%% file: thesis.tex
\documentclass[11pt,a4paper]{book}

\usepackage{amssymb, amsmath, amsfonts, type1cm}
\usepackage[usenames, dvipsnames]{color}
\usepackage{cite, graphicx, verbatim, wrapfig, setspace}

\usepackage{appendix}

\usepackage{sectsty}
\partfont{\centering}

\newcommand{\be}{\begin{equation}}
\newcommand{\ee}{\end{equation}}
\newcommand{\bea}{\begin{eqnarray}}
\newcommand{\eea}{\end{eqnarray}}
\newcommand{\beqar}{\begin{eqnarray*}}
\newcommand{\eeqar}{\end{eqnarray*}}
\newcommand{\lp}{\left(}
\newcommand{\rp}{\right)}

\def\ii{\textrm{i}}
\def\Y{\mathbb{Y}}
\def\H{{\cal H}}
\def\M{{\cal M}}
\def\CD{{ \cal D }}

\def\pd{\partial}
\def\dr{\frac{\partial}{\partial r}}
\newcommand{\PP}[4]{P^{#1\phantom{\underline{#2}}#3\phantom{\underline{#4}}}_{\phantom{#1}\underline{#2}\phantom{#3}\underline{#4}}}
\newcommand{\QQ}[3]{Q^{\phantom{\underline{#1}}#2\phantom{\underline{#3}}}_{\underline{#1}\phantom{#2}\underline{#3}}}

\newcommand{\der}[2]{\frac{\partial{#1}}{\partial{#2}}}
\newcommand{\dder}[2]{\frac{\partial{}^2 #1 }{ {\partial{#2}}^2}}
\newcommand{\dderf}[3]{\frac{\partial{}^2 #1 }{ {\partial{#2} \partial{#3}}}}

\numberwithin{equation}{chapter}

\begin{document}


\include{Coverpage}

\pagenumbering{roman}

\include{Acknowledgements}

\include{Declaration}

\include{Summary}

\tableofcontents

\pagenumbering{arabic}


\part{Introduction} \label{part:Introduction}

    \include{Introduction}


\part{Black hole partition functions} \label{part:partitionfunctions}

    \include{PFtheory}
    \include{Kerrads}
    \include{Charged}


\part{Stability of higher-dimensional black holes} \label{part:higherdim}

    \include{Classthermo}
    \include{MPsingle}
    \include{MPeq}


\part{Conclusion} \label{part:conclusion}

    \include{Conclusion}
    \include{Spectralmethod}


         
\bibliographystyle{utphys}
\bibliography{bibliography}

\end{document}

%% file: Coverpage.tex

\thispagestyle{empty}

\null\vfill
\begin{center}

  {\fontsize{28pt}{28pt}\selectfont \bf Classical and thermodynamic \\ stability of black holes \par}

\vspace*{1cm}
  {\Large \sc Ricardo Jorge Ferreira Monteiro \par
\vspace*{6ex}}
  {\large {{Dissertation submitted for the Degree of} \par}
          {{Doctor of Philosophy} \par}
          {{at the University of Cambridge} \par}}

%

%

\vspace*{0.5cm}
   \includegraphics[height=30mm]{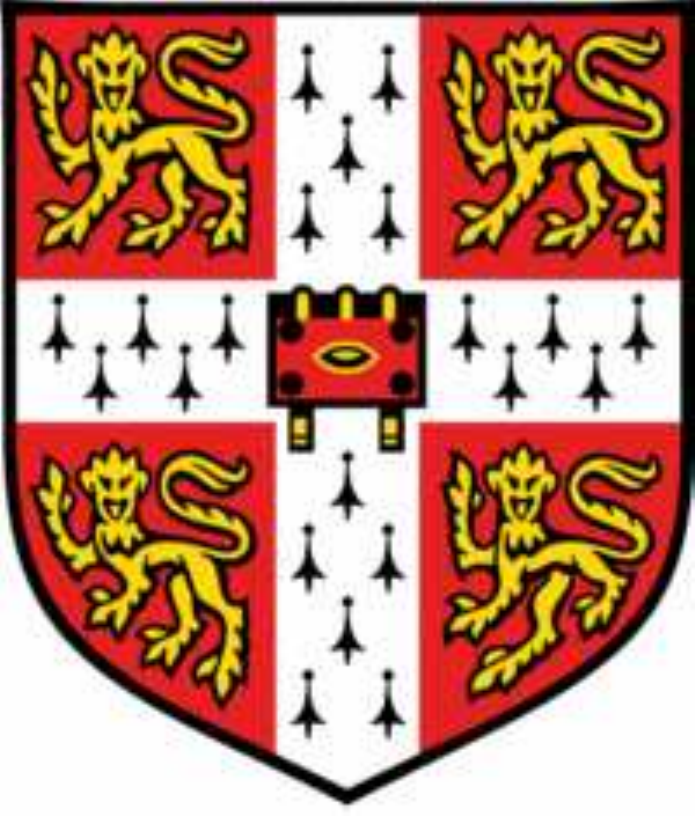} \par

  {\large \it Department of Applied Mathematics and Theoretical Physics \\
              \& Hughes Hall\\
              University of Cambridge \par}
\vspace*{6ex}
  {\large May 2010}

\end{center}
\null\vfill


\newpage
\thispagestyle{empty}
\qquad
\newpage

%% file: Acknowledgements.tex
\newpage

\begin{center}
  {\Large {\bfseries Acknowledgements}}
\end{center}
\addcontentsline{toc}{chapter}{Acknowledgements}

\vspace{0.5cm}

\noindent I am very thankful to my supervisor Stephen Hawking for his support and inspiration, for being always available in spite of the many solicitations on his time, and for his interest in my independent work. Thanks are also due to his personal assistant Judith Croasdell for her help with all sorts of paperwork burdens.

My graduate work is indebted to the talent and enthusiasm of Jorge Santos. Our initial side project turned into the core of our research. Special thanks are reserved for \'Oscar Dias and Pau Figueras for our fruitful and immensely enjoyable collaboration. I am also very grateful to my collaborators Roberto Emparan, Malcolm Perry and Harvey Reall for all that they taught me. I have benefitted from discussions with Tim Clunan, Gary Gibbons, Hadi Godazgar, Mahdi Godazgar, Steven Gratton, Mihalis Dafermos, Carlos Herdeiro, Gustav Holzegel, Veronika Hubeny, Alexander Kaus, Miguel Paulos and Claude Warnick, among others. Steven Gratton and Malcolm Perry are due an extra acknowledgement for their advice in difficult times. For comments on drafts of this thesis, I am very thankful to Gustav Holzegel, Mette Iversen and most especially \'Oscar Dias. Finally, I would like to thank Mette Iversen for her love and for guaranteeing my sanity while finishing the thesis.

This work was supported by the Funda\c c\~ao para a Ci\^encia e Tecnologia (Portugal), throught the grant SFRH\-/BD\-/22211\-/2005, and, in its final stages, by the Cambridge Philosophical Society.

This thesis is dedicated to my parents Jorge and Isabel, and to my siblings Maria Jo\~ao and Rui, for their love and support at all times during my education.


\newpage
\quad
\newpage

%% file: Declaration.tex
\newpage

\begin{center}
  {\Large {\bfseries Declaration}}
\end{center}
\addcontentsline{toc}{chapter}{Declaration}

\vspace{0.5cm}

\noindent This dissertation is the result of my own work and includes nothing which is the outcome of work done in collaboration except where specifically indicated in the text. The research described in this dissertation was carried out in the Department of Applied Mathematics and Theoretical Physics at the University of Cambridge between October 2006 and May 2010. Except where reference is made to the work of others, all the results are original and based on the following works of mine:

\begin{itemize}
\item \emph{Negative modes and the thermodynamics of Reissner-Nordstr\"om black holes}, R. Monteiro and J. E. Santos, Phys.~Rev.~D 79, 064006 (2009), arXiv:0812.1767
\item \emph{Thermodynamic instability of rotating black holes}, R. Monteiro, M. J. Perry and J. E. Santos, Phys.~Rev.~D 80, 024041 (2009), arXiv:0903.3256
\item \emph{Semiclassical instabilities of Kerr-AdS black holes}, R. Monteiro, M. J. Perry and J. E. Santos, Phys.~Rev.~D 81, 024001 (2010), arXiv:0905.2334
\item \emph{Instability and new phases of higher-dimensional rotating black holes}, O. Dias, P. Figueras, R. Monteiro, J. E. Santos and R. Emparan, Phys.~Rev.~D 80, 111701 (R) (2009), arXiv:0907.2248
\item \emph{An instability of higher-dimensional black holes}, O. Dias, P. Figueras, R. Monteiro, H. S. Reall and J. E. Santos, arXiv:1001.4527
\end{itemize}

\noindent None of the original works contained in this dissertation has been submitted by me for any other degree, diploma or similar qualification.

\phantom{a}

\noindent Signed: ........................................................... (Ricardo Jorge Ferreira Monteiro) \\
Date: ...........................................................


\newpage
\quad
\newpage

%% file: Summary.tex
\newpage

\begin{center}
  {\Large {\bfseries \underline{Classical and thermodynamic stability of black holes}}} \\
  \vspace{0.2cm}
  {\large {\bfseries Ricardo Jorge Ferreira Monteiro}} \\
  \vspace{0.6cm}
  {\Large {\bfseries Thesis summary}}
\end{center}
\addcontentsline{toc}{chapter}{Summary}

\vspace{0.4cm}

We consider the stability of black holes within both classical general relativity and the semiclassical thermodynamic description. In particular, we study linearised perturbations and their contributions to the gravitational partition function. Exploring the connection between classical and thermodynamic stability, we find classical instabilities and new families of vacuum black holes.

We start by studying negative modes of black hole partition functions, which represent pathologies in the one-loop quantum corrections. In particular, we extend this study to charged black holes (Reissner-Nordstr\"om), using a method based on gauge-invariant perturbations, and to rotating black holes (Kerr-AdS), where a numerical technique is employed. In the both cases, we find a negative mode in the region where local thermodynamic stability fails, as expected.

We then present the first examples of linearised classical instabilities of vacuum asymptotically flat black holes. We analyse numerically perturbations of Myers-Perry solutions, both in the single spin and in the equal spins (odd $D$) cases. For sufficiently high rotation, in the so-called ultraspinning regime, new negative modes of the partition function may arise whose threshold marks both the onset of a classical instability of the black \emph{hole} (not just of the associated black branes) and the bifurcation to a new family of black hole solutions. In the case of singly-spinning solutions, we find the threshold stationary modes signalling the instabilities, confirming a conjecture by Emparan and Myers. In the case of solutions with equal spins, we are able to find perturbations that grow exponentially in time in $D=9$ (we believe that this extends to higher odd $D$). Furthermore, the new family of solutions bifurcating at the onset of the instability should have a \emph{single} rotational symmetry, saturating the rigidity theorem.



%% file: Introduction.tex
\newpage

\chapter{Black holes} \label{chap:blackholes}

Black holes are arguably the most interesting objects in theoretical physics. Understanding their dynamics forces us to fit together two widely accepted theories of Nature: general relativity (Einstein's classical theory of gravity) and quantum mechanics, a goal that has eluded theoretical efforts so far, despite encouraging successes. Black hole thermodynamics is at the crossroad between the classical and the quantum pictures. In this thesis, we will study the behaviour of black holes as seen both from classical general relativity and from their quantum thermodynamic properties.

This Chapter includes a basic introduction to black holes, a summary of their thermodynamic properties, and a review of black hole solutions in spacetimes of different dimensionality. Only later will we focus on our main subject of research: black hole stability.


\section{Introduction} \label{sec-bh:intro}

In Newtonian gravity, a massive body can have an escape velocity greater than the ve\-lo\-ci\-ty of light. The analogue of such a ``black planet" in general relativity is a black hole. However, the analogy does not go far since black holes are intrinsically relativistic objects highlighting two basic features of Einstein's theory: causal horizons and spacetime singularities.

According to general relativity, the gravitational force is caused by the curvature of spacetime. Spacetime consists of three spatial dimensions and a time dimension put together in a geometrical way. (In higher-dimensional gravity, additional spatial dimensions are considered.) It is a pseudo-Riemannian manifold $(\M,g_{ab})$ of Lorentzian signature whose dynamical metric $g_{ab}$ obeys the Einstein field equations. Let $R_{ab}$ be the Ricci curvature and $R$ the scalar curvature. The Einstein equations are
\be
\label{einstein}
R_{ab}-\frac{1}{2} R g_{ab} + \Lambda g_{ab} = \frac{8 \pi G}{c^4} T_{ab} \,.
\ee
The parameter $\Lambda$, the cosmological constant, denotes the contribution from vacuum energy, and, on the right-hand side, $T_{ab}$ is the energy-momentum tensor of all the matter fields. It is clear that the dynamics of spacetime is affected by any field content. On the other hand, any field is affected by gravity since it lives on a curved spacetime, with massive particles moving along timelike geodesics and massless particles moving along null geodesics.

Solid evidence for general relativity has been provided by the corrections to Newtonian dynamics in the Solar system, and the indirect detection of gravitational waves from binary pulsars \cite{Will:2005va}.

A crucial distinction from Newtonian gravity is that the ``action-at-a-distance" is substituted by a built-in causality structure in Einstein's theory. The initial-value formulation of general relativity splits the 10 Einstein equations into 6 evolution equations and 4 constraint equations \cite{Wald:1984rg}. The latter constrain what is suitable initial data on a Cauchy spacelike surface. The former consist of a system of hyperbolic quasilinear equations that evolves the initial data in time. The resulting causality structure looks locally like the light-cone structure of special relativity. However, since the spacetime is dynamical, there is the possibility that the Cauchy evolution of smooth geometry and matter data on a spacelike surface may lead to a singularity due to a cathastrophic event, such as the gravitational collapse of a massive body or a high energy collision. By singularity, we mean that the spacetime is geodesically incomplete for timelike or null geodesics, and that geometric invariants constructed with the metric curvature may diverge. The causal propagation of such a spacetime pathology could challenge the physical significance of general relativity, in spite of the experimental successes.

Singularities indeed arise in Einstein's theory as shown by the theorems of Penrose and Hawking \cite{Hawking:1973uf}. Their standard interpretation is that general relativity breaks down for curvatures of the order of the Planck scale, e.g. $R_{abcd}R^{abcd} \sim c^3 / \hbar G$ where $\hbar$ is Planck's constant, giving way to a quantum description of spacetime. Since general relativity is non-renormalisable when treated as a quantum field theory of gravitons, a more fundamental quantum theory of spacetime is required, and general relativity should be viewed as an effective low energy theory. The physical significance of classical general relativity relies on a protection mechanism from singularities known as \emph{cosmic censorship} \cite{Penrose:1969pc}. This hypothesis (in the ``weak" formulation) states that no naked singularities exist, i.e. singularities arising from realistic matter (except perhaps at the Big Bang) can only form behind a surface, called an \emph{event horizon}, from within which no information may reach observers outside that surface.

A \emph{black hole} is the region contained inside an event horizon. More precisely, for spacetimes with an asymptotic conformal structure, a black hole is the region of spacetime that does not lie in the causal past of future null infinity, and its boundary in the full spacetime $\M$ is called the future event horizon $\H^+$ (to distinguish it from the past event horizon $\H^-$, the boundary of communication of past null infinity, present in time-symmetric solutions but absent for dynamically formed black holes). The singularity theorems mentioned above then imply, under certain assumptions for the matter content, that there is a singularity in the black hole.

The formation of black holes through the gravitational collapse of massive objects has been studied analytically and numerically (see reviews e.g. in \cite{Christodoulou:2008nj} and \cite{Stergioulas:2009zz}), and recently the formation through high energy collisions has also been addressed numerically \cite{Choptuik:2009ww}. Using semi-realistic matter, an event horizon forms in agreement with cosmic censorship. In fact, no matter is necessary for the formation of a black hole, as the focusing of incoming gravitational waves may be sufficient \cite{Christodoulou:2008nj}. The results support the expectation that a black hole will form whenever a given amount of energy is contained in a sufficiently small region of space, as proposed by the hoop conjecture \cite{Thorne}. What is truly remarkable is that the spacetime (at least in the four-dimensional asymptotically flat case) settles down to a unique stationary black hole solution, regardless of the details of the initial matter distribution.

The uniqueness theorems (see \cite{Heusler} for a review) show that four-dimensional vacuum black holes which are asymptotically flat, stationary and have a connected horizon are described by the Kerr metric \cite{Kerr:1963ud}. This solution depends on two parameters only: the mass $M$ and the angular momentum $J$ of the black hole, which are defined asymptotically as conserved charges of the spacetime. They satisfy the extremality bound $|J| \leq G M^2$, whose saturation gives a degenerate horizon (zero surface gravity), while otherwise the spacetime is nakedly singular. The Kerr-Newman black hole \cite{Newman:1965my} generalises the Kerr solution to include electric and magnetic charges, the uniqueness for a connected horizon still holding in the Einstein-Maxwell theory. The connectedness assumption can be dropped in the static limit, except in the charged case with degenerate horizons where there exist multi-black hole solutions of the Majumdar-Papapetrou type \cite{Hartle:1972ya}. In the case of solutions with a cosmological constant, uniqueness has not been proven for asymptotically de Sitter (dS, $\Lambda>0$) or anti-de Sitter (AdS, $\Lambda<0$) spacetimes but the only localised black hole known in four dimensions is the generalisation of the Kerr case \cite{Carter:1968ks}. We shall later discuss how this simple picture changes drastically in higher-dimensional gravity, where there is a great variety of rotating black hole solutions.

Observational evidence for the existence of black holes in the universe is indirect only, since no signal comes out of them according to general relativity. X-ray signals do come from accretion disks around black hole candidates, sometimes in binary systems. The geodesic motion of nearby bodies can also indicate the presence of a black hole. The size, mass and angular momentum of the candidates is inferred from that data, and the evidence for black holes is abundant. A supermassive black hole is thought to exist at the centre of most galaxies \cite{Magorrian:1997hw}, and even evidence for near-extremal black holes in binary systems has been reported, for instance $|J| > 0.98 \, G M^2$ in \cite{McClintock:2006xd}. Quantum evaporation of black holes, to be discussed in the next Section, opens the possibility of direct evidence, but the effects are too small for large astrophysical black holes. Very small astrophysical black holes, which would provide a definite signature, have not been detected.

Hereafter, we shall use natural units $c=G=\hbar=k_B=1$, where $k_B$ is Boltzmann's constant, implicit in the discussion of thermodynamics.


\section{Quantum description} \label{sec-bh:quantum}

\subsection{Black hole thermodynamics}

The laws of black hole mechanics have a close analogy with the common laws of thermodynamics \cite{Bekenstein:1973ur,Bardeen:1973gs,Bekenstein:1974ax}. This observation and the discovery of Hawking radiation \cite{Hawking:1974rv,Hawking:1974sw} paved the way for progress in the quantum understanding of spacetime.

A small variation in the mass $M$ of a rotating black hole with angular momentum $J$, and say electric charge $Q$, satisfies
\be
\label{1stlaw4d}
d M = \frac{\kappa}{8 \pi} dA + \Omega dJ + \phi dQ \,,
\ee
where $\kappa$ is the surface gravity, $A$ is the area of the event horizon, $\Omega$ is the angular velocity of the horizon and $\phi$ is the electric potential on the horizon. The analogy with the common \emph{first law} of thermodynamics is clear since $M$ is the conserved charge associated with the time-translation symmetry, i.e. the energy. $J$ and $Q$ are identified with ``particle numbers", and $\Omega$ and $\phi$ are identified with ``chemical potentials", which are constant on the horizon, as required by equilibrium. We are left with the identifications of the temperature and the entropy of the black hole, $T=\alpha \kappa / 8 \pi$ and $S = A/\alpha$, respectively, where $\alpha$ is a positive constant. The \emph{zeroth law} of thermodynamics is the statement that the surface gravity $\kappa$ is constant on the horizon. The \emph{second law} is the area law of classical general relativity, stating that the event horizon area $A$ never decreases \cite{Hawking:1971tu}. When matter is considered, both the black hole entropy and the total entropy -- black hole plus matter outside -- must be non-decreasing classically. (Hawking radiation, a quantum effect, may cause the horizon area $A$ to decrease, but this is compensated by the entropy of the radiation so that the total entropy never decreases.) To complete the four laws of thermodynamics, the \emph{third law} says that it is not possible to make $\kappa$ vanish through a finite-time physical process.

Although $M$, $A$, $J$ and $Q$ play the role of the extensive thermodynamic variables in the common applications of the first law, they have different scaling dimensions. For asymptotically flat solutions in four spacetime dimensions, the first law and the scaling of those quantities 
imply the Smarr relation \cite{Smarr:1972kt},
\be
\label{smarr4d}
M = 2 \left( \frac{\kappa A}{8 \pi} + \, \Omega J \right) + \phi Q\,.
\ee
The relation fails for solutions with a cosmological constant, since an independent length scale is introduced.

The laws of black hole mechanics result from the structure of the Einstein equations and from the fact that event horizons of stationary black holes are Killing horizons \cite{Bardeen:1973gs}. A Killing horizon is generated by a Killing vector $K$ which is null on the horizon. For a four-dimensional black hole, $K=\pd_t + \Omega\, \pd_\varphi\,$, where $\pd_t$ and $\pd_\varphi$ are the stationarity and axisymmetry Killing vectors, respectively. We will later discuss specific aspects of higher dimensions. The first law has actually been shown to hold for any covariant Lagrangian theory of gravity (e.g. arising as a higher-derivative curvature correction to general relativity) if the Bekenstein-Hawking entropy is substituted by the so-called Wald entropy \cite{Wald:1993nt,Iyer:1994ys}. The corresponding second law, however, remains an open problem. Let us also point out the result of Jacobson \cite{Jacobson:1995ab} stating that, while the first law is a consequence of the equations of motion, the converse is also true if we formulate the first law conveniently. Indeed, the Einstein equations can be derived by requiring that energy-momentum fluxes $\delta {\mathcal Q}$ across all local Rindler horizons (causal horizons of uniformly accelerated observers) through each spacetime point are reversible, i.e. $\delta {\mathcal Q} = \kappa \,dA/ 8 \pi$.

The considerations above are purely classical. The temperature of a black hole, which emits nothing classically, can only be understood quantum-mechanically. The breakthrough by Hawking \cite{Hawking:1974rv,Hawking:1974sw} was to use quantum field theory on a curved background to show that black holes behave like black bodies in the usual thermodynamic sense, emiting radiation with a thermal spectrum. The Hawking temperature is given by $T= \kappa /2 \pi$, i.e. $\alpha=4$. The classical limit $\hbar \to 0$ is clear if we do not use natural units: $k_B T= \hbar \kappa / 2 \pi c \to 0$. A stationary observer detects a thermal bath with temperature $T/V$, where $V=\sqrt{-\xi^a \xi_a}$ is the redshift factor ($\xi=\pd_t$). The radiation is the effect of acceleration, characteristic of flat space too: Rindler observers, i.e. observers with uniform acceleration $a$, detect Unruh radiation with $T=a/2\pi$ \cite{Unruh:1976db}.

One of the great puzzles in this picture is what happens to the information of an object that enters the black hole. Since the radiation emitted by the black hole is thermalised, this information seems to be irremediably lost, violating unitarity \cite{Hawking:1976ra}. The rate of evaporation, i.e. mass loss through Hawking radiation, is given by the Stefan-Boltzmann law, $dM/dt \sim -A\, T^4 \sim -M^{-2}$, since $A \sim M^2$ and $T \sim M^{-1}$. Evaporation is thus unimportant for large astrophysical black holes, but it dominates the behaviour of very small black holes, and it would provide a distinct observational signature in case they would be produced in accelerator experiments. In the final stages of evaporation, as the black hole approaches the Planck size, a full quantum description is unavoidable. Such a description must solve both the information paradox and the singularity problem.

The discussion above implies that the entropy of a black hole, known as the Beken\-stein-\-Hawking entropy, is given by $S=A/4$, which strongly suggests that the quantum degrees of freedom of the black hole are effectively distributed over a surface, rather than a volume. This crucial observation is the basis of the holographic principle, proposed by 't Hooft \cite{'tHooft:1993gx} and Susskind \cite{Susskind:1994vu}. This principle says that quantum gravity in a given volume should be described by a theory on the boundary of that volume, analogously to a common planar hologram that generates a 3D image. The number of degrees of freedom is thus drastically smaller than the na\"ive expectation.

The Euclidean path integral approach to black hole thermodynamics, on which this thesis heavily relies, will be reviewed in Chapter~\ref{cha:negmodes}.

\subsection{String theory and the AdS/CFT correspondence}

Only a theory of quantum gravity can identify the microscopic degrees of freedom which give rise to the Bekenstein-Hawking entropy. Moreover, according to the holographic principle, this theory should admit two equivalent formulations: one as a bulk theory, in which semiclassical general relativity arises explicitly as a low energy limit, and one as a boundary theory, in which the description of gravity is not explicit but the distribution of degrees of freedom is more conventional.

String theory, a promising candidate for a theory of quantum gravity, has made progress on these two challenges coming from black hole thermodynamics: (i) it has provided a counting of microscopic states for specific classes of black holes leading to the Bekenstein-Hawking entropy, first reported in \cite{Strominger:1996sh}, and (ii) it has provided a concrete realisation of the holographic principle, which is the anti-de Sitter / conformal field theory (AdS/CFT) correspondence \cite{Maldacena:1997re,Aharony:1999ti}. We shall make a brief comment although this is not a required background for the research described in this thesis.

The idea of string theory is that the universe is composed of small strings, which resemble point particles at low energies, and that the worldsheet of a string (the (1+1)-surface spanned by its time evolution) is quantised. The consistency of the quantum field theory of the worldsheet has drastic consequences: the introduction of supersymmetry (to avoid the existence of tachyons and naturally include both bosons and fermions in the spectrum) and, more remarkably, the existence of ten spacetime dimensions (to ensure Lorentz invariance). Agreement with the commonly perceived four dimensions requires that six spatial dimensions are compactified on a very small scale. To offer in return, the theory: contains a graviton-type string state in its spectrum; requires that, at low energies and curvatures, the spacetime background on which the worldsheet moves satisfies the supergravity equations of motion (to ensure conformal invariance of the worldsheet theory); and has a natural high energy cutoff scale (the string length) since interactions are not point-like, which solves the non-renormalisability problem of general relativity. The drastic consequences mentioned above can also be seen in a different light: supersymmetric field theories have appealing properties, and the compactification of the extra-dimensions provides a geometric framework to understand how the various parameters of the Standard Model of particle physics can arise from a simpler fundamental theory. There are five different types of consistent string theories (types I, IIA, IIB, heterotic $SO(32)$ and heterotic $E_8 \times E_8$), some of which have been shown to be related by dualities making them equivalent. They have all been conjectured to arise as special cases of an eleven-dimensional theory called M-theory.

The quantum understanding of black holes provided by string theory comes mainly from the introduction of D$p$-branes, which are extended objects with $p$ spatial dimensions on which the endpoints of open strings are restricted to move. A gauge theory arises as the low energy description of these open strings. From a dual gravity perspective, D$p$-branes are extremal black branes (i.e. black holes with infinitely extended horizons along $p$ directions). The wrapping of D-branes on compact spaces can reproduce black holes in lower dimensions, whose quantum properties can then be derived from those of the D-branes. Strominger and Vafa \cite{Strominger:1996sh} counted the degeneracy of D-brane states corresponding to microstates of a five-dimensional class of extremal black holes, leading to the first microscopic derivation of the Bekenstein-Hawking entropy. We should point out that, while D-branes are intrinsically string theory objects, it has been argued that the computation of the entropy may actually not rely on the string theory input, but rather on the symmetry of the near-horizon geometry or of the low energy wave equation \cite{Guica:2008mu,Castro:2010fd} (see \cite{Dias:2009ex} for potential issues).

An outstanding development brought about by the understanding of D-branes is the AdS/CFT correspondence proposed by Maldacena \cite{Maldacena:1997re}. The original conjecture states that type IIB string theory with $AdS_5 \times S^5$ boundary conditions is dual to ${\cal N}=4$ $SU(N)$ super-Yang-Mills theory defined on $R \times S^{3}$. This is a realisation of the holographic principle, since we have a gravity theory (string theory) in a five-dimensional spacetime ($AdS_5$, after compactification on $S^5$) proposed to be equivalent to a four-dimensional field theory without gravity. In particular, the semiclassical supergravity limit on the gravity side corresponds to the strongly coupled limit on the field theory side. Black hole physics in AdS can describe strongly coupled gauge theories at finite temperature (the Hawking temperature of the black hole) and vice-versa. The conjecture is expected to extend to other cases of string theory or M-theory with AdS (or ``almost AdS") boundary conditions, the dual field theory being defined on the AdS boundary.


\section{Higher-dimensional solutions} \label{sec-bh:highdim}

In this Section, we motivate the study of higher-dimensional gravity and review briefly the literature on the existence and properties of black hole solutions, leaving the important issue of stability for the next Chapter. Our main focus will be on vacuum solutions to general relativity, with particular attention to Myers-Perry black holes, which will be analysed in the last part this thesis. See \cite{Emparan:2008eg} for a comprehensive review, although some developments are more recent.

String theory is the original and most important motivation for the study of higher-dimensional gravity. Black holes in spacetimes with up to ten dimensions are part of string theory (or eleven dimensions, for M-theory). We mentioned in the last Section that one of the theory's major successes, the microscopic derivation of the Bekenstein-Hawking entropy, was obtained for a class of five-dimensional black holes \cite{Strominger:1996sh}. The AdS/CFT correspondence \cite{Maldacena:1997re,Aharony:1999ti} and its applications further motivate this study. The correspondence implies that the dynamics of $D$-dimensional field theories at finite temperature and that of ($D+1$)-dimensional black holes are equivalent. Gravity solutions have been constructed \emph{\`a la carte} to describe strong-coupling features of field theories, in an effort to learn something about quantum chromodynamics \cite{Gubser:2009md} and condensed matter physics \cite{Hartnoll:2009qx}. Other scenarios, which try to explain the hierarchy problem (the weakness of 4D gravity when compared to the Standard Model interactions) by introducing extra-dimensions, possibly in string theory embeddings, open the possibility that black hole production in particle collisions is within future experimental reach, as reviewed in \cite{Kanti:2004nr}.

Higher-dimensional gravity is also important in its own right. A new insight is gained by studying gravity using the spacetime dimensionality $D$ as a parameter. Indeed, many properties of black hole solutions are specific to four dimensions.

In Newtonian gravity, the attractive gravitational force is supressed with the radial distance as $r^{-(D-2)}$, while the repulsive centrifugal force is supressed as $r^{-3}$ (for given mass and angular momentum) independently of $D$ since it acts on a plane of rotation. This is why stable planetary orbits are not possible for $D>4$. One might expect that the physics of black holes, especially when there is rotation, will also be different. In fact, there will be black holes with different horizon topology and also disconnected horizons. Black holes with an arbitrarily large angular momentum for a given mass can also be found. These are closely related to the existence of solutions with extended horizons (absent in four dimensions without cosmological constant).

Another obvious contrast is the number of rotation planes. The Cartan subgroup, i.e. the maximal Abelian subgroup or maximal torus, of $SO(3)$ is $U(1)$, which means that there is a single independent rotation plane in $D=4$. The associated rotational Killing vector $\xi=\pd_\phi$ defines the angular momentum as an asymptotic conserved charge, $J \sim \int_{S^2_\infty} \ast d\xi \,$. However, the Cartan subgroup of $SO(D-1)$ is $U(1)^n$, so that there are $n= \lfloor (D-1)/2 \rfloor$ (where $\lfloor \rfloor$ stands for the smallest integer part) independent rotation planes in $D-1$ spatial dimensions. These allow for the definition of $n$ independent angular momenta, $J_i \sim \int_{S^{D-2}_\infty} \ast d\xi_i \,$. (Notice that only one of the $\xi_i$'s must be a Killing vector of the spacetime according to the rigidity theorem, to be discussed below.)

As we mentioned in Section~\ref{sec-bh:intro}, the uniqueness of the Kerr black hole among stationary asymptotically flat vacuum black holes with a connected horizon is a remarkable feature of $D=4$. In higher dimensions, the generalisation of the Schwarzschild solution, obtained by Tangherlini \cite{Tangherlini:1963bw}, is also the unique static vacuum black hole \cite{Gibbons:2002bh}. Uniqueness extends to the higher-dimensional Reissner-Nordstr\"om black hole for the Einstein-Maxwell theory in the static non-degenerate case \cite{Gibbons:2002av}, while in the degenerate case there exist the higher-dimensional Majumdar-Papapetrou solutions, as in four dimensions. It is the inclusion of rotation that changes this simple picture.

\subsection{Myers-Perry solutions \label{subsec:MPsolution}}

The Myers-Perry solution \cite{Myers:1986un} is the natural generalisation of the Kerr black hole to higher dimensions, and (so far) the only exactly known rotating solution in $D>5$. It is a rather non-trivial generalisation, due to the possibility of rotation in different planes. Fortunately, as the Kerr solution, it can be written in the so-called Kerr-Schild form,
\be
g_{\mu \nu} = \eta_{\mu \nu} + H(x^\lambda) k_\mu k_\nu\,,
\ee
where $\eta_{\mu \nu}$ is the Minkowski spacetime metric and $k_\mu$ is a null vector with respect to $g_{\mu \nu}$. The fact that $k_\mu$ is then null also with respect to $\eta_{\mu \nu}$ allows for a simplification of the Einstein equations similar to linearisation.

We present the solution here in more detail because it will be useful in this thesis. Let us take $D=2 n+1+\epsilon$, where $n= \lfloor (D-1)/2 \rfloor$ is the number of angular momenta and $\epsilon= (D-1) \{\text{mod 2}\}\,$ is 0 ($D$ odd) or 1 ($D$ even). There are $n$ azimuthal angles $\phi_i$ and $n+\epsilon$ direction cosines $\mu_i$ obeying $\sum_{i=1}^{n+\epsilon} \mu_i^2=1$. In Boyer-Lindquist coordinates, the metric can be written as
\bea
\label{fullMP}
ds^2 &=& - dt^2
 + \frac{2m}{U}\Bigl(dt - \sum_{i=1}^n {a_i\, \mu_i^2\, d\phi_i}\Bigr)^2
 + \sum_{i=1}^n (r^2 + a_i^2)\,(d\mu_i^2+ \mu_i^2\, d\phi_i^2) \nonumber\\
&&
 + \frac{U\, dr^2}{V-2m}
 + {r^2}\, d\mu_{n+\epsilon}^2 \,,
\eea
where the last term is present only for even $D$, and
\bea
U(r,\mu_i) \equiv  r^{\epsilon}\, \sum_{i=1}^{n+\epsilon} \frac{\mu_i^2}{r^2 + a_i^2}\,
\prod_{j=1}^n (r^2 + a_j^2)\,,\qquad
V(r) \equiv r^{\epsilon-2}\,
   \prod_{i=1}^n (r^2 + a_i^2)\,.
\eea
The event horizon is located at $r=r_+$, where $r_+$ is the largest root of $V(r)-2m=0$. The Kerr black hole case ($D=4$) is included as $n=\epsilon=1$.

The solution is parametrised by $n+1$ length scales: the horizon radius $r_+$ and the $n$ rotation parameters $a_i$. The horizon area is 
\be
A = \frac{{\cal A}_{D-2}}{r_+^{1-\epsilon}}\, \prod_{i=1}^n (r_+^2 + a_i^2) = 2{\cal A}_{D-2}\, m\, r_+ \,,\qquad \text{where} \quad {\cal A}_{D-2} = \frac{2 \,\pi^{(D-1)/2}}{\Gamma[(D-1)/2]} \quad
\ee
is the volume of a unit-radius $(D-2)$-sphere. The surface gravity $\kappa$ and the angular velocities on the horizon $\Omega_i$ are given by
\be
\kappa = r_+\, \sum_{i=1}^n \frac{1}{r_+^2 + a_i^2} -\frac{2-\epsilon}{2\,r_+}\,,\,\qquad \Omega_i = \frac{a_i}{r_+^2 + a_i^2} \,.
\ee
The asymptotic charges, the mass $M$ and the angular momenta $J_i$, which uniquely specify a solution, are given by
\be
M = \frac{(D-2){\cal A}_{D-2}}{8 \pi}\,m \,,\,\qquad J_i = \frac{2}{D-2}\, a_i\, M \,.
\ee

Extremality, for which the temperature $T=\kappa/2\pi$ vanishes, occurs for
\bea
\label{extremal}
\sum_{i=1}^n \frac{1}{1 + (8\pi\,J_i/A)^2} = \frac{2-\epsilon}{2}
= \left\{  \begin{array}{ll} \;1 & \;\;\text{$D$ odd} \\ \;1/2 & \;\;\text{$D$ even} \end{array} \right. \,.
\eea
In $D=4$, this introduces the bound $J \leq M^2$ on the angular momentum, beyond which the Kerr solution is nakedly singular. Does such a bound exist in higher dimensions? Let us pick up a direction through the origin in the space $\{J_i\}$ for a fixed mass. We can readily see from \eqref{extremal} that a direction with regular extremal limit requires $J_i \neq 0\;\; \forall i$. Otherwise:

$\bullet\,$ \underline{$D$ even}: There is no bound when one or more of the angular momenta vanish, since it is impossible to satisfy \eqref{extremal}, which leaves all the remaining angular momenta unbounded for a given mass.

$\bullet\,$ \underline{$D$ odd}: There is no bound when \emph{two} or more of the angular momenta vanish, leaving all the remaining angular momenta unbounded for a given mass. However, there is a bound corresponding to \eqref{extremal} when a single angular momentum vanishes. The saturation of the bound gives a nakedly singular solution, rather than an extremal black hole, since it is required that $A \to 0$ (no event horizon).

\begin{figure}[t]
\centerline{ \large{$D=5$:} \hspace{3.5cm} \large{$D=6$:} $\phantom{aaaaaaaaaaaaaaaaaaaaa}$ }
\centerline{ \includegraphics[width=.3\textwidth]{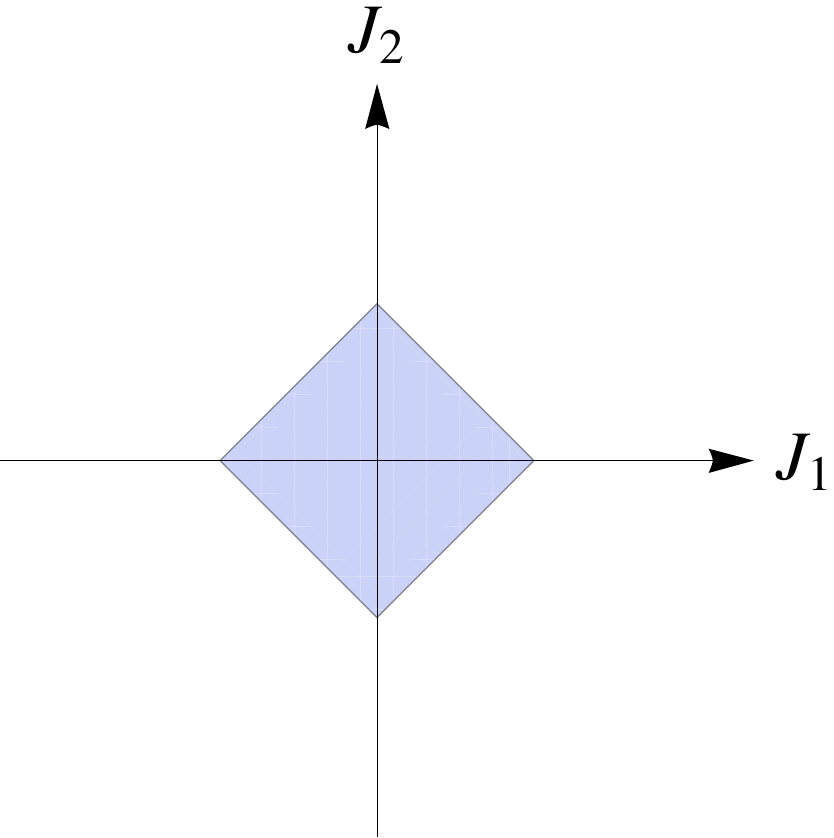}
\hspace{1.3cm} \includegraphics[width=.3\textwidth]{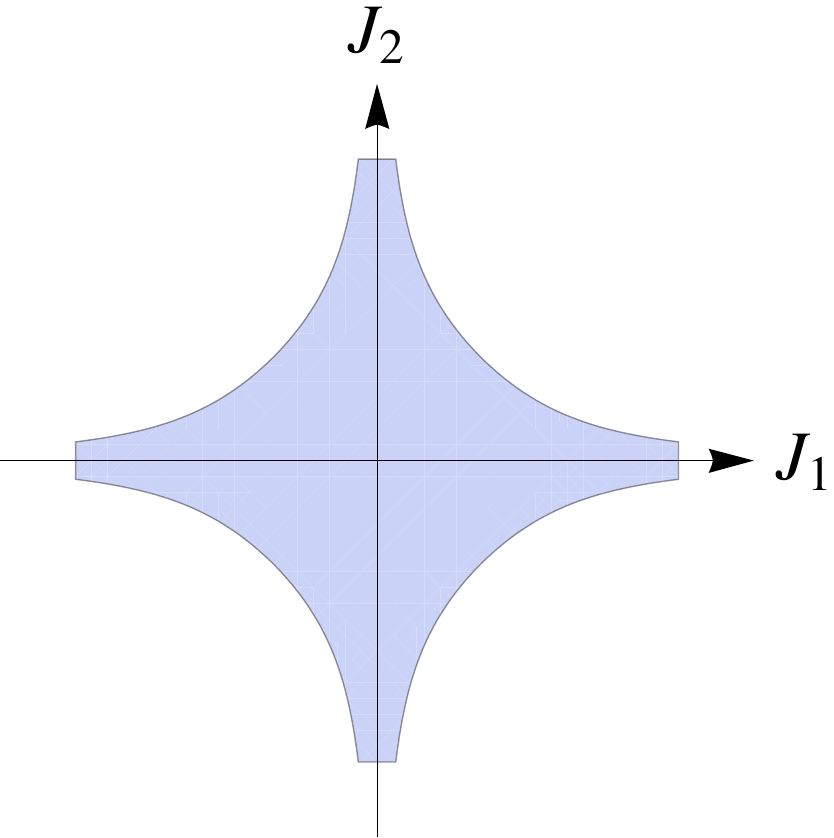}}
\centerline{ \large{$D=7$:} \hspace{3.5cm} \large{$D=8$:} $\phantom{aaaaaaaaaaaaaaaaaaaaa}$ }
\centerline{ \hspace{-0.8cm} \includegraphics[width=.35\textwidth]{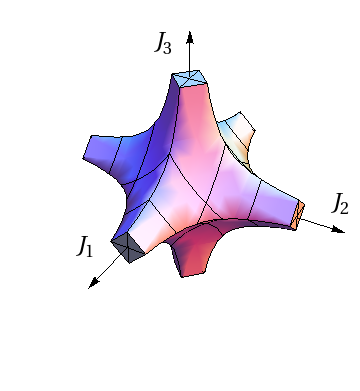}
\hspace{0.5cm} \includegraphics[width=.35\textwidth]{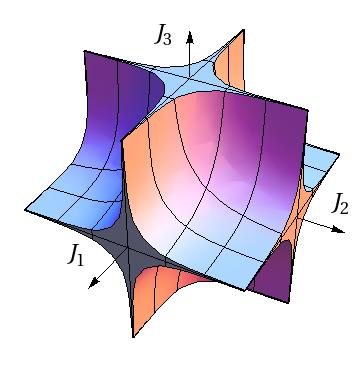}}
\vspace{-1cm}
\caption{Shape of the parameter space of angular momenta for fixed mass in Myers-Perry black holes for $D=5,6,7,8$.}
\label{fig:parameter5d6d7d8d}
\end{figure}

Figure~\ref{fig:parameter5d6d7d8d} presents some examples. In $D=5$, there are two angular momenta and, if one of them vanishes, there is a bound on the other for a given mass, the saturation of which gives a naked singularity (the four corners in Figure~\ref{fig:parameter5d6d7d8d}). In $D=6$, there are two angular momenta and, if one of them vanishes, there is no bound on the other. In $D=7$, there are three angular momenta and two cases occur: if only one vanishes, there is a singular bound (the intersections of the extremality curve with the $J_{1,2,3}=0$ planes in Figure~\ref{fig:parameter5d6d7d8d}); but if two vanish, there is no bound (along the axes). In $D=8$, there are three angular momenta and, if any of them vanish, there is no bound on the other two.

The fact that there are solutions with arbitrarily high angular momentum for a given mass allows for different regimes, e.g. in the $D>5$ singly-spinning we can have $|J|/M^{\frac{D-2}{D-3}} \ll 1$ but also $|J|/M^{\frac{D-2}{D-3}} \gg 1$. This hints at a rich phenomenology, as we shall confirm.

The isometry group of the Myers-Perry family is $\mathbb{R} \times U(1)^n$ (up to discrete factors), where $\mathbb{R}$ corresponds to the time translations and $U(1)^n$ to the rotational symmetries. In Chapters~\ref{cha:MPsingle} and \ref{cha:MPequal}, we will consider particular cases of the Myers-Perry family with enhanced symmetry, namely solutions with a single spin (say $J_1 \neq 0$, $J_i = 0\;\; \forall i>1$), which have isometry group $\mathbb{R} \times U(1) \times SO(D-3)$, and solutions with equal spins ($J_i=\bar{J}\;\; \forall i$), which have isometry group $\mathbb{R} \times U(n)$.

The Myers-Perry family has been extended to asymptotically (A)dS spacetimes in \cite{Hawking:1998kw} ($D=5$) and \cite{Gibbons:2004uw,Gibbons:2004js} ($D>5$). It is the only exactly known family of localised black holes with that asymptotic behaviour.

\subsection{Plethora of black holes}

There are no solutions with extended event horizons in four-dimensional general relativity unless a cosmological constant is present.\footnote{Although black hole spacetimes in $D=4$ which asymptote to AdS in all spatial directions are restricted to have spherical event horizon topology \cite{Galloway:1999bp}, solutions with extended cylindrical, planar and hyperbolic horizons exist \cite{Lemos:1994xp,Lemos:1994fn,Lemos:1995cm,Mann:1997jb}.} A reason for this is that no asymptotically flat black holes exist in $D<4$. Notice that the direct product of two Ricci-flat manifolds is also a Ricci-flat manifold, so that solutions with extended horizons are trivially constructed from lower dimensional black holes. Consider the metric of a ($D+N$)-dimensional vacuum black brane which uniformly extends a $D$-dimensional asymptotically flat vacuum black hole with metric $g_{ab}$ along $\mathbb{R}^N$,
\bea
\label{blackbrane}
ds_{\text{brane}}^2 = g_{ab}dx^a dx^b + d\vec{z} \cdot d\vec{z}\,.
\eea
The simplest example is the Schwarzschild string, $Schwarz_D \times \mathbb{R}$. Black branes need not be uniform and need not be obtained from a direct product. We will discuss how families of non-uniform branes bifurcate from uniform branes due to linear instabilities with stationary threshold.

The existence of black strings is suggestive. Take a segment of a black string and bend it to make it circular like a ring, introducing also rotation on the plane of the circle so that the centrifugal force may balance the gravitational pull. Emparan and Reall showed in \cite{Emparan:2001wn} that there is indeed an asymptotically flat \emph{black ring} solution in $D=5$ (generalised to have two angular momenta by Pomeransky and Sen'kov \cite{Pomeransky:2006bd}). The topology of the black ring event horizon is $S^1 \times S^2$. Topological restrictions, such as Hawking's proof \cite{Hawking:1971vc} that the horizon topology must be spherical in $D=4$, are much weaker in higher dimensions (see e.g. \cite{Hollands:2006rj}).

\begin{figure}[t]
\centerline{\includegraphics[width=.4\textwidth]{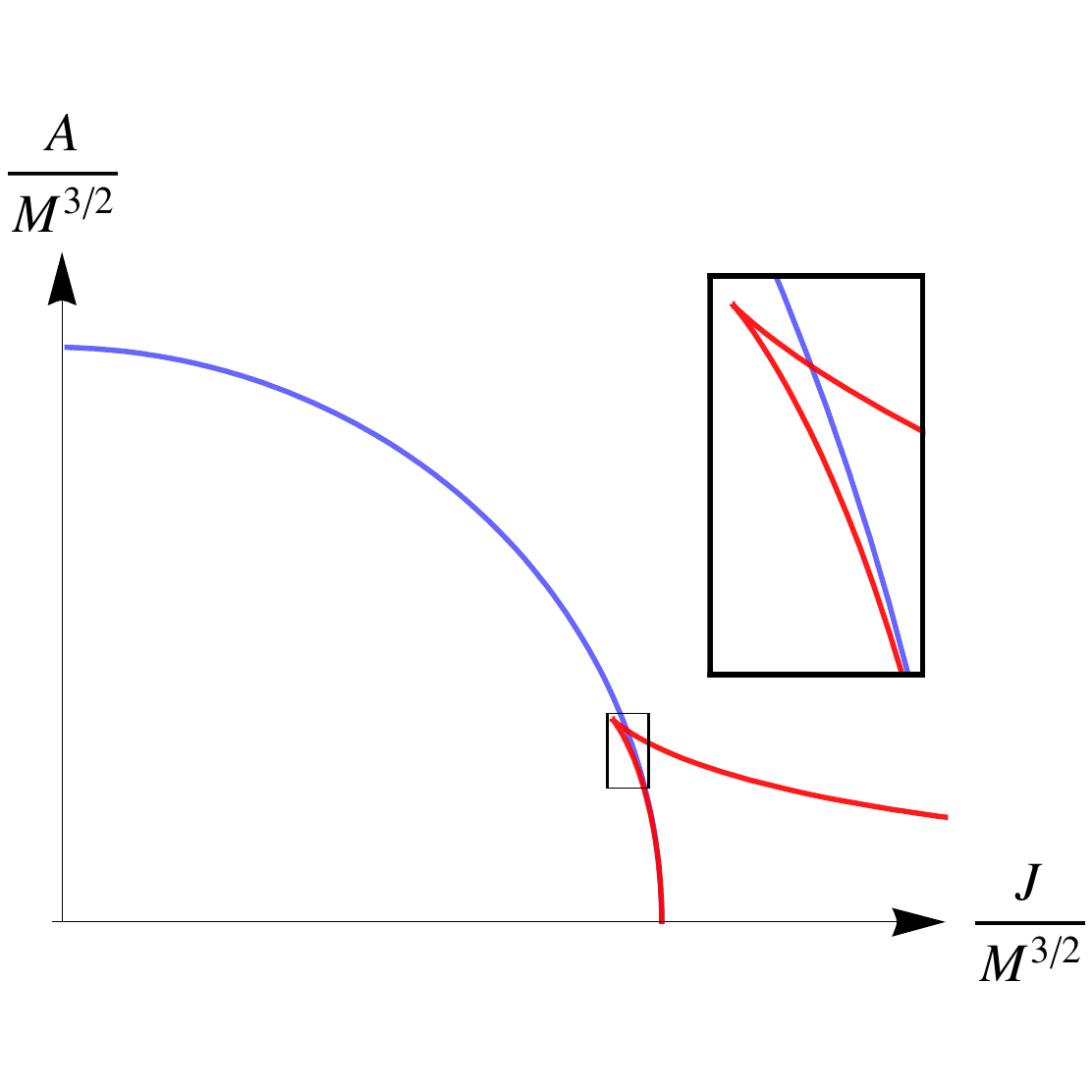}}
\caption{Entropy $S$ versus angular momentum $J$ for fixed mass $M$, for the singly-spinning black rings (red) and the $D=5$ Myers-Perry black hole (blue). Notice the zoomed detail in the top-right corner, where the fat and the thin ring solutions coincide at the cusp.}
\label{fig:nonunique}
\end{figure}

\begin{figure}[t]
\centerline{\includegraphics[width=.7\textwidth]{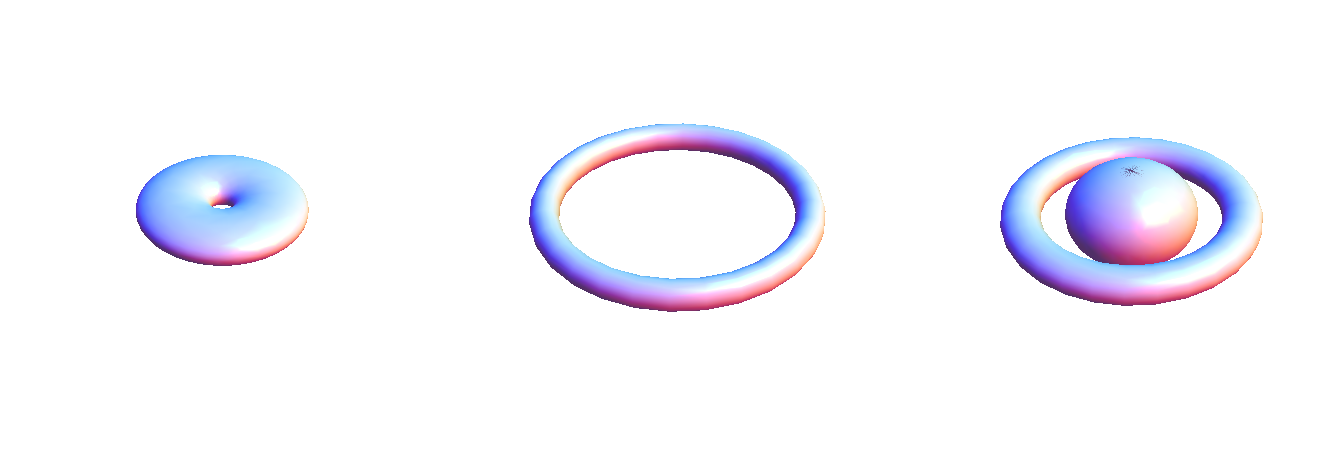}}
\caption{Lower-dimensional pictures of the event horizons of a fat ring, a thin ring and a black Saturn (left to right).}
\label{fig:ringssat}
\end{figure}

The existence of black rings shows that uniqueness is lost generically for rotating black holes. There is a region in the parameter space of $D=5$ black holes, near the corners of the Myers-Perry parameter space of Figure~\ref{fig:parameter5d6d7d8d}, where three solutions with connected horizons exist: the Myers-Perry black hole and \emph{two} black ring solutions. Let us concentrate on the case with a single spin, shown in Figure~\ref{fig:nonunique}. There are two black ring solutions meeting at a cusp in the graph: the \emph{fat ring} has smaller entropy and finite $J$ range, and the \emph{thin ring} can have arbitrarily large $J$.
The existence of these two solutions shows that uniqueness fails even for a given horizon topology. In the plot, the Myers-Perry black hole and the fat ring meet at a zero-area point where both cases give the same nakedly singular solution. Figure~\ref{fig:ringssat} presents lower-dimensional pictures of a fat ring and a thin ring.

One may also wonder whether stationary vacuum solutions with disconnected horizon components can exist, which is believed to be impossible in $D=4$. Take a black ring (BR), put it around a Myers-Perry (MP) black hole, and then make the black ring rotate faster to balance the gravitational pull, in a Saturn-like configuration as in Figure~\ref{fig:ringssat}. Elvang and Figueras \cite{Elvang:2007rd} constructed the singly-spinning \emph{black Saturn} solution, which has a disconnected event horizon: an $S^3$ component and an $S^1 \times S^2$ component. There are regions of the parameter space where the solution can have the same asymptotic charges as a Myers-Perry black hole or a black ring. However, the solution possesses a continuous type of non-uniqueness given the asymptotic charges\footnote{Continuous non-uniqueness was first obtained for black rings with a dipole charge \cite{Emparan:2004wy}.}: masses and angular momenta can be assigned to each horizon component by Komar integrals, and they can vary while keeping the asymptotic mass $M=M_\text{`MP'}+M_\text{`BR'}$ and angular momentum $J=J_\text{`MP'}+J_\text{`BR'}$ fixed. The case $J_\text{`MP'}=-J_\text{`BR'}$ shows that non-static $J=0$ solutions are possible. Furthermore, there are interesting frame dragging effects between the two components. Solutions have also been constructed which represent two concentric black rings rotating on the same plane (\emph{di-rings}) \cite{Iguchi:2007is} or on perpendicular planes (\emph{bi-rings}) \cite{Izumi:2007qx,Elvang:2007hs}.

These solutions were obtained with powerful solution generating techniques for $D=5$, reviewed in \cite{Emparan:2008eg}, which assume the existence of two rotational Killing symmetries. There is uniqueness in this class, given the asymptotic charges and the so-called rod structure that generates the solution \cite{Hollands:2007aj}. The failure to construct an analogous scheme in $D>5$ is the reason why Myers-Perry solutions are the only rotating solutions known exactly. However, thin black rings with large rotation in $D>5$ have been constructed using approximate methods \cite{Emparan:2007wm}. These methods were later put into a more systematic framework, where black holes with quasi-extended horizons are given approximate solutions called \emph{blackfolds} \cite{Emparan:2009cs,Emparan:2009at,Emparan:2009vd}. The approach allows for the construction of a variety of solutions with different horizon topology. This is important since, as we mentioned, topological restrictions are much weaker in higher dimensions.

An important recent development is related to the rigidity theorem, generalised from $D=4$ \cite{Hawking:1973uf} to higher dimensions in \cite{Hollands:2006rj,Moncrief:2008mr}, which requires stationary (non-extremal) black holes to have at least one rotational symmetry. Until recently, not a single black hole had been found with only one rotational symmetry, though such solutions had been conjectured to exist \cite{Reall:2002bh}. The blackfold approach gave the first example, as it allowed for the approximate construction of \emph{helical} black rings in any $D>4$ which possess a single rotational symmetry \cite{Emparan:2009vd}. The perturbative numerical work of \cite{Dias:2010eu}, presented in Chapter~\ref{cha:MPequal} of this thesis, provided the first evidence of such solutions with spherical horizon topology, in $D=9$ (and all higher odd dimensions, we expect). A single rotational symmetry was also obtained in all even $D \geq 8$ for near-horizon geometries which give an infinite class of distinct horizon topologies and are consistent with the requirements of asymptotically flat or AdS extremal black holes \cite{Kunduri:2010vg}.

The goal is to uncover the phase diagram analogous to Figure~\ref{fig:nonunique} but for all solutions in any $D>4$. Notice how the fat ring and the Myers-Perry solution coincide at one point in that Figure. The black Saturn in thermodynamic equilibrium (horizon components with the same temperature and angular velocity) appears in this plot in a similar way to the black ring, also joining the Myers-Perry solution at the singular point; see Figure 2 of \cite{Elvang:2007hg}. In $D>5$, this singularity is ``resolved" since the angular momentum is unbounded for the singly-spinning Myers-Perry black hole (recall Figure~\ref{fig:parameter5d6d7d8d}). The expectation \cite{Emparan:2003sy,Emparan:2007wm} is that the connection between phases at the singular bound in $D=5$ will correspond in $D>5$ to bifurcations of regular solutions which interpolate between Myers-Perry black holes and phases such as the black ring and the black Saturn. We shall elaborate on this in Chapter~\ref{cha:MPsingle}, which is based on the numerical work of \cite{Dias:2009iu}. The results imply non-uniqueness in $D>5$ given spherical horizon topology.

It would be very interesting to see how the picture described here would change with the inclusion of a cosmological constant, especially a negative one having in mind the AdS/CFT correspondence. We mentioned that the Myers-Perry solution is the only one to have been extended to asymptotically (A)dS spacetimes. 
However, black rings have been constructed approximately in AdS \cite{Caldarelli:2008pz} and there is no reason why a variety of solutions should not exist in this case too.

\subsection{Thermodynamics}

The existence and -- as we shall see -- the stability of black hole solutions depend critically on the dimensionality of spacetime. However, the laws of black hole thermodynamics are a consequence of the structure of Einstein's equations in any dimensions.

For charged black holes, the first law in higher dimensions is
\be
\label{1stlawcomp}
dM = \frac{\kappa^{(\alpha)}}{8 \pi} \, dA^{(\alpha)} + \Omega^{(\alpha)}_i dJ^{(\alpha)}_i + \phi^{(\alpha)}_\ell dQ^{(\alpha)}_\ell \,,
\ee
where we have a sum over different planes of rotation -- index $i$ --, a sum over charges -- index $\ell$ -- and a sum over different event horizon components -- index $(\alpha)$. This extension of \eqref{1stlaw4d} can be derived with the standard procedure \cite{Bardeen:1973gs}, by considering that each component is a Killing horizon of $K^{(\alpha)} = \pd_t + \Omega^{(\alpha)}_i \pd_{\varphi_i} $. Notice that the individual angular momenta $J^{(\alpha)}_i$ and charges $Q^{(\alpha)}_\ell$ are not asymptotically defined conserved quantities in the case of disconnected horizon components. Even in the case of a connected horizon, a dipole charge $Q^{\text{dip}}_\ell$ has no net asymptotic contribution, despite being included in the first law \cite{Emparan:2004wy,Copsey:2005se}.

The Smarr relation, valid for asymptotically flat spacetimes, reads
\be
\label{smarrcomp}
(D-3)\, M = (D-2)\,\left( \frac{\kappa^{(\alpha)} A^{(\alpha)}}{8 \pi} + \Omega^{(\alpha)}_i J^{(\alpha)}_i \right) + \sigma_\ell \,\phi^{(\alpha)}_\ell dQ^{(\alpha)}_\ell \,,
\ee
where $\sigma_\ell$ is the scaling dimension of the charge (it differs for dipoles and conserved charges).

Thermodynamic equilibrium between components of the horizon requires
\be
\kappa^{(\alpha)}=\kappa\,, \qquad \Omega^{(\alpha)}_i=\Omega_i\,, \qquad \phi^{(\alpha)}_\ell=\phi_\ell\,,
\ee
which reduces the space of solutions, e.g. the contributions $J^{(\alpha)}_i$ from different horizon components to the total conserved angular momenta $J_i$ are fixed. The first law becomes simply
\be
\label{1stlawc}
dM = \frac{\kappa}{8 \pi} \, dA + \Omega_i dJ_i + \phi_\ell dQ_\ell \,.
\ee


\chapter{Stability of black holes} \label{chap:stab}

The construction of a stationary black hole spacetime satisfying the equations of motion is not enough to appreciate the physical significance of the solution. In a ``realistic" scenario, we have to know whether or not the spacetime is robust for small perturbations of the geometry and matter fields. If not, a probe, say a particle moving on the black hole background, may cause a disruptive backreaction; and the possibility of dynamically forming such a black hole through a physical process, such as gravitational collapse, is put in doubt.

Nevertheless, unstable solutions are not devoid of physical significance. Instabilities have characteristic timescales, and a solution may be ``sufficiently stable" for shorter timescale phenomena. Furthermore, the manifestation of an instability is itself of great interest. The onset may indicate a phase transition through a bifurcation, as we shall see. Information on the timescales is very useful especially in the cases where different instabilities are present, since it may give hints on what the final state is, given that the full time evolution is often beyond our technology.

The complete description of black holes  -- stability included -- must take into account their quantum thermodynamic properties. It is well known that the Schwarzschild black hole is classically stable at the linear mode level and yet it has a negative specific heat, which signals a (local) thermodynamic instability. Standard thermodynamic arguments can also identify (global) phase transitions between different spacetimes.

In this Chapter, we will discuss both the classical and the thermodynamic stability of black holes. It is one of the goals of this thesis to explore their connection.


\section{Classical stability} \label{sec:classstab}

We are concerned here with the stability of stationary black hole spacetimes as solutions to a system of differential equations of motion, namely the Einstein equations, or extensions. The literature often consists of linear stability studies. Just as happened for uniqueness, stability does not always hold for asymptotically flat vacuum black holes in $D>4$ when rotation is considered. In fact, we shall argue that uniqueness and stability are intimately related. Is there uniqueness of classically stable black holes? This remains a major open problem.

\subsection{Statement}

Classical stability is based on the initial value problem of general relativity \cite{Wald:1984rg}. Brushing aside the intricate technical aspects, the stability problem can be put as follows. Take a Cauchy surface $\Sigma$ of a stationary solution $(\M,g_{ab})$ and perturb the initial data $C$ on $\Sigma$. That is, instead of the initial data $C$ that develops into the stationary solution from $\Sigma$, $(\M_\Sigma,g_{ab})$, take smooth initial data $\widetilde{C}$ which is ``close" to $C$ and satisfies the Einstein constraint equations. Now consider the future development $(\M_\Sigma,\widetilde{g}_{ab})$ of $\widetilde{C}$. If, for any ``small" perturbation ($\widetilde{C}$ close enough to $C$), the future development $(\M_\Sigma,\widetilde{g}_{ab})$ is ``close" to -- or actually ``approaches" -- $(\M_\Sigma,g_{ab})$, then the stationary solution $(\M,g_{ab})$ is stable. For instance, starting with the Kerr spacetime, the perturbed solution $(\M_\Sigma,\widetilde{g}_{ab})$ could become a Kerr solution at late times, but with slightly different mass and angular momentum. Otherwise, the stationary spacetime is unstable, and the time evolution in $(\M_\Sigma,\widetilde{g}_{ab})$ will lead ``away" from $(\M_\Sigma,g_{ab})$ for a class of perturbations. The inclusion of matter fields $\Phi_s$ presents no problem of principle, the initial value problem being set for the solution $(\M,g_{ab},\Phi_s)$.

For black hole spacetimes, only the development of the data on $\Sigma$ between the future event horizon $\H^+$ and spatial infinity is important, as the evolution inside the horizon cannot affect the evolution outside. We are therefore interested only in the stability of the black hole exterior.

The non-linearity of the Einstein equations makes the stability problem very challenging. A rigorous proof of non-linear stability for small perturbations has been achieved only for the simplest of spacetimes, Minkowski space, by Christodoulou and Klainerman \cite{Christodoulou:1993uv}. For black holes, apart from numerical approaches, only the linear problem has been studied. In this case, one considers perturbations $h_{ab} = \widetilde{g}_{ab} - g_{ab}$ of a black hole background $g_{ab}$, subject to boundary conditions to ensure the regularity of the perturbed spacetime $\widetilde{g}_{ab}$. The perturbations satisfy the \emph{linearised} Einstein equations,
\bea
\label{linearEinstein}
&& (\Delta_L h)_{ab} - \nabla_a \nabla_b h^c_{\phantom{c}c} + 2 \nabla_{(a} \nabla^c h_{b)c}
+ g_{ab}\big(-\nabla^c \nabla^d h_{cd} + \Box h^c_{\phantom{c}c} + R^{cd} h_{cd}\big) \nonumber\\
&&
\phantom{aaaaaaaaa}+ (2\Lambda - R) h_{ab} \;\; = \;\; 16 \pi \delta T_{ab}\,,
\eea
where $\Delta_L$ is the Lichnerowiz operator, defined as
\be
\label{deflichn}
(\Delta_L h)_{ab} \equiv -\nabla^c \nabla_c h_{ab} -2 R_{a\phantom{c}b}^{\phantom{a}c\phantom{b}d} h_{cd} + 2 R^c_{(a} h_{b)c} \,,
\ee
and $\delta T_{ab}$ is the linear perturbation of the energy-momentum tensor. The gauge ambiguity in $h_{ab}$ must be dealt with, either by fixing the gauge or by considering gauge-invariant quantities. Stability requires that these perturbations remain bounded in their time evolution, the bound depending on the initial data. This usually means that an appropriate positive definite functional ${\mathcal F}[h](t)$ of the perturbation, defined on a time slice $\Sigma_t$ outside the horizon, obeys ${\mathcal F}[h](t) \leq C {\mathcal F}[h](t_0)$, where the initial data is defined on $\Sigma_{t_0}$, $t>t_0$, and $C$ is a constant depending on the parameters of the background geometry. It may also be possible to show ``pointwise" boundedness, e.g. for the perturbation as a function of the radius in a spherically symmetric spacetime, as opposed to a functional over a time slice. When considering the implications to the non-linear problem, stronger conditions of time decay -- and not just boundedness -- are typically required at the linear level (this is because some integrals over time should also be bounded). Notice, however, that the solution approached at late times by the time decay of perturbations will have small variations in its parameters, such as mass and angular momenta, with respect to the initial solution. The time decay of the scalar wave equation is an important first step in understanding the decay of metric perturbations.

The symmetries of the black hole background can simplify the problem considerably. For stationary backgrounds $g_{ab}$, a linear perturbation is a superposition of Fourier modes $h_{ab}^{(\omega)} \sim e^{-\ii \omega t}$, and stability implies that, when analysing each mode separately, only modes with $\text{Im}(\omega)<0$ are allowed. Spatial symmetries of the background also allow for decompositions, e.g. into spherical harmonics when there is spherical symmetry. Mode analyses are much simpler and very effective, as we shall see in this thesis. In the case that no unstable modes are found, the approach falls short of the mathematical rigour required to establish linear stability.\footnote{Because an infinite sum of modes which decay in time may not decay in time. Furthermore, there are examples, such as coloured black holes \cite{Bizon:1991nt}, where regular initial data can be constructed from unstable modes which are irregular per se, and thus usually discarded.} However, the detection of an \emph{instability} is clearer, since the initial data may be fine-tuned to reproduce a particular unstable mode.

\subsection{Review} \label{sec:classstabreview}

Let us start with the static case (for a more detailed review, see \cite{Kodama:2004gz}). In $D=4$, the linear mode stability of the Schwarzschild black hole was established by \cite{Vishveshwara:1970cc,Price:1971fb,Wald:1979,Kay:1987ax}. Spherical symmetry allows for a decomposition of linear perturbations into an even-parity class (constructed from scalar harmonics of $S^2$) and an odd-parity class (constructed from vector harmonics of $S^2$), as shown by Regge and Wheeler \cite{Regge:1957td}. Each class of perturbations is governed by a master equation: the Regge-Wheeler equation in the odd-parity case \cite{Regge:1957td} and the Zerilli equation in the even-parity case\cite{Zerilli:1970se}. After a Fourier decomposition of the time dependence, each becomes a Schr\"odinger-type ODE allowing for no unstable regular modes. A rigorous approach avoiding that decomposition is put forward in \cite{Wald:1979,Kay:1987ax}, where the boundedness of linear perturbations is shown. Analyses of Schr\"odinger-type master equations have established the mode stability of the non-degenerate Reissner-Nordstr\"om black hole \cite{Chandrasekhar}, using the Newman-Penrose formalism \cite{Newman:1961qr}, and the Schwarzschild-(A)dS black hole \cite{Cardoso:2001bb,Kodama:2003jz}. In the latter case, rigorous studies of the boundedness and decay of solutions to the scalar wave equation are available, see e.g. the review \cite{Dafermos:2007jd} for dS and \cite{Holzegel:2009ye} for AdS, where a scalar field mass obeying the Breitenlohner-Freedman bound \cite{Breitenlohner:1982jf} is considered.\footnote{In asymptotically AdS spacetimes, the presence of a Cauchy horizon demands more care in setting boundary conditions at spatial infinity, in order to have a well-defined initial value problem \cite{Breitenlohner:1982jf,Ishibashi:2004wx}. Negative mass-squared scalar fields can have a well-defined dynamics if the mass obeys the Breitenlohner-Freedman bound \cite{Breitenlohner:1982jf}.}

In higher dimensions, the procedure has been applied with success to establish the mode stability of Schwarzschild black holes for arbitrary $D$ \cite{Gibbons:2002pq,Ishibashi:2003ap}. Stability extends to the $D=5$ Reissner-Nordstr\"om case, but no such proof is available for $D>5$. Schwarzschild-dS black holes have been shown to be stable for $D \leq 6$. While Reissner-Nordstr\"om-dS black holes are stable for $D \leq 5$, an instability was found for sufficiently large charge in $D>6$ \cite{Konoplya:2008au,Cardoso:2010rz}. In the asymptotically AdS case, not even the stability of Schwarzschild-AdS black holes is established for $D>4$.

Now we consider rotating asymptotically flat black holes, starting with $D=4$. For the Kerr spacetime, which is codimension-2, Teukolsky obtained the remarkable result that the linearised Einstein equations can also be reduced to a single separable master equation \cite{Teukolsky:1972my}, using the Newman-Penrose formalism \cite{Newman:1961qr}.\footnote{The Teukolsky equation considers radiative modes, thus excluding stationary modes, which consist only of trivial mass and angular momentum variations, as shown by Wald \cite{Wald:1973}.} Whiting showed that the Teukolsky equation does not allow for unstable Fourier modes \cite{Whiting:1988vc}. Such a separability of the equations has not been achieved in the Kerr-Newman case, which remains an open problem. Ref.~\cite{Dafermos:2008en} reviews the rigorous study of the scalar wave equation for slowly-rotating ($|J| \ll M^2$) small-charge ($|Q| \ll M$) Kerr-Newman black holes, avoiding the use of separability properties. That approach has provided proofs of the boundedness and decay of solutions to the scalar wave equation, and has also strengthened the results of \cite{Wald:1979,Kay:1987ax} for the Schwarzschild case. Let us also point out that numerical studies of black hole formation (e.g. \cite{Stergioulas:2009zz,Choptuik:2009ww}) provide important evidence for stability in the full non-linear regime. Astrophysical black holes should be unique and stable.

\begin{figure}[t]
\centerline{\includegraphics[width=0.4\textwidth]{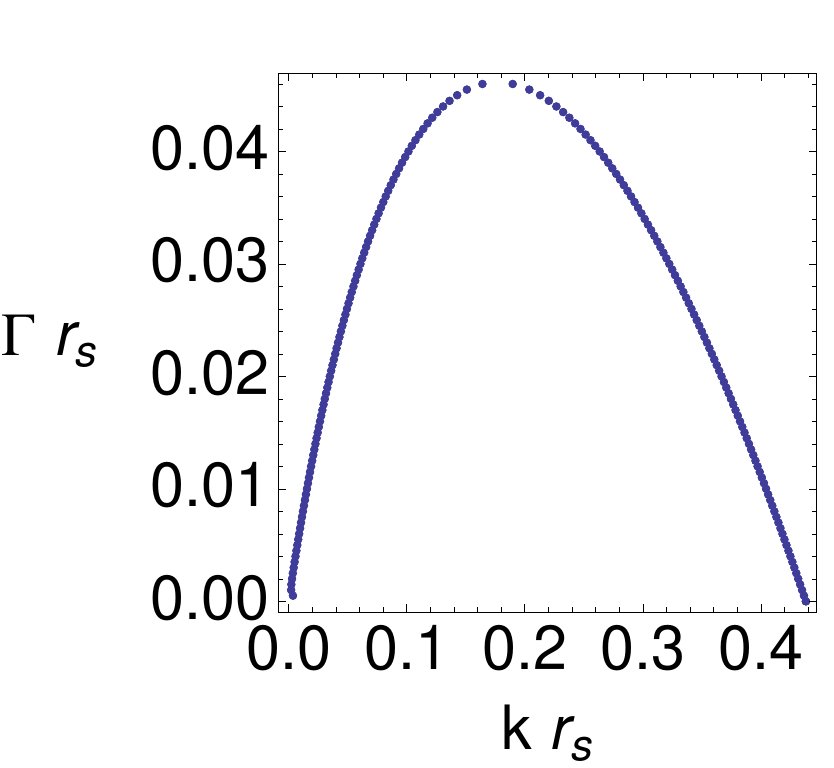}}
\caption{Profile $\Gamma(k)$ for the Gregory-Laflamme unstable mode of a Schwarzschild brane ($D_\text{Schwarz}=4$), where $r_s$ stands for the Schwarzschild radius.}
\label{fig:4dgl}
\end{figure}

The picture of stability for higher-dimensional rotating black holes is very different. The first discovery of an instability in higher dimensions concerned solutions which, despite being static, possess an extended horizon. Gregory and Laflamme showed numerically that black branes of the type \eqref{blackbrane} constructed with a $D$-dimensional Schwarzschild black hole are unstable \cite{Gregory:1993vy}. Consider a transverse and traceless (TT) linear perturbation of the metric \eqref{blackbrane} of the form
\be
\label{glpert}
ds^2_\mathrm{brane} = g_{ab}(x)dx^a dx^b + d\vec{z} \cdot d\vec{z}\qquad \to \qquad ds^2_\mathrm{brane}+e^{\ii\, \vec{k}\cdot\vec{z}} h_{ab} (x) dx^a dx^b\,,
\ee
The condition is equivalent to $h_{ab}$ being TT with respect to the $D$-dimensional black hole metric $g_{ab}$,
\be
\label{eqn:TTgauge}
h^a_{\phantom{a}a}=0\,, \qquad \nabla^a h_{ab} =0\,.
\ee
The linearised Einstein equations \eqref{linearEinstein} reduce to
\be
\label{lichn}
(\Delta_L h)_{ab} = -k^2 h_{ab}\,,
\ee
where $\Delta_L$ is the Lichnerowicz operator for the black hole background, and $k^2 \equiv \vec{k} \cdot \vec{k}$. Hence perturbations with non-zero $k$ correspond to {\it negative modes} of $\Delta_L$. The boundary conditions are that $h_{ab}$ should be regular on the future horizon and vanishing at infinity. Gregory and Laflamme studied time dependent modes $h_{ab} \propto e^{\Gamma t}$ which preserve the spherical symmetry of the Schwarzschild $g_{ab}$ background.\footnote{The equations \eqref{lichn} reduce to a single ODE. In a Fourier decomposition $h_{ab} \propto e^{-\ii\, \omega t}$, the hermiticity of the ODE operator guarantees a real eigenvalue $\omega^2$, so that $\Gamma=\ii\, \omega$ is purely real (and positive) for an instability.} They found exponentially growing solutions to \eqref{lichn} for $k<k_\ast$, as shown in Figure~\ref{fig:4dgl}.
The instability pinches the branes along the extra-directions $\vec{z}$, breaking the translational symmetry. The final point of the time evolution remains an open problem \cite{Horowitz:2001cz}. The stability properties may be improved with the inclusion of charge (see the review \cite{Harmark:2007md}), according to the Gubser-Mitra conjecture, to be discussed in the next Section.

\begin{figure}[t]
\centerline{\includegraphics[width=1\textwidth]{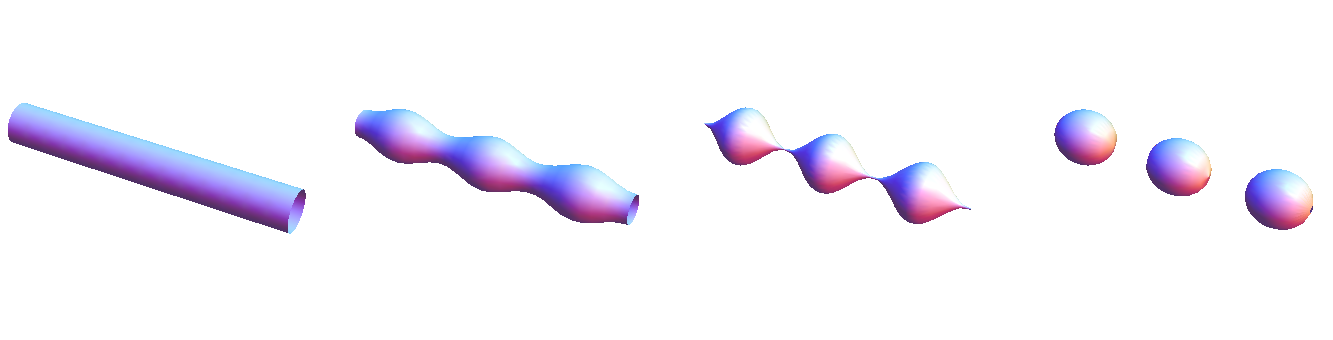}}
\caption{Family of solutions from a uniform black string to an array of black holes.}
\label{fig:nonuniform}
\end{figure}

Notice that, while the mode $(k,\Gamma)=(0,0)$ is a gauge mode \cite{Gregory:1994bj}, the mode $(k,\Gamma)=(k_\ast,0)$ is a static linear perturbation signaling the existence of a one-parameter family of static non-uniform branes. The existence of such a family of black strings was shown first in \cite{Gubser:2001ac}, at the perturbative level, and then in \cite{Wiseman:2002zc}, with a non-linear numerical analysis. Along the family of non-uniform solutions, the pinches become greater and greater, supposedly until the string splits into an array of black holes; see \cite{Kol:2004ww,Harmark:2007md} for reviews and \cite{Headrick:2009pv} for more recent work. The idea is represented in Figure~\ref{fig:nonuniform}. The same reasoning should apply to the rotating non-uniform strings discovered in \cite{Kleihaus:2007dg} and \cite{Dias:2010eu}, also associated with instabilities as verified numerically in \cite{Dias:2010eu}.

The fact that the event horizon is infinite along the $\vec{z}$ directions but finite along the black hole directions is the cause of the instability.\footnote{Ref.~\cite{Cardoso:2006ks} proposed an analogy with the Rayleigh-Plateau instability of fluid tubes, which was made more precise in the context of the fluid-gravity correspondence \cite{Caldarelli:2008mv,Camps:2010br}.} Now, in higher dimensions, black hole horizons can be characterised by very different length scales. Say we have a horizon scale $L_x$ along certain directions $x$ and a horizon scale $L_y \ll L_x$ along other directions $y$. A linear perturbation along the $x$ directions with a length scale $\lambda$ satisfying $L_y<\lambda<L_x$ may behave like a Gregory-Laflamme unstable mode. Recall that the horizon of a Kerr black hole is a deformed sphere, but that there is a limit on the deformation which is set by extremality, $|J|/M^2 \leq 1$. We already mentioned that such a bound is absent for singly-spinning Myers Perry black holes in $D>5$: we have the regime $|J|/M^{\frac{D-2}{D-3}} \ll 1$, for which the horizon is approximately spherical, but also the regime $|J|/M^{\frac{D-2}{D-3}} \gg 1$, for which some directions along the horizon spread, becoming quasi-extended. In the latter regime, the black hole is expected to be unstable, because the metric near the axis of rotation takes the form of a black-brane metric and should thus possess a Gregory-Laflamme-type instability \cite{Emparan:2003sy}. Similarly, very thin black rings, satisfying $|J|/M^{\frac{3}{2}} \gg 1$, have a horizon with topology $S^1 \times S^2$ for which the $S^1$ radius is much larger than the radius of the squashed $S^2$. Locally, a segment of the very thin ring looks like a boosted black string and is expected to be unstable to Gregory-Laflamme-type modes \cite{Emparan:2001wn}.

In Chapters~\ref{cha:MPsingle} and \ref{cha:MPequal}, based on \cite{Dias:2009ex,Dias:2009iu}, we will analyse this type of instability numerically for Myers-Perry black holes with a single spin and with equal spins. Our results for a single spin are consistent with the expectations of \cite{Emparan:2003sy,Emparan:2007wm} regarding connections between black hole families in $D>5$. These connections arise in an analogous way to the bifurcation of non-uniform branes from uniform branes, i.e. through the stationary thres\-hold mode $(k,\Gamma)=(k_\ast,0)$ of an instability, as in Figure~\ref{fig:nonuniform}. In the case of Myers-Perry black holes with equal spins, we find a similar instability in $D=9$ near the extremality bound. No clear geometrical understanding of the instability, by analogy with the Gregory-Laflamme case, has been proposed for equal spins due to the presence of the extremality bound. However, a definition of an \emph{ultraspinning} regime will be proposed, based on a connection with thermodynamics, and conjectured to be a necessary condition for this type of instability. The thermodynamic argument is only expected to hold for perturbations which preserve the stationarity $\pd_t$ and axisymmetry $\Omega_i \pd_{\varphi_i}$ Killing vectors.

Also very recently, Shibata and Yoshino detected an instability of rapidly-rotating singly-spinning Myers-Perry black holes, using a non-linear numerical method \cite{Shibata:2009ad,Shibata:2010wz}. The instability occurs for non-axisymmetric deformations. Although the approach is non-linear, the instability probably holds at the linear level, but it should not have a stationary and axisymmetric threshold mode. Hence the thermodynamic argument for our ultraspinning conjecture does not apply. Indeed, no $D=5$ Myers-Perry black hole falls into our ultraspinning class. Ref.~\cite{Murata:2008yx} found no evidence of instability for $D=5$ Myers-Perry black holes with equal spins.

We mentioned above that very thin black rings should be unstable to Gregory-Laflamme-type modes along the ring circumference. Ref.~\cite{Elvang:2006dd} points out the possibility that this affects the entire thin ring branch, since all thin rings can fit several dangerous modes in their circumference. Furthermore, evidence is provided for the instability of fat rings for radial perturbations. Fat rings had been argued to be unstable in \cite{Arcioni:2004ww}, based on the thermodynamic turning-point method.\footnote{Such a connection between classical stability and thermodynamics is of a different nature than the one we will explore in this thesis, which relies on local thermodynamic stability in the spirit of the Gubser-Mitra conjecture.} The question remains whether there is a portion of the thin ring branch with $J/M^{\frac{3}{2}} \approx \mathcal{O}(1)$ which is stable. More generally, among the plethora of higher-dimensional asymptotically flat vacuum black holes, are Myers-Perry solutions with moderate rotation the only stable ones?

To conclude, let us consider rotating asymptotically (A)dS black holes. In dS, the full stability problem remains open even in $D=4$. In the Kerr-AdS/Myers-Perry-AdS case, however, it is known that an instability for non-axisymmetric perturbations occurs when $\exists i\;|\Omega_i| \ell>1$, where $\ell$ is the AdS curvature radius. The instability is caused by superradiant modes which are amplified near the horizon and reflected back to it by the AdS gravitational potential \cite{Hawking:1999dp,Cardoso:2004nk,Cardoso:2004hs,Kunduri:2006qa,Kodama:2009rq}. When the rotation is slow, $|\Omega_i| \ell<1 \; \forall i$, these black holes should be stable \cite{Hawking:1999dp}. Ref.~\cite{Holzegel:2009ye} provides a rigorous study showing the boundedness of solutions to the scalar wave equation for slowly rotating Kerr-AdS black holes, when the scalar field mass obeys the Breitenlohner-Freedman bound \cite{Breitenlohner:1982jf}.


\section{Thermodynamic stability} \label{sec:thermalstab}

The thermodynamic description of black holes, outlined in Section~\ref{sec-bh:quantum}, accounts for their quantum properties at the semiclassical level. Stability with respect to quantum effects can be studied, within the semiclassical approximation, using the well-established criteria of thermodynamic stability.

Thermodynamic stability tells us how a system in thermodynamic equilibrium responds to fluctuations of energy, temperature and other thermodynamic parameters. The conditions for stability are derived from the second law of thermodynamics. If they are violated, which often happens for black holes, the system is not in a preferred configuration. We distinguish between \emph{global} and \emph{local} stability. Because black holes, as other self-gravitating systems, are not extensive some care is required in interpreting these conditions.

For definiteness, we will focus on vacuum black holes in thermodynamic equilibrium, for which the first law reads
\be
\label{1stlaw}
dM= T dS + \Omega_i dJ_i\,.
\ee

\subsection{Global stability} \label{sec:globalstab}

Global stability is concerned with the phase of a system corresponding to the global maximum of the total entropy.

Consider a system in equilibrium with a thermodynamic reservoir at temperature $T$ and angular velocities $\Omega_i$. If we allow for mass and angular momenta exchanges ($\Delta M \neq 0$, $\Delta J_i\neq0$) between the system and the reservoir, the relevant ensemble to describe the system is the \emph{grand-canonical}. The preferred phase of the system is the one that minimises the Gibbs free energy,
\be
G=M-TS-\Omega_i J_i\,,
\ee
given $T$ and $\Omega_i$. Therefore, for two possible phases -- 1 and 2 -- satisfying $G_1(T,\Omega_i) > G_2(T,\Omega_i)$, phase 1 is unstable. Starting with phase 1, thermodynamic fluctuations will eventually lead to a phase transition to phase 2. To see this from the second law, the crucial point is that the reservoir is assumed to be very large so that no exchange affects its internal equilibrium. Hence, exchanges are reversible for the reservoir: $\Delta M_\text{res} = T \Delta S_\text{res} + \Omega_i \Delta J_{i\,\text{res}}$. Conservation of mass and angular momentum requires $\Delta M_\text{res} = - \Delta M$ and $\Delta J_{i\,\text{res}} = - \Delta J_i$. The second law then favours fluctuations obeying
\be
\label{gcanglobal}
\Delta S + \Delta S_\text{res} \geq 0 \quad \Leftrightarrow \quad \Delta S - \frac{\Delta M- \Omega_i \Delta J_i}{T} \geq 0 \quad \Leftrightarrow \quad \Delta G \leq 0\,.
\ee

If we restrict the exchanges such that $\Delta J_i=0$, but $\Delta M \neq 0$, then the relevant ensemble is the \emph{canonical}. The preferred phase is the one that minimises the Helmholtz free energy,
\be
\label{canglobal}
F=M-TS\,,
\ee
given $T$ and $J_i$, since the second law implies that $\Delta F \leq 0$ is favoured.

An isolated system ($\Delta M=\Delta J_i=0$) is described by the \emph{micro-canonical} ensemble. The second law is direct: $\Delta S \geq 0$, the preferred phase of the system being the one that maximises the entropy $S$ given $M$ and $J_i$.

In higher-dimensions, there are several black hole phases so that we may expect to find phase transitions due to the global instabilities. For instance, Figure~\ref{fig:nonunique} represents three phases in a certain parameter range. However, as we will show later, following \cite{Dias:2010eu}, asymptotically flat vacuum black holes are \emph{locally} thermodynamically unstable. The most celebrated example of a global phase transition between locally stable phases occurs in asymptotically AdS spacetimes: it is the Hawking-Page phase transition between thermal AdS and a large asymptotically AdS black hole \cite{Hawking:1982dh}. According to the AdS/CFT corresponce such a semiclassical phase transition between asymptotically AdS spacetimes has a corresponding ``confinement/deconfinement" phase transition in the free energy of the dual field theory, and that is indeed what Witten found \cite{Witten:1998qj}.

\subsection{Local stability} \label{sec:localstab}

Local stability is concerned with whether a certain phase in equilibrium is a local maximum of the total entropy. This depends on how the system responds to small fluctuations of its thermodynamic parameters. The phase is locally stable if the response counteracts the effect of the fluctuation leading back to equilibrium. The standard analysis is based on linear response functions, such as specific heats.

Let us consider the general condition of local thermodynamic stability. As happened for global stability, it is a consequence of the second law of thermodynamics. Take a black hole in equilibrium with a reservoir at temperature $T$ and angular velocities $\Omega_i$, as described by the \emph{grand-canonical} ensemble. The inequality \eqref{gcanglobal} gives the preferred evolution of the system. The important point is that, if a phase is locally stable, then small fluctuations must be entropically suppressed. That is, we change the sign of the middle inequality \eqref{gcanglobal} so that small fluctuations $\{\Delta M, \Delta J_i\}$ obey
\be
\label{deltaSpm}
\Delta S - \frac{\Delta M - \Omega_i \Delta J_i}{T} < 0 \,.
\ee
The fluctuations take the black hole temporarily out of equilibrium with the reservoir. For small fluctuations $\Delta M$ and $\Delta J_i$, the change in the entropy of the black hole is
\be
\Delta S = \der{S}{x_\mu}\, \Delta x_\mu + \frac{1}{2} {\dderf{S}{x_\mu}{x_\nu}} \, \Delta x_\mu \Delta x_\nu + \ldots \,, \qquad \; x_\mu=(M,J_i)\,.
\ee
Notice that, although the thermodynamic fluctuations are off-equilibrium, the coefficients in the expansion above are still determined by the equilibrium phase, because one may take $x_\mu=x_\mu(T,\Omega_i)$. Using the first law of thermodynamics,
\be
dS = \frac{1}{T} \, dM - \frac{\Omega_i}{T} \, dJ_i\,,
\ee
the inequality \eqref{deltaSpm} becomes
\be
\Delta S - \frac{\Delta M- \Omega_i \Delta J_i}{T} = \frac{1}{2} {\dderf{S}{x_\mu}{x_\nu}} \Delta x_\mu \Delta x_\nu + \ldots \leq 0 \,.
\ee
For small arbitrary fluctuations $\Delta x_\mu$, this is true if and only if
\be
\label{localstabS}
-\,{\dderf{S(x_\lambda)}{x_\mu}{x_\nu}} \,, \qquad \; x_\mu=(M,J_i)\,, \qquad \text{is positive definite.}
\ee
This is the condition for local thermodynamic stability. If it is not satisfied, the black hole will drop out of its equilibrium phase.

We can express these conditions in several different ways. In Appendix~\ref{app:localstab}, we prove that each of the following statements (the list is not exhaustive) is equivalent to condition \eqref{localstabS}:
\bea
\label{localstabG}
\bullet & \displaystyle{-\,{\dderf{G(y_\lambda)}{y_\mu}{y_\nu}}} \,, \qquad \; y_\mu=(T,\Omega_i)\,, \qquad \text{is positive definite.} \\
\label{localstabM}
\bullet & \displaystyle{\phantom{i}\,{\dderf{M(\tilde{x}_\lambda)}{\tilde{x}_\mu}{\tilde{x}_\nu}}} \,, \qquad \; \tilde{x}_\mu=(S,J_i)\,, \qquad \text{is positive definite.} \\
\label{localstabW}
\bullet & \displaystyle{\phantom{aaa..}\,{\dderf{W(\tilde{y}_\lambda)}{\tilde{y}_\mu}{\tilde{y}_\nu}}} \,, \qquad \; \tilde{y}_\mu=(\beta,-\beta \Omega_i)\,, \qquad \text{is positive definite.}
\eea
We introduced here $\beta \equiv 1/T$ and the Legendre transform of the entropy,
\be
W=S-\beta M + \beta \Omega_i J_i = - \beta G \,.
\ee
The formulation \eqref{localstabG} is more convenient to relate the stability condition to the usual linear response functions, like the specific heat. Explicitly,
\be
\label{localstabGmatrix}
-{\dderf{G(y_\lambda)}{y_\mu}{y_\nu}} = \left(
\begin{array}{cc}
\beta C_\Omega & \eta_j \\
\eta_i & \epsilon_{ij}
\end{array}
\right)\,,
\ee
where
\be
C_\Omega = T \left( {\der{S}{T}} \right)_\Omega
\ee
is the specific heat at constant angular velocities (all $\Omega_i$ fixed). The isothermal differential moment of inertia tensor is
\be
\epsilon_{ij} = \left( {\der{J_i}{\Omega_j}} \right)_T = \epsilon_{ji}.
\ee
There is also the vector
\be
\eta_{i} = \left( {\der{S}{\Omega_i}} \right)_T = \left( {\der{J_i}{T}} \right)_\Omega,
\ee
where the second equality, given by the symmetry of the Hessian matrix, corresponds to a Maxwell relation. The specific heat at constant angular momenta $C_J$, defined as
\be
C_J = T \left( {\der{S}{T}} \right)_J,
\ee
satisfies the identity:
\be
\label{cjco}
C_J = C_\Omega - T (\epsilon^{-1})_{ij} \,\eta_{i} \,\eta_{j},
\ee
which follows simply from the chain and cyclic rules for partial derivatives. We show in Appendix~\ref{app:localstab} that the usual stability conditions,
\be
\label{localstabusual}
C_J >0 \quad \qquad \text{and} \quad \qquad \epsilon_{ij} \quad \text{is positive definite,}
\ee
amount to the statement \eqref{localstabG}. The identity \eqref{cjco} then implies that $C_\Omega > C_J$.

Let us also obtain these conditions starting from the partition function. We can write the grand-canonical partition function $Z = e^{-\beta G}$ as
\be
Z(\tilde{y}_\mu) = \int (\Pi d x_\nu) \, \rho(x_\sigma) \, e^{-\tilde{y}_\lambda x_\lambda}\,,
\ee
where $\rho(x_\sigma)$ is the density of states with mass or angular momenta $x_\sigma$. The mean value of these variables in the ensemble is given by
\be
\langle x_\mu \rangle = - \der{\ln Z}{\tilde{y}_\mu}\,.
\ee
If the partition function is well-defined, the statistical variance of a linear combination of the $x_\mu$ variables must be positive:
\be
\langle (c_\mu x_\mu - \langle c_\nu x_\nu \rangle)^2 \rangle = c_\mu c_\nu (\langle x_\mu x_\nu \rangle - \langle x_\mu \rangle \langle x_\nu \rangle) > 0\,,
\ee
where $c_\mu$ are real coefficients, one of which, at least, is non-zero. This clearly implies that the matrix
\be
\label{goodpartfunc}
\langle x_\mu x_\nu \rangle - \langle x_\mu \rangle \langle x_\nu \rangle = {\dderf{\ln Z}{\tilde{y}_\mu}{\tilde{y}_\nu}} = 
{\dderf{(-\beta G)}{\tilde{y}_\mu}{\tilde{y}_\nu}} = {\dderf{W}{\tilde{y}_\mu}{\tilde{y}_\nu}}
\ee
is positive definite, which is simply the statement \eqref{localstabW}. We will introduce in Chapter~\ref{cha:negmodes} the gravitational partition function of Euclidean quantum gravity, and confirm that it is ill-defined when local thermodynamic stability fails.

Consider now that the fluctuations are restricted to $\Delta J_i=0$. The conditions are relaxed so that $C_J >0$ is enough to ensure stability for exchanges $\Delta M\neq 0$ with the reservoir. Notice, for instance, that the matrices in \eqref{localstabS} and \eqref{localstabM} give the appropriate condition if restricted to the $\{0,0\}$ component:
\be
-\lp {\dder{S}{M}} \rp_J = \beta^3 \, \lp {\dder{M}{S}} \rp_J = \beta^2 \, C_J^{-1}\,.
\ee
The canonical partition function being well-defined requires, analogously to \eqref{goodpartfunc},
\be
\langle M^2 \rangle - \langle M \rangle^2 = {\dder{(-\beta F(T,J_i))}{\beta}}= T^2 C_J >0 \,.
\ee

What about the local thermodynamic stability of a black hole as an isolated system, i.e. in the micro-canonical ensemble? The common derivation of the stability condition does not make reference to a reservoir, which we assumed above, but only to the \emph{internal} equilibrium of the system, where there is no net variation of mass and angular momentum. However, those systems are extensive. Suppose that a system, which we take to be in internal equilibrium in the phase $S(M,J_i)$, is extensive. Then it can be divided into two separate subsystems with equivalent properties, each with entropy $S(M/2,J_i/2)=S(M,J_i)/2$. The suppression of fluctuations within the total system requires
\bea
S(M+\Delta M,J_i+\Delta J_i) + S(M-\Delta M,J_i-\Delta J_i) < 2 \,S(M,J_i)\,,
\eea
which for small fluctuations implies \eqref{localstabS}, without reference to a reservoir.

This fails for gravitational systems because gravity is a long-range interaction. We cannot assign a mass to different parts of the spacetime since the energy/mass is non-local. It is well known that the Schwarzschild solution is classically stable (at the linear level), and yet its specific heat is negative\footnote{Negative specific heats are common in self-gravitating systems \cite{LyndenBell:1998fr}, e.g. a virialised self-gravitating ideal gas with $N_p$ particles in four dimensions has energy $E=E_\text{kin}+E_\text{pot}=-E_\text{kin} =-\frac{3}{2}N_p T$.}: $C=-8\pi M^2$ since $T=1/8\pi M$. In fact, we will show in Chapter \ref{cha:classthermo}, based on \cite{Dias:2010eu}, that \emph{all} asymptotically flat vacuum black holes are unstable according to the condition of local thermodynamic stability.

For extensive systems, the thermodynamic limit gives a direct connection between classical and thermodynamic stability.\footnote{This is the gas limit $U/V \to u(N_p/V,T)$ for large volume $V$ and number of particles $N_p$, where $U$ is an extensive quantity, say the energy. The statistical variance in an ensemble becomes negligible in the thermodynamic limit making the ensembles equivalent, e.g. for the energy in the canonical or the grand-canonical ensembles: $(\langle E^2 \rangle -\langle E \rangle^2 )/\langle E \rangle^2 \to 0$. A quantum statistical system then behaves classically for quasi-stationary long-wavelength processes.} This clearly fails for black holes, as shown by the Schwarzschild case. However, there is also a well-known relation, the Gubser-Mitra conjecture \cite{Gubser:2000ec,Gubser:2000mm}, which states that \emph{black branes with a non-compact translational symmetry are classically stable if and only if they are locally thermodynamically stable}. The reason for this is discussed in Chapter~\ref{cha:classthermo} for vacuum black branes of the type \eqref{blackbrane}. However, we can readily notice that these branes are extensive along the $\vec{z}$ directions, since we may take a partition of the spacetime by periodically identifying each of the $\vec{z}$ coordinates, $z=z+L$. The thermodynamic limit in the $\vec{z}$ directions is exact because $U/L^N$ (where $U$ can be the entropy, the mass, the angular momenta or the charge) is independent of $L$.

How can the internal equilibrium conditions be derived for black \emph{holes}? The works \cite{Dias:2009ex,Dias:2009iu}, on which Chapters~\ref{cha:classthermo}--\ref{cha:MPequal} are based, give a partial answer to this question using the gravitational partition function of Euclidean quantum gravity. New pathologies in the partition function correspond to instabilities which are also classical. They correspond to the failure of \eqref{goodpartfunc} being positive definite, not for fluctuations of the mass and angular momenta, but for additional degrees of freedom that are not captured by the usual thermodynamic description.

\begin{subappendices}
\section{Appendix: Equivalent statements of local stability \label{app:localstab}}

In this Appendix, we will connect some different ways of stating the condition for local thermodynamic stability. Let us start by showing the equivalence of the statements \eqref{localstabS}, \eqref{localstabG}, \eqref{localstabM} and \eqref{localstabW}. We denote
\be
x_\mu=(M,J_i)\,, \quad y_\mu=(T,\Omega_i)\,, \quad \tilde{x}_\mu=(S,J_i)\,, \quad \tilde{y}_\mu=( \beta,-\beta \Omega_i ) \,.
\ee
The first law of thermodynamics \eqref{1stlaw} can be written as
\be
dS= \tilde{y}_\mu dx_\mu \,, \qquad \text{or} \qquad dG= -\tilde{x}_\mu dy_\mu\,.
\ee
The Hessian matrices are
\bea
\label{Gmat}
{\mathbf S}_{\mu\nu} \equiv {\dderf{S(x_\lambda)}{x_\mu}{x_\nu}} = \der{\tilde{y}_\nu}{x_\mu} = \der{y_\sigma}{x_\mu} \der{\tilde{y}_\nu}{y_\sigma}
= (\mathbf{\pd_x y} \cdot \mathbf{\pd_y \tilde{y}})_{\mu\nu} \,, \phantom{()^{-1}--.} \nonumber \\
{\mathbf G}_{\mu\nu} \equiv {\dderf{G(y_\lambda)}{y_\mu}{y_\nu}} = -\der{\tilde{x}_\nu}{y_\mu} = -\der{x_\sigma}{y_\mu} \der{\tilde{x}_\nu}{x_\sigma}
= -( (\mathbf{\pd_x y})^{-1} \cdot \mathbf{\pd_x \tilde{x}})_{\mu\nu} \,,
\eea
which implies that
\be
{\mathbf S} = - (\mathbf{\pd_x \tilde{x}}) \cdot {\mathbf G}^{-1} \cdot (\mathbf{\pd_y \tilde{y}})\,.
\ee
Since
\be
{\mathbf U}_{\mu\nu} \equiv (\mathbf{\pd_y \tilde{y}})_{\mu \nu} = -\beta \, \left(
\begin{array}{cc}
\beta & - \beta \Omega_j \\
0 & \delta_{ij}
\end{array}
\right) =- \beta \, (\mathbf{\pd_x \tilde{x}})_{\nu \mu} \,,
\ee
we find that
\be
\label{SUGU}
{\mathbf S} = T \; {\mathbf U}^{\text{T}} \cdot {\mathbf G}^{-1} \cdot {\mathbf U} \,.
\ee
Now, the matrix $(\mathbf{-S})$ is positive definite -- therefore satisfying the statement \eqref{localstabS} -- if and only if
\be
\label{localstabSdef}
\forall \vec{v} \neq \vec{0}\, \qquad \vec{v}^{\text{T}} \cdot (\mathbf{-S}) \cdot \vec{v} >0 \,.
\ee
From Eq.~\eqref{SUGU},
\be
\vec{v}^{\text{T}} \cdot (\mathbf{-S}) \cdot \vec{v} = T \, ({\mathbf U} \vec{v})^{\text{T}} \cdot (\mathbf{-G})^{-1} \cdot ({\mathbf U} \vec{v})\,.
\ee
Since $|\text{det}({\mathbf U})|=1/T$, we conclude that \eqref{localstabSdef} is equivalent to
\be
\label{quoteforproof}
\forall \vec{v} \neq \vec{0}\, \qquad \vec{v}^{\text{T}} \cdot (\mathbf{-G})^{-1} \cdot \vec{v} >0 \,,
\ee
so that $(\mathbf{-G})^{-1}$ is positive definite. Its inverse must also be positive definite, which is the statement \eqref{localstabG}. We can easily see that the statements \eqref{localstabM} and \eqref{localstabW} are also equivalent: since
\bea
dM=y_\mu d\tilde{x}_\mu \,, \qquad dW= -x_\mu d\tilde{y}_\mu\,,
\eea
we have
\bea
({\mathbf M})_{\mu\nu} \equiv {\dderf{M(\tilde{x}_\lambda)}{\tilde{x}_\mu}{\tilde{x}_\nu}} = \der{y_\nu}{\tilde{x}_\mu} = (- \mathbf{G^{-1}})_{\mu\nu}\,,
 \phantom{i} \nonumber \\
({\mathbf W})_{\mu\nu} \equiv {\dderf{W(\tilde{y}_\lambda)}{\tilde{y}_\mu}{\tilde{y}_\nu}} = -\der{x_\nu}{\tilde{y}_\mu} = (- \mathbf{S^{-1}})_{\mu\nu} \,,
\eea
from \eqref{Gmat}.

Let us now express these statements in terms of the usual linear response functions. Recall the standard linear algebra result:\\
---------------------------------------\\
\noindent {\bf Theorem}: Let $L$ be a symmetric real matrix denoted by 
\be
{L}_{\mu\nu} = \left(  \begin{array}{cc}
A & B_j \\
B_i & C_{ij}
\end{array}
\right)\,,
\ee
where the matrix $C_{ij}$ is invertible. Then (i) the following determinant identity holds:
\be
\label{detidentity}
\mathrm{det}({L}) = \big( A - (C^{-1})_{kl} B_k B_l \big) \;\mathrm{det}(C_{ij})\,;
\ee
and (ii) $L$ is positive definite if and only if
\be
A - (C^{-1})_{kl} B_k B_l >0  \qquad \text{and} \qquad C_{ij} \quad \text{is positive definite}\,.
\ee
\phantom{aa} \\
\noindent {\bf Proof}: Notice that $L=V^{\text{T}}\, \tilde{L}\, V$, where
\be
{\tilde{L}}_{\mu\nu} = \left(  \begin{array}{cc}
A - (C^{-1})_{kl} B_k B_l & 0 \\
0 & C_{ij}
\end{array}
\right)
\quad \textrm{and} \quad
V_{\mu\nu} = \left(  \begin{array}{cc}
1 & 0 \\
(C^{-1})_{ik} B_k & \delta_{ij}
\end{array}
\right)\,.
\ee
The statement (i) holds since $\mathrm{det}({V})=1$, and the right-hand side of \eqref{detidentity} is $\mathrm{det}({\tilde{L}})$. The statement (ii) follows from steps analogous to Eqs.~\eqref{localstabSdef}-\eqref{quoteforproof}, showing that $L$ is positive definite if and only if $\tilde{L}$ is positive definite.\\
---------------------------------------\\
\noindent If we apply this theorem to the matrix \eqref{localstabGmatrix}, for which
\be
C_{ij} = \epsilon_{ij}\, \qquad \text{and} \qquad A - (C^{-1})_{kl} B_k B_l = \beta \big( C_\Omega - (\epsilon^{-1})_{kl} \,\eta_{k} \,\eta_{l} \big)= \beta \, C_J\,,
\ee
we see that the conditions \eqref{localstabusual} are equivalent to \eqref{localstabG}.

To conclude, notice that these conditions can also be obtained using the Helmholtz free energy $F(T,J_i)$. Local stability holds if $F$ is convex with respect to the temperature and concave with respect to the angular momenta:
\be
\lp {\dder{F}{T}} \rp_J = - \beta\, C_J\,, \qquad \lp {\dderf{F}{J_i}{J_j}} \rp_T = (\epsilon^{-1})_{ij} \,.
\ee 

\end{subappendices}

%% file: PFtheory.tex
\chapter{Thermodynamic negative modes} \label{cha:negmodes}

The thermodynamic description of black holes reviewed earlier holds at the semiclassical level. Quantum effects are incorporated by using thermal quantum field theory on a classical spacetime background. Therefore, a semiclassical effective theory of quantum gravity -- and not necessarily a complete theory -- should be able to identify the low energy degrees of freedom of thermodynamic instabilities. This is indeed the case for Euclidean quantum gravity.

We saw in the last Chapter that a local thermodynamic instability must be signalled by a pathology in the partition function. For Euclidean quantum gravity, the pathology consists of a divergent mode in the one-loop quantum corrections of the gravitational partition function. It is commonly referred to as a \emph{negative mode}, since it has the wrong sign for convergence.

\section{Introduction}

Shortly after the advent of black hole thermodynamics, the Euclidean path integral methods of quantum field theory at finite temperature were extended to semiclassical quantum gravity \cite{Hartle:1976tp,Gibbons:1976pt}. Gibbons and Hawking \cite{Gibbons:1976ue} proposed a construction for the partition functions of black holes. These were defined as path integrals with given boundary conditions, which correspond to fixing the temperature, through the periodicity in imaginary time, and possibly other quantities such as angular velocities. A Euclideanised black hole solution is seen as a saddle point of the path integral -- an \emph{instanton} -- its action being related to the thermodynamic free energy.

The semiclassical approach to the path integral allows for more than that. It is possible to go beyond the instanton approximation, corresponding to the classical black hole, and analyse small thermal (quantum) fluctuations around it. This was made possible after a better understanding of the path integral for gravitational perturbations, through the works of Gibbons, Hawking and Perry \cite{Gibbons:1978ac,Gibbons:1978ji}. They found that conformal perturbations of the metric, which always decrease the Euclidean action and seem to render the path integral divergent, are an unphysical artifact that can be eliminated by choosing a suitable integration contour and by applying a standard gauge-fixing procedure.

The first application of those methods was the work of Gross, Perry and Yaffe \cite{Gross:1982cv}. They found that the Schwarzschild instanton possesses a non-conformal radial negative mode in the path integral for perturbations, which renders the partition function ill-defined. This was expected because the Hawking temperature formula for the Schwarzschild black hole, $T=1/8 \pi M$, corresponds to a negative specific heat, so that there is no local thermodynamic stability. The authors further interpreted the instability as the possibility of spontaneous nucleation of black holes in hot flat space.\footnote{In de Sitter space, the cosmological horizon gives rise to the Gibbons-Hawking temperature \cite{Gibbons:1977mu}, and a black hole in thermal equilibrium must match this temperature. Ginsparg and Perry found that the corresponding instanton possesses a negative mode \cite{Ginsparg:1982rs}.} The physics of black hole nucleation was clarified when York \cite{York:1986it} considered the partition function with boundary conditions at finite radius. Such a cavity, with fixed temperature on the wall, allows for two black hole solutions. The smaller radius solution is unstable and tends to the usual Schwarzschild case when the radius is taken to infinity. The larger radius solution is stable and its nucleation is thermodynamically allowed, since its free energy is inferior to that of the hot flat space in the cavity.

A non-trivial test of the correspondence between local thermodynamical stability and a well-defined gravitational partition function was given by Prestidge \cite{Prestidge:1999uq}. He analysed the Schwarzschild-AdS instanton, soon after the AdS/CFT correspondence was proposed, and found numerically that the negative mode disappears when the specific heat of the black hole becomes positive. The curvature of AdS simulates the finite cavity in the asymptotically flat case. The correspondence was put on a firmer footing when Reall \cite{Reall:2001ag} showed, for a certain class of black holes, that a negative specific heat implies the existence of a negative mode. The proof of the converse result, however, remained elusive. Refs. \cite{Dias:2009iu,Dias:2010eu}, on which Chapters~\ref{cha:classthermo}--\ref{cha:MPequal} are based, clarified this question by showing that negative modes could arise which were not connected with the standard conditions for local thermodynamic stability.

Work on Euclidean negative modes of black holes includes also higher-dimensional solutions \cite{Kol:2006ga,Asnin:2007rw}, the Taub-NUT and Taub-bolt instantons \cite{Young:1983dn} also with cosmological constant \cite{Warnick:2006ih}, and connections to Ricci-flow \cite{Headrick:2006ti,Holzegel:2007zz}. We shall review in Chapter~\ref{cha:classthermo} the correspondence between the classical stability of black strings or branes and the existence of thermodynamic negative modes.

This Chapter is organised as follows. In Section~\ref{sec:comp}, we review the partition function of Euclidean quantum gravity, paying special attention to the conformal factor problem and the relevance of traceless-transverse perturbation modes. In Section~\ref{sec:negmodestab}, we connect the existence of local thermodynamic instabilities to negative modes of the action.

\section{\label{sec:comp}The gravitational partition function}

\subsection{\label{subsec:EQG}Euclidean quantum gravity}

The path integral of Euclidean quantum gravity,
\be
\label{pathintegral}
Z = \int d[g] e^{-I[g]}\,,
\ee
is a sum over imaginary time manifolds whose boundaries match a prescribed geometry, which is the argument of the gravitational partition function. The periodicity of the imaginary time coordinate on the boundary is the inverse temperature $\beta=1/T$, as is common in thermal field theory, but the periodicity will now also include the rotation angles, fixing the angular velocities $\Omega_i$. The action for Euclidean metrics is
\be
\label{action}
I[g] = -\frac{1}{16 \pi} \int_{\mathcal M} d^D x \sqrt{g}\, (R-2 \Lambda) -\frac{1}{8 \pi} \int_{\partial {\mathcal M}} d^{D-1} x \sqrt{g^{(D-1)}}\, K - I_0\,.
\ee
The first term is the usual Einstein-Hilbert action and the second is the York-Gibbons-Hawking boundary term \cite{York:1972sj,Gibbons:1976ue}, where $K$ is the trace of the extrinsic curvature on $\partial \mathcal M$. This term is required for non-compact manifolds $\mathcal M$, such as the ones we will study, in order that the boundary condition on $\partial \mathcal M$ is a fixed induced metric, and not fixed derivatives of the metric normal to $\partial \mathcal M$. The inclusion of such a boundary term in the action specifies the thermodynamic ensemble, in this case the grand-canonical (fixed temperature $T$ and angular velocities $\Omega_i$).

The term $I_0$ can depend only on $g^{(D-1)}_{ab}$, the induced metric on $\partial \mathcal M$, and not on the bulk metric $g_{ab}$, so that it can be absorbed into the measure of the path integral. However, since we are interested in the partition functions of black holes, it is convenient to choose it so that $I=0$ for the background spacetime that the black hole solution approaches asymptotically. For asymptotically flat black holes \cite{Gibbons:1976ue}, the Einstein-Hilbert term is zero and the action becomes
\be
-\frac{1}{8 \pi} \int_{\partial {\mathcal M}} d^{D-1} x \sqrt{g^{(D-1)}}\, ( K - K_0 )\,,
\ee
where $K_0$ is the trace of the extrinsic curvature of the flat spacetime matching the black hole metric on the boundary $\partial \mathcal M$ at infinity. This subtraction renders the action of the black hole finite. For asymptotically AdS black holes \cite{Hawking:1982dh,Gibbons:2004ai}, the boundary terms cancel when the background subtraction is performed, but the bulk volume integral diverges and requires an analogous subtraction that sets the action of AdS space to zero. See \cite{Henningson:1998gx,Balasubramanian:1999re,Kraus:1999di,deHaro:2000xn,Olea:2005gb,Olea:2006vd} for the AdS/CFT interpretation of $I_0$ as a counterterm in the dual conformal field theory. In fact, that prescription for regularising the gravitational action is preferable since it is not possible in general to embed an arbitrary boundary geometry in the reference spacetime, as the subtraction method requires.

It should be emphasised that the gravitational path integral (\ref{pathintegral}) is known to be non-renormalisable and is considered here as a low energy (and low-curvature) effective theory. A different issue is that the action (\ref{action}) can be made arbitrarily negative so that the path integral appears to be always divergent. As we shall see, this problem can be addressed in the semiclassical approximation, where the path integral is dealt with by saddle-point methods. We consider a saddle-point $g_{ab}$, i.e. a non-singular solution of the equations of motion,
\be
\label{rab}
R_{ab}= \frac{2 \Lambda}{D-2}\, g_{ab}\,,
\ee
usually referred to as a gravitational instanton. While the boundary conditions in the partition function may admit the existence of several instantons, which are different phases of the system, we are here insterested here in the local stability of a single black hole phase. A black hole instanton is defined by the analytic continuation $t=-\ii \tau$ of the Lorentzian solution. Regularity at the bolt (instanton horizon) requires the periodic identifications of imaginary time and rotation angles: $(\tau,\phi_i) \sim (\tau,\phi_i + 2\pi) \sim (\tau + \beta, \phi_i - \ii\, \Omega_i \beta)$. Such an identification of the rotation angles, which fixes the angular velocities $\Omega_i$, leads to an instanton metric for rotating black holes which is not Euclidean, having complex components.

To go beyond the leading order instanton contribution, we treat as a quantum field the normalisable perturbations $h_{ab}$ about the saddle-point, $g_{ab} \to g_{ab} + h_{ab}$. This leads to a perturbative expansion of the action,
\be
I[g+h] = I[g] + I_2[h;g] + {\mathcal O}(h^3)\,.
\ee
The first order action $I_1$ vanishes since $g_{ab}$ obeys the equations of motion, while the second order action $I_2$, which gives the one-loop correction, is the action for the quantum field $h_{ab}$ on the background geometry $g_{ab}$. The effective field theory requires that the background geometry $g_{ab}$ has a curvature far below the Planck scale. The partition function is
\be
\label{1loopPI}
Z_\mathrm{1-loop} = e^{-I[g]} \int d[h] (\mathrm{G.F.})\,  e^{-I_2[h;g]}\,,
\ee
where $(\mathrm{G.F.})$ denotes all contributions induced by fixing the gauge in the path integral. The second order action is given by
\be
\label{actioninitial}
I_2[h;g] = - \frac{1}{16 \pi} \int d^D x \sqrt{g}\, \left[ - \frac{1}{4} \bar{h} \cdot Gh + \frac{1}{2} (\delta \bar{h})^2  \right]\,,
\ee
where $\cdot$ denotes the metric contraction of tensors. We have defined\footnote{The operator $G$ is related to the Lichnerowicz operator $\Delta_L$ defined in \eqref{deflichn} by $G= \Delta_L - 4 \Lambda / (D-2)$.}
\be
\bar{h}_{ab} = h_{ab} - \frac{1}{2} g_{ab} h^c_{\phantom{i}c}
\ee
and
\be
(G h)_{ab} = - \nabla^c \nabla_c h_{ab} -2 R_{a\phantom{c}b}^{\phantom{a}c\phantom{b}d} h_{cd}\,.
\ee
We have also introduced the following operations on tensors $T$:
\begin{subequations}
\be
(\delta T)_{b \dots c} = - \nabla^a T_{ab \dots c}\,,
\ee
\be
(\alpha T)_{ab \dots c} = \nabla_{(a} T_{b \dots c)}\,,
\ee
\end{subequations}
following \cite{Gibbons:1978ji}.

\subsection{\label{subsec:conffactor}The conformal factor problem and physical instabilities}

We mentioned that the instantons of rotating black holes are not real Euclidean geometries. Although the action $I[g]$ is real, difficulties arise when dealing with quantum corrections. We will ignore these subtleties for now and leave them for the next Chapter, where we consider Kerr-AdS black holes. In the following, we assume the instanton to be real Euclidean.

Consider the action for a conformal normalisable perturbation $h_{ab} = \varphi g_{ab}$,
\be
\label{conf}
I_2[\varphi g;g] = -\frac{(D-1)(D-2)}{64 \pi} \int d^D x \sqrt{g}\, (\pd \varphi)^2 \,.
\ee
This action is always negative, so that the path integral in \eqref{1loopPI} seems irremediably ill-defined. This is the perturbative manifestation of the fact that, if the path integral is taken without further prescription, there are geometries for which the action \eqref{action} can be made arbitrarily negative.

This challenge, known as the conformal factor problem, was addressed by Gibbons, Hawking and Perry \cite{Gibbons:1978ac}, who proposed that the integration contour over the conformal direction in the space of metrics should be imaginary rather than real.\footnote{This prescription is very suggestive because it corresponds to making the timelike direction in the Wheeler-DeWitt metric into a spacelike one, becoming a positive definite metric on the space of spacetime metrics. It is thus analogous to the imaginary time prescription in common Euclidean path integrals.} We follow here the more detailed procedure of \cite{Gibbons:1978ji}, straightforwardly extended to higher dimensions. We will decompose the second order action, applying a standard gauge fixing procedure, and show that the conformal divergent modes are unphysical and do not contribute to the one-loop partition function.

The second order action $I_2[h;g]$ is invariant for the diffeomorphism transformations
\be
\label{hgauge}
h_{ab} \to h_{ab} + \nabla_{a} V_{b} + \nabla_{b} V_{a}  = (h + 2 \alpha V)_{ab}\,.
\ee
Following the Feynman-DeWitt-Faddeev-Popov gauge fixing method,
\be
\label{fpgf}
(\mathrm{G.F.}) = (\mathrm{det}\, C) \,\delta(C_a[h] - w_a)\,.
\ee
We consider the linear class of gauges
\be
\label{wa}
C_b[h]= \nabla^a \left( h_{ab}- \frac{1}{\beta} g_{ab} h^c_{\phantom{i}c} \right)\,,
\ee
where $\beta$ is an arbitrary constant, so that the Fadeev-Popov determinant $(\mathrm{det}\, C)$ is given by the spectrum of the operator
\be
(C V)_a = - \nabla^b \nabla_b V_a - R_{ab} V^b +\left( \frac{2}{\beta} -1 \right) \nabla_a \nabla_b V^b\,.
\ee
To study the spectrum, let us consider the Hodge-de Rham decomposition\footnote{Notice that, while the instanton background is not compact for asymptotically flat/AdS black holes, the off-shell perturbations $h_{ab}$ in the path integral should be normalisable.} of the gauge vector $V$ into harmonic (H), exact (E) and coexact (C) parts,
\be
V = V_\mathrm{H} + V_\mathrm{E} + V_\mathrm{C}\,.
\ee
This induces a decomposition of the action of $C$, which we denote by $C_\mathrm{H}$ for harmonic vectors, $C_\mathrm{E}$ for exact vectors and $C_\mathrm{C}$ for coexact vectors.

The harmonic part satisfies $d V_\mathrm{H} = 0$ and $\delta V_\mathrm{H} = 0$. We can check that
\be
\label{spch}
C V_\mathrm{H} = -\frac{4 \Lambda}{D-2} V_\mathrm{H}\,.
\ee
The spectrum is positive for $\Lambda<0$ and zero for $\Lambda=0$, with multiplicity given by the number of linearly independent harmonic vector fields. For $\Lambda > 0$, the background solution satisfying (\ref{rab}) does not allow for harmonic vector fields if assumed to be compact and orientable \cite{Yano}. Thus, the spectrum of $C_\mathrm{H}$ is never negative.

The exact part is such that $V_\mathrm{E} = d \chi$, where $\chi$ is a scalar. We can show that
\be
\label{Ceop}
\mathrm{spec}\;C_\mathrm{E} = \mathrm{spec}\; \left( 2 \left[ \left(\frac{1}{\beta} -1 \right) \Box -\frac{2 \Lambda}{D-2} \right] \right)\,,
\ee
where the operator on the right-hand side acts on scalars, and $\Box$ is the Laplacian. For $\Lambda < 0$, the operator is positive for $\beta>1$, being positive semi-definite for $\Lambda=0$. For $\Lambda > 0$, the Lichnerowicz-Obata theorem tells us that the spectrum of the Laplacian on a compact and orientable manifold satisfying (\ref{rab}) is bounded from above by $- 2D \Lambda (D-1)^{-1} (D-2)^{-1}$, the saturation of the bound corresponding to the sphere \cite{Yano}. This implies that, for $\Lambda > 0$, the spectrum of $C_\mathrm{E}$ is positive for $\beta > D$.

The coexact part is such that $\delta V_\mathrm{C} = 0$. Hence
\be
\label{spcc}
C V_\mathrm{C} = 2 \delta \alpha V_\mathrm{C}
\ee
and the spectrum of $C_\mathrm{C}$ can be shown to be positive semi-definite,
\be
\label{ineqCc}
\int d^D x \sqrt{g}\, [ V_\mathrm{C} \cdot C V_\mathrm{C} ] = 2 \int d^D x \sqrt{g}\, [ \alpha V_\mathrm{C} \cdot \alpha V_\mathrm{C} ] \geq 0\,,
\ee
with equality for coexact Killing vectors.

The Faddeev-Popov determinant contribution to the partition function is then
\be
\label{detC0}
\mathrm{det}\, \tilde{C} \sim (\mathrm{det}\, \tilde{C}_\mathrm{E}) (\mathrm{det}\, \tilde{C}_\mathrm{C})\,,
\ee
the tilde denoting that the zero modes have been projected out. The harmonic contribution is not explicitly considered because, if it exists ($\Lambda<0$), it is a positive factor dependent only on $\Lambda$ and on the dimension of the space of harmonic vector fields, as mentioned above; it will not be relevant to our discussion. The contribution from the exact part is fundamental since it will cancel the divergent conformal modes.

In order to make the results independent of the arbitrary vector $w$ in the gauge fixing (\ref{fpgf}), the 't Hooft method of averaging over gauges is adopted. The arbitrariness is then expressed in terms of a constant $\gamma$ introduced by the weighting factor of the averaging. The final result will be independent of $\gamma$, as required. The unconstrained effective action for the perturbations is given by
\begin{align}
& \phantom{aaaaaaaaaaaa} I_2^\mathrm{eff}[h;g] = I_2[h;g] + \frac{\gamma}{32 \pi}  \int d^D x \sqrt{g}\, C^a[h] C_a[h] = \nonumber \\
- \frac{1}{16 \pi} &  \int d^D x \sqrt{g}\, \left[ - \frac{1}{4} \bar{h} \cdot Gh + \frac{1}{2} (1-\gamma) (\delta \bar{h})^2  +\frac{\gamma}{2} \left( 1- \frac{2}{\beta} \right) \delta \bar{h} \cdot d \hat{h} - \frac{\gamma}{8} \left( 1- \frac{2}{\beta} \right)^2 (d \hat{h})^2  \right],
\end{align}
where we denote $\hat{h} \equiv h^c_{\phantom{i}c}$.

We now decompose the quantum field $h_{ab}$ into a traceless-transverse (TT) part, a traceless-longitudinal (TL) part, built from a vector $\eta$, and a trace (i.e. conformal) part,
\be
h_{ab} = h_{ab}^{TT} + h_{ab}^{TL} + \frac{1}{D} g_{ab} \hat{h}\,,
\ee
with
\be
h_{ab}^{TL} = 2 (\alpha \eta)_{ab} + \frac{2}{D} g_{ab} \delta \eta\,.
\ee

The constant $\beta$, unspecified in the gauge condition (\ref{wa}), can be chosen so that the trace $\hat{h}$ and the longitudinal vector $\eta$ decouple. This requires
\be
\label{beta}
\beta = 2 \left( 1- \frac{D-2}{D} \frac{ \gamma-1 }{ \gamma } \right)^{-1}\,.
\ee
The effective action becomes
\begin{align}
I_2^\mathrm{eff}[h;g] & = - \frac{1}{16 \pi} \int d^D x \sqrt{g}\, \Bigg[ - \frac{1}{4} h^{TT} \cdot G h^{TT} - \alpha \eta \cdot \alpha \Delta_1 \eta - \frac{1}{D} \delta \eta \Box \delta \eta +
\nonumber \\
& + 2 (1-\gamma) \left( \delta \alpha \eta \cdot \delta \alpha \eta + \frac{1}{D^2} \alpha \delta \eta \cdot \alpha \delta \eta - \frac{2}{D} \delta \alpha \eta \cdot \alpha \delta \eta \right) +
\nonumber \\
& +\frac{4}{D-2} \Lambda \left( \alpha \eta \cdot \alpha \eta  - \frac{1}{D} (\delta \eta)^2 \right) +\frac{1}{2} \hat{h} F \hat{h}  \Bigg]\,,
\end{align}
where the operator $F$ is given by
\be
F = - \frac{D-2 }{ 4D} \left( 1+ \frac{D-2}{ D} \frac{\gamma-1}{\gamma} \right) \Box - \frac{1}{D} \Lambda\,.
\ee
Recalling the choice of $\beta$ (\ref{beta}), we find that the operator on the right-hand side of the expression (\ref{Ceop}) is given by $4D F /(D-2)$. The contribution of the ghosts (\ref{detC0}) can be recast as
\be
\label{detC1}
\mathrm{det}\, \tilde{C} \sim (\mathrm{det}\, \tilde{F}) (\mathrm{det}\, \tilde{C}_\mathrm{C})\,.
\ee

For the vector $\eta$, as we did for $V$ in the ghost part, we perform a Hodge-de Rham decomposition into harmonic, coexact and exact parts,
\be
\eta = \eta_\mathrm{H} + \eta_\mathrm{C} + \eta_\mathrm{E}\,,
\ee
respectively. Using $\eta_\mathrm{E} = d \chi$, the result for the effective action is then
\begin{align}
\label{I2eff}
I_2^{\mathrm{eff}}[h;g] = & -\frac{1}{16\pi} \int d^D x \sqrt{g}\, \Big[ - \frac{1}{4} h^{TT} \cdot G h^{TT} + \frac{1}{2} \hat{h}F\hat{h} + \nonumber \\ & + \frac{4}{D-2} \gamma \Lambda \alpha \eta_\mathrm{H} \cdot \alpha \eta_\mathrm{H} - \gamma \alpha \eta_\mathrm{C} \cdot \alpha C_\mathrm{C} \eta_\mathrm{C} - \frac{4D}{D-2} \gamma {\mathcal D}\chi \cdot {\mathcal D}F\chi \Big]\,,
\end{align}
where we defined the operator
\be
{\mathcal D}_{ab} = \nabla_a \nabla_b - \frac{1}{D} g_{ab} \Box\,.
\ee
Notice that the Hodge-de Rham decomposition of $\eta$ in harmonic, coexact and exact parts gives, for $h^{TL}_{ab}$, a decomposition in $2\alpha \eta_\mathrm{H}$, $2\alpha \eta_\mathrm{C}$ and $2{\mathcal D}\chi$, respectively.

Finally, we can evaluate the Gaussian integrals in the partition function. The prescription of \cite{Gibbons:1978ac} corresponds to taking the imaginary contour for $\hat{h}$. The dependence of the partition function is
\begin{align}
\label{Zfinal}
Z_\mathrm{ 1-loop} & \sim (\mathrm{det}\, \tilde{C}) (\mathrm{det}\,\tilde{G})^{-1/2}  (\mathrm{det}\, \tilde{F})^{-1/2} (\mathrm{det}\, \tilde{C}_\mathrm{C})^{-1/2} (\mathrm{det}\,\tilde{F})^{-1/2} 
\nonumber \\
& \sim (\mathrm{det}\,\tilde{G})^{-1/2} (\mathrm{det}\, \tilde{C}_\mathrm{C})^{1/2}\,.
\end{align}
Again, the tilde on the operators denotes that the zero modes have been projected out.\footnote{Zero modes can be dealt with by the standard collective coordinates method \cite{Jevicki:1976kd}.} The Gaussian integrals are subject to $\zeta$-function regularisation \cite{Gibbons:1978ji}. It is understood that the spectrum of $G$ here is restricted to traceless-transverse normalisable modes.

Let us summarise the treatment of the conformal factor problem. Conformal modes make the action for perturbations negative. However, these modes do not contribute to the path integral. The two factors $(\mathrm{det}\,F)^{-1/2}$ arising from the Gaussian integrals in $\hat{h}$ (taken through an imaginary contour) and $\chi$ cancel with the $\mathrm{det}\,F$ factor arising from the exact part of the Fadeev-Popov determinant. This makes the unphysical character of the divergence obvious, at least in perturbation theory.

The relevant operators are then $C_\mathrm{C}$ and $G$. For a real metric, the operator $C_\mathrm{C}$ is positive semi-definite, as we have shown above. Once its zero modes are projected out, it contributes a positive factor to the final result. Physical instabilities are only possible if there are negative eigenvalues of the operator $G$,
\be
\label{negmodeG}
(G h^{TT})_{ab} = \lambda h^{TT}_{ab}\,,
\ee
in which case the eigenmodes $h^{TT}_{ab}$ are called negative modes. Notice that, for $\lambda \neq 0$, there is no gauge ambiguity \eqref{hgauge} because $G\alpha V=0$.

\section{\label{sec:negmodestab}Negative modes and local thermodynamic stability}

We saw how to identify a physical pathology of the partition function through the spectrum of the operator $G$. The contribution of a negative mode is to lower the action $I[g+h]$. Now, we wish to connect local thermodynamic instabilities to the existence of negative modes. The discussion follows Ref.~\cite{Dias:2009ex}. The idea, taken from Ref.~\cite{Reall:2001ag}, is to consider a family of off-shell geometries for the Euclidean path integral for which one can show that the Euclidean action decreases in a certain direction if the black hole is locally thermodynamically unstable. It then follows that there must exist a Euclidean stationary negative mode.

Let us point out first that the negative modes $h^{TT}_{ab}$ considered before are regular TT fields on the instanton background. This is different from saying that the perturbed metric $g_{ab}+h^{TT}_{ab}$ is regular, since we are demanding that $g_{ab}$ and $h^{TT}_{ab}$ are regular \emph{separately}. Regularity of $g_{ab}$ requires the association of $T$ and $\Omega_i$ to the periodic identification of imaginary time and rotation angles. A field $h^{TT}_{ab}$ on the background $g_{ab}$ must respect the same periodic identifications in order to be regular. Therefore, the off-shell metric $g_{ab}+h^{TT}_{ab}$ leaves $T$ and $\Omega_i$ unchanged to first order.

Consider a black hole instanton $\mathcal{B}(x)$ uniquely specified by parameters $x^\alpha$. Let $T(x)$, $\Omega_i(x)$, $M(x)$, etc. denote the temperature, angular velocities, mass, etc. of this solution. We can construct an off-shell generalization $\mathcal{B}(x,\hat{x})$, specified by parameters $\hat{x}^\alpha$ as follows \cite{Brown:1990di,Brown:1990fk}. Assume that the Killing isometries $\pd_\tau$ and $\pd_{\phi_i}$, where $\phi_i$ are the rotation angles, of $\mathcal{B}(x)$ are preserved. Perform an ADM decomposition of the metric, using the imaginary time coordinate $\tau$. Take the spatial geometry of $\mathcal{B}(x,\hat{x})$ to be the same as that of $\mathcal{B}(\hat{x})$. Now choose the lapse function and shift vector so that (i) $\mathcal{B}(x,\hat{x})$ has the same asymptotics as $\mathcal{B}(x)$; (ii) $\mathcal{B}(x,\hat{x})$ is regular everywhere, in particular at the bolt, subject to the identifications $(\tau,\phi_i) \sim (\tau,\phi_i + 2\pi) \sim (\tau + \beta(x), \phi_i - \ii\, \Omega_i (x) \beta(x))$; (iii) $\mathcal{B}(x,x)=\mathcal{B}(x)$. Note that (ii) implies that $\mathcal{B}(x,\hat{x})$ is a configuration in the Euclidean path integral defined for temperature $T(x)$ and angular velocities $\Omega_i(x)$, for which the saddle point is $\mathcal{B}(x)$. Calculating the Euclidean action of $\mathcal{B}(x,\hat{x})$ using the Hamiltonian formalism gives
\be
I(x,\hat{x})=\beta(x) M(\hat{x}) - S(\hat{x})-\beta(x) \Omega_i(x) J_i(\hat{x}).
\ee
Condition (iii) implies that the geometry with $\hat{x}=x$ satisfies the equations of motion and hence the first derivative of the action with respect to $\hat{x}^\alpha$ must vanish for $\hat{x}^\alpha=x^\alpha$. This is a consequence of the fact that the black hole satisfies the first law of thermodynamics,
\be
\label{firstlaw}
dM=TdS + \Omega_i J_i\,.
\ee
The second derivative of the action, i.e. the Hessian of the action, now reduces to
\be
\left( \frac{ \partial^2 I}{\partial \hat{x}^\alpha \partial \hat{x}^\beta} \right)_{\hat{x}=x} =  \left(\beta  \frac{ \partial^2 M}{\partial x^\alpha \partial x^\beta} - \frac{ \partial^2 S}{\partial x^\alpha \partial x^\beta} - \beta \Omega_i \frac{ \partial^2 J_i}{\partial x^\alpha \partial x^\beta} \right),
\ee
where the right-hand side is evaluated at $x$. If the charges $M$ and $J_i$ uniquely parameterize the solution, we can choose $x^\alpha=(M,J_i)$. We then have
\be
\left( \frac{ \partial^2 I}{\partial \hat{x}^\alpha \partial \hat{x}^\beta} \right)_{y=x} =  -S_{\alpha\beta}(M,J)\,.
\ee
Therefore, if $-S_{\alpha\beta}$ fails to be positive definite for a black hole (i.e. if condition \eqref{localstabS} for local thermodynamic stability fails), then the Euclidean action decreases in some direction and hence the black hole must admit a negative mode.\footnote{Note that we have not constructed the negative mode explicitly by this argument: the linearisation of $\mathcal{B}(x,\hat{x})$ around $\hat{x}=x$ will give a superposition of eigenfunctions of $G$. The point is that, since the action decreases in some direction, this must involve a negative mode.} Given that our off-shell geometries are stationary, this negative mode must also be stationary.\footnote{The operator $G$ commutes with $\pd_\tau$ so one can work with simultaneous eigenfunctions of these operators. Eigenfunctions with different eigenvalues of the latter will be orthogonal.} In general, there must be at least as many negative modes as there are negative eigenvalues of this Hessian.

We shall refer to a negative mode whose existence is predicted by this thermodynamic argument as a {\it thermodynamic negative mode}. We shall see in Chapters~\ref{cha:classthermo}--\ref{cha:MPequal} that there are some negative modes whose existence cannot be predicted in this way. The latter modes found in \cite{Dias:2009iu,Dias:2010eu} are relevant not just for the quantum stability of the black hole but also for its classical stability.

%% file: Kerrads.tex
\chapter{Kerr-AdS negative mode} \label{cha:KerrAdS}

In this Chapter, based on \cite{Monteiro:2009ke}, we will analyse the semiclassical stability of Kerr-AdS black holes. In particular, we will see that a stationary and axisymmetric negative mode exists only when local thermodynamic stability fails. This test of the gravitational partition function is remarkable also because rotating black holes have complex instanton metrics, whose subtleties we briefly discuss.

The lack of symmetry of the Kerr-AdS solution makes the problem much harder to solve than the spherically symmetric cases. We address this by applying a spectral numerical method to solve linear coupled partial differential equations. The method is reviewed in the Appendix at the end of the thesis.

\section{Quasi-Euclidean instantons}

The gravitational partition function is defined as the path integral \eqref{pathintegral}, a sum over geometries with imaginary time $\tau=\ii\, t$. However, while static geometries remain real for this analytical continuation, the same does not hold for stationary non-static geometries. In the canonical formalism, where $\gamma_{ij}$ is the metric on a constant time slice, $N$ is the lapse function and $N^i$ is the shift vector required for rotating spacetimes, we have
\be
\label{ds2i}
d s^2 = N^2 d \tau^2 + \gamma_{ij}(d x^i-\ii N^i d \tau)(d x^j-\ii N^j d \tau).
\ee
Regularity at the bolt (instanton horizon) requires the periodic identifications of imaginary time and rotation angles: $(\tau,\phi_i) \sim (\tau,\phi_i + 2\pi) \sim (\tau + \beta, \phi_i - \ii\, \Omega_i \beta)$. These geometries have been called \emph{quasi-Euclidean}.\footnote{Note that, although the metric is complex, the manifold is real. Real coordinates can be defined by setting $\tilde{\phi_i} = \phi_i + \ii\,\Omega_i \tau$, so the identifications are $(\tau,\tilde{\phi}_i) \sim (\tau,\tilde{\phi}_i + 2\pi) \sim (\tau+\beta,\tilde{\phi}_i)$.} This seems to pose a difficulty for the path integral because we expect physical quantities to be real. Notice also that the procedure applied in the last Chapter to decompose the metric perturbations and deal with the conformal factor problem assumed the instanton to be (real) Euclidean.

Nevertheless, we share the view of \cite{Brown:1990di,Brown:1990fk} that quasi-Euclidean instantons pose no problem of principle. The instanton action is real and gives the physical free energy (divided by the temperature). Notice that, while one might be tempted to take imaginary lapse functions, which would make the line element \eqref{ds2i} real, the resulting geometry would bear no relation to the Lorentzian black hole, e.g. the bolt radius would be different from the event horizon radius and there would be no ergosphere. Furthermore, although this can be done for Myers-Perry black holes \eqref{fullMP} by analytically continuing the rotation parameters $a_i$, it is impossible for black rings because not all conical singularities can be removed in the would-be real instanton \cite{Astefanesei:2005ad,bella}.

The treatment of quantum corrections about a quasi-Euclidean instanton is more subtle. However, Refs.~\cite{Brown:1990di,Brown:1990fk} show that the action is real not just for the instanton, but also for a family of off-shell geometries -- the one used in Section~\ref{sec:negmodestab} to connect negative modes to local thermodynamic stability. The problem of thermodynamic negative modes should then be posed in terms of real quantities.

The procedure leading to the expression (\ref{Zfinal}) for the one-loop quantum corrections assumed a Euclidean instanton, while we now want to consider the quasi-Euclidean case. However, that expression may still hold for an appropriate complex contour of integration in the space of perturbations (and ghosts), specified as usual by the steepest descent method. This is our \emph{assumption}. It would be important to construct such a contour explicitly.

The numerical technique applied here differs from the analytical but perhaps simplistic first approach to the problem in \cite{Monteiro:2009tc}, which could only account for the effect of a single direction in the perturbation space. That single direction was provided by an easily constructed traceless-transverse perturbation, which kept the second order action real but was not an eigenmode. The difficulty with that approach, which may explain the small discrepancy in the final result for the Kerr-AdS negative mode, is that it is not clear if such a direction lies on the steepest descent path, despite the second order action being real. The steepest descent path is here infinite-dimensional, spanned by the normalised eigenmodes of $G$, which we can only determine numerically.

The family of off-shell geometries related to thermodynamic stability in Section~\ref{sec:negmodestab} preserves the Killing isometries $\pd_\tau$ and $\pd_{\phi_i}$, where $\phi_i$ are the rotation angles. Therefore, we will focus on stationary and axisymmetric negative modes \eqref{negmodeG} of the Kerr-AdS instanton. Since this problem is independent of the time and the rotation angle, which are responsible for the instanton metric being non-real, it is equivalent to the Lorentzian problem, and it reduces to a set of explicitly real differential equations.

In the next Section, we review the Kerr-AdS solution and its thermodynamic stability. In Section~\ref{sec:KAdSeigen}, we outline our implementation of the eigenvalue problem for negative modes. The results are presented and discussed in Section~\ref{sec:KAdSresults}.

\section{The Kerr-AdS solution}

\subsection{Instanton geometry}

The instanton geometry is obtained through the analytical continuation of the black hole solution:
\begin{eqnarray}
ds^2 \!\!&=&\!\!
\frac{\Delta(r)}{\Sigma^2(r,\theta)}\left(
d\tau-\ii\, \frac{a}{\Xi} \sin^2\theta \,d\phi \right)^2 +
\frac{\Delta_\theta(\theta) \sin^2\theta}{\Sigma^2(r,\theta)}\left(
\frac{r^2+a^2}{\Xi}d\phi +\ii\,a\,dt\right)^2 \nonumber\\
&&
+\frac{\Sigma^2(r,\theta)}{\Delta(r)}\left(dr+r_0 \delta\chi\,
\sin\theta \,d\theta \right)^2+
\Sigma^2(r,\theta)\,d\theta^2\,,
\label{eq:kerrAdS}
\end{eqnarray}
where $\ell$ is the curvature radius of AdS, $\ell^2 = - 3/\Lambda$, and
\begin{subequations}
\begin{eqnarray}
\Delta (r) &=& (r^2+a^2)(1+r^2\ell^{-2}) -r_0 r, \\
\Sigma (r,\theta)^2 &=& r^2 +a^2 \cos^2\theta, \\
\Delta_\theta (\theta) &=& 1-a^2\ell^{-2} \cos^2{\theta}, \\
\Xi &=& 1-a^2\ell^{-2}.
\end{eqnarray}
\end{subequations}
The bounds on the parameter space of the instanton are given by the extremality condition and by the requirement that $|a|<\ell$, since the limit $|a| \to \ell$, for which the 3-dimensional Einstein universe at infinity rotates at the speed of light, is singular \cite{Hawking:1998kw}. The avoidance of a conical singularity at the bolt, located at $r= r_{_+}$, the largest root of $\Delta$, requires the coordinate identification $(\tau,\phi)= (\tau+ \beta, \phi + \ii \beta (\Omega -a \ell^{-2}))$. Here,
\be
\beta = \frac{4\pi(r_{_+}^2 +a^2)}{r_{_+}(1+a^2\ell^{-2} + 3 r_{_+}^2 \ell^{-2} - a^2r_{_+}^{-2})}
\ee
is the inverse temperature and
\be
\Omega = \frac{a(1+ r_{_+}^2\ell^{-2})}{r_{_+}^2 + a^2}
\ee
is the angular velocity in a reference frame non-rotating at infinity, the quantity that satisfies the first law of thermodynamics \cite{Gibbons:2004ai}. It differs from the angular velocity in the reference frame of \eqref{eq:kerrAdS}, which is $\Omega'=\Omega-a \ell^{-2}$. The mass $M$, the angular momentum $J$ and the Bekenstein-Hawking entropy $S$ are given by
\be
M = \frac{r_0}{2 \,\Xi}\,, \qquad \quad J = \frac{a\, r_0}{2 \,\Xi}\,, \qquad \quad S = \frac{\pi(r_+^2+a^2)}{\Xi} \,.
\ee

\subsection{Thermodynamic stability}

We analyse here the Kerr-AdS black hole not only because of the usual theoretical motivations (AdS/CFT correspondence) but also because there is a phase transition to local thermodynamic stability for large enough black holes in the grand-canonical ensemble. Our purpose is to verify that this criterion for stability matches the one given by the quantum corrections to the gravitational partition function.

Local thermodynamic stability in the grand-canonical ensemble requires that both the specific heat at constant angular momentum $C_J$ and the isothermal moment of inertia $\epsilon$ are positive \eqref{localstabusual}. These are given by
\be
C_J=-\frac{2 \pi \ell^2 \, \Xi^{-1} \, (r_+^2 + a^2)^2  \, [ r_+^2(3 r_+^2+ a^2) + \ell^2 (r_+^2-a^2)]}{r_+^4\ell^4 - 3 r_+^2 \ell^2 -a^2 (9 r_+^6 + 23 r_+^4 \ell^2 + 6 r_+^2 \ell^4) -a^4 (6 r_+^4 + 13 r_+^2 \ell^2 + 3 \ell^4) -a^6 (r_+^2 + \ell^2)}
\ee
and
\be
C_J\;\epsilon =-\frac{\pi (r_+^2+a^2)^3 \, [r_+^2(3 r_+^2+ a^2) + \ell^2 (r_+^2-a^2)]}{r_+ \,\Xi^4\,[r_+^2 \ell^2- 3 r_+^4 +(r_+^2+\ell^2) a^2]} \,.
\ee
An interesting feature of the Kerr-AdS black hole, as opposed to the black ring for instance, is that the specific heat at constant angular velocity $C_\Omega$ is sufficient to describe the stability in the grand-canonical ensemble.
\begin{figure}
\centering{ \includegraphics[width = 6cm]{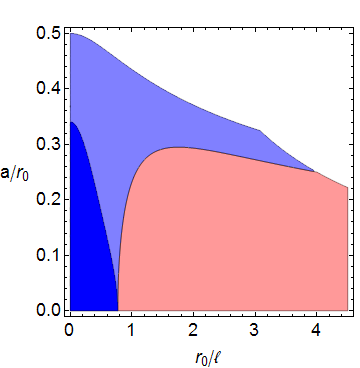} \hspace{1cm} \includegraphics[width = 6cm]{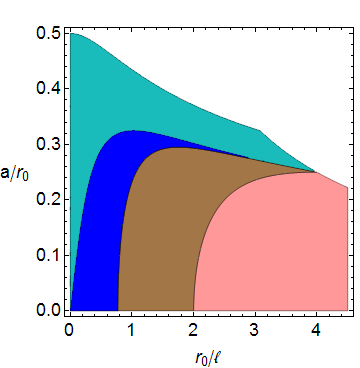}}
\caption{\label{fig:kerradstermo}Phase diagrams of the Kerr-AdS black hole. The parameter space is bounded above by extremality and by the singular limit $a=\ell$. \underline{Left plot}: $C_J<0$, $\epsilon >0$ and $C_\Omega <0$ for the dark blue region; $C_J>0$, $\epsilon <0$ and $C_\Omega <0$ for the light blue region; local thermodynamical stability for the red region ($C_J,\epsilon,C_\Omega >0$). \underline{Right plot}: thermodynamic stability in the red region; global thermodynamic instability ($r_+ < \ell$) in the brown, blue and green regions; local thermodynamic instability ($C_\Omega <0$) in the blue and green regions; classical superradiant instability ($\Omega \ell>1$) in the green region.}
\end{figure}
It is positive when both $C_J$ and $\epsilon$ are positive, and negative when one of them is negative; $C_J$ and $\epsilon$ are never simultaneously negative and complement the region of instability, as we can see in the left plot of Figure~\ref{fig:kerradstermo}. Therefore, the local thermodynamic stability criterium is
\be
\label{localstabkerrads}
\beta C_\Omega= -\frac{8 \pi^2  r_+ \left(r_+^2 + a^2\right)}{\Xi \left(1 -3 r_+^2 \ell^{-2} + a^2 r_+^{-2} + a^2 \ell^{-2} \right)} > 0\,.
\ee
The vanishing of the denominator in the right-hand side identifies the line of critical stability for each value of $\ell$. Across this line, $C_\Omega$ diverges and changes sign. In the static case, this occurs for $r_+=\ell/\sqrt{3}$, or $r_0 \approx 0.77\, \ell$.

In the right plot of Figure~\ref{fig:kerradstermo}, we present a phase diagram including information on thermodynamic stability (local and global) and also on classical instability. For $\Omega \ell>1$, in the light blue region, the black hole suffers from the classical superradiant instability \cite{Hawking:1999dp,Cardoso:2004nk,Cardoso:2004hs}, the only known instability of these black holes in the absence of matter. In the two blue regions, the black hole is locally thermodynamically unstable. In those regions and also in the brown region, which all satisfy $r_+ < \ell$, the black hole is globally thermodynamically unstable because its Gibbs free energy with respect to thermal AdS space is positive,
\be
G=M - T S - \Omega J = \frac{\left(\ell^2-r_+^2\right)\left(r_+^2+ a^2\right)}{4 \,r_+\, \ell^2\, \Xi}\,.
\ee
In the light red region, the black hole is thermodynamically stable. The transition to this region is the Hawking-Page phase transition \cite{Hawking:1982dh}. The three critical curves meet at the point $r_+=r_0/4=a= \ell$.

\section{The eigenvalue problem \label{sec:KAdSeigen}}

\subsection{Ansatz}

We will look at eigenvalues of the operator $G$ in search of stationary axisymmetric negative modes, $G h^{TT} =  \lambda \, h^{TT}$, with $\lambda <0$. Our ansatz for the perturbed metric is given by
\begin{eqnarray}
ds^2 &=&
\frac{\Delta(r)}{\Sigma^2(r,\theta)}\,e^{\delta\nu_0}\left(
d\tau-\ii\, \frac{a}{\Xi} \sin^2\theta \,e^{\delta\omega}\,d\phi \right)^2 \nonumber\\
&&
+ \frac{\Delta_\theta(\theta) \sin^2\theta}{\Sigma^2(r,\theta)}\,e^{\delta\nu_1}\left(
\frac{r^2+a^2}{\Xi}d\phi +\ii\,a\, e^{-\delta\omega}\,dt\right)^2 \nonumber\\
&&
+\frac{\Sigma^2(r,\theta)}{\Delta(r)}\,e^{\delta\mu_0}\left(dr+r_0 \delta\chi\,
\sin\theta \,d\theta \right)^2+
\Sigma^2(r,\theta)\,e^{\delta\mu_1}\,d\theta^2\,,
\label{eq:kerrAdS1}
\end{eqnarray}
where the functions $\delta\nu_0$, $\delta\nu_1$, $\delta\omega$, $\delta\mu_0$, $\delta\mu_1$, $\delta\chi$ are small perturbations which are functions of $(r,\theta)$. The imposition of the TT conditions on the linearised perturbation $h_{ab}$,
\be
h^a_{\phantom{a}a}=0\,, \qquad \nabla^a h_{ab} =0\,,
\ee
gives three constraints, which are solved by explicit relations
\be
\Upsilon = \Upsilon(\;
\pd_r \delta\mu_0,\pd_r \delta\mu_1,\pd_r \delta\chi,\quad
\pd_\theta \delta\mu_0,\pd_\theta \delta\mu_1,\pd_\theta \delta\chi,\quad
\delta\mu_0,\delta\mu_1,\delta\chi\;),
\ee
where $\Upsilon = \{\delta\nu_0,\delta\nu_1,\delta\omega\}$. Substituting these three relations in the metric perturbation, we get the general expression for $h^{TT}_{ab}$ respecting the isometries of the instanton.\footnote{The inclusion of the terms $h_{\tau r}$, $h_{\tau \theta}$, $h_{\phi r}$, $h_{\phi \theta}$ in the ansatz \eqref{eq:kerrAdS1} leads to the conclusion that these must vanish when imposing the TT conditions. Hence we are working with the most general ansatz for stationary and axisymmetric perturbations.} Notice that this expression includes first derivatives of the functions $\delta\mu_0$, $\delta\mu_1$, $\delta\chi$ in the components $h_{\tau\tau}^{TT}$, $h_{\tau\phi}^{TT}$, $h_{\phi\phi}^{TT}$.

Now we look at the eigenvalue problem $E = G h^{TT} - \lambda \, h^{TT}=0$. There are six equations to be solved, corresponding to the non-zero components $E_{\tau\tau}$, $E_{\tau\phi}$, $E_{\phi\phi}$, $E_{rr}$, $E_{r\theta}$, $E_{\theta\theta}$. Since $G$ is a second order differential operator, and $h^{TT}$ already includes first derivatives, the components $E_{\tau\tau}$, $E_{\tau\phi}$, $E_{\phi\phi}$ are third order equations. The components $E_{rr}$, $E_{r\theta}$, $E_{\theta\theta}$ are second order only, and they imply (by taking their derivatives) that the third order components are automatically satisfied, as expected. Hence we have the three coupled partial differential equations $E_{rr}$, $E_{r\theta}$, $E_{\theta\theta}$ for the three unknown functions $\delta\mu_0$, $\delta\mu_1$, $\delta\chi$.

One thing we should emphasise is that the subtlety of having a quasi-Euclidean metric, as opposed to Euclidean, has vanished now. The equations obtained are real equations for real perturbation functions. Unfortunately, it will not be possible to present the three equations explicitly here. Even in the asymptotically flat limit $\ell \to \infty$, they are too cumbersome.

Let us first consider the boundary conditions for the perturbations. The region of integration for the differential equations is an infinite strip $r_+ \leq r < \infty$, $0 \leq \theta \leq \pi$. The perturbations must vanish at infinity, $r \to \infty$. The boundary conditions on the horizon are obtained by considering a regular basis. This procedure is described explicitly in Chapter~\ref{cha:MPsingle} and we will not repeat it here. We find that $\delta\chi(r,\theta) \propto (r-r_+)$ near the horizon, while $\delta\mu_0 (r_+,\theta)$ and $\delta\mu_1 (r_+,\theta)$ must simply be finite. Regularity at the poles $\theta = 0 ,\pi$ implies only that the functions are finite there.

\subsection{Implementation}

It is convenient to make rescalings such that only adimensional quantities are involved in the problem,
\begin{subequations}
\begin{eqnarray}
y = \frac{r}{r_+} -1\,, &&  y_\ast = \frac{r_+}{r_0}\,, \\
\ell_\ast = \frac{\ell}{r_0}\,, &&  \lambda_\ast = \lambda \, r_0^2\,,
\end{eqnarray}
and notice that
\be
\frac{y_\ast - y_\ast^2 - y_\ast^4 \ell_\ast^{-2}}{1+ y_\ast^2 \ell_\ast^{-2}} = \left( \frac{a}{r_0} \right)^2\,. 
\ee
\end{subequations}
We will use the coordinate
\be
x=\cos \theta\,,
\ee
so that the numerical integration will be performed on a rectangle $0 \leq y \leq Y$, $-1 \leq x \leq 1$, where $Y \gg y_\ast$ must be sufficiently large. Let us rescale the perturbation functions too,
\begin{subequations}
\label{eq:KAdSbc}
\begin{eqnarray}
\delta\mu_0(r,\theta) &=& \frac{q_1(y,x)}{y(1-x^2)} , \\
\delta\chi(r,\theta) &=& \frac{q_2(y,x)}{1-x^2} , \\
\delta\mu_1(r,\theta) &=& \frac{q_3(y,x)}{y(1-x^2)} ,
\end{eqnarray}
\end{subequations}
so that the boundary conditions discussed before are simply $q_i=0$ ($i=1,2,3$) along the edges $x=\pm 1$ and $y=0,Y$. It is convenient to combine our previous equations $rr$, $r\theta$ and $\theta\theta$, now in terms of the redefined quantities, into the form
\be
\partial^2_y q_i (y,x) + \ldots + \lambda_\ast f(y,x;y_\ast,\ell_\ast) q_i (y,x) =0,
\ee
which turns out to be possible. Notice that the last term above gives the only dependence of the equations on $\lambda_\ast$. We are now ready to implement the spectral numerical method \cite{Trefethen}, which is briefly described in the Appendix at the end of this thesis.

\section{Results \label{sec:KAdSresults}}

\begin{figure}
\centering
\includegraphics[width = 6 cm]{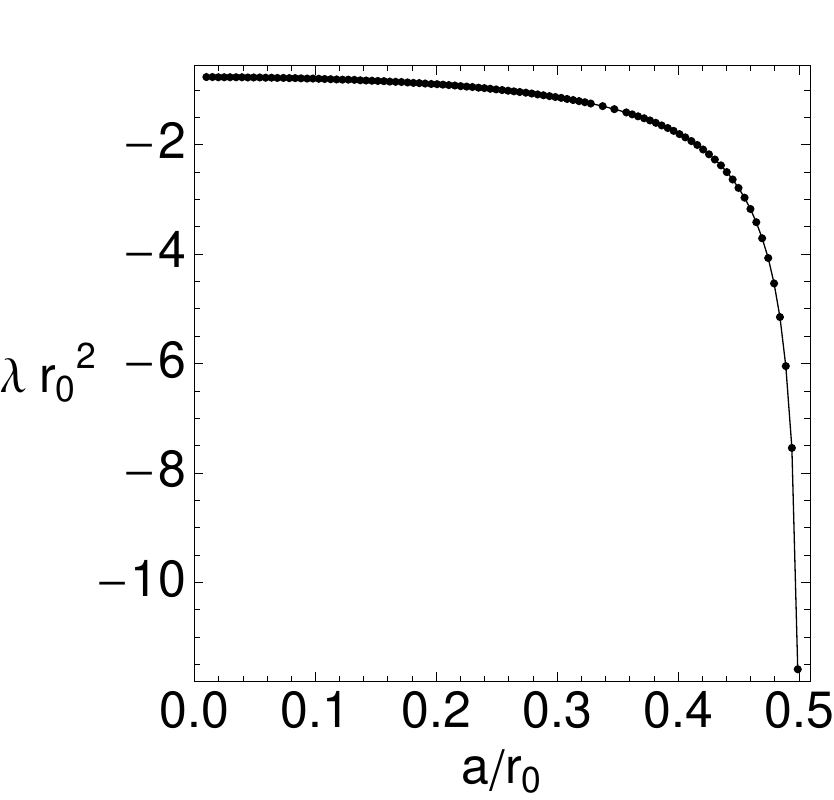}
\caption{\label{fig:kerr}For the Kerr instanton, $\lambda_*$ is negative, decreasing monotonically away from $a=0$ and evaluating to a finite value at extremality $|a| = r_0/2$.}
\end{figure}

Let us first consider the Kerr case, i.e. the limit $\ell \to \infty$. The results are represented in Fig.~\ref{fig:kerr}. We find that there is a single negative mode for $|a| \leq r_0/2$, monotonically increasing in magnitude with the angular momentum. Surprisingly, our probe perturbation method \cite{Monteiro:2009tc} approached the value of the negative eigenvalue now found to within 10\%.
One way to understand the increase in magnitude is to recall the connection between the black hole thermodynamic negative mode and the classical Gregory-Laflamme instability of the respective black string/brane \cite{Reall:2001ag}, which we shall review in greater detail in Chapter~\ref{cha:classthermo}. The threshold wavenumber $k= (\vec{k} \cdot \vec{k})^{1/2}$ for the Gregory-Laflamme instability corresponds to the four dimensional stationary solution of $G h^{TT} = -k^2 \, h^{TT}$, with the appropriate boundary conditions \cite{Gregory:1993vy}. This is exactly the problem addressed here if we identify $\lambda=-k^2$. The fact that we are dealing with a quasi-Euclidean geometry rather than a Lorentzian geometry is irrelevant since time plays no role in the solutions to the perturbation functions defined in (\ref{eq:kerrAdS1}). The curve in Fig.~\ref{fig:kerr} thus implies that the Gregory-Laflamme instability of the Kerr string persists up to extremality. The larger in magnitude is the negative mode, the smaller is the threshold length scale $k^{-1}$ for the instability. We expect on physical grounds that the centrifugal force caused by the rotation will favour the instability of ripples along the string (Ref.~\cite{Caldarelli:2008mv} presents a fluid dual analogy) thus decreasing their threshold length scale and explaining the stronger negative mode of the black hole. See \cite{Kleihaus:2007dg,Dias:2009iu,Dias:2010eu} for analogous results in higher dimensions.

\begin{figure}
\centering
\includegraphics[width = 10 cm]{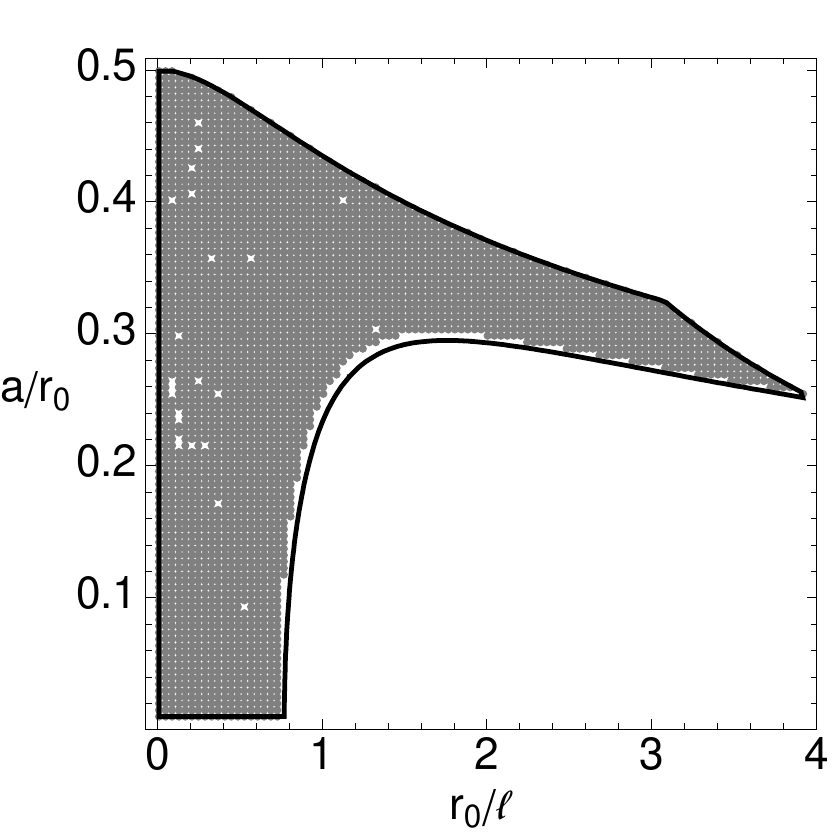}
\caption{\label{fig:kerr_ads}Phase diagram of the Kerr-AdS black hole. The points represent the parameter region where we find a negative mode and the line represents the change of sign of $C_\Omega$, which is negative in the Kerr limit $\ell \to \infty$.}
\end{figure}

We now turn to the Kerr-AdS case. If we look at Fig.~\ref{fig:kerr_ads}, the agreement between the existence of a negative mode and the condition \eqref{localstabkerrads} is striking.
We find a negative mode only when $C_\Omega$ is negative. Unfortunately the numerical method does not allow us to safely zoom in the line of critical stability. Nevertheless, it is clear that the gravitational partition function reproduces the thermodynamics of the system beyond the instanton approximation, even for this non-static black hole with a quasi-Euclidean instanton.

%% file: Charged.tex
\chapter{Reissner-Nordstr\"om negative mode} \label{cha:RN}

In this Chapter, based on \cite{Monteiro:2008wr}, we analyse the problem of thermodynamic negative modes of the Reissner-Nordstr\"om black hole in four dimensions. We find analytically that a negative mode disappears when the specific heat at constant charge becomes positive. The sector of perturbations analysed here is included in the canonical partition function of the magnetically charged black hole. The result obeys the usual rule that the partition function is only well-defined when there is local thermodynamical equilibrium.

We emphasise the challenge of quantising Einstein-Maxwell theory, even as a low energy effective theory, because the conformal factor problem is more intricate than in the vacuum case, which was discussed in Section~\ref{subsec:conffactor}. We circumvent this difficulty by considering a dimensional reduction from five to four dimensions. The method allows us to decouple the divergent gauge volume and treat the metric perturbations sector in a gauge-invariant way.

\section{\label{sec:RNintro}Introduction}

The inclusion of matter presents considerable challenges to the study of black hole negative modes. This is because the standard decomposition of the perturbations \cite{Gibbons:1978ac,Gibbons:1978ji}, reviewed in Section~\ref{subsec:conffactor}, which singles out the unphysical conformal modes, was performed for the Einstein theory only (with or without cosmological constant). In particular, the presence of a background gauge field invalidates the procedure even for metric perturbations only, i.e. when the gauge field perturbations can be consistently set to zero.

One way out to address this problem is suggested by the alternative procedure of Gratton, Lewis and Turok \cite{Gratton:2000fj,Gratton:2001gw} for cosmological instantons. They look for well-behaved gauge-invariant perturbations which allow for a decoupling in the action between the relevant modes and the unphysical ones, which are integrated out. See Kol \cite{Kol:2006ux} for a general and systematic discussion of the same basic idea, which he calls ``power of the action,'' following its initial application to the Schwarzschild black hole and black string \cite{Kol:2006ga}. This corresponds to the procedure we adopt in this work. There is a further complication though. The problem of radial perturbations requires two gauge conditions (as the traceless and transverse conditions in pure gravity) for the metric perturbations. The method is inconsistent because the radial ansatz for the perturbation of the action only has a radial diffeomorphism invariance (one gauge choice), and including time does not allow for the construction of gauge-invariant quantities. The work of Kol \cite{Kol:2006ga} on the Schwarzschild black hole gives a solution to this problem. Considering a lift to one higher dimension, along which there is translational invariance, it is possible to construct gauge-invariant quantities, i.e. perturbations which are invariant for infinitesimal diffeomorphisms along the radial direction and along the extra dimension. If the action for the zero modes (infinite wavelength) along the extra-dimension reproduces the lower-dimensional action for perturbations, then the higher-dimensional action can be decomposed, and the long wavelength limit can be taken when a simple reduced action is available.

Using this Kaluza-Klein method, we are able to study the four-dimensional magnetic Reissner-Nordstr\"om black hole in a sector of perturbations corresponding to the canonical ensemble, \emph{i.e.} fixed charge $Q$. We find that one negative mode still exists if the charge is small compared with the mass $M$, as expected, but disappears for $|Q| \geq \sqrt{3}M/2$, exactly when the specific heat becomes positive. This supports the validity of the canonical partition function of Euclidean quantum gravity \cite{Gibbons:1976ue} even when gauge fields are present.

One might be worried by the fact that such a theory is not renormalisable at one loop, which is indeed the case of Einstein-Maxwell theory \cite{Deser:1974cz}, so that we cannot compute quantum corrections. However, we can take an effective field theory approach \cite{Donoghue:1994dn}, where Einstein-Maxwell theory is regarded as the low energy limit of an underlying fundamental theory. The effective theory is valid up to a cut-off scale, after which ultraviolet completion effects become important. As long as the energies of the fields involved are nowhere near that scale, the perturbative quantisation of the non-renormalisable effective theory is meaningful. It may help to recall that Einstein-Maxwell theory corresponds to the bosonic sector of four-dimensional $\mathcal{N}=2$ supergravity, which can be embedded into string theory.

We mentioned that Einstein-Maxwell theory had not been considered for the perturbative path integral quantisation on black hole backgrounds. Miyamoto and Kudoh \cite{Miyamoto:2006nd,Miyamoto:2007mh} have analysed the classical stability of magnetically charged branes and verified that they are stable when there is local thermodynamic stability. This result is related to the Gubser-Mitra conjecture mentioned in Sections~\ref{sec:localstab} and \ref{subsec:GMconjecture}. The issue we address here is whether the partition function at one loop, defined as a saddle point approximation to a Euclidean path integral, conforms to the usual thermodynamic stability criterion. It is also worth mentioning that Einstein-Maxwell theory in four dimensions cannot be the result of dimensionally reducing Einstein-Maxwell theory in five dimensions, since there is then an extra scalar field. The fact that the Reissner-Nordstr\"om black hole is not straightforwardly related to a string will force us to be careful in our Kaluza-Klein action method, so that we make sure the correct quantum theory is obtained.

This Chapter is organised as follows. In Section~\ref{sec:EM}, we discuss the problems with quantising Einstein-Maxwell theory, stressing the differences with the pure Einstein case, and recall the relation between the boundary conditions of the path integral and the corresponding thermodynamic ensemble. In Section~\ref{sec:KK}, we explain the Kaluza-Klein action method to analyse the second-order action. In Section~\ref{sec:app0}, we describe the application of the method to the magnetic Reissner-Nordstr\"om black hole. We start by presenting an appropriate lift to five dimensions. We then construct the gauge-invariant quantities and obtain the reduced action. We find that the action possesses a negative mode when the specific heat at constant charge is negative. Finally, in Section~\ref{sec:conc}, we present the conclusions.

\section{\label{sec:EM}The Einstein-Maxwell path integral}

We wish to consider gravity coupled to electromagnetism in asymptotically flat spacetimes. The partition function is given by the path integral
\be
\label{pathintegralA}
Z=\int d[g]d[A] e^{-I[g,A]},
\ee
constructed from the Euclidean action
\be
\label{euclidaction}
I[g,A]=-\int_{\mathcal{M}} d^d x \sqrt{g}(R-F_{ab}F^{ab})-2\int_{\partial \mathcal{M}} d^{d-1} x \sqrt{h} (K-K_0),
\ee
twhere the first term corresponds to the usual Einstein-Hilbert action and the second term is the Maxwell action. As before, the third term is the Gibbons-Hawking-York boundary term \cite{Gibbons:1976ue,York:1972sj}, required if the configurations summed over have a prescribed induced metric on $\partial \mathcal{M}$. $K$ represents the trace of the extrinsic curvature on $\partial \mathcal{M}$ and $K_0$ is the trace of the extrinsic curvature of flat spacetime, matching the black hole metric at infinity, necessary to render the on-shell action of asymptotically flat black hole solutions finite. Recall that $F_{ab}$ are the components of the Maxwell field strength, the exterior derivative $F=d A$ of the gauge 1-form potential $A$.

As discussed in Section~\ref{sec:comp}, the purely gravitational action can be made arbitrarily negative for conformal transformations that obey the boundary conditions of the path integral, i.e. by geometries included in the sum. In the absence of electromagnetic sources, this apparent divergence of the functional integration, called the conformal factor problem, is circumvented by choosing an appropriate complex integration contour \cite{Gibbons:1978ac}, at least at the one loop level of the semiclassical quantisation. Furthermore, it was later shown in \cite{Gibbons:1978ji} that an orthogonal decomposition of the metric perturbations into a trace, a longitudinal-traceless and a transverse-traceless part, complemented by an appropriate gauge choice to deal with the diffeomorphism invariance of the action, leads to the complete decoupling of these components in the second-order action. This procedure was detailed in Section~\ref{subsec:conffactor}. It is a key requirement that the gauge choice kills the interaction between the trace and the longitudinal traceless parts. The final result confirms the prescription of \cite{Gibbons:1978ac}, and the decoupled trace can be integrated by choosing a suitable complex contour. Moreover, the scalar parts of the partition function, comprising the contributions from the tracelike (conformal) perturbations and from the scalar parts of the vector modes, both in the longitudinal traceless part of the metric and in the ghost vectors, cancel. This shows that the apparent non-positivity of the action for perturbations, and the consequent divergence of the path integral, are fixed by projecting out this contribution. We would like to address this problem in the presence of electromagnetism.

\subsection{\label{sec:EM_DEC}The second-order action}

In this section, we will perturb the action about a saddle point $(g,A)$, which we define here as a non-singular solution of the equations of motion
\begin{subequations}
\be
R_{ab}-\frac{1}{2}\,R\, g_{ab}=2\left(F_{a}^{\phantom{a}c}F_{bc}-\frac{1}{4}\,g_{ab}\,F^{pq}F_{pq}\right)
\ee
and
\be
\nabla_a F^{ab}=0\,.
\ee
\end{subequations}
Small perturbations $(h_{ab},a_b)$ about this solution,
\be
\begin{array}{c@{\hspace{1 cm}}c@{\hspace{1 cm}}c}
g_{ab}\;\to\;g_{ab}+h_{ab} & \text{and} & A_b\;\to\;A_b + a_b\,,
\end{array}
\ee
are treated as quantum fields living on the saddle point background. We then perturb the action to second order,
\be
I[g+h,A+a]=I[g,A]+I_2[h,a;g,A]+\mathcal{O}(h^3,h^2 a,h a^2, a^3)\,,
\ee
so that the partition function can be approximated by a saddle point functional integral
\be
Z\simeq e^{-I[g,A]}\int d[h]d[a](\mathrm{G.F.})\,e^{-I_2[h,a;g,A]}\equiv e^{-I[g,A]} Z_{(2)}\,,
\ee
where $(\mathrm{G.F.})$ denotes all contributions induced by fixing the gauge. The first-order term is absent because the instanton solution satisfies the equations of motion. The second-order action is given by
\begin{multline}
I_2 [h,a]= \frac{1}{4}\int d^D x \sqrt{g}\Big\{h^{ab}\Big[\Delta_L h_{ab}+2 \nabla_a \nabla^c h_{b c}-2 g_{ab}\nabla^m\nabla^n h_{mn}+g_{ab}\Box \hat{h}+4 F_a^{\phantom{a}m}F_b^{\phantom{b}n}h_{mn}\\
+\frac{2 F^{mn}F_{mn}}{D-2}\Big(h_{ab}-\frac{g_{ab}}{2}\hat{h}\Big)\Big]+16 a_c\Big(F_{ab}\nabla^a h^{bc}+\nabla^a F^{bc}h_{ab}+F^{ac}\nabla^b h_{ab}-\frac{1}{2}F^{ac}\nabla_a \hat{h}\Big)\\
-8\Big(a_b\Box a^b-a_a a_b R^{ab}+\nabla_a a^a \nabla_b a^b\Big)\Big\},
\label{eq:EM_DEC_1}
\end{multline}
where we have defined $\hat{h} \equiv g^{ab}h_{ab}$. $\Delta_L$ is the Lichnerowicz operator defined in \eqref{deflichn}. This action can be checked to be invariant under the following gauge transformations
\be
\left\{
\begin{array}{c}
h_{ab}\to h_{ab}
\\
a_a \to a_a+\nabla_a \chi
\end{array}
\right.,
\ee
and
\be
\left\{
\begin{array}{c}
h_{ab}\to h_{ab}+\nabla_a V_b+\nabla_b V_a
\\
a_a \to a_a+V^b F_{ba}
\end{array}
\right. .
\ee
The first set of transformations corresponds to the $U(1)$ invariance associated with the electromagnetic field, and the second to the invariance under diffeomorphisms, conveniently mixed with the $U(1)$ symmetry. The gauge ambiguity from both these transformations should be fixed in the path integral.

Let us point out a property of \eqref{eq:EM_DEC_1} which will be crucial in this work. If we focus on spherically symmetric four-dimensional instantons, the metric perturbations can be expanded into odd and even perturbations \cite{Regge:1957td}. Substituting this general expansion in \eqref{eq:EM_DEC_1}, in the background of the magnetic Reissner-Nordstr\"om solution, the cross term gives zero for any value of the vector field perturbation $a_b$. If we further assume that whatever gauge choice that makes this problem tractable in four dimensions does not involve the mixture between metric perturbations and gauge potential perturbations, then the metric sector completely decouples from the gauge potential sector. In the electric Reissner-Nordstr\"om case, this decoupling does not occur, and a complete analysis of the problem requires the inclusion of such a term.

We further decompose the metric perturbations into a traceless and a trace part:
\be
h_{ab}=\phi_{ab}+\frac{1}{D}\,g_{ab}\,\hat{h} \,.
\ee
After this expansion, the second-order action can be written as
\begin{multline}
I_2[\phi_{ab},\hat{h},a_a] =\frac{1}{4}\int d^D x\sqrt{g} \\
\times \Big\{\phi^{ab}\Big[\Delta_L \phi_{ab}+4\Big(F_a^{\phantom{a}m}F_b^{\phantom{b}n}-\frac{g_{ab}}{D}F^{mc}F^{n}_{\phantom{n}c}\Big)\phi_{mn}+\frac{2}{D-2}F^{mn}F_{mn}\phi_{ab}\Big]\\
+\frac{8}{D}\phi_{ab}F^{ac}F^b_{\phantom{b}c}\hat{h}+\left(\frac{D^2-3 D+2}{D^2}\right)\hat{h}\Box \hat{h}-\frac{D-4}{D^2}F^{ab}F_{ab}\hat{h}^2 \phantom{aa}\\
-2 \nabla_a \phi^a_{\phantom{a}c}\nabla_b \phi^{bc}+\frac{2(D-2)}{D}\nabla_a \phi^{ab}\nabla_b \hat{h} \phantom{aaaaaaaaaaaaaaaaaaa}\\
+16 a_c\Big[F_{ab}\nabla^a \phi^{bc}+\nabla^a F^{bc}\phi_{ab}+F^{ac}\nabla^b \phi_{ab}-\frac{D-4}{2D}F^{ac}\nabla_a \hat{h}\Big] \phantom{i}\\ -8\Big(a_b\Box a^b-a_a a_b R^{ab}+\nabla_a a^a \nabla_b a^b\Big)\Big\} \phantom{aaaaaaaaaaaaaaaaaaaaaaai} \,.
\label{EM_DEC_2}
\end{multline}
The first term in the third line of \eqref{EM_DEC_2} is the term that makes the study of Einstein-Maxwell instantons more involved. It couples the traceless part of the metric with the trace part. Hence one cannot simply associate the trace with the unphysical divergent modes, as in the pure-gravity conformal factor problem. There may be several ways of tackling this. One might try to make a particular gauge choice for the metric perturbations. This is exactly what one does in the purely gravitational case to remove the last term in the second line, by choosing a gauge of the form
\be
C_a[h]=\nabla^b\left(h_{ab}-\frac{1}{\beta}\,g_{ab}\,\hat{h}\right) \,,
\label{eq:123}
\ee
where $\beta$ is a constant to be conveniently fixed \cite{Gibbons:1978ji}. However, we were not able to find such a gauge when electromagnetism is introduced without making a shift in the gauge potential $a_a$. This shift results in a new coupling between the trace and the vector potential, invalidating once more the pure gravity interpretation of the trace. Furthermore, this new choice of gauge requires a new decomposition of the metric perturbations instead of the usual longitudinal/transverse decomposition corresponding to the choice (\ref{eq:123}).

An alternative approach to the problem would be to shift the trace component of the metric by a term proportional to $\Box^{-1}(\phi_{ab}F^{ac}F^b_{\phantom{b}c})$. This choice removes the problematic term, but also makes the second-order action dependent on the inverse box operator, rendering any further computations unpractical.

A third method, based on \cite{Kol:2006ga}, is to attempt to solve the problem by using Kaluza-Klein techniques, and will be the one followed here.

\subsection{\label{sec:EM_BC}Boundary conditions}

The partition function is defined as the path integral (\ref{pathintegralA}) with appropriate boundary conditions. These boundary conditions specify the thermodynamic quantities which are held fixed in the ensemble. Here, we recall the discussion of Hawking and Ross \cite{Hawking:1995ap} for the Reissner-Nordstr\"om case.

As usual, the 3-metric on the boundary $\partial {\mathcal M} = S^1 \times S^2_{\infty}$ fixes the temperature $T=\beta^{-1}$, where $\beta$ is the periodicity of imaginary time. The boundary condition on the electromagnetic field at infinity typically fixes either the charge $Q$ or the potential $\Phi =Q/{r_+}$, as we shall discuss. Imposing a periodicity $\beta$, the leading order approximation to the path integral is the Reissner-Nordstr\"om instanton,
\be
d s^2 = f(r)d \tau^2+\frac{{d r^2}}{f(r)}+r^2d \Omega^2 \,,
\ee
where $f(r) = 1-2 M/r+Q^2/r^2$, $d \Omega^2=d\theta^2+\sin^2\theta d \phi^2$, and $M$ and $Q$ are the black hole mass and charge, respectively. Since an instanton solution is required to be non-singular, the mass and the charge are fixed by the boundary data and by the condition of regularity at $r=r_+=M +\sqrt{M^2 -Q^2}$, the location of the outer horizon in the Lorentzian solution and of the ``bolt'' in the Euclidean solution. That condition is the formula for the Hawking temperature:
\be
T = \frac{r_+ -M}{2 \pi r_+^2} \,.
\ee

However, the action of the instanton depends on the boundary terms which make the fixing of quantities on $\partial {\mathcal M}$ consistent. Suppose one fixes the potential $A$ at a surface of very large $r=R$, where $R$ will be taken to infinity. In the magnetic case, this corresponds to specifying the charge
\be
Q= \frac{1}{4 \pi} \int_{S^2_{\infty}} F \,,
\ee
which is determined by integrating the magnetic field strength $F_{\theta \phi}$ over $S^2_{\infty}$, which in turn is determined by $A$ on the boundary alone. But, in the electric case, the charge is computed using the dual of the electric field strength,
\be
Q= \frac{i}{4 \pi} \int_{S^2_{\infty}} \star F \,,
\ee
i.e. it requires fixing $F_{\tau r}$ (the $i$ is due to the use of imaginary time). The canonical ensemble, for which the charge $Q$ is fixed, includes the configurations for which derivatives of $A$ normal to the boundary are fixed. The thermodynamic quantity associated with specifying $A$ on $\partial {\mathcal M}$, in the electric case, is the electric potential at infinity $\Phi = -i A_\tau = Q/{r_+}$ (the gauge choice $A= -i (Q/r-\Phi) d\tau$ ensures that $A$ is regular on the horizon $r=r_+$). Fixing the potential $\Phi$ corresponds to the grand-canonical ensemble. For the canonical ensemble, the action (\ref{euclidaction}) must include a boundary term appropriate to the variational problem in question,
\be
- 4 \int_{\partial {\mathcal M}} d^3 x \sqrt{h} F^{ab} n_a A_b \,,
\ee
i.e. to the fixing of $F_{\tau r}$ and thus of the electric charge $Q$. This term ensures that the Helmholtz free energies of the electric and magnetic black holes coincide \cite{Hawking:1995ap}:
\be
F_{\text{Helmh}} = M-T S = -T \ln{Z_{\mathrm{canonical}}(\beta,Q)} \,,
\ee
where $S = \pi r_+^2$ is the black hole entropy.

In the present work, we look at a sector of metric perturbations about the instanton such that the metric is fixed at the boundary. We consider the magnetic case, leaving the electromagnetic potential $A$ unperturbed, which can be done consistently for $SO(3)$ symmetric backgrounds, as mentioned in the previous Section. The sector is thus included in the canonical ensemble.

\section{\label{sec:KK}The Kaluza-Klein action method}

We described in Section~\ref{sec:EM} the difficulties in quantising Einstein-Maxwell theory. The standard decomposition of the perturbations around the instanton \cite{Gibbons:1978ji} does not apply. However, a decomposition that explicitly decouples the divergent modes, the conformal modes in the case of pure gravity, is essential. We therefore look for a different approach.

In \cite{Kol:2006ga}, Kol addresses the problem of the negative mode of the Schwarzschild black hole by looking at the ``dynamical'' part of the action. This procedure, useful if the problem has a single non-homogeneous dimension (the radial one here), was formalised in \cite{Kol:2006ux}. Instead of the treatment of Gross, Perry and Yaffe \cite{Gross:1982cv}, which looks at the Lichnerowicz operator $\Delta_L$ acting on transverse-traceless metric perturbations, an auxiliary extra dimension $z$ is added. The extended space of metric perturbations and the dependence on the extra dimension allow for the construction of several gauge-invariant quantities. These decouple into two sectors, a ``dynamical'' part, which is the relevant reduced action, and a ``non-dynamical'' part, which takes away the divergent modes. The action of the five-dimensional zero modes, i.e. the $k=0$ modes in a Fourier decomposition along $z$, or at least a sector of it, is the four-dimensional action. Notice that, in four dimensions, the radial problem has a single gauge transformation, $\xi_r$. The appropriate fixing of the gauge freedom requires a second condition (as in the traceless and transverse gauge), but including time does not allow for the construction of gauge-invariant quantities. This is provided by the $\xi_z$ component if we use the auxiliary extra dimension instead.

A different way of looking at the extra dimension is to relate the black hole to a black string/brane, as we shall discuss in the next Chapter.\footnote{A straightforward correspondence of this type will not hold in our case because of the non-trivial factor in front of the second term in (\ref{nega_10}).} The difficulty in the charged black hole case is that the standard traceless-transverse gauge does not decouple the divergent modes. However, the Kaluza-Klein action method (``power of the action'', as Kol prefers) is still available. It requires:

\begin{itemize}
\item[(i)] A ``maximally general ansatz'' for the perturbation of the fields; i.e. the ansatz must reproduce all of the background field equations by variation of the metric, and must be closed for the relevant group of gauge transformations ($\xi_r$ and $\xi_z$).
\item[(ii)] The lift must be such that the five-dimensional action for the $k=0$ perturbation modes is equivalent to the action for the perturbations around the four-dimensional black hole ins\-tan\-ton. Ac\-tual\-ly, it suf\-fi\-ces that a particular sector of the five-di\-men\-sio\-nal action satisfies this. We must then restrict ourselves to that sector of the path integral.
\end{itemize}

The lift we consider in this work, along a \emph{timelike} direction $z$, is a ``magnetic string'' of a theory with electromagnetism and Chern-Simons term. We will need to restrict to a sector of the path integral by introducing a Delta functional on the space of perturbations. The restriction will ensure that we obtain the four-dimensional action when we look at the $k=0$ modes.

Before we start, let us make two clarifications. First, how do we know that the divergent modes are being decoupled from the path integral? The decomposition of the path integral into ``dynamical'' and ``non-dynamical'' parts leads to a ``non-dynamical action'' simply composed of squares of gauge-invariant quantities. One of them will have a minus sign, meaning that a rotation to the imaginary line is needed in order for the Gaussian path integral to converge. This is the exact analogy of the prescription of \cite{Gibbons:1978ac}. Second, the particular eigenvalue obtained from this method is not the same, in general, as the one obtained from the standard decomposition. The negative eigenvalue in \cite{Kol:2006ga} is quantitatively different from the one in \cite{Gross:1982cv} because the decomposition of the action is different. But the positivity properties of the action, i.e. whether a negative eigenvalue exists or not, must be the same.

\section{\label{sec:app0}Magnetic Reissner-Nordstr\"om analysis}

\subsection{\label{sec:app} Lift to five dimensions}

In this Section, we will apply the method described in Section~\ref{sec:KK} to the magnetic Reissner-Nordstr\"om black hole. The first non-trivial step is to find a five-dimensional system of gravity, possibly coupled to some fields, that reduces to four-dimensional Einstein-Maxwell theory upon a Kaluza-Klein reduction on a circle. This truncation was first discussed in \cite{LozanoTellechea:2002pn} in the context of Lorentzian signature. Here we straightforwardly extend it to Euclidean signature spacetimes.

We start with minimal five-dimensional supergravity with the action given by
\be
I^{(5)} = -\int d x^5 \sqrt{\hat{g}}\Big(\hat{R}-\hat{F}^{\hat{a}\hat{b}}\hat{F}_{\hat{a}\hat{b}}-\frac{2}{3\sqrt{3}}\frac{\hat{\varepsilon}^{\hat{a}_1\ldots\hat{a}_5}}{\sqrt{\hat{g}}}\hat{F}_{\hat{a}_1\hat{a}_2}\hat{F}_{\hat{a}_3\hat{a}_4}\hat{A}_{\hat{a}_5}\Big)\,,
\label{eq:app_1}
\ee
where $\;\hat{}\;$ represents quantities in five dimensions, $\hat{g}=|\det{g_{\hat{a}\hat{b}}}|$ and $\hat{F}=d \hat{A}$. Our dimensional reduction ansatz will take the generic form
\be
d s_5^2 = \Delta^{\gamma}d s_4^2+\epsilon \Delta (d z+2 \tilde{A}_a d x^a)^2\,,
\label{eq:app_2}
\ee
where $\gamma$ is a constant and $\Delta$ is related to the exponential of the dilaton field. Here $\epsilon=\pm1$, according to the signature of the five-dimensional spacetime. The signature of the four-dimensional space is chosen to be Euclidean. Also, $\partial_z$ is a Killing vector of the five-dimensional spacetime, meaning that $\Delta$, $g_{ab}$ and $\tilde{A}_a$ do not depend on $z$.

The reduction of the Maxwell field parallels that of the metric, and in particular we expand it as
\be
\hat{A}=A+\Sigma(d z+2 \tilde{A})\,,
\label{eq:app_3}
\ee
where both $A$ and $\tilde{A}$ only have components in the four-dimensional space, and again do not depend on $z$. Choosing $\gamma=-1/2$, setting $\Delta=\exp{(-4\phi/\sqrt{3}})$ and substituting in the action (\ref{eq:app_1}) yields the following form of the action:
\begin{multline}
I^{(5)}=-V \int d^4 x \sqrt{g}\Big[R-2\partial_a\phi \partial^a \phi-e^{-2\sqrt{3}\phi}\epsilon \tilde{F}^{ab}\tilde{F}_{ab}-e^{-\frac{2}{\sqrt{3}}\phi}H^{ab}H_{ab}-e^{\frac{4}{\sqrt{3}}\phi}\partial_a \Sigma \partial^a \Sigma \\
-\frac{2}{\sqrt{3}}\Sigma \frac{\varepsilon^{a_1\ldots a_4}}{\sqrt{g}}\big(H_{a_1a_2}-4 \tilde{A}_{a_1}\partial_{a_2}\Sigma\big)\big(H_{a_3a_4}-4 \tilde{A}_{a_3}\partial_{a_4}\Sigma\big)\Big]\,,
\label{app_4}
\end{multline}
where $V$ is the volume of the circle along which we are performing our dimensional reduction and $H = F+2\Sigma \tilde{F}$. Varying this action with respect to $g_{ab}$, $\tilde{A}_a$, $A_a$, $\phi$ and $\Sigma$ yields the following equations of motion
\begin{subequations}
\bea
R_{ab}=2 \partial_a \phi \partial_b \phi+e^{\frac{4}{\sqrt{3}}\phi}\partial_a \Sigma \partial_b \Sigma \hspace{8cm} \nonumber \\
+2 e^{-2 \sqrt{3}\phi}\varepsilon \Big(\tilde{F}_{ac}\tilde{F}_b^{\phantom{b}c}-\frac{g_{ab}}{4}\tilde{F}_{mn}\tilde{F}^{mn}\Big)+2 e^{-\frac{2}{\sqrt{3}}\phi}\Big(H_{ac}H_b^{\phantom{b}c}-\frac{g_{ab}}{4}H_{mn}H^{mn}\Big)\,,
\label{eq:app_a_1}
\eea
\be
\epsilon \nabla_a(e^{-2 \sqrt{3}\phi}\tilde{F}^{ab})+2\nabla_a (\Sigma e^{-\frac{2}{\sqrt{3}}\phi}H^{ab})+\frac{4}{\sqrt{3}}\frac{\varepsilon^{a_1 a_2 a_3 b}}{\sqrt{g}}H_{a_1 a_2} \Sigma \partial_{a_3}\Sigma = 0\,,
\label{eq:app_a_2}
\ee
\be
\nabla_a(e^{-\frac{2}{\sqrt{3}}\phi}H^{ab})+\frac{2}{\sqrt{3}}\frac{\varepsilon^{a_1 a_2 a_3 b}}{\sqrt{g}}H_{a_1 a_2} \partial_{a_3}\Sigma=0\,,
\label{eq:app_a_3}
\ee
\be
\Box \phi+\frac{\sqrt{3}}{2}\epsilon \tilde{F}_{ab}\tilde{F}^{ab}+\frac{1}{2 \sqrt{3}}e^{-\frac{2}{\sqrt{3}}\phi}H_{a b}H^{ab}-\frac{1}{\sqrt{3}}e^{\frac{4}{\sqrt{3}}\phi}\partial^a \Sigma \partial_a \Sigma =0\,,
\label{eq:app_a_4}
\ee
\begin{multline}
\nabla_a(e^{\frac{4}{\sqrt{3}}\phi}\nabla^a \Sigma)-\frac{1}{\sqrt{3}}\frac{\varepsilon^{a_1\ldots a_4}}{\sqrt{g}}\big(H_{a_1a_2}-4 \tilde{A}_{a_1}\partial_{a_2}\Sigma\big)\big(H_{a_3a_4}-4 \tilde{A}_{a_3}\partial_{a_4}\Sigma\big)\\
+4 \frac{\varepsilon^{a_1\ldots a_4}}{\sqrt{g}} H_{a_1 a_2}\tilde{A}_{a_3}\partial_{a_4}\Sigma-2e^{-\frac{2}{\sqrt{3}}\phi}H^{ab}\tilde{F}_{ab}=0\,.
\label{eq:app_a_5}
\end{multline}
\label{eq:app_5}
\end{subequations}
We now search for a consistent truncation in which the two fields $\phi$ and $\Sigma$ are set to zero. We are then left with two constraints on $\tilde{F}$ and $F$ coming from both Eqs. (\ref{eq:app_a_4}) and (\ref{eq:app_a_5})
\be
\begin{array}{c@{\hspace{1 cm}}c@{\hspace{1 cm}}c}
3 \epsilon \tilde{F}^{ab}\tilde{F}_{ab}+F^{ab}F_{ab}=0 & \text{and} &\star F+\sqrt{3}\tilde{F}=0\,,
\end{array}
\label{eq:app_6}
\ee
where $\star$ is the Hodge dual with respect to the four-dimensional geometry. If the four-dimensional manifold has Lorentzian signature, the second condition solves the first if $\epsilon =1$. However, if the four-dimensional manifold is assumed to have the Euclidean signature, one must require $\epsilon=-1$. We thus from now on choose $\epsilon=-1$. The equations of motion (\ref{eq:app_5}) then reduce to
\bea
R_{ab}=-2\Big(\tilde{F}_{ac}\tilde{F}_b^{\phantom{b}c}-\frac{g_{ab}}{4}\tilde{F}_{mn}\tilde{F}^{mn}\Big)+2\Big(F_{ac}F_b^{\phantom{b}c}-\frac{g_{ab}}{4}F_{mn}F^{mn}\Big)\,, \nonumber \\
\nabla_a F^{ab}=\nabla_a \tilde{F}^{ab}=0\,, \hspace{3cm}
\label{eq:neg_7}
\eea
subject to the second constraint in (\ref{eq:app_6}). In order to further simplify the equations of motion and to explicitly solve the constraint, we change from $F$ and $\tilde{F}$ to $F^{(1)}$ and $F^{(2)}$ given by
\be
\begin{array}{c@{\hspace{1 cm}}c@{\hspace{1 cm}}c}
F^{(1)}=\frac{1}{2}\star \tilde{F}-\frac{\sqrt{3}}{2}F & \text{and} & F^{(2)}=\frac{\sqrt{3}}{2}\star \tilde{F}+\frac{1}{2}F\,.
\end{array}
\ee
The constraint is solved by setting $\star F^{(2)}=0$, that is, $F^{(2)}=0$. Substitution in (\ref{eq:neg_7}) yields
\be
\begin{array}{ccc}
R_{ab}=2\Big(F^{(1)}_{ac}{F^{(1)}}_b^{\phantom{b}c}-\frac{g_{ab}}{4}F^{(1)}_{mn}{F^{(1)}}^{mn}\Big)\,, \qquad \quad \nabla_a({F^{(1)}}^{ab})=0\,.
\end{array}
\ee
These are precisely the equations of motion that we were searching for, i.e. the Einstein-Maxwell equations in four dimensions. Note that both these equations can be deduced from the four-dimensional action
\be
I^{(4)} = -\int d x^4 \sqrt{g} \Big(R-F^{(1)}_{ab}{F^{(1)}}^{ab}\Big)\,.
\ee
The four-dimensional quantities are related to the five-dimensional quantities via
\be
\begin{array}{c@{\hspace{1 cm}}c@{\hspace{1 cm}}c@{\hspace{1 cm}}c}
d s^2_5 =d s^2_4-(d z+2 \tilde{A}_a d x^a)^2\,, & \displaystyle{\tilde{F} = \frac{\star F^{(1)}}{2}} & \text{and} & \displaystyle{\hat{A} = -\frac{\sqrt{3}}{2}A^{(1)}\,.}
\end{array}
\label{eq:app_8}
\ee
We further remark that this truncation is only valid at the level of the equations of motion. In fact, directly substituting the ansatz (\ref{eq:app_8}) in \eqref{app_4} gives
\be
-V \int d^4 x \sqrt{g}\Big(R -\frac{1}{2}F^{(1)}_{ab}{F^{(1)}}^{ab}\Big)\neq V I^{(4)}\,.
\label{app_11}
\ee

We can now write the five-dimensional lift of both the magnetic and electric Reissner-Nordstr\"om instantons, respectively as
\be
\begin{array}{ccc}
\displaystyle{d s^2_5 = f(r)d \tau^2+\frac{{d r^2}}{f(r)}+r^2d \Omega^2-\Big(d z-\frac{Q}{r} d\tau \Big)^2}\,,& \quad & \displaystyle{A = -\frac{\sqrt{3}}{2}Q \cos\theta d\phi}\,,
\end{array}
\label{app_9}
\ee
and
\be
\begin{array}{ccc}
\displaystyle{d s^2_5 = f(r)d \tau^2+\frac{{d r^2}}{f(r)}+r^2d \Omega^2-\Big(d z-\ii\, Q \cos\theta d\phi \Big)^2}\,, & \quad  & \displaystyle{A = -\ii\,\frac{\sqrt{3}Q}{2 r} d\tau}\,,
\label{eq:app_10}
\end{array}
\ee
where $f(r) = 1-2 M/r+Q^2/r^2$, $d \Omega^2=d\theta^2+\sin^2\theta d \phi^2$, and $M$ and $Q$ are the black hole mass and charge, respectively. We are now ready to use the ``power of action''.


\subsection{Reducing the action}

The ``maximally general ansatz'' for the magnetic case can be written as
\bea
d s^2_5 = e^{2 A(r,z)}[d\tau-\kappa(r,z)]^2+e^{2 B(r,z)}d r^2+e^{2 C(r,z)}d \Omega^2 \nonumber \\
-e^{2\beta(r,z)}[d z-\Gamma(r,z)d \tau-\alpha(r,z)d r]^2\,, \hspace{1.6cm}
\label{nega_0}
\eea
and $A = -(\sqrt{3}Q/{2})\cos\theta d\phi$. With this ansatz, it is consistent, in terms of gauge-invariance, not to perturb the electromagnetic potential. This conforms to what we have seen before, because in the magnetic case the metric perturbations decouple from the vector potential perturbations, and the former have a positive definite action. For the electric case, one would have to find a form of perturbing the solution (\ref{eq:app_10}), maintaining all variables involved only dependent on $r$ and $z$. However, the authors were not able to accomplish this. From now on, we will only consider the magnetic case, for which the metric is explicitly codimension one. We then perturb the quantities in \eqref{nega_0} as
\be
\begin{array}{cc}
\displaystyle{A(r,z)=A_0(r)+a(r,z)}\,, & \displaystyle{B(r,z)=B_0(r)+b(r,z)}\,,
\\
\\
\displaystyle{C(r,z)=C_0(r)+c(r,z)}\,, & \displaystyle{\Gamma(r,z) = \Gamma_0(r)+\gamma(r,z)}\,, 
\end{array}
\label{nega_1}
\ee
where all lower case letters are perturbations, and thus absent in the background solution, which is given by
\be
\begin{array}{c@{\hspace{1 cm}}c@{\hspace{1 cm}}c@{\hspace{1 cm}}c}
\displaystyle{A_0=-B_0=\frac{1}{2} \log f}, & C_0(r)=\log r & \text{and} & \displaystyle{ \Gamma_0(r)=\frac{Q}{r}}\,.
\end{array}
\ee
At zero-th order, we exactly recover \eqref{app_9}. We then substitute \eqref{nega_1} into the action (\ref{eq:app_1}) and expand it to second order,
\be
I^{(5)}_2=\int d r d z (P^{a\phantom{I}b\phantom{J}}_{\phantom{a}I\phantom{b}J}\partial_a u^I\partial_b u^J+Q^{\phantom{I}a\phantom{J}}_{I\phantom{a}J}u^I\partial_a u^J+V_{IJ}u^Iu^J)\,,
\label{nega_2}
\ee
where $u=\{a,b,c,\alpha,\beta,\gamma,\kappa\}$, $I\in\{\underline{1},\ldots,\underline{7}\}$ and the independent non-vanishing components of the tensors $P^{a\phantom{I}b\phantom{J}}_{\phantom{a}I\phantom{b}J}$, $Q_{I\phantom{a}J}^{\phantom{I}a\phantom{J}}$ and $V_{IJ}$ are given in Appendix~\ref{app:1}. We can now expand all the fields in Fourier modes, take the limits $k=0$ and $\beta=0$, and compare the five-dimensional quadratic action with the four-dimensional counterpart. As we predicted in \eqref{app_11}, the two actions are not equal. Instead, they differ by the following term:
\be
I^{(5)}_{2,k=0,\beta=0}-I^{(4)}_2 = -\int d r\frac{\{r^2\partial_r\gamma(r)+Q [a(r)+b(r)-2 c(r)]\}^2}{2 r^2}\,.
\label{eq:diff5d4d}
\ee
Note that here we are perturbing the four-dimensional action only in the metric sector, and with an ansatz equal to the $l=0$ ansatz for the Schwarzschild perturbations used in \cite{Gross:1982cv}. We will see later how to deal with this difference in actions.

The second-order action (\ref{nega_2}) is invariant under the following gauge transformations
\be
\begin{array}{l}
\delta a = e^{-2 B_0}\xi_r A_0'+e^{-2 A_0}\Gamma_0^2\partial_z\xi_z
\\
\delta b = e^{-2 B_0}(\partial_r \xi_r-B_0' \xi_r)
\\
\delta c = e^{-2 B_0}\xi_r C_0'
\\
\delta \alpha = \partial_z \xi_r+\partial_r \xi_z-e^{-2 A_0}\xi_z \Gamma_0 \Gamma_0'
\\
\delta \beta = -\partial_z \xi_z
\\
\delta \gamma = e^{-2 B_0}\xi_r \Gamma_0'+2 \Gamma_0 \partial_z \xi_z
\\
\delta \kappa = -e^{-2 A_0}[\xi_z\Gamma_0'+\Gamma_0(\partial_z \xi_r +\partial_r \xi_z -2 \xi_z A_0')]\,,
\end{array}
\label{eq:nega_3}
\ee
where $'$ denotes differentiation with respect to $r$ in zero-th-order quantities. These correspond to infinitesimal diffeomorphisms along the 1-form $\xi = \xi_r d r+\xi_z d z$.

The quadratic action (\ref{nega_2}) can be cast in a different form, in which $\alpha$ and $\kappa$ only appear as $\partial_z \alpha$ and $\partial_z \kappa$, if we integrate by parts. This was expected because, in the $k=0$ sector, one does not need to set $\alpha$ or $\kappa$ to zero, as one does for $\beta$. Let us then consider the following gauge independent quantities
\be
\begin{array}{l}
\displaystyle{q^1 = a-\left(\frac{c A_0'}{C_0'}-e^{-2 A_0}\Gamma_0^2 \beta\right)}
\\
\displaystyle{q^2 = b+\left[\frac{c B_0'}{C_0'}-e^{-2 B_0}\partial_r\left(\frac{e^{2 B_0}c}{C_0'}\right)\right]}
\\
\displaystyle{q^3 = \partial_z \kappa-e^{-2 A_0}\left(\beta \Gamma_0'-2 \Gamma_0 \beta A_0'-\frac{e^{2 B_0}\Gamma_0}{C_0'}\partial^2_z c+\Gamma_0 \partial_r\beta\right)}
\\
\displaystyle{q^4 = \partial_z \alpha-e^{-2 A_0}\left[\beta \Gamma_0 \Gamma_0'+\frac{e^{2 (A_0+B_0)}}{C_0'}\partial^2_zc-e^{2 A_0}\partial_r \beta\right]}
\\
\displaystyle{q^5 = \gamma-\left(\frac{c \Gamma_0'}{C_0'}-2 \Gamma_0 \beta\right)}\,.
\label{eq:nega_4}
\end{array}
\ee
Plugging in the expressions for the $u^I$'s, as a function of the $q^i$'s, in \eqref{nega_2}, we obtain a quadratic action which only depends on the $q^i$'s. As expected, the action can be entirely written in terms of gauge-invariant quantities. We now proceed to integrate the ``non-dynamical'' quantities $q^2$, $q^3$ and $q^4$, which appear with no derivatives in the action. The action for the perturbations can be written as
\be
I^{(5)}_2 = I^{(5)}_{2,q^1,q^5}+\int d r d z\left[q^{\tilde{i}}L_{\tilde{i}\tilde{j}}q^{\tilde{i}}-2 R_{\tilde{i}}q^{\tilde{i}}\right]\,,
\label{eq:nega_5}
\ee
where $\tilde{i}\in\{2,3,4\}$. $L$ does not depend on $q^1$ or $q^5$ and $R$ depends on $q^1$, $q^5$ and their derivatives. Explicitly,
\be
L = \left[
\begin{array}{ccc}
\displaystyle{ \frac{3 Q^2}{2 r^2}-2} & \displaystyle{\frac{3 f Q}{2}} & \displaystyle{\frac{1}{2} (1+3 f) r}
\\
\\
\displaystyle{ \frac{3 f Q}{2}} & \displaystyle{-\frac{1}{2} f^2 r^2} & \displaystyle{-\frac{1}{2} f Q r}
\\
\\
\displaystyle{ \frac{1}{2} (1+3 f) r }& \displaystyle{-\frac{1}{2} f Q r} & \displaystyle{-\frac{Q^2}{2}}
\end{array}
\right]\,,
\label{eq:nega_6}
\ee
\be
R=\left[
\begin{array}{c}
\displaystyle{\frac{1}{2} \left(\frac{q^1 Q^2}{r^2}-\frac{2 r \partial^2_z q^5 Q}{f}+\partial_r q^5 Q+2 r^2 \partial^2_z q^1-4 r f\partial_r q^1\right)}
\\
\\
\displaystyle{\frac{(Q^2-r^2) q^5+r f[r (q^5+r\partial_r q^5)-Q q^1]}{2 r}}
\\
\\
\displaystyle{\frac{1}{2} \left[\left(r-r f-\frac{2 Q^2}{r}\right) q^1+2 Q q^5+r (2 r f \partial_r q^1-Q \partial_r q^5)\right]}
\end{array}
\right]
\label{eq:nega_7}
\ee
and
\be
I^{(5)}_{2,q^1,q^5}=-\frac{(Q q^1+r^2 \partial_r q^5)^2}{2 r^2}\,.
\label{eq:nega_8}
\ee
We can thus integrate the $q^{\tilde{i}}$ by constructing the following shifted variables
\be
\tilde{q}^{\tilde{i}} = q^{\tilde{i}}-(L^{-1})^{\tilde{i}\tilde{j}}R_{\tilde{j}}\,,
\label{eq:nega_9}
\ee
in terms of which the action (\ref{eq:nega_5}) can be rewritten as
\be
I^{(5)}_2=I^{(5)}_{2,q^1,q^5}-\int d r d z R_{\tilde{i}}(L^{-1})^{\tilde{i}\tilde{j}}R_{\tilde{j}}+I_{ND}\,,
\ee
where
\be
I_{ND}=\int d r d z \tilde{q}^{\tilde{i}}L_{\tilde{i}\tilde{j}}\tilde{q}^{\tilde{i}}\,.
\ee
The ``non-dynamical'' $\tilde{q}^{\tilde{i}}$'s have now decoupled from the $q^1$ and $q^5$ dependent part of the action. The new effective action for the variables $q^1$ and $q^5$ is
\be
I_{\mathrm{eff}}\equiv I^{(5)}_{2,q^1,q^5}-\int d r d z R_{\tilde{i}}(L^{-1})^{\tilde{i}\tilde{j}}R_{\tilde{j}}\,.
\ee
This last expression is too cumbersome to be shown. However, we still have freedom to rotate $q^1$ and $q^5$. If one sets
\be
\begin{array}{l}
q^1 = a_{11} q+a_{15} \tilde{q}^5
\\
q^5 = a_{51} q+a_{55}\tilde{q}^5
\end{array},
\ee
where
\bea
\displaystyle{a_{11}=\frac{r^2+3 f r^2-5 Q^2}{2 \sqrt{6} f r^2}}\,, \quad \displaystyle{a_{15} = \frac{Q a_{55}}{r f}}\,, \quad \displaystyle{a_{51}=-\frac{Q}{\sqrt{6}r}}\,, \quad \displaystyle{\partial_r a_{55} = -\frac{Q^2a_{55}}{r^3}}\,,
\eea
then $\tilde{q}^5$ disappears from the effective action, which reduces to the simple expression
\be
I_{\mathrm{eff}} = \int d r d z r^2\left[f (\partial_r q)^2-\left(1-\frac{Q^2}{r^2 f}\right)(\partial_z q)^2+V q^2\right]\,,
\label{nega_10}
\ee
where $V$ is given by
\be
V=-\frac{6 M^2}{r^2(3 M-2 r)^2}\left[1-\frac{4}{3}\left(\frac{Q}{M}\right)^2\right]\,.
\ee

We are now in a position to explain why the Kaluza-Klein method works in this specific case. The path integral for perturbations in five dimensions $Z^5_{(2)}$ reduces to the one in four dimensions $Z^4_{(2)}$ for $k=0$ modes if we only integrate over directions in which the difference \eqref{eq:diff5d4d} between the two actions is zero, i.e. if we tune $\gamma$. So, one has a particular sector of $Z^5_{(2)}$,
\be
\tilde{Z}^5_{(2)} =\lim_{k=0} \int d [u^I]e^{-I^{(5)}_2[u^I]}\delta[Q(a+b-2c)+r^2\partial_r \gamma]\delta[\beta] = \tilde{Z}^4_{(2)}\,,
\ee
where $\delta[.]$ is a Dirac delta functional. Above, $\tilde{Z}^4_{(2)}$ denotes the restriction of the four-dimensional path integral to the sector of radial metric perturbations (we showed before that it is consistent, in the magnetic case, to leave the electromagnetic potential unperturbed). Introducing the gauge-invariant variables gives the following form
\begin{align}
\tilde{Z}^4_{(2)} & = \lim_{k=0} \int d [q^i]d[c]d[\beta]e^{-I^{(5)}_2 [q^i]}\delta\left[Q(q^1+q^2)+r^2\partial_r q^5-\frac{Q(Q^2-2 r^2 f)\beta}{r^2 f}+2 Q r \partial_r \beta \right]\delta[\beta] \nonumber\\
    & = \lim_{k=0} \int d [q^i]e^{-I^{(5)}_2 [q^i]}\delta[Q(q^1+q^2)+r^2\partial_r q^5] \nonumber\\
    & = \lim_{k=0} \int d [q]d[\tilde{q}^5]d[\tilde{q}^{\tilde{i}}]e^{-I_{\mathrm{eff}}[q]-I_{ND}[\tilde{q}^{\tilde{i}}]}\delta\Big[\sqrt{\frac{2}{3}} Q \left(1-\frac{Q^2}{f r^2}+\frac{2 r^2}{r^2+3 f r^2-3 Q^2}\right)q \nonumber \\
    & \hspace{6.5cm} +Q \tilde{q}^{\tilde{2}}+a_{55} r^2 \partial_r \tilde{q}^5\Big]\nonumber \\
    & = \lim_{k=0} \int d [q]e^{-I_{\mathrm{eff}}[q]} \,,
\end{align}
where each equality holds up to an infinite constant. It is the crucial fact that the final form of the action, $I_{\mathrm{eff}}[q]$, does not depend on $\tilde{q}^5$ that makes the two path integrals equivalent when we look at the $k=0$ sector. The coefficients $a_{ij}$ were chosen in such a way that the argument of the last Dirac delta functional depends on $\partial_r \tilde{q}^5$ and not on both $\tilde{q}^5$ and $\partial_r \tilde{q}^5$ simultaneously.


\subsection{The negative mode}

In this section, we will compute the negative mode of the action by directly studying the eigenvalue associated with \eqref{nega_10}, when the $z$ dependence drops out ($k=0$). The action reduces to
\be
I_{\mathrm{eff},k=0} = \int d r r^2\left[f(\partial_r q)^2+V q^2\right].
\label{nega_1_1}
\ee
We first remark that, if $|Q|/M\geq\sqrt{3}/2$, the potential $V$ becomes positive, and as a result the action does not have a negative mode beyond this value of the charge. This is in exact agreement with the thermodynamic prediction, as we shall see.

\begin{figure}
\centering
\includegraphics[width = 8 cm]{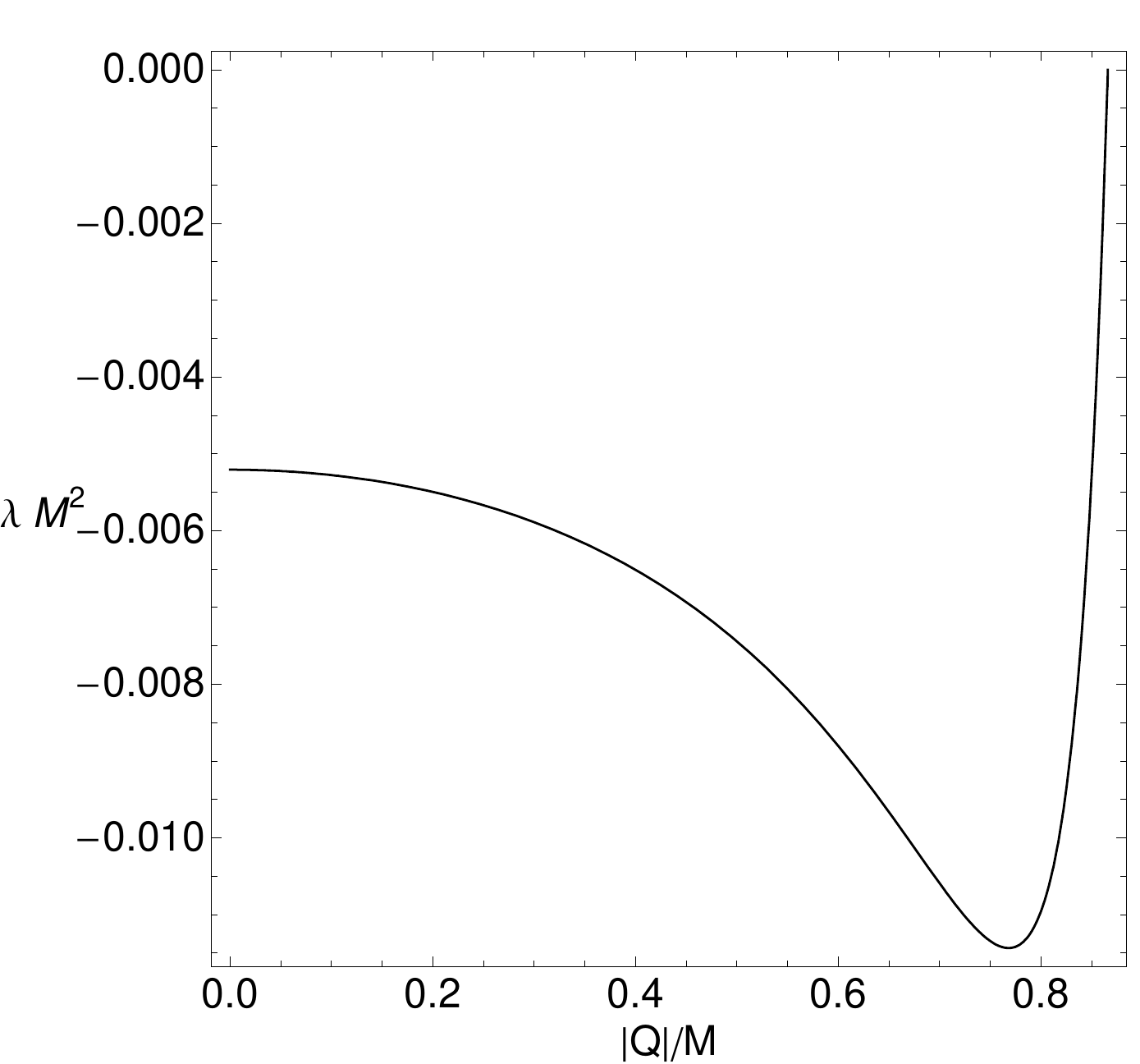}
\caption{\label{fig:nega_1}Evolution of the negative mode with $|Q|/M$.}
\end{figure}

To proceed, we change variables in \eqref{nega_1_1}, in such a way that the resulting eigenvalue problem reduces to a one-dimensional Schr\"odinger equation. A convenient change of variables is
\be
\begin{array}{c@{\hspace{1 cm}}c@{\hspace{1 cm}}c}
\displaystyle{u=q \sqrt{\mathcal{T}}} & \text{ and } & \displaystyle{y = -\frac{\mathcal{T}}{r_{_+}-r_{_-}}\log\left(\frac{r-r_{_+}}{r-r_{_-}}\right),}
\end{array}
\label{nega_1_1_a}
\ee
where $r_{\pm}=M\pm\sqrt{M^2-Q^2}$ are the locations of the black hole outer and inner horizons, respectively, and $\mathcal{T}$ is a constant chosen for later convenience. The action (\ref{nega_1_1}) reduces to
\be
I_{\mathrm{eff},k=0}=\int d y \left[(\partial_y u)^2+\tilde{V} u^2\right],
\label{eq:nega_1_2}
\ee
where
\be
\tilde{V}=-\frac{2 \varepsilon (r_{_+}-r_{_-})^2 }{\mathcal{T}^2}\frac{e^{\frac{r_{_+}-r_{_-}}{\mathcal{T}}y}}{\left[1+\varepsilon e^{\frac{r_{_+}-r_{_-}}{\mathcal{T}}y}\right]^2}
\ee
and $\varepsilon=(r_{_+}-3 r_{_-})/(3 r_{_+}-r_{_-})$. Here we choose $\mathcal{T}=(r_{_+}-r_{_-})^2/\sqrt{\varepsilon}$, making the potential negative definite. This choice is only valid for $\varepsilon>0$, that is, $|Q|/M<\sqrt{3}/2$. The eigenvalue problem can now be formulated as
\be
-\partial^2_{y} u+\tilde{V} u=\lambda u.
\label{eq:nega_1_4}
\ee
The change of coordinates (\ref{nega_1_1_a}) maps $r=r_{_+}$ to $y=+\infty$ and $r=+\infty$ to $y=0$. As a result, the boundary conditions are now changed to $u$ being regular at $y=+\infty$, and integrable near $y=0$. Equation (\ref{eq:nega_1_4}) can be analytically solved for these boundary conditions and one finds the unique normalised bound state
\be
u=\frac{\sqrt{2}\varepsilon^{3/4}\sqrt{r_{_+}+r_{_-}}}{(r_{_+}-r_{_-})}\frac{e^{-\frac{(r_{_+}+r_{_-})y}{4 \mathcal{T}}}}{(\varepsilon-1)+(\varepsilon+1)\coth\left[\frac{(r_{_+}-r_{_-})y}{2 \mathcal{T}}\right]},
\ee
corresponding to
\be
\lambda=-\frac{2\sqrt{1-\eta^2}-1}{64 M^2 (1-\eta^2)^2(2\sqrt{1-\eta^2}+1)^2},
\label{eq:nega_1_5}
\ee
where $\eta=|Q|/M$. As expected, there is only one negative mode, and it disappears for $\eta\geq\sqrt{3}/2$ (Fig.~\ref{fig:nega_1}).


\subsection{Local thermodynamic stability}

Let us recall the thermodynamics of Reissner-Nordstr\"om black holes. The partition functions correspond to the different ensembles according to the boundary conditions of the path integral, as explained in Section \ref{sec:EM_BC}. In the magnetic case, fixing the temperature $T=\beta^{-1}$, by imposing the periodicity in imaginary time, and fixing the electromagnetic potential $A_a$ at infinity, which gives the magnetic charge $Q$, corresponds to the canonical ensemble.

The thermodynamic stability condition for the Reissner-Nordstr\"om black hole is stated in Section~\ref{sec:localstab} if we substitute $\Omega \to \Phi=Q/r_+$ and $J \to Q$. In the canonical ensemble, local stability requires that the specific heat at constant charge,
\be
C_Q = T \left( \der{S}{T} \right)_Q= - \frac{2 \pi r_+^2  (r_+ - r_-) }{r_+ - 3 r_-}\,,
\ee
is positive. This occurs for $\sqrt{3}M/2 < |Q| < M $, which is exactly the range we found previously for the disappearance of the negative mode in the partition function.

Had we studied the grand-canonical ensemble, we would expect a negative mode to persist. The stability condition is the positivity of not just $C_Q$ but also the isothermal permittivity $\epsilon$, really the capacitance here, which is given by
\be
\epsilon = \left( {\der{Q}{\Phi}} \right)_T = \frac{T \eta^2}{C_\Phi - C_Q} = \frac{r_+ \left(r_+ - 3r_-\right)}{r_+ - r_-}\,.
\ee
Since $C_Q$ and $\epsilon$ have opposite signs, the grand-canonical ensemble is always unstable and the negative mode is not expected to disappear.

\section{\label{sec:conc}Conclusions}

In this Chapter, we have studied the problem of negative modes of the Euclidean section of the magnetic Reissner-Nordstr\"om black hole in four dimensions. Solving this problem within four dimensions seems very difficult, as the identification of the unphysical perturbations which render the partition function divergent becomes considerably more intricate than in the vacuum case.

Following \cite{Kol:2006ga}, we devised a method to study this problem by lifting the magnetic Reissner-Nordstr\"om solution to five dimensions. The five-dimensional action is equal to the four-dimensional action, up to a quadratic term that can be set to zero by a suitable constraint, imposed by a Dirac delta functional, on the five-dimensional path integral. Furthermore, in five dimensions, the action can be solely written in terms of gauge-invariant variables, which in turn can be divided into two decoupled sectors: ``non-dynamical'' and ``dynamical''. The former is algebraic, and can thus be readily integrated out. The final form of the five-dimensional action depends on a single gauge-invariant variable, and the study of its negative modes in the long wavelength limit, corresponding to four dimensions, is now accessible.

We found complete agreement between the local stability of the canonical ensemble and the existence of the negative mode. We analytically determined the eigenmode as a function of the black hole mass and magnetic charge, from which we concluded that the negative mode ceased to exist for $|Q|/M\geq\sqrt{3}/2$. This is the range for which the specific heat at constant charge $C_Q$ becomes positive.

The Kaluza-Klein action method is a practical procedure to determine the quantum stability of gravity coupled to electromagnetism. However, a standard treatment of the metric and electromagnetic potential perturbations along the lines of \cite{Gibbons:1978ji} remains most desirable. The clarification of the divergent modes problem, analogous to the conformal factor problem of pure Einstein theory, would possibly lead to a better understanding of perturbative quantum gravity coupled to matter.

\begin{subappendices}

\section{\label{app:1}Appendix: Components of the tensors $P$, $Q$ and $V$}

The independent components of $P^{a\phantom{I}b\phantom{J}}_{\phantom{a}I\phantom{b}J}$ in the expression \eqref{nega_2} are given by
\be
\begin{array}{l}
\begin{array}{cccc}
\PP{1}{1}{1}{3}=\PP{1}{3}{1}{3}=\PP{1}{3}{1}{5}=\PP{1}{3}{2}{4}=4 \PP{1}{6}{2}{7}, & \PP{1}{1}{1}{5}=\PP{1}{1}{2}{4}=2\PP{1}{6}{2}{7}, & \PP{1}{3}{2}{7}=\PP{1}{5}{2}{7},
\end{array}
\\
\begin{array}{cccccc}
\PP{1}{5}{2}{7}=2 \PP{2}{4}{2}{7}, & \PP{2}{1}{2}{3}=2\PP{2}{1}{2}{2}, & \PP{2}{2}{2}{3}=\PP{2}{3}{2}{3}, & \PP{2}{3}{2}{5}=2\PP{2}{5}{2}{5}, & \PP{2}{3}{2}{6}=2 \PP{2}{6}{2}{6},
\end{array}
\\
\;\,\PP{2}{1}{2}{2}=-2 \PP{1}{6}{1}{6}, \\
\begin{array}{cccccc}
\PP{1}{6}{2}{7}=-\frac{r^2 f}{2}, & \PP{1}{6}{1}{6}=-\frac{r^2}{2}, & \PP{1}{6}{2}{4}=\frac{Q r}{2}, & \PP{2}{3}{2}{3}=2r^2\left(1-\frac{Q^2}{r^2 f}\right), & \PP{2}{2}{2}{5} = -\frac{Q^2}{f},
\end{array}
\\
\begin{array}{cccc}
\PP{2}{2}{2}{6}=-\frac{Q r}{f}, & \PP{2}{4}{2}{4}=-\frac{Q^2}{2}, & \PP{2}{4}{2}{7}=-\frac{Q r f}{2}, & \PP{2}{7}{2}{7}=-\frac{r^2 f^2}{2},
\end{array}
\end{array}
\ee
where all the other non-vanishing components can be obtained via the symmetry $P^{a\phantom{I}b\phantom{J}}_{\phantom{a}I\phantom{b}J}=P^{b\phantom{J}a\phantom{I}}_{\phantom{b}J\phantom{a}I}$. The terms that only involve one derivative were written using the auxiliary tensor $Q_{I\phantom{a}J}^{\phantom{I}a\phantom{J}}$, whose non-zero components are given by
\be
\begin{array}{l}
\begin{array}{ccc}
\QQ{1}{1}{1}=\QQ{5}{1}{1}=-2 \QQ{2}{1}{1}, & \QQ{1}{1}{3}=\QQ{5}{1}{3}=-2 \QQ{2}{1}{3}, & \QQ{1}{1}{5}=\QQ{5}{1}{5}=-2 \QQ{2}{1}{5},
\end{array}
\\
\begin{array}{ccccc}
\QQ{5}{1}{6}=-3 \QQ{1}{1}{6}, & \QQ{1}{1}{6}=\QQ{2}{1}{6}, & \QQ{3}{1}{1}=-2\QQ{2}{1}{1}, & \QQ{3}{1}{3}=-2\QQ{2}{1}{3}, & \QQ{3}{1}{5}=-2\QQ{2}{1}{5},
\end{array}
\\
\begin{array}{ccc}
\QQ{3}{1}{6}=-\QQ{6}{2}{4}=-2\QQ{2}{1}{6}, & \QQ{6}{2}{4}=2 \QQ{2}{1}{6}, & \QQ{2}{2}{7}=3 \QQ{1}{2}{7},
\end{array}
\\
\begin{array}{cccc}
\QQ{1}{2}{4}=2\frac{Q^2}{r}-r+r f, & \QQ{1}{2}{7}= Q f, & \QQ{2}{1}{1}= 4 r f, & \QQ{2}{1}{6}=-Q,
\end{array}
\\
\begin{array}{ccc}
\QQ{2}{1}{3}=-2\frac{Q^2}{r}+2 r+2 r f, & \QQ{2}{1}{5}=-\frac{Q^2}{r}+r+3 r f, & \QQ{2}{2}{4}=r(1+3 f),
\end{array}
\\
\begin{array}{ccc}
\QQ{3}{2}{4}= -2\left(\frac{Q^2}{r}+2 r f\right), & \QQ{3}{2}{7}=2 Q-2 \frac{Q^3}{r^2}-4 Q f, & \QQ{5}{2}{4}=-\frac{3 Q^2}{r},
\end{array}
\\
\begin{array}{cc}
\QQ{5}{2}{7}=Q \left(1-\frac{Q^2}{r^2}\right), & \QQ{6}{2}{7}=-\frac{Q^2}{r}+r-r f.
\end{array}
\end{array}
\ee
Finally, the potential $V_{IJ}$ takes the following form:
\be
V_{IJ}=\left[
\begin{array}{ccccccc}
 \frac{3 Q^2-4 r^2}{2 r^2} & -\frac{Q^2}{2 r^2} & \frac{Q^2-2 r^2}{r^2} & 0 & \frac{5 Q^2-4 r^2}{2 r^2} & 0 & 0 \\
 -\frac{Q^2}{2 r^2} & \frac{3 Q^2-4 r^2}{2 r^2} & \frac{4 r^2-6 Q^2}{2 r^2} & 0 & \frac{Q^2}{2 r^2} & 0 & 0 \\
 \frac{Q^2-2 r^2}{r^2} & \frac{4 r^2-6 Q^2}{2 r^2} & \frac{6 Q^2-4 r^2}{r^2} & 0 & -\frac{Q^2+2 r^2}{r^2} & 0 & 0 \\
 0 & 0 & 0 & 0 & 0 & 0 & 0 \\
 \frac{5 Q^2-4 r^2}{2 r^2} & \frac{Q^2}{2 r^2} & -\frac{Q^2+2 r^2}{r^2} & 0 & -\frac{Q^2+4 r^2}{2 r^2} & 0 & 0 \\
 0 & 0 & 0 & 0 & 0 & 0 & 0 \\
 0 & 0 & 0 & 0 & 0 & 0 & 0
\end{array}
\right].
\ee

\end{subappendices}

%% file: Classthermo.tex
\chapter{Classical stability and thermodynamics} \label{cha:classthermo}

In this Chapter, we will discuss the connection between the local thermodynamic instability and the classical stability of black branes and black holes, focusing on vacuum solutions without cosmological constant. The connection arises from the fact that stationary zero-modes of the Euclidean action, which mark the onset of pathologies of the partition function, are also classical stationary zero-modes. These may be thermodynamic in origin, being predicted by the thermodynamic stability Hessian, in which case they identify the threshold of new Gregory-Laflamme-type instabilities of black branes, but not of black holes. Or they may correspond to additional degrees of freedom in the partition function, not captured by the usual thermodynamic description, in which case they may indeed correspond to classical instabilities of black holes.

When connected to the classical instability of a black brane or a black hole, a stationary zero-mode corresponds to the bifurcation to a new family of solutions. When connected to the thermodynamic instability of a black hole, the zero-mode corresponds simply to a change in the asymptotic charges within the same family of solutions.

The discussion here, taken mainly from Refs.~\cite{Dias:2009iu,Dias:2010eu}, provides the explanation for the numerical results in the following two Chapters.

\section{Negative modes and black brane stability}

We start with a review of the Gubser-Mitra conjecture. Then we will show, based on \cite{Dias:2010eu}, that this conjecture implies the classical instability of any black brane constructed by trivially extending a vacuum black hole along extra dimensions.

\subsection{The Gubser-Mitra conjecture} \label{subsec:GMconjecture}

The relevance of black branes in string theory motivated the study of the stability of these objects. Gregory and Laflamme showed that Schwarzschild black branes are classically unstable for perturbation modes with a large wavelength along the extended directions \cite{Gregory:1993vy}. As we discussed in Section~\ref{sec:classstabreview}, the linearised Einstein equations for those modes in the traceless-transverse gauge are
\be
\label{Cllichn}
(\Delta_L h)_{ab} = -k^2 h_{ab}\,.
\ee
Here, $k$ is the wavenumber of perturbations of the type $e^{\ii\, \vec{k}\cdot\vec{z}} h_{ab}$, and $\Delta_L$ is the Lichnerowicz operator on the black \emph{hole} background. The instability occurs for modes with $|k|<k_\ast$, where the critical wavelength $k_\ast$ corresponds to the threshold stationary mode. The inclusion of charge may improve the stability properties. Recall Chapter~\ref{cha:RN}, where we saw how the charge can make a thermodynamic instability disappear. Indeed, a connection between the classical and thermodynamic stability of black branes was found. It is known as the Gubser-Mitra conjecture \cite{Gubser:2000ec,Gubser:2000mm}, which states that \emph{black branes with a non-compact translational symmetry are classically stable if and only if they are locally thermodynamically stable}. We alluded to this conjecture in Section~\ref{sec:localstab}, when we discussed local thermodynamic stability; see the review \cite{Harmark:2007md} and references therein.

The problem was clarified by Reall \cite{Reall:2001ag}, who noticed that Eq.~\eqref{Cllichn} for the stationary and axisymmetric mode marking the onset of the black brane instability, $k=k_\ast$, is the same as Eq.~\eqref{negmodeG} for traceless-transverse negative modes of the Euclidean partition function of black holes, if we identify $\lambda=-k_\ast^2$. Recall also Section~\ref{sec:negmodestab}, where we show how negative modes arise from local thermodynamic instabilities. If there is a negative eigenvalue of the thermodynamic stability Hessian, there will be a negative mode $h_{ab}$ of the black hole partition function. Hence, there will also be a threshold for a classical Gregory-Laflamme instability of the associated black branes.

The stationary perturbation mode with $k=k_\ast$ is also the perturbative signal of the bifurcation to a new family of non-uniform black brane solutions \cite{Gubser:2001ac,Wiseman:2002zc}. Recall in particular Figure~\ref{fig:nonuniform} in Section~\ref{sec:classstabreview}.

\subsection{Rotating black branes are unstable} \label{subsec:stringinstab}

We shall now prove, based on \cite{Dias:2010eu}, that the condition for local thermodynamic stability fails for \emph{all} asymptotically flat vacuum black holes. According to the discussion above, this implies that \emph{all} black branes of the type \eqref{blackbrane} constructed with such a black hole are classically unstable.

The proof goes as follows. Consider the Legendre transform of the entropy,
\be
W = S- \beta M+ \beta \Omega_i J_i\,,
\ee
in terms of which the first law is expressed as $dW=-Md\beta+\beta\Omega_idJ_i$.
Recall from Section~\ref{sec:thermalstab}, expression \eqref{localstabW}, that local thermodynamic stability exists if and only if
\begin{equation}
\label{eq:ClW}
W_{\alpha\beta} \equiv {{\dderf{W(\tilde{y}_\gamma)}{\tilde{y}_\alpha}{\tilde{y}_\beta}}} \,, \qquad \; \tilde{y}_\alpha=(\beta,-\beta \Omega_i)\,, \qquad \text{is positive definite.}
\end{equation}
A sufficient condition for $W_{\alpha\beta}$ not being positive definite is
\be
\label{w00i}
W_{00} = \frac{\partial^2 W(\tilde{y}_\alpha)}{\partial \beta^2} = - \left( \frac{\partial M}{\partial \beta} \right)_{\beta \Omega} <0\,.
\ee

Now, the Smarr relation, valid for asymptotically flat vacuum black holes, reads
\be
\label{smarr}
\frac{D-3}{D-2}\, M = T S - \Omega_i J_i  \qquad \Leftrightarrow \qquad M = -(D-2) WT\,.
\ee
Hence we have for such black holes, using the first law again,
\be
\label{w00f}
W_{00} = -(D-3) MT <0\,,
\ee
which implies that local thermodynamic stability fails. It then follows that any vacuum black hole solution must admit a thermodynamic negative mode, and that any black brane based on such a black hole solution must always be classically unstable. This explains our results for the Kerr black hole (Chapter~\ref{cha:KerrAdS}) and for Myers-Perry black holes with a single spin (Chapter~\ref{cha:MPsingle}) and equal spins (Chapter~\ref{cha:MPequal}). We find that at least one continuous negative mode exists for any value of the rotation parameter, so that the black branes are unstable. In fact, we find for Myers-Perry black holes, which may exhibit more than one negative mode, that this is the \emph{most negative}.

Notice that this negative mode must be present for all values of the rotation, and thus there is no critical rotation for which it reduces to a {\it zero}-mode. Hence there cannot be a classical instability of the black hole associated with this particular mode.

\section{Zero-modes and black hole stability \label{sec:zeromodeultraspin}}

In this Section, we will argue that a stationary zero-mode of the black hole can correspond: ($i$) to a change in the parameters of the solution, if the zero-mode can be predicted by the thermodynamic Hessian, or ($ii$) to the threshold of a classical instability of the black hole, not just the black brane. We will also conjecture that classical instabilities associated with stationary zero-modes can only appear in a regime which we call \emph{ultraspinning}. We analyse this regime for the Myers-Perry family of solutions.

\subsection{The ultraspinning regime \label{subsec:ultraspin}}

In Ref.~\cite{Emparan:2003sy}, rapidly-rotating Myers-Perry black holes with a single spin in $D\geq6$ were conjectured to be unstable for Gregory-Laflamme-type modes. Such black holes have quasi-extended event horizons for large values of the angular momentum compared to the mass ($|J|/M^{\frac{D-2}{D-3}} \gg 1$), acquiring some properties of black branes, which are Gregory-Laflamme-unstable. An order of magnitude estimate for the threshold of the black hole instability was given in \cite{Emparan:2003sy} by considering the thermodynamic behaviour. For small rotations, the temperature decreases with the rotation for fixed mass, as it happens for the Kerr black hole, which is expected to be classically stable. However, after a critical value of the rotation, the temperature actually increases, as it happens for black branes. Classical stability was conjectured to fail roughly after the critical rotation.

The critical rotation is actually a thermodynamic zero-mode of the Hessian \eqref{eq:ClW}, beyond which the black hole possesses \emph{two} thermodynamic instabilities. To understand this, recall our discussion in Section~\ref{sec:negmodestab}. Since a traceless-transverse negative mode $h_{ab}$ is a regular tensor on the black hole background $g_{ab}$, it follows that $T$ and $\Omega_i$ are left unchanged by $h_{ab}$. A negative mode, which is an off-shell perturbation in the path integral, occurs when the thermodynamic Hessian has a negative eigenvalue. However, this argument also applies if a negative mode is continuously connected through the variation of the black hole parameters to a \emph{zero}-mode ($\lambda=0$), which is a classical perturbation of the black hole. This will be the case if one of the eigenvalues of the thermodynamic Hessian changes from positive to negative continuously, marking the appearance of a new local thermodynamic instability. It is now more convenient to use the thermodynamic Hessian
\be
-S_{\alpha\beta}\equiv -\,{\dderf{S(x_\gamma)}{x_\alpha}{x_\beta}} \,, \qquad \; x_\alpha=(M,J_i)\,,
\ee
which is simply the inverse matrix of $W_{ab}$ (recall Appendix~\ref{app:localstab}). If the zero-mode $h_{ab}$ preserves $T$ and $\Omega_i$ then, using the first law, it preserves $(\partial S/\partial M)_J=1/T$ and $(\partial S/\partial J_i)_M=-\Omega_i/T$, that is, it preserves $\partial S/\partial x^\alpha$. Hence it must correspond to an eigenvector of $S_{\alpha \beta}$ with eigenvalue zero:
\be
0= \delta(\partial_\alpha S) = \delta x^\beta\partial_\beta \partial_\alpha S = S_{\alpha \beta} \delta x^\beta.
\ee
Notice that the corresponding thermodynamic instability occurs only for a certain range of the rotation, beyond the zero-mode, while the thermodynamic instability shown to exist for all vacuum black holes in the last Section occurs for any value of the rotation. These instabilities correspond to distinct negative eigenvalues of the thermodynamic Hessian, and hence to distinct Gregory-Laflamme instabilities of the black branes. This is a refinement of the Gubser-Mitra conjecture.

Ref.~\cite{Dias:2009iu} confirmed numerically this picture and the conjecture of Ref.~\cite{Emparan:2003sy}, showing that a new negative mode appears at the critical rotation and, more importantly, that additional negative modes which are not thermodynamic in origin occur for higher rotations. These are thresholds for \emph{classical} instabilities of the black hole, as explicitly verified for the equal spin case in \cite{Dias:2010eu}, on which Chapter~\ref{cha:MPequal} is based. Notice that the zero-modes which are thermodynamic in origin can be identified by the simpler Hessian matrix
\be
H_{ij}\equiv -\lp{\dderf{S}{J_i}{J_j}}\rp_M= -S_{ij}\,,
\ee
due to the identity
\be
\mathrm{det}(-S_{\alpha\beta}) = -\frac{1}{(D-3) MT}\; \mathrm{det}(H_{ij})\,,
\ee
valid for asymptotically flat vacuum black holes; for a proof of this identity, see Appendix~\ref{app:determinants}. It follows that, for a black hole parameterised by $(M,J_i)$, additional negative eigenvalues of $-S_{\alpha\beta}$ correspond precisely to negative eigenvalues of $H_{ij}$.

In the Myers-Perry case, for fixed $M$, the eigenvalues of $H_{ij}$ are all positive for small enough angular momenta. However, as some or all of the angular momenta are increased, an eigenvalue of $H_{ij}$ may become negative. If we consider the space parameterised by $J_i$ (for fixed $M$), there is some region containing the origin in which $H_{ij}$ is positive definite. We define the boundary of this region to be the {\it ultraspinning surface}. Following Ref.~\cite{Dias:2009iu}, we shall say that a given black hole is ultraspinning if it lies outside the ultraspinning surface. From the above arguments, we know that as one crosses the ultraspinning surface, the black hole will develop a new negative mode, and the associated black branes will develop a new classical instability. This is in addition to the instability already present at low angular momenta. Furthermore, on the ultraspinning surface, the new negative mode must reduce to a stationary zero-mode that corresponds to a variation of parameters within the Myers-Perry family of solutions, since they are identified by the Myers-Perry equation of state $S(M,J_i)$.

Ref.~\cite{Dias:2009iu} conjectured that classical instabilities whose threshold is a stationary and axisymmetric zero-mode occur only for rotations higher than a thermodynamic zero-mode, i.e. in the ultraspinning regime. We emphasise that our conjecture gives a necessary condition for an instability, not a sufficient one.

The intuition leading to the conjecture is that modes of lower symmetry are usually the most unstable ones. For instance, the original Gregory-Laflamme instability occurs for the ``s-wave'' of the transversal black hole. An additional classical instability will arise after a critical value of the rotation, and it will correspond to a ``p-wave'' of the transversal black hole. As the rotation is increased, higher order waves may become unstable. Now, if we consider a black hole, instead of a black brane, the ``s-wave'' and the ``p-wave'' are associated with the asymptotic charges, mass and angular momenta. Therefore they are associated to purely thermodynamic instabilities. Higher order waves, on the other hand, may become classically unstable as the rotation is increased, starting with the ``d-wave''. Notice that these waves do not affect the asymptotic charges.\footnote{Recall that, in Section~\ref{sec:localstab}, we discussed how the thermodynamic Hessian gave the stability with respect to mass/angular momenta exchanges with a ``reservoir''. If the asymptotic charges are unaltered, no ``reservoir'' is invoked, and it is the internal stability of the black hole which is being probed.} The results of Refs.~\cite{Dias:2009iu,Dias:2010eu}, described in the next two Chapters, clarify the harmonic structure underlying this problem.

Recall also that the thresholds of the classical instabilities should be associated with bifurcations to different black hole families, highlighting the connection between stability and uniqueness.

\subsection{Zero-modes of the Myers-Perry family \label{subsec:ultraspinMP}}

\begin{figure}[t]
\centerline{\includegraphics[width=.45\textwidth]{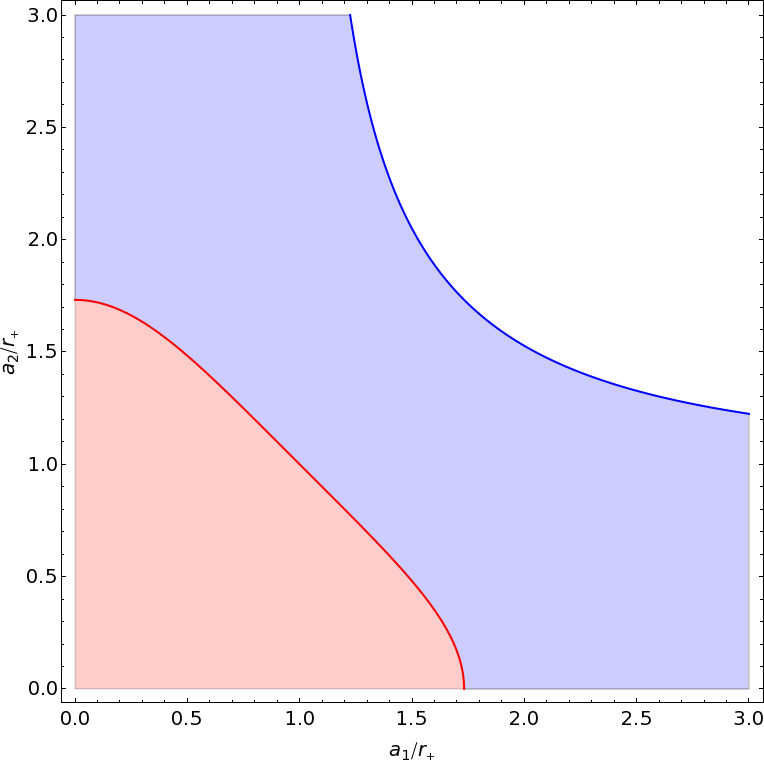}
\hspace{2cm}\includegraphics[width=.45\textwidth]{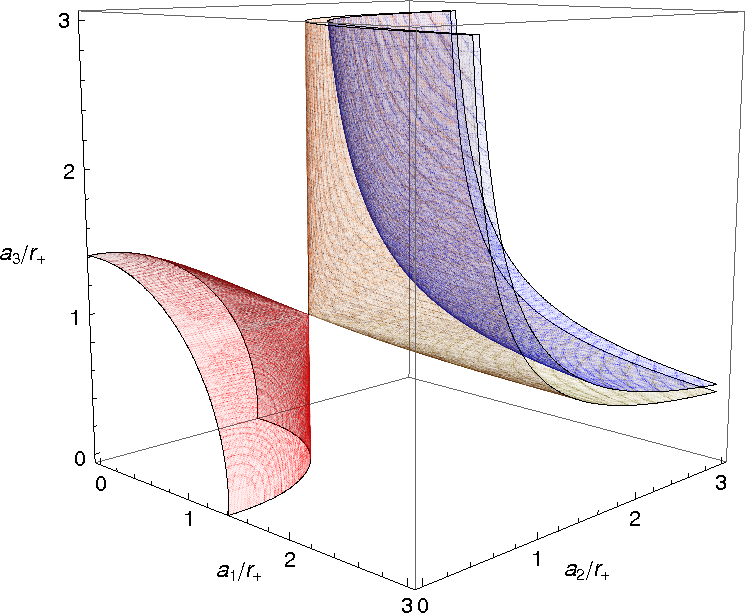}}
\caption{Parameter space for MP black holes with $D=6$ (left) and $D=7$ (right). The black hole is labelled by the horizon radius $r_+$ and the spin parameters $a_i$ which we take to be positive for clarity. For $D=6$, the blue curve corresponds to extreme black holes. In the red region, both eigenvalues of $H_{ij}$ are positive. In the blue region, corresponding to ultraspinning black holes, one eigenvalue is positive and the other is negative. For $D=7$, the blue surface corresponds to extreme black holes. The ultraspinning surface is the red surface near the origin. Inside this surface, $H_{ij}$ is positive definite. The orange surface is where another eigenvalue of $H_{ij}$ vanishes. Between the red and orange surfaces, two eigenvalues of $H_{ij}$ are positive and one is negative. Between the orange and blue surfaces, one eigenvalue of $H_{ij}$ is positive and two are negative. Ultraspinning black holes correspond to points between the red and blue surfaces. Note that the ``cusp" where the red and orange surfaces meet has equal spins.}
\label{fig:ultraspinning}
\end{figure}

Let us now examine the particular form of $H_{ij}$ in the case of Myers-Perry solutions. For $D=5$, there is no ultraspinning region, i.e. $H_{ij}$ is always positive definite. For $D=6,7$, we find a more interesting behaviour, as shown in Figure~\ref{fig:ultraspinning}. We parameterise the black hole by the horizon radius $r_+$ and the spin parameters $a_i \sim J_i/M$ (defined in Section \ref{subsec:MPsolution}), rather than $M$ and $J_i$, in order to produce clearer figures.
For $D=6$, $H_{ij}$ is a $2 \times 2$ matrix, and the ultraspinning ``surface" is a closed curve that encloses the origin. Inside this curve, both eigenvalues of $H_{ij}$ are positive but outside one is positive and one is negative. Notice that all extreme black holes are ultraspinning.

For $D=7$, $H_{ij}$ is a $3 \times 3$ matrix. The ultraspinning surface encloses the origin. However, now there is another surface on which a second eigenvalue of $H_{ij}$ changes from positive to negative. This surface lies between the ultraspinning surface and the surface corresponding to extreme black holes, and touches the ultraspinning surface at a point with equal $a_i$. For generic $a_i$, if one gradually scales up the $a_i$ then one eigenvalue of $H_{ij}$ becomes negative as the ultraspinning surface is crossed, and another eigenvalue becomes negative as the other surface is crossed. As each surface is crossed, the black hole should develop a new stationary negative mode, and the associated black branes should develop a new instability.

Some $D=7$ black holes are special, e.g. solutions for which one of the angular momenta vanishes do not intersect the second surface. For instance, consider the singly-spinning case. All such black holes must possess the thermodynamic negative mode predicted in Section~\ref{subsec:stringinstab}, and a second negative mode should appear at the ultraspinning surface, but there should be no further thermodynamic negative modes. This is consistent with the results of Ref.~\cite{Dias:2009iu}, where it was found that a stationary negative mode does indeed appear at the ultraspinning surface. As we shall see in the next Chapter, new and infinitely many stationary {\it non-thermodynamic} negative modes appear at larger angular momenta, the first of which marks the threshold of instability of the black hole.

Using the expressions in Section \ref{subsec:MPsolution}, we can actually derive an explicit expression for the reduced thermodynamic Hessian,
\bea
H_{ij} &=& \frac{(D-2)\pi}{M \kappa} \Big\{ \frac{r_+^2-a_i^2}{(r_+^2+a_i^2)^2} \; \delta_{ij} \nonumber \\
&+& 2 \; \frac{\Omega_i \Omega_j}{\kappa} \left[ \frac{r_+}{(r_+^2+a_i^2)^2} + \frac{r_+}{(r_+^2+a_j^2)^2} - \frac{1}{2r_+}+ \frac{\tilde{\Omega}^2}{\kappa} \right]
\Big\}\,,
\eea
where there is no sum over $i,j$ and we have defined $\tilde{\Omega}^2 \equiv \sum_i \Omega_i^2$. The matrix is positive definite in the static case, $a_i=0$.

In the singly-spinning case, say $a_1 \neq 0$, we have
\bea
H_{11} &=& \frac{2(D-2)(D-3)\pi\,r_+(r_+^2+a^2)}{M\,[(D-3)r_+^2+(D-5)a^2]^3} \,[(D-3)r_+^2-(D-5)a^2]\,, \nonumber \\
H_{ij} &=& \frac{(D-2)\pi}{M \kappa\, r_+^2} \delta_{ij} \qquad \textrm{for} \quad (i,j)\neq(1,1) \,.
\eea
There is a single zero-mode in $D>5$, as we see in Figure~\ref{fig:ultraspinning}, which occurs for
\be
\label{MPsingleacrit}
\lp \frac{a}{r_+} \rp^2 = \frac{D-3}{D-5}\,.
\ee
The associated eigenvector is $\delta_{i1}$, so that the angular momenta which vanish in the background solution are not excited by the perturbation, i.e. the zero-mode keeps the black hole within the singly-spinning Myers-Perry family. The ultraspinning regime occurs for rotations higher than \eqref{MPsingleacrit}.

In the equal spins case, we have
\bea
H_{ij} = \frac{(D-2)\pi}{M \kappa} \Big\{ \frac{r_+^2-a^2}{(r_+^2+a^2)^2} \; \delta_{ij} + 2 \; \frac{\Omega^2}{\kappa} \left[ \frac{2 r_+}{(r_+^2+a^2)^2} - \frac{1}{2r_+}+ \frac{n \Omega^2}{\kappa} \right]\,Q_{ij}
\Big\}\,,
\eea
where $Q_{ij} =1\; \forall i,j$. An eigenvector $V_i$ of $H_{ij}$ must then be an eigenvector of $Q_{ij}$, which leaves only two options: the eigenvector is such that $V_i=V\, \forall i$, or is such that $\sum_i V_i=0$. In the former case, there can be no zero-mode, since this would require
\be
\frac{r_+^2-a^2}{(r_+^2+a^2)^2} + 2 \; \frac{\Omega^2}{\kappa} \left[ \frac{2 r_+}{(r_+^2+a^2)^2} - \frac{1}{2r_+}+ \frac{n \Omega^2}{\kappa} \right] n=0\,,
\ee
which can be simplified to $a^2 = - (D-3)r_+^2/(2-\epsilon)$, and thus has no real solution for $a$ (recall that $\epsilon=0,1$ for odd and even $D$, respectively). Hence no instability occurs for modes that preserve the equality between the spins, which is consistent with the results of \cite{Kleihaus:2007dg}. However, the eigenvectors satisfying $\sum_i V_i=0$ do change sign once at $|a|=r_+$.\footnote{Notice that, in Chapter~\ref{cha:MPequal}, we will use the radial variable of Ref.~\cite{Kunduri:2006qa}, which is related to the variable presently used by $\tilde{r}^2 = r^2+a^2$, so that the ultraspinning surface is at $|a|=\tilde{r}_+/\sqrt{2}$.} There is one such eigenvalue in $D=6$, and in $D=7$ there are two, as shown in Figure~\ref{fig:ultraspinning}. The associated eigenvectors break the symmetry between the spins, so that the perturbed black hole is a Myers-Perry solution which is not in the equal spins sector. Due to the presence of the extremality bound, it is not clear whether the rotation can be sufficiently high to excite classical instabilities in the ultraspinning regime. However, we shall see in Chapter~\ref{cha:MPequal} that this indeed happens for $D=9$ (and, we believe, for higher $D$, at least in the odd $D$ case for which the solutions are codimension-1).

\begin{subappendices}
\section{Appendix: Thermodynamic determinants \label{app:determinants}}

We wish to show that
\be
\mathrm{det}(-S_{\alpha\beta}) = -\frac{1}{(D-3) MT} \,\mathrm{det}(H_{ij})\,,
\ee
where $S_{\alpha\beta} \equiv \partial^2 S / \partial x^\alpha \partial x^\beta$, with $x^\alpha=(M,J_i)$, and $H_{ij} \equiv \partial^2 (-S) / \partial J_j \partial J_j = -S_{ij}$. This identity follows directly from the linear algebra result \eqref{detidentity}, if we take $C_{ij}=H_{ij}$. The proportionality coefficient is given by
\be
\sigma = -S_{00} - H^{-1}_{ij} S_{0i} S_{0j} = - \left(\frac{\partial \beta}{\partial M}\right)_J + \left(\frac{\partial (\beta \Omega_j)}{\partial J_i}\right)_M^{-1} \left(\frac{\partial (\beta \Omega_i)}{\partial M}\right)_J \left(\frac{\partial (\beta \Omega_j)}{\partial M}\right)_J\,,
\ee
where we have used the first law of thermodynamics, $dS= \beta dM -\beta \Omega_i dJ_i$. Since 
\be
\left(\frac{\partial (\beta \Omega_j)}{\partial J_i}\right)_M^{-1} = \left(\frac{\partial J_j}{\partial (\beta \Omega_i)}\right)_M \qquad \mathrm{and} \qquad
\left(\frac{\partial J_j}{\partial M}\right)_{\beta \Omega} = \left(\frac{\partial J_j}{\partial (\beta \Omega_i)}\right)_M \left(\frac{\partial (\beta \Omega_j)}{\partial M}\right)_J\,,
\ee
we get
\be
\sigma= - \left(\frac{\partial \beta}{\partial M}\right)_J + \left(\frac{\partial J_j}{\partial M}\right)_{\beta \Omega} \left(\frac{\partial (\beta \Omega_j)}{\partial M}\right)_J\,.
\ee
The identity
\be
\left(\frac{\partial \beta}{\partial M}\right)_J = \left(\frac{\partial \beta}{\partial M}\right)_{\beta \Omega} - \left(\frac{\partial \beta}{\partial J_j}\right)_M \left(\frac{\partial J_j}{\partial M}\right)_{\beta \Omega}
\ee
and the $S_{0i}=S_{i0}$ ``Maxwell relation'', $\left(\partial (\beta \Omega_j) / \partial M \right)_J = \left(\partial \beta / \partial J_j \right)_M $, further imply that
\be
\sigma= - \left(\frac{\partial \beta}{\partial M}\right)_{\beta \Omega} = - \left(\frac{\partial M}{\partial \beta}\right)_{\beta \Omega}^{-1} \,.
\ee
Following the steps taken between Eqs.~\eqref{w00i} and \eqref{w00f}, we finally obtain
\be
\sigma = -\frac{1}{(D-3) MT} <0\,.
\ee
These steps require that the Smarr relation \eqref{smarr} is valid, i.e. they apply only to asymptotically flat vacuum black holes.

\end{subappendices}

%% file: MPsingle.tex
\chapter{Singly-spinning Myers-Perry instability} \label{cha:MPsingle}

In this Chapter, based on \cite{Dias:2009iu}, we will confirm numerically the conjecture that Myers-Perry black holes with a single spin become unstable at a sufficiently large value of the rotation, and that new black holes with pinched horizons appear at the threshold of the instability. We search numerically, and find, the stationary axisymmetric perturbations that mark the onset of the instability and the appearance of the new black hole phases.

\section{Introduction}

In higher-dimensional spacetimes a vast landscape of novel black holes has begun to be uncovered, as discussed in Section~\ref{sec:classstabreview}. Its layout -- i.e. the connections between different classes of black holes in the space of solutions -- hinges crucially on the analysis of their classical stability: most novel black hole phases are conjectured to branch-off at the threshold of an instability of a known phase. Showing how this happens was an outstanding open problem first addressed in Ref.~\cite{Dias:2009iu}.

The best known class of higher-dimensional black holes, discovered by Myers and Perry (MP) in \cite{Myers:1986un} and described in Section~\ref{subsec:MPsolution}, appears in many respects as the natural generalisation of the Kerr solution. In particular, the event horizon is topologically spherical. However, the actual shape of the horizon can differ markedly from the four-dimensional one, which is always approximately round with a radius parametrically $\sim M$. This is not so in $D\geq 6$. Considering for simplicity the case where only one spin $J$ is turned on (of the $n=\lfloor (D-1)/2 \rfloor$ independent angular momenta available), it is possible to have black holes with arbitrarily large $J$ for a given mass $M$. These black holes fall into the \textit{ultraspinning regime} defined in the last Section. The horizon of these black holes spreads along the rotation plane out to a radius $a\sim J/M$ much larger than the thickness transverse to this plane, $r_+\sim (M^3/J^2)^{1/(D-5)}$.

\begin{figure}[t]
\centerline{\emph{A }: \hspace{.6\textwidth} \hspace{2cm}}
\centerline{\emph{B }: \includegraphics[width=.6\textwidth]{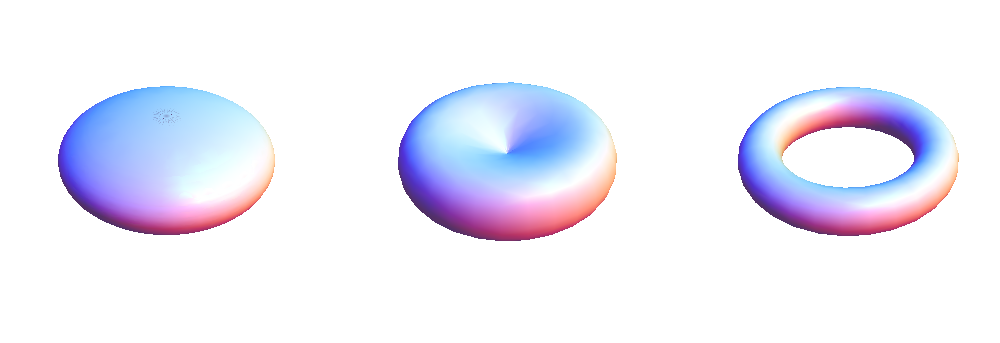} \hspace{2cm}}
\centerline{\emph{C }: \includegraphics[width=.6\textwidth]{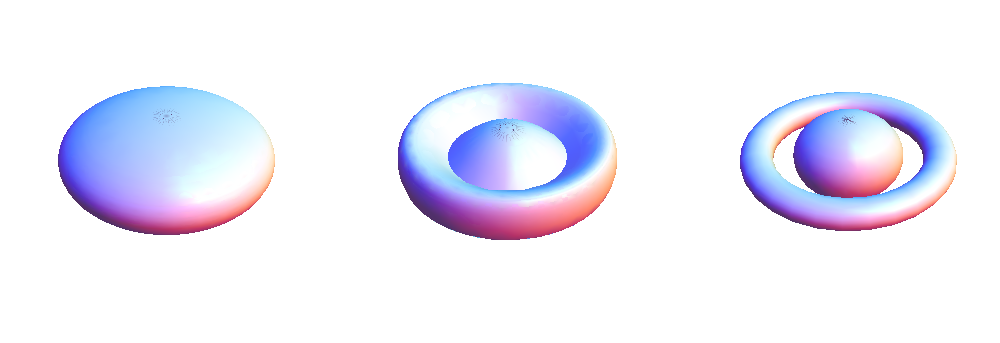} \hspace{2cm}}
\centerline{\phantom{\emph{A }:} \hspace{1cm} \includegraphics[width=.82\textwidth]{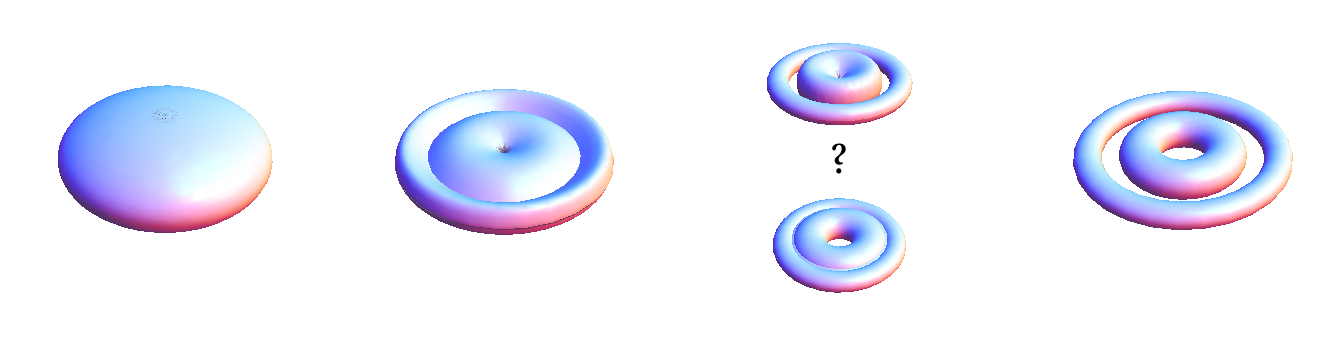}}
\caption{Conjectured connections between a $D\geq 6$ singly-spinning MP black hole and a black ring ($A$), a black Saturn ($B$) or a di-ring ($C$), through a family of pinched solutions. The bifurcation from the MP family occurs at the points $A$, $B$ and $C$, respectively, in the phase diagram of Figure~\ref{fig:phases}.}
\label{fig:pinchbhs}
\end{figure}

\begin{figure}
\centering
\includegraphics[width = 7 cm]{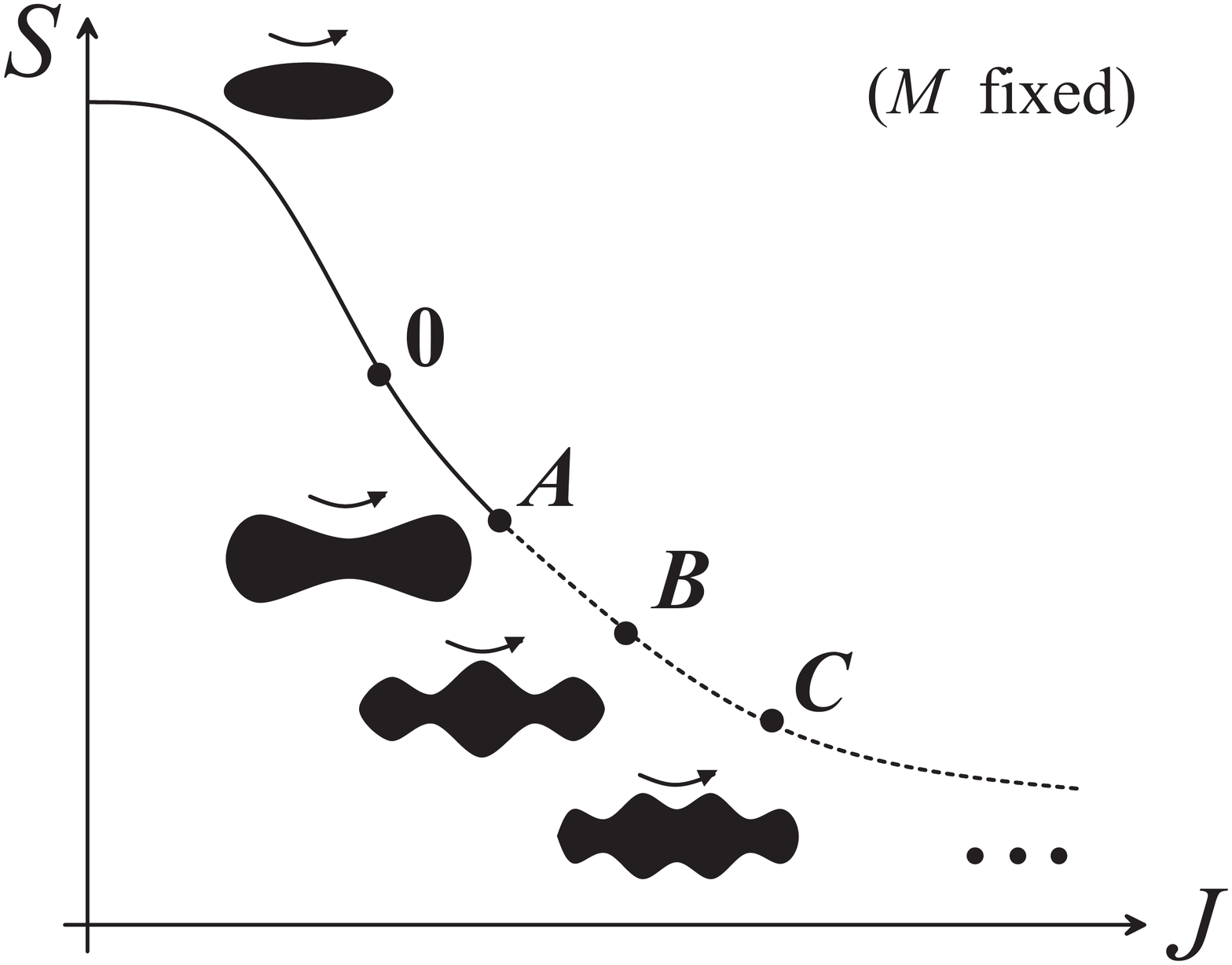}
\caption{\label{fig:phases} Diagram of entropy versus angular momentum, at fixed mass, for singly-spinning MP black holes in $D\geq 6$ illustrating the conjecture of \cite{Emparan:2003sy} (see also \cite{Emparan:2007wm}): at sufficiently large spin the MP solution becomes unstable, and at the threshold of the instability a new branch of black holes with a central pinch appears ($A$). As the spin grows new branches of black holes with further axisymmetric pinches ($B,C,\dots$) appear. We determine numerically the points where the new branches appear, but it is not yet known in which directions they run. We also indicate that at the inflection point ($0$), where $(\partial^2 S/\partial J^2)_M=0$, there is a stationary perturbation that should not correspond to an instability nor a new branch, but rather to a thermodynamic zero-mode that moves the solution along the curve of singly-spinning MP black holes.}
\end{figure}

This fact was picked out by Emparan and Myers \cite{Emparan:2003sy} as an indication of an instability and a connection to novel black hole phases. In more detail, in the limit $a\to\infty$ with $r_+$ fixed, the geometry of the black hole in the region close to the rotation axis approaches that of a black brane. Black branes are known to exhibit classical instabilities \cite{Gregory:1993vy}, at whose threshold a new branch of black branes with inhomogeneous horizons appears \cite{Gubser:2001ac,Wiseman:2002zc}. Ref.~\cite{Emparan:2003sy} conjectured that this same phenomenon should be present for MP black holes at finite but sufficiently large rotation: they should become unstable beyond a critical value of $a/r_+$, and the marginally stable solution should admit a stationary, axisymmetric perturbation signalling a new branch of black holes pinched along the rotation axis. Simple estimates suggested that in fact $(a/r_+)_\mathrm{crit}$ should not be much larger than one. As $a/r_+$ increases, the MP solutions should admit a sequence of stationary perturbations, with pinches at finite latitude, giving rise to an infinite sequence of branches of `pinched black holes'. Ref.~\cite{Emparan:2007wm} argued that this structure is indeed required in order to establish connections between MP black holes and the black ring and black Saturn solutions more recently discovered. See the idea in Figure~\ref{fig:pinchbhs}, which is analogous to Figure~\ref{fig:nonuniform} for the non-uniform branes, in Section~\ref{sec:classstabreview}.
See also how this picture arises in the black hole phase diagram of Figure~\ref{fig:phases}.

Our main result is a numerical analysis that verifies the conjecture of Ref.~\cite{Emparan:2003sy}.

\section{Background solution}

The solution for a MP black hole rotating in a single plane is
\begin{eqnarray}
ds^2 \!\!&=&\!\!
-\frac{\Delta(r)}{\Sigma^2(r,\theta)}\left[ dt+a\sin^2\theta \,d\phi \right]^2 +
\frac{\sin^2\theta}{\Sigma^2(r,\theta)}\left[ (r^2+a^2)d\phi-a\,dt\right]^2 \nonumber\\
&&
+\frac{\Sigma^2(r,\theta)}{\Delta(r)}dr^2+
\Sigma^2(r,\theta)\,d\theta^2
+r^2\cos^2\theta\, d\Omega^2_{(D-4)}\,,\label{mpbh}
\end{eqnarray}
\begin{equation}
\Sigma(r,\theta)^2=r^2+a^2\cos^2\theta\,,\qquad \Delta(r)=r^2+a^2-\frac{r_m^{D-3}}{r^{D-5}}\,.
\end{equation}
The parameters here are the mass-radius $r_m$ and the rotation-radius $a$,
\be
r_m^{D-3}=\frac{16\pi GM}{(D-2)\Omega_{D-2}}\,,\qquad a=\frac{D-2}{2}\frac{J}{M}\,.
\ee
The event horizon lies at the largest real root $r=r_+$ of $\Delta$.

The linearised perturbation theory of the Kerr black hole ($D=4$) was disentangled in \cite{Teukolsky:1973ha} using the Newman-Penrose
formalism. Attempts to extend this formalism to decouple a master equation for the linear perturbations of MP black holes have not yet been successful; see \cite{Durkee:2010xq} for recent progress. Moreover, even though some subsectors of perturbations for some classes of MP black holes had been decoupled before the work described here \cite{Dias:2009iu}, e.g. \cite{Ishibashi:2003ap,Kunduri:2006qa,Murata:2008yx,Oota:2008uj,Kodama:2009bf}, none of them had shown signs of any instability and they do not contain the precise kind of perturbations we are interested in.\footnote{Notice that \cite{Shibata:2009ad,Shibata:2010wz} and \cite{Dias:2010eu}, on which the next Chapter is based, are posterior to \cite{Dias:2009iu}.} Thus we take a more frontal numerical approach to a full set of coupled partial differential equations (PDEs).

\section{The eigenvalue problem}

We intend to solve for a stationary linearised perturbation $h_{ab}$ around the background \eqref{mpbh}. Choosing the traceless-transverse (TT)
gauge, $h^a_{\phantom{a}a}=0$ and $\nabla^a h_{ab}=0$, the equations to
solve are
\begin{equation} (\triangle_L h)_{ab}= 0\,,
\label{Lichnerowicz}
\end{equation}
where $\triangle_L$ is the Lichnerowicz operator defined in \eqref{deflichn}. Actually, we solve the more general eigenvalue problem
\begin{equation} \label{eigenh}
(\triangle_L h)_{ab} =-k^2 h_{ab}\,.
\end{equation}
As we discussed in the previous Chapter, this problem appears in two contexts. Eq.~\eqref{eigenh} determines the stationary perturbations of a black
brane obtained by adding flat directions $\vec{z}$ to the line element \eqref{mpbh}, for perturbations with a profile $e^{\ii\, \vec{k}\cdot\vec{z}} h_{ab}$. Thus such modes with $k>0$ correspond to the threshold of the Gregory-Laflamme instability of black branes \cite{Gregory:1993vy}. The same equations also describe the negative modes of quadratic quantum corrections to the gravitational Euclidean partition function \cite{Gibbons:1978ji,Gross:1982cv}. In Chapter~\ref{cha:KerrAdS}, based on \cite{Monteiro:2009ke}, we studied this problem for the Kerr black hole, and showed the existence of a branch of solutions extending the negative Schwarzschild mode ($k_\mathrm{Sch}\neq 0$) \cite{Gross:1982cv} to finite rotation, with $k$ growing as the rotation increases towards the extremality bound.

If the ultraspinning instability is present for MP black holes in $D\geq 6$, then, in addition to the analogue of the branch studied in \cite{Monteiro:2009ke}, a new branch of negative modes extending to $k=0$ must appear. The eigenvalue $k=0$ corresponds to a (perturbative) stationary solution with a slightly deformed horizon. In fact, as explained above, we expect an infinite sequence of such branches that reach $k=0$ at increasing values of the rotation. The solutions for $k>0$ imply new kinds of Gregory-Laflamme instabilities and inhomogeneous phases of ultraspinning black branes (see also \cite{Kleihaus:2007dg}).

Another reason to consider \eqref{eigenh} instead of trying to solve directly for $k=0$ is that there exist powerful numerical methods for eigenvalue problems that give the eigenvalues $k$ together with the eigenvectors, i.e. the metric perturbations; see the Appendix at the end of this thesis.

The modes we seek preserve the $SO(D-3)\times SO(2)$ rotational symmetries of the singly-spinning MP solution and depend only on the radial and polar
coordinates, $r$ and $\theta$ \cite{Emparan:2003sy}. Thus we take the ansatz
\begin{eqnarray}
ds^2 \!\!&=&\!\!
-\frac{\Delta(r)}{\Sigma^2(r,\theta)}\,e^{\delta\nu_0}\left[
dt+a\sin^2\theta \,e^{\delta\omega}\,d\phi \right]^2 +
\frac{\sin^2\theta}{\Sigma^2(r,\theta)}\,e^{\delta\nu_1}\left[
(r^2+a^2)d\phi-a\, e^{-\delta\omega}\,dt\right]^2 \nonumber\\
&+&
\frac{\Sigma^2(r,\theta)}{\Delta(r)}\,e^{\delta\mu_0}\left[dr+\delta\chi\,
\sin\theta \,d\theta \right]^2+
\Sigma^2(r,\theta)\,e^{\delta\mu_1}\,d\theta^2
+r^2\cos^2\theta\,e^{\delta\Phi}\,
d\Omega^2_{(D-4)}\,,\label{ansatz}
\end{eqnarray}
where $\{\delta\nu_0,\delta\nu_1,\delta\mu_0,\delta\mu_1,\delta\omega,\delta\chi,\delta\Phi\}$, which are functions of $(r,\theta)$, describe our perturbations. These functions will be determined numerically by solving the eigenvalue problem \eqref{eigenh} subject to appropriate boundary conditions. After imposing the TT gauge, Eq.~\eqref{eigenh} reduces to four coupled PDEs for $\delta\mu_0$, $\delta\mu_1$, $\delta\chi$ and $\delta\Phi$ (the TT conditions then give $\delta\nu_0$, $\delta\nu_1$ and $\delta \omega$), in a procedure equivalent to the one described in Chapter~\ref{cha:KerrAdS}, based on \cite{Monteiro:2009ke}, for the Kerr-AdS black hole.

The boundary conditions are discussed in detail in the Appendix \ref{app:mpsingleBC}. For the numerical implementation, it is convenient to use the new
radial and polar variables
 \be y=1-\frac{r_+}{r}\,,\qquad x=\cos\theta\,, \ee
which are dimensionless and take values in the interval $[0,1]$. Dirichlet boundary conditions are simpler to implement, and so we will use the following perturbation functions:
\be
\label{MPs:newfunc} 
\begin{array}{l}
\displaystyle{q_1(y,x) =  \left( 1-\frac{r_+}{r} \right) x(1-x)} \,\delta\mu_0(y,x)\,, \qquad \qquad
\displaystyle{q_2(y,x) = r_m^{-1} (1-x)}\,\delta\chi(y,x) \,, \\
\displaystyle{q_3(y,x) =  \left( 1-\frac{r_+}{r} \right)x(1-x)} \,\delta\mu_1(y,x)\,,  \qquad \qquad
\displaystyle{q_4(y,x) =  \left( 1-\frac{r_+}{r} \right) x(1-x)} \,\delta\Phi(y,x) \,, \\
\end{array}
\ee
which vanish at the boundaries $r=r_+$, $r=\infty$, $x=0$ and $x=1$ when the boundary conditions determined in the Appendix \ref{app:mpsingleBC} are enforced.

It is important to ensure that the eigenmodes we find are not pure gauge, $h_{ab}=\nabla_{(a}\xi_{b)}$. This is trivial when $k \neq 0$, since $\triangle_L \nabla_{(a}\xi_{b)}=0$, but not for zero-modes. We can prove that in the TT gauge, where the equations for the gauge parameter $\xi_{a}$ are separable, pure gauge perturbations within our ansatz necessarily diverge at the horizon or at infinity (the proof is along the lines of the equal spins MP case, in the Appendix~\ref{sec:NoPureGauge}). Thus, with our boundary conditions, the zero-modes are never pure gauge.

\section{Results and discussion}

We use again, as in Chapter~\ref{cha:KerrAdS} describing \cite{Monteiro:2009ke}, a spectral numerical method to solve the eigenvalue problem; see the Appendix at the end of the thesis. We have carried out the calculations for $D=7,8,9$. The cases $D=5$ (where there is no ultraspinning regime) and
$D=6$ were left out in \cite{Dias:2009iu} due to numerical difficulties which have since been solved.\footnote{Ref.~\cite{Dias:2010ma} was published after the submission of this thesis.} For $D=5$, there is no instability of this type. For $D=6$, the results conform to our expectations, being similar to the ones for $D=7,8,9$ discussed here.

\begin{figure}
\centering{ \includegraphics[width = 6cm]{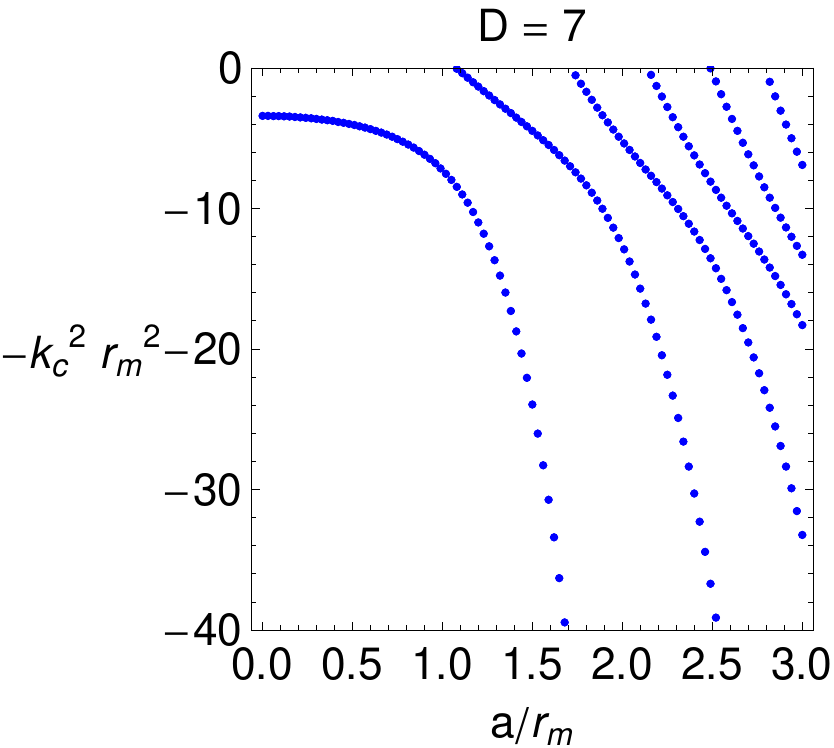} \hspace{1cm} \includegraphics[width = 6cm]{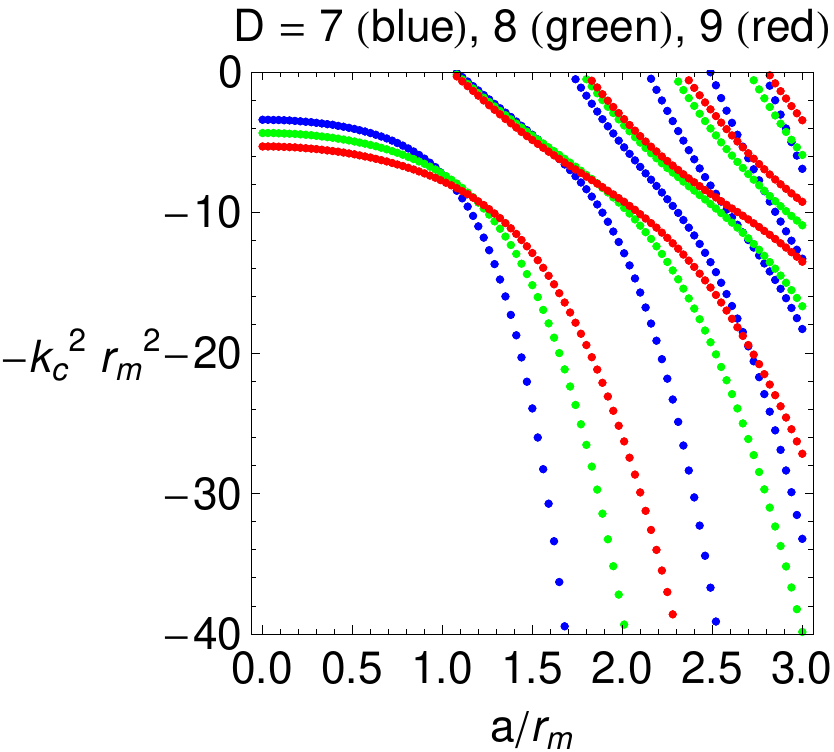}}
\caption{\label{fig:negmode789d} Negative eigenvalues for the MP black hole in $D=7$ (left) and, for comparison, in $D=7,8,9$ (right).}
\end{figure}
The results for $D=7,8,9$ are displayed in Figure~\ref{fig:negmode789d}. The negative eigenvalue $-k^2$ is plotted as a function of the rotation parameter $a$. We normalise $k$ and $a$ relative to the mass-radius $r_m$, which is equivalent to plotting their values for fixed mass (or mass per unit
length, in the black string interpretation). As described above, the leftmost curve, which does not reach $k=0$, is the higher-dimensional counterpart of the Kerr negative mode, and the eigenvalues $k$ are the wavenumbers of the Gregory-Laflamme threshold modes at rotation $a$. As we consider larger rotation, new branches of negative modes appear that intersect $k=0$ at finite $a/r_m$. We label these successive branches with an integer $\ell=1,2,3,\dots$, and refer to them as `harmonics'. The values of $a/r_m$ at which the stationary perturbations appear are listed in
Table~\ref{Table:critRot}.
\begin{table}[ht]
\begin{eqnarray}
\nonumber
\begin{array}{|c|c|c|c|}\hline
 D & (a/r_m)|_{\ell=1} & (a/r_m)|_{\ell=2} &
 (a/r_m)|_{\ell=3} \\ \hline\hline
 7  & 1.075  & 1.714 & 2.141 \\
\hline
 8  & 1.061 & 1.770 & 2.275 \\
 \hline
9  & 1.051  & 1.792 & 2.337 \\
\hline
\end{array}
\end{eqnarray}
\caption{Values of the rotation $a/r_m$ for the first three zero-modes ($k=0$). The estimated numerical error is $\pm 3\times 10^{-3}$ in $D=7$ and $\pm 5\times 10^{-3}$ in $D=8,9$.}
\label{Table:critRot}
\end{table}

It is important to note that the eigenmode $k=0$ of the harmonic $\ell=1$ does {\it not} correspond to a new stationary solution. Instead it is a  thermodynamic zero-mode that marks the onset of a new local thermodynamic instability, and takes the solution to an infinitesimally nearby one along the family of MP black holes. The location of this zero-mode is predicted by the reduced thermodynamic Hessian $H_{ij}$, introduced in Section~\ref{sec:zeromodeultraspin}, and corresponds to the inflection point of the curve $S(J)$ at fixed $M$ (point $0$ in Figure~\ref{fig:phases}). According to the expression \eqref{MPsingleacrit}, this point is given by
\be
\label{EMestimate}
 \left(\frac{a}{r_m}\right)^{D-3} \Big|_{\ell=1} =
\frac{D-3}{2(D-4)}\left(\frac{D-3}{D-5}\right)^{\frac{D-5}{2}}\,.
\ee
For rotations larger than this value, the black holes are in the \emph{ultraspinning regime}, as defined in the last Chapter. The values of $(a/r_m)$ given by \eqref{EMestimate} for $D=7,8,9$ agree with the central values of the numerically-determined rotations $(a/r_m)$ for $\ell=1$ (first column in Table~\ref{Table:critRot}) up to the third decimal place. This is a very good check of the accuracy of our numerical method.

The $k=0$ eigenmodes of the higher harmonics, $\ell\geq 2$, do not admit this interpretation as perturbations along the MP family of solutions and thus correspond to genuinely new (perturbative) black hole solutions with deformed horizons. Their appearance conforms perfectly to the predictions in \cite{Emparan:2003sy} and \cite{Emparan:2007wm}. It is then natural to expect, although our approach does not prove it since it only captures zero-frequency perturbations, that the harmonic $\ell=2$ signals the onset of the instability conjectured in \cite{Emparan:2003sy}. The $k=0$ eigenmodes for higher harmonics confirm the appearance of the sequence of new black hole phases as the rotation grows.

\begin{figure}
\centering
\includegraphics[width = 6 cm]{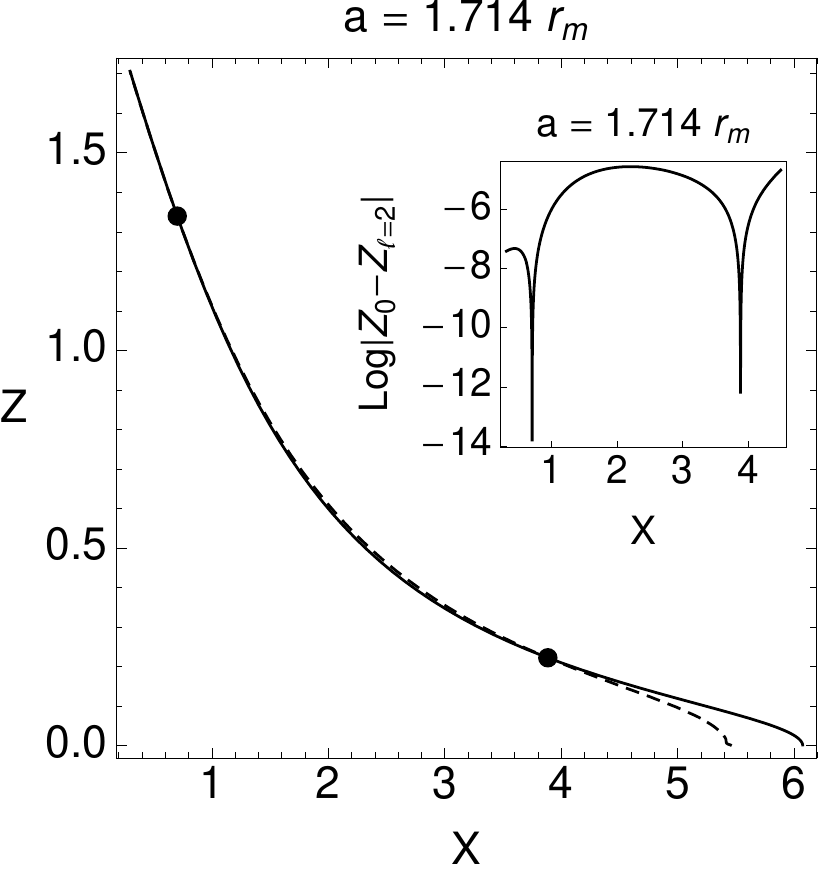} \\
\includegraphics[width = 6 cm]{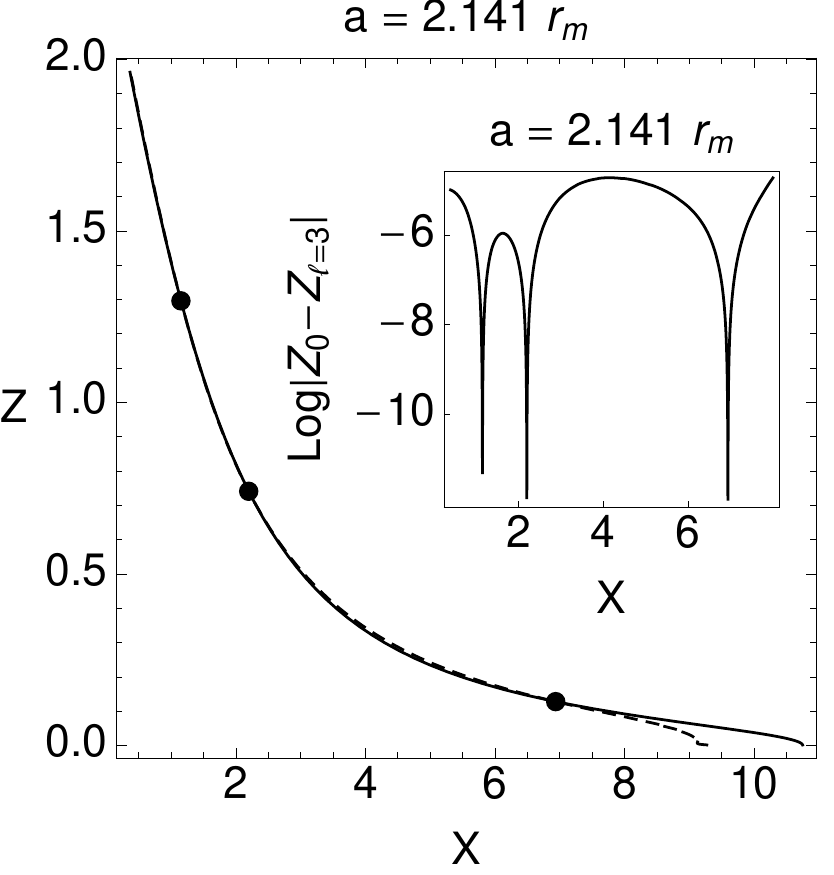} \hspace{1cm}
\includegraphics[width = 6 cm]{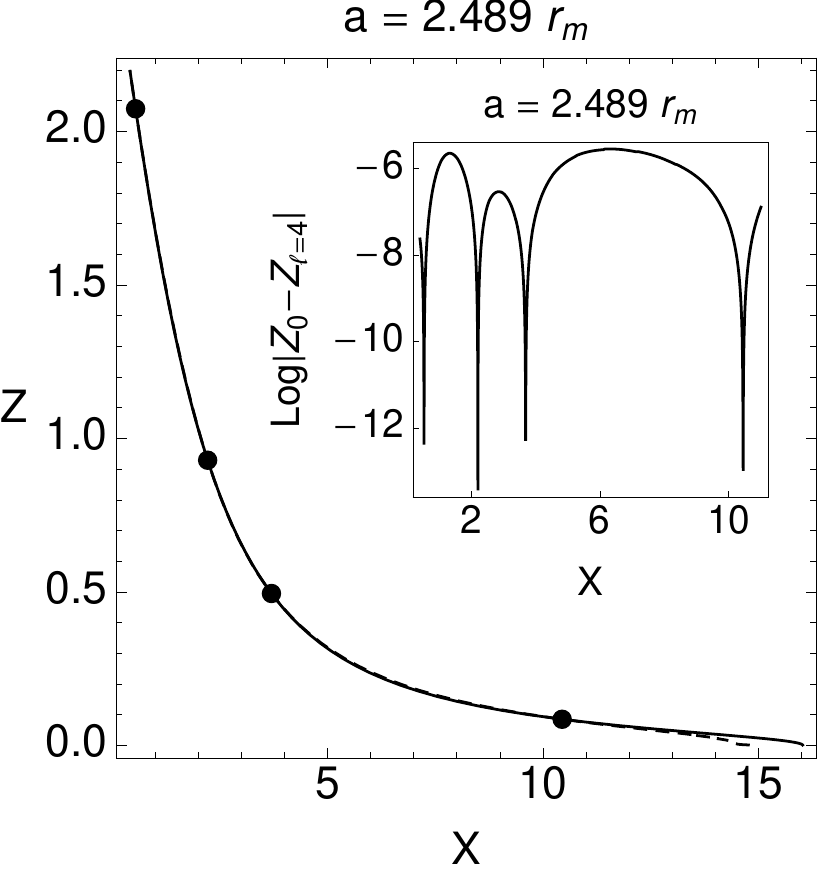}
\caption{\label{fig:embeddingl} \underline{Top plot}: Embedding diagram at $(a/r_m)|_{\ell=2}$ of the $D=7$ black hole horizon, unperturbed (solid), and with the first unstable harmonic perturbation ($\ell=2$, $k=0$) (dashed). The embedding Cartesian coordinates $Z$ and $X$ lie respectively along the rotation axis $\theta=0$ and the rotation plane $\theta=\pi/2$. We also show the logarithmic difference between the embeddings of the perturbed ($Z_{\ell=2}$) and unperturbed ($Z_0$) horizons. The spikes represent the points where the two embeddings intersect. The perturbation has two nodes, so the horizon squeezes around the rotation axis, then bulges out, and squeezes again at the equator, as in the conjectured shape $A$ in Figures~\ref{fig:pinchbhs} and \ref{fig:phases}. \underline{Bottom-left plot}: for $\ell=3$, between the first two nodes of the perturbation, the horizon has a pinch, like shape $B$. \underline{Bottom-right plot}: for $\ell=4$, the four nodes deform the horizon into shape $C$.}
\end{figure}

To visualise the effect on the horizon of the perturbations that give new solutions, and provide further confirmation of our interpretation, we draw an embedding diagram of the unperturbed MP horizon and compare it with the deformations induced by the ultraspinning harmonics $\ell\geq 2$. This is best done using the embedding proposed in \cite{Frolov:2006yb}, which has the advantage of allowing one to embed the horizon along the entire range $0\leq
\theta\leq \pi/2$ for any rotation, although at the cost of stretching the pole region, which acquires a conical profile. We do it for the $\ell=2,3,4$ ultraspinning harmonics in Figure~\ref{fig:embeddingl}. In spite of the distortion created by the embedding, the effect of the perturbations is clear: $\ell=2$ modes create a pinch centred on the rotation axis $\theta=0$; $\ell=3$ modes have a pinch centred at finite latitude $\theta$; $\ell=4$ modes pinch the horizon twice: around the rotation axis and at finite latitude. These are the kind of deformations depicted in Figures~\ref{fig:pinchbhs} and \ref{fig:phases}. To better identify the number of times that the perturbed horizon crosses the unperturbed solution, in these figures we also plot the logarithmic difference between the two embeddings.

Ref.~\cite{Emparan:2003sy} gave several arguments to the effect that critical values $a/r_m$ close to 1 were to be expected. In particular, it was pointed out that the change in the behavior of the black hole from `Kerr-like' to `black brane-like' could be pinpointed to the value of the spin where the temperature (i.e. surface gravity) reaches a minimum for fixed mass, which is the same, for solutions with a single spin, as the inflection point of
$S(J)$. As we have argued, the zero-mode at this solution should not signal an instability. The $\ell=2$ mode at the threshold of the actual instability instead appears at larger rotation, well within the ultraspinning regime defined in the last Chapter. We conjecture this to be true in general: instabilities of MP black holes which have stationary threshold modes can only occur in the ultraspinning regime.

We have identified the points in the phase diagram where the new branches must appear, but we cannot determine in which direction these run. This requires calculating the area, mass and spin of the perturbed solutions. However, for any $k\neq 0$ --- and numerically we can never obtain an exact zero --- the linear perturbations decay exponentially in the radial direction, and thus the mass and spin, measured at asymptotic infinity, are not corrected. Moreover, the modes $\ell\geq 2$ should not affect the asymptotic charges, which are usually associated with the lowest modes (we will confirm this explicitly in the next Chapter for MP black holes with equal spins). It seems that in order to obtain the directions of the new branches one has to go beyond our level of approximation or adopt a different approach.

The new $\ell\geq 1$ branches extend to non-zero eigenvalues $k$. These imply a new ultraspinning Gregory-Laflamme instability for black branes, in which the horizon is deformed not only along the extended directions of the branes, but also along the polar direction of the transverse black hole. Observe that, even if the $\ell=1$ mode does not signal a classical instability of the MP black hole, the modes $\ell=1$, $k>0$ are expected to correspond to thresholds of instabilities of MP black branes. At a given rotation, modes with larger $\ell$ have longer wavelength $k^{-1}$ and so the branch $\ell=1$ is expected to dominate the instability. The growth of $k$ with $a$ can be understood heuristically, since as $a$ grows the horizon becomes thinner in directions transverse to the rotation plane and hence it can fit into a shorter compact circle.

After this work was released \cite{Dias:2009iu}, Shibata and Yoshino \cite{Shibata:2009ad,Shibata:2010wz} were able to find an instability for \emph{non-axisymmetric} perturbations of singly-spinning MP black holes, using a non-linear numerical method. This instability seems to kick in for lower values of the rotation than the mode $\ell=2$ found here. In particular, it also occurs in the $D=5$ case, where no instability of the type we studied occurs. Since the perturbations are not axisymmetric, the thermodynamics arguments on which our conjecture is based do not apply. Furthermore, if the threshold mode is not stationary, it is not associated with new families of stationary black hole solutions bifurcating from the MP family.

To finish, we mention that pinched phases of rotating plasma balls, dual to pinched black holes in Scherk-Schwarz compactifications of AdS, have been found \cite{Lahiri:2007ae}, as well as new kinds of deformations of rotating plasma tubes \cite{Caldarelli:2008mv} and rotating plasma ball instabilities \cite{Cardoso:2009bv,Cardoso:2009nz}. The relation of our results to these and other phenomena of rotating fluids is an interesting problem.

\begin{subappendices}
\section{Appendix: Boundary conditions \label{app:mpsingleBC}}

\vspace{1cm}
\noindent {\bf Boundary conditions at the event horizon}
\vspace{.3cm}


We want to determine the conditions for the metric perturbations $h_{ab}$ to be regular at the event horizon. As we discussed in Section~\ref{sec:negmodestab}, requiring that the traceless-transverse perturbation is a regular 2-tensor on the black hole background $g_{ab}$ is more restrictive than requiring that the perturbed geometry is regular. In particular, such a perturbation cannot change the temperature $T$ and angular velocity $\Omega$ of the background solution. From the Euclidean perspective, if $h_{ab}$ does not obey the periodic identifications of the imaginary time and rotation angles of the background geometry, it clearly cannot be a regular tensor on that geometry.

Let us first discuss the unperturbed background geometry \eqref{mpbh}. Near $r\sim r+$, we can write $\Delta(r)=\Delta'(r_+)(r-r_+)+O[(r-r_+)^2]$, with $\Delta'(r_+)>0$, and the near horizon geometry of \eqref{mpbh} reads
\begin{eqnarray}
ds^2{\bigr |}_{r\sim r+} &\simeq& \frac{\Sigma ^2\left(r_+,\theta
\right) \Delta'\left( r_+ \right)
\left(r-r_+\right)}{\left(r_+^2+a^2 \right)^2}\,dt^2 + \frac{\Sigma
^2\left(r_+,\theta \right)}{\Delta'\left( r_+ \right)
\left(r-r_+\right)}\,dr^2 \nonumber \\
&& + \Sigma^2\left(r_+,\theta \right)
d\theta^2+\frac{\left(r_+^2+a^2\right)^2 \sin^2\theta }{\Sigma
^2\left(r_+,\theta
\right)}\left(d\phi-\frac{a}{r_+^2+a^2}dt\right)^2
+r_+^2\cos^2\theta\, d\Omega^2_{(D-4)}\,.\nonumber \\
\label{mpbhHorizonAUX}
\end{eqnarray}
This suggests the introduction of a new azimuthal coordinate
\begin{equation}
\widetilde{\phi}=\phi-\Omega_H t\,,\qquad
\Omega_H=\frac{a}{r_+^2+a^2}\,, \label{AngVel}
\end{equation}
with period $\Delta\widetilde{\phi}=2\pi$. Now we perform a Wick rotation into Euclidean time $\tau$ and define a new radial coordinate $\rho$ according to
\begin{eqnarray}
&& t=-\ii \,\tau\,,\qquad
\tau=\frac{\widetilde{\tau}}{2\pi T} \qquad \hbox{with} \quad
T=\frac{\Delta'(r_+)}{4\pi\lp r_+^2+a^2 \rp}\,,  \nonumber\\
&& r= r_+ +\frac{\Delta'(r_+)}{4} \rho^2\,.
\end{eqnarray}
The near horizon geometry is then given by
\begin{eqnarray}
ds^2{\bigr |}_{r\sim r+} &\simeq&
 \Sigma^2\left(r_+,\theta \right)\left[ \rho^2d\widetilde{\tau}^2 +d\rho^2 \right]
+ \Sigma^2\left(r_+,\theta \right) d\theta^2 \nonumber \\
&+& \displaystyle{\frac{\left(r_+^2+a^2\right)^2 \sin^2\theta }{\Sigma^2\left(r_+,\theta \right)}}\,d\widetilde{\phi}^2
+r_+^2\cos^2\theta\, d\Omega^2_{(D-4)}\,.
\label{mpbhHorizon}
\end{eqnarray}
The periodic identifications required to ensure regularity are
\be
(\widetilde{\tau},\widetilde{\phi}) \sim (\widetilde{\tau}+2\pi,\widetilde{\phi}) \sim (\widetilde{\tau},\widetilde{\phi}+2\pi)\,.
\ee
These correspond to the already mentioned identifications which give the temperature $T=\beta^{-1}$ and angular velocity $\Omega$ of the solution,
\be
(\tau,\phi) \sim (\tau+\beta,\phi- \,\ii \Omega \beta) \sim (\tau,\phi+2\pi) \,.
\ee

Let us now introduce Cartesian coordinates in the $(\widetilde{\tau},\rho)$ plane by taking $\widetilde{\tau}=\hbox{Arctan}(y/x)$ and $\rho=\sqrt{x^2+y^2}$. The following 1-forms are manifestly regular:
\be \label{SmoothForm1}
E^{\widetilde{\tau}}=\rho^2 d\widetilde{\tau}=x\,dy-y\,dx\,,\qquad E^{\rho}=\rho\,
d\rho=x\,dx+y\,dy\,.
\ee
In terms of these 1-forms, the metric perturbation reads
\begin{equation}
\begin{aligned}
& h_{ab}\,dx^a\,dx^b{\bigr |}_{r\sim r_+} \simeq \\
&~ \Sigma^2\left(r_+,\theta\right) \left[ \delta\nu_0-2 a^2
\sin^2\theta \left(\frac{\Sigma^2\left(r_+,\theta\right)
\Delta'(r_+) +2 r_+ \left(r_+^2+a^2\right)}{\Sigma^4\left(r_+,\theta
\right)\Delta'\left(r_+\right)}\right)\delta\omega \right] \rho^2d\widetilde{\tau}^2
\\
&~ +\Sigma^2\left(r_+,\theta \right) \delta\mu_0\, d\rho^2
+\frac{4\Sigma^2\left(r_+,\theta \right)\sin \theta }
{\Delta'(r_+)}\,\frac{\delta\chi}{\rho^2}\, E^{\rho} d\theta
 -\frac{4 \ii \, a\left(r_+^2+a^2\right)^2 \sin^2\theta}{\Sigma^2\left(r_+,\theta \right)\Delta'\left(r_+\right)}
 \, \frac{ \delta\omega}{\rho^2} E^{\widetilde{\tau}} d\widetilde{\phi}
 \\
&~  +\Sigma^2\left(r_+,\theta \right) \delta\mu_1\, d\theta^2
+\frac{\left(r_+^2+a^2\right)^2 \sin^2\theta}
{\Sigma^2}\,\delta\nu_1\, d\widetilde{\phi}^2 +r_+^2\cos^2\theta^2\,
\delta\Phi\, d\Omega_{(D-4)}^2 \,.
\end{aligned}
\label{eqn:Regularhmetric:r+}
\end{equation}
Regularity then requires that
\begin{equation}
\begin{aligned}
&\delta\chi{\bigr |}_{r=r_+}=0\,,\qquad \delta\omega{\bigr
|}_{r=r_+}=0\,, \qquad \delta\nu_0{\bigr |}_{r=r_+} =
\delta\mu_0{\bigr |}_{r= r_+}\,, \quad \hbox{and}\\
& \delta\mu_0{\bigr |}_{r= r_+}\,,\:\:  \delta\mu_1{\bigr
|}_{r=r_+}\,,\:\:  \delta\nu_1{\bigr |}_{r=r_+}\,,\:\:
\delta\Phi{\bigr |}_{r=r_+} \,\quad \hbox{are finite}\,.
\end{aligned}
\label{eqn:BC:r+}
\end{equation}
The first and second conditions eliminate irregular contributions arising respectively from the terms $E^{\rho} d\theta$ and $E^{\widetilde{\tau}}
d\widetilde{\phi}$ in \eqref{eqn:Regularhmetric:r+}. The second and third conditions guarantee that there is no conical singularity in the $(\widetilde{\tau},\rho)$ plane, since the first two terms in \eqref{eqn:Regularhmetric:r+} then read simply $\Sigma^2\,\delta\mu_0\lp \rho^2d\widetilde{\tau}^2 +d\rho^2 \rp$, which is manifestly regular.

\vspace{1cm}
\noindent {\bf Boundary conditions at the $\theta=0$ equator }
\vspace{.3cm}

We follow here the same strategy. Let us focus on $\theta=0$, and introduce the coordinate
\begin{equation}
\cos\theta=1-\frac{1}{2}\,\chi^2\,. \label{ZoomPolar}
\end{equation}
The background geometry \eqref{mpbh} near $\chi=0$ is given by
\begin{align}\label{mpbhAxisPhi}
ds^2{\bigr |}_{\theta\sim 0} \simeq & -\frac{\Delta(r)}{r^2+a^2}\,dt^2+2 a\lp 1-\frac{\Delta(r)}{r^2+a^2}\rp dt d\phi \\
 &+\left(r^2+a^2\right)\left(d\chi^2+\chi^2 d\phi^2\right)+\frac{r^2+a^2}{\Delta(r)}dr^2+r^2 d\Omega_{(D-4)}^2\,,
\end{align}
which is manifestly regular given that $\phi$ has period $2 \pi$.

Cartesian coordinates on the $(\chi,\phi)$ plane can be chosen by taking $\phi=\hbox{Arctan}(y/x)$ and $\chi=\sqrt{x^2+y^2}$, and the following 1-forms are manifestly regular:
\be \label{SmoothForm2}
E^{\chi}=\chi\, d\chi=x\,dx+y\,dy\,,\qquad E^{\phi}=\chi^2
d\phi=x\,dy-y\,dx\,.
\ee
The metric perturbation then reads
\begin{equation}
\begin{aligned}
h_{ab}\,dx^a\,dx^b{\bigr |}_{\theta\sim 0} &\simeq
-\frac{\Delta(r)}{r^2+a^2}\,\delta\nu_0\, dt^2+\frac{r^2+a^2}
{\Delta(r)}\,\delta\mu_0\,dr^2 +\frac{r^2+a^2}{\Delta(r)}\,
\delta\chi \,E^{\chi} dr \\
&~~~
+\frac{2a}{r^2+a^2}\left[\left(r^2+a^2+\Delta(r)\right)\delta\omega
-\left(r^2+a^2\right)\delta\nu_1+\Delta(r)\delta\nu_0 \right] E^{\phi}dt\\
&~~~ +\left(r^2+a^2\right)\left[\delta\mu_1 d\chi^2
+\delta\nu_1\chi^2 d\phi^2\right] +r^2\, \delta \Phi
\,d\Omega_{(D-4)}^2 \,,
\end{aligned}
\label{eqn:Regularhmetric:x1}
\end{equation}
and regularity requires that
\begin{equation}
\begin{aligned}
& \delta\nu_1{\bigr |}_{\theta=0}=\delta\mu_1{\bigr|}_{\theta=0}\,, \qquad \textrm{and} \\ 
& \delta\chi{\bigr |}_{\theta=0}\,,\:\:
\delta\omega{\bigr |}_{\theta=0}\,, \:\: \delta\mu_0{\bigr|}_{\theta=0}\,,\:\: \delta\nu_0{\bigr |}_{\theta=0}\,,\:\: \delta\Phi{\bigr |}_{\theta=0} \,\quad \hbox{are
finite}\,.
\label{eqn:BC:x1}
\end{aligned}
\end{equation}

\vspace{1cm}
\noindent {\bf Boundary conditions at the $\theta=\pi/2$ equator }
\vspace{.3cm}

We introduce the new coordinate $x=\cos\theta$. The geometry \eqref{mpbh} in the neighbourhood of the rotation plane $\theta=\pi/2$ is given by
\begin{equation}\label{mpbhAxisSphere}
\begin{aligned}
ds^2{\bigr |}_{\theta\sim \pi/2} &\simeq
-\frac{\Delta(r)-a^2}{r^2}\,dt^2-\frac{2a}{r^2}\left[r^2+a^2
-\Delta(r)\right]dt d\phi+\frac{\left(r^2+a^2\right)^2-a^2
\Delta(r)}{r^2}\,d\phi^2\\
&~~~ +\frac{r^2}{\Delta(r)}\,dr^2+r^2\left[dx^2+ x^2
d\Omega_{(D-4)}^2\right]\,,
\end{aligned}
\end{equation}
which is manifestly regular given that $d\Omega_{(D-4)}^2$ is the line element of an $S^{(D-4)}$.

Introducing the manifestly regular and smooth 1-forms,
\be \label{SmoothForm3}
E^{x}=x\, dx\,,\qquad E^{\Omega}=x^2 d\Omega_{(d-4)}\,,
\ee
the metric perturbation reads
\begin{equation}
\begin{aligned}
& h_{\mu\nu}\,dx^\mu\,dx^\nu{\bigr |}_{\theta\sim \pi/2} \simeq
-\frac{\left[\Delta(r)-a^2\right]\delta\nu_0 +2a^2\delta\omega}{r^2}\,dt^2 \\
&~~~ +\frac{\left(r^2+a^2\right)^2 \delta\nu_1 -a^2 \Delta(r) \left(\delta\nu_0+2\delta\omega \right)}{r^2}\,d\phi^2  \\
&~~~ +\frac{2a}{r^2}\left[\left(r^2+a^2+\Delta(r)\right)\delta\omega
-\left(r^2+a^2\right)\delta\nu_1+\Delta(r)\delta\nu_0 \right] dtd\phi\\
&~~~
+\frac{r^2}{\Delta(r)}\,\delta\mu_0\,dr^2-\frac{r^2}{\Delta(r)}\left(
\frac{\delta\chi}{x}- \partial_x \delta\chi \right)E^x dr+r^2
\left(\delta\mu_1 \,dx^2+x^2\,\delta\Phi
\,d\Omega_{(d-4)}^2\right)\,.
\end{aligned}
\label{eqn:Regularhmetric:x0}
\end{equation}
Regularity requires that
\begin{equation}
\begin{aligned}
&\delta\chi{\bigr |}_{\theta=\frac{\pi}{2}}=0\,,\:\: \delta\Phi{\bigr |}_{\theta=\frac{\pi}{2}}=\delta\mu_1{\bigr|}_{\theta=\frac{\pi}{2}}\,\quad \hbox{and}\\
&
 \delta\omega{\bigr |}_{\theta=\frac{\pi}{2}}\,, \:\:
 \delta\mu_0{\bigr |}_{\theta=\frac{\pi}{2}}\,,\:\:
 \delta\nu_0{\bigr |}_{\theta=\frac{\pi}{2}}\,,\:\:
 \delta\nu_1{\bigr |}_{\theta=\frac{\pi}{2}}\,\quad \hbox{are finite}\,.
\end{aligned}
\label{eqn:BC:x0}
\end{equation}

\vspace{1cm}
\noindent {\bf Boundary conditions at the asymptotic region $r \to \infty$}
\vspace{.3cm}

At spatial infinity, $r\rightarrow\infty$, the solutions to Eq.~\eqref{eigenh} behave as $h_{ab} \propto e^{\pm |k|r}$. Therefore, regular perturbations must vanish at infinity.

\end{subappendices}

%% file: MPeq.tex
\chapter{Equal-spinning Myers-Perry instability} \label{cha:MPequal}

In this Chapter, based on Ref.~\cite{Dias:2010eu}, we present the first example of a linearised gravitational instability of an asymptotically flat vacuum black hole. We study perturbations of a Myers-Perry black hole with equal angular momenta in an odd number of dimensions. We find no evidence of any instability in five or seven dimensions, but in nine dimensions, for sufficiently rapid rotation, we find perturbations that grow exponentially in time. The onset of instability is associated with the appearance of time-independent perturbations which generically break all but one of the rotational symmetries. This is interpreted as evidence for the existence of a new 70-parameter family of black hole solutions with only a single rotational symmetry. We also present results for the Gregory-Laflamme instability of rotating black branes, demonstrating that rotation makes black branes more unstable.

\section{Introduction}

In the previous Chapter, based on Ref.~\cite{Dias:2009iu}, we presented very strong evidence for the existence of an ultraspinning instability of singly-spinning Myers-Perry black holes. However, no actual instability, i.e.\ a perturbation growing in time, was demonstrated. In this Chapter, based on Ref.~\cite{Dias:2010eu}, we shall demonstrate that some Myers-Perry (MP) black holes do admit perturbations that grow exponentially in time, thereby providing the first example of a linearised gravitational instability of an asymptotically flat vacuum black hole solution.\footnote{Ref.~\cite{Shibata:2009ad}, where an instability of singly-spinning Myers-Perry black holes for non-axisymmetric perturbations was found (see also \cite{Shibata:2010wz}), was released shortly before \cite{Dias:2010eu}. However, the numerical analysis in \cite{Shibata:2009ad,Shibata:2010wz} is fully non-linear, despite dealing with small perturbations. It would be very interesting to analyse the instability found there at the linear level, so that the threshold mode can be better studied.}

We shall exploit the idea introduced in Ref.~\cite{Kunduri:2006qa} of considering MP solutions with enhanced symmetry. The generic MP solution (Section~\ref{subsec:MPsolution}) has isometry group $R \times U(1)^n$ where $R$ corresponds to time translations and $n=\lfloor (D-1)/2 \rfloor$. However, this is enhanced when some of the angular momenta coincide. In particular, for odd $D$, the MP solution with all angular momenta equal ($J_i=J/n\;\; \forall i$) has a much larger $R \times U(N+1)$ isometry group, where $D=2N+3$. Furthermore, the solution is cohomogeneity-1, i.e. it depends only on a single coordinate. The metric involves a fibration over complex projective space $CP^N$. Gravitational perturbations of this solution can be decomposed into scalar, vector and tensor types according to how they transform under isometries of $CP^N$. The tensors, which exist only for $N \ge 2$ ($D \ge 7$), were studied in Ref.~\cite{Kunduri:2006qa} and no evidence of any instability was found. The special case of $D=5$, for which only scalar perturbations exist, was studied in Ref.~\cite{Murata:2008yx}. Again, no evidence of any instability was found.

In this Chapter, we shall study scalar-type perturbations of these cohomogeneity-1 MP black holes. The symmetries enable the problem to be reduced to coupled linear ordinary differential equations (ODEs) which we solve numerically. We find no evidence of any instability for $D=5$ (consistent with Ref.~\cite{Murata:2008yx}) or $D=7$. However, for $D=9$, when $J$ exceeds a certain critical value $J_{\rm crit}$, there is a perturbation that grows exponentially in time, i.e. an instability. We believe that such an instability will exist for all (odd) $D \ge 9$ although we have demonstrated this only for $D=9$.

As expected, the onset of instability is associated with the appearance of a stationary zero-mode of the MP solution with $J=J_{\rm crit}$. This zero-mode is interesting for another reason. It has been proven that a stationary, rotating black hole must admit a rotational isometry (i.e. a $U(1)$ isometry) \cite{Hollands:2006rj,Moncrief:2008mr}. This is known as the rigidity theorem. However, all known higher-dimensional black holes have {\it multiple} rotational isometries (e.g. MP black holes have $\lfloor (D-1)/2 \rfloor$ commuting $U(1)$ isometries), i.e. more symmetry than one expects on the basis of general arguments. Therefore, it has been conjectured that there exist solutions with less symmetry than any known solution, specifically solutions with a single rotational symmetry \cite{Reall:2002bh}.

It was proposed in Ref.~\cite{Reall:2002bh} that one could seek evidence for the existence of such solutions in the same way that the first evidence was obtained for the existence of non-uniform black string solutions. For black strings, the static zero-mode associated with the onset of the Gregory-Laflamme instability was conjectured to describe the ``branching off" of a new family of non-uniform black string solutions from the already known branch of uniform solutions \cite{Horowitz:2001cz}. Perturbative \cite{Gubser:2001ac} and numerical \cite{Wiseman:2002zc} work subsequently confirmed that this was correct. For rotating black holes, the idea proposed in Ref.~\cite{Reall:2002bh} is to look for a stationary zero-mode of a MP solution. By analogy with the black string example, this could be interpreted as the branching off of a new family of solutions. If the zero-mode preserves only a single rotational symmetry then this would be evidence for the existence of new black holes with just one rotational symmetry. 

In our case, the stationary zero-mode generically preserves only a single rotational symmetry. Therefore we conjecture that there exists a family of stationary black hole solutions with just a single rotational symmetry, that bifurcates from the cohomogeneity-1 MP black hole solution at $J=J_{\rm crit}$. In fact, we find not just one stationary zero-mode, but a large family, corresponding to all scalar harmonics on $CP^N$ with a certain eigenvalue. For $D=9$, we shall argue that the new family of solutions will involve 70 independent parameters, considerably more than the 5 parameters required to specify the MP solution! If correct, this implies that any hope of specifying higher-dimensional black holes uniquely using just a few parameters is bound to fail. Of course, the new black hole solutions may turn out to be unstable themselves.

Recently, Ref.~\cite{Emparan:2009vd} has provided other evidence for the existence of higher-dimensional black holes with a single rotational symmetry. Using the ``blackfold" approach of Refs.~\cite{Emparan:2009cs,Emparan:2009at}, approximate solutions were constructed for $D \ge 5$ that describe ``helical" black rings. The blackfold approximation is based on the fact that higher-dimensional black holes can have widely separated horizon scales. The results of the present Chapter concern black holes that lie outside the regime of validity of this approximation. Our results complement those of Ref.~\cite{Emparan:2009vd} because we are presenting evidence for topologically spherical black holes with a single rotational symmetry, whereas Ref.~\cite{Emparan:2009vd} considered black rings.

A difference between the results to be discussed here and the results in the previous Chapter concerns the nature of the stationary zero-mode. In the present case, the unstable perturbation breaks all but one rotational symmetry. However, the stationary zero-modes found in the previous Chapter for singly-spinning black holes preserved the isometries of the background, as Ref.~\cite{Emparan:2003sy} had conjectured. Singly-spinning black holes exhibit symmetry enhancement for $D \ge 6$. They are cohomogeneity-2 with isometry group $R \times U(1) \times SO(D-3)$ where $SO(D-3)$ has $S^{D-4}$ orbits.\footnote{One can decompose metric perturbations of these solutions into scalar, vector and tensor types using this $SO(D-3)$ symmetry. The tensors, which exist only for $D \ge 7$ have been studied previously in Ref.~\cite{Kodama:2009bf} and show no evidence of any instability. However, tensor perturbations arise from deformations of the $S^{D-4}$ part of the metric, whereas the expected ultraspinning instability should arise from perturbations of the metric transverse to $S^{D-4}$. The results of Ref.~\cite{Dias:2009iu} (Chapter~\ref{cha:MPsingle}) indicate that the instability should be a scalar-type perturbation.} While these isometries are preserved by the zero-modes found previously, the results in this Chapter raise the question of whether singly-spinning MP black holes might admit further stationary zero-modes that break some of their symmetry, and provide further evidence for new black hole solutions with reduced symmetry.

A corollary of our approach is the first data for the Gregory-Laflamme instability \cite{Gregory:1993vy} for rotating black {\it branes}. We consider black branes obtained as the product \eqref{blackbrane} of a cohomogeneity-1 MP black hole with flats directions. We argue that such solutions are always classically unstable. Our numerical results demonstrate that the branes become more unstable as the angular momentum increases: the instability becomes stronger (i.e. it occurs on a shorter time scale) and the critical wavelength of unstable modes decreases, as the angular momentum increases.

An important feature of cohomogeneity-1 MP black holes is that they exhibit an upper bound on their angular momentum (Section~\ref{subsec:MPsolution}). Solutions saturating this bound are extreme black holes. Because of this upper bound, it is not obvious that such black holes should exhibit the type of behaviour discussed in Ref.~\cite{Emparan:2003sy}. However, we were motivated by the observation in Section~\ref{subsec:ultraspinMP} that, for $D \geq 7$, cohomogeneity-1 MP black holes do satisfy the \emph{ultraspinning criterion} once $J$ exceeds a critical value $J_{\rm ultra}$ (for given $M$). This criterion says that a classical instability with a stationary threshold can occur only if the Hessian matrix $H_{ij} = (\partial^2 (-S)/\partial J_i \partial J_j)_M$, where $S$ is the black hole entropy, fails to be positive definite. Hence, there {\it might} be an ultraspinning instability for $J > J_{\rm crit}$ where $J_{\rm ultra} < J_{\rm crit} < J_{\rm extreme}$. Our results show that there is no instability for $D=7$ but an instability does occur for $D=9$ and, we believe, for (odd) $D>9$.

For $D=5$ MP black holes, $H_{ij}$ is always positive definite \cite{Dias:2009iu} so such black holes are never ultraspinning. However, for $D \ge 6$, singly spinning MP black holes with large enough angular momentum are, of course, ultraspinning and the numerical results of Ref.~\cite{Dias:2009iu} supply strong evidence that the instability appears only when $H_{ij}$ fails to be positive definite.

This Chapter is organized as follows. Section \ref{sec:background} describes the cohomogeneity-1 black holes that we shall study. In Section~\ref{sec:strategyresults}, we explain our approach and discuss the results. The technical details of our work are presented in the later Sections~\ref{sec:decomposition} and \ref{sec:eigenvalue}, and in the Appendices.

\section{Cohomogeneity-1 Myers-Perry black holes} \label{sec:background}

The Kerr solution was extended to higher dimensions by Myers and Perry \cite{Myers:1986un}. The Myers-Perry family can be parameterized by a mass-radius parameter $r_M$ and $\lfloor (D-1)/2 \rfloor$ angular momentum parameters $a_i$. In the particular case of equal angular momenta, $a_i = a$, the solution in odd dimensions $D=2 N+3$ is cohomogeneity-1. The metric can be written as:\footnote{The radial coordinate used here can be related to the standard Boyer-Lindquist radial coordinate of \cite{Myers:1986un} through $r^2 \to r^2+a^2$. In this Chapter, Latin indices will be used for coordinates on $CP^{N}$, whereas Greek indices will be used for the black hole spacetime coordinates, unlike the previous Chapters.}
\be
\label{background}
ds^2 = -f(r)^2dt^2 +g(r)^2dr^2 + h(r)^2[d\psi +A_a dx^a - \Omega(r)dt]^2 + r^2 \hat{g}_{ab} dx^a dx^b\,,
\ee
\bea
g(r)^2 = \left(1- \frac{r_M^{2N}}{ r^{2N}} + \frac{r_M^{2N}a^2}{r^{2N+2}}\right)^{-1} &,& \quad h(r)^2 = r^2\left( 1+ \frac{r_M^{2N}a^2}{ r^{2N+2}} \right)\,, \nonumber \\
f(r) = \frac{r}{g(r)h(r)} &,& \quad \Omega(r) = \frac{r_M^{2N}a}{ r^{2N} h^2} \,, \nonumber
\eea
where $\hat{g}_{a b}$ is the Fubini-Study metric on $CP^{N}$ with Ricci tensor $\hat{R}_{ab} =2(N+1) \hat{g}_{ab}\,$, and $A = A_a dx^a\,$ is related to the K\"ahler form $J$ by $dA=2J$. Surfaces of constant $t$ and $r$ have the geometry of a homogeneously squashed $S^{2N+1}$, written as an $S^1$ fibre over $CP^{N}$. The fibre is parameterized by the coordinate $\psi$, which has period $2\pi$. Explicit expressions for the metric $\hat{g}_{a b}$ and K\"ahler potential $A$ of $CP^N$ can be obtained through the iterative Fubini-Study construction summarized in Appendix~\ref{cpnappendix}.

The spacetime metric satisfies $R_{\mu\nu} =0$ and the solution is asymptotically flat. The event horizon is located at $r=r_+$ (the largest real root of $g^{-2}$) and it is a Killing horizon of $\xi = \partial_t+\Omega_H \partial_\psi\,$, where the angular velocity of the horizon is given by:
\be
\label{angvel}
\Omega_H = \frac{r_M^{2N} a}{r_+^{2N+2}+r_M^{2N}a^2}.
\ee
The mass $M$ and angular momentum $J$, defined with respect to $\partial_{\psi}$, are~\cite{Gibbons:2004ai}
\be 
\label{EJmpeq}
M = \frac{A_{2N+1}}{8\pi G}r_M^{2N}\left (N+\frac{1}{2} \right)\,, \qquad J  = \frac{A_{2N+1}}{8\pi G} (N+1) r_M^{2N}a\,,
\ee
where $A_{2N+1}$ is the area of a unit ($2N+1$)-sphere.

There is an extremality bound on the angular momentum which can be expressed as
\be
\left( \frac{a}{r_+} \right)^2 \leq \left( \frac{a_{\rm ext}}{r_+} \right)^2 = N\,, \qquad \mathrm{or} \qquad \lp \frac{a}{r_M} \rp^2 \le \left( \frac{a_{\rm ext}}{r_M} \right)^2 = \frac{N}{(N+1)^{(N+1)/N}}\,.
\ee
The solution saturating this bound has a regular, but degenerate, horizon. For fixed $r_+$, or $r_M$, ultraspinning behaviour (Section~\ref{subsec:ultraspinMP}) occurs for 
\be
\label{eqn:a1}
\lp \frac{a}{r_+}\rp^2  > \lp \frac{a_1}{r_+}\rp^2 \equiv \frac{1}{2}\,, \qquad \mathrm{or} \qquad \lp \frac{a}{r_M}\rp^2 >  \lp \frac{a_1}{r_M}\rp^2  \equiv \frac{1}{2^{(N+1)/N}}\,.
\ee
Note that the range $a_1 < a \le a_{\rm ext}$ (for fixed $r_+$) for which the black hole is ultraspinning becomes larger as $N$ increases, and that $a_1=a_{\rm ext}$ if $N=1$, so there is no ultraspinning behaviour for $D=5$.

\section{Strategy and Results} \label{sec:strategyresults}

\subsection{Strategy}

Ref.~\cite{Monteiro:2009ke} (Chapter~\ref{cha:KerrAdS}) introduced new numerical techniques for determining negative modes of rotating black holes. In Ref.~\cite{Dias:2009iu} (Chapter~\ref{cha:MPsingle}), these techniques were exploited to construct the stationary zero-mode expected to indicate the onset of an ultraspinning instability of a singly-rotating MP black hole. In the present Chapter, based on \cite{Dias:2010eu}, we shall determine the stationary zero-mode indicating the onset of instability for cohomogeneity-1 black holes. However, our main achievement is to generalise these methods to demonstrate the existence of perturbations that grow exponentially in time. 

Our approach is explained in Chapter~\ref{cha:classthermo}. We consider the eigenvalue problem
\be
\label{lichnMPeq}
(\Delta_L h)_{\mu\nu} = -k^2 h_{\mu\nu}\,,
\ee
where $\Delta_L$ is the Lichnerowicz operator \eqref{deflichn} for the MP background, and $h_{\mu\nu}$ are traceless-transverse perturbations $(h^\mu_{\phantom{\mu}\mu}=\nabla^\mu h_{\mu\nu} =0)$ of the MP black hole. This problem arises for Gregory-Laflamme-type perturbations \eqref{glpert} of uniform black branes. Perturbations with non-zero $k$ correspond to negative modes of $\Delta_L$, which, in the stationary case, may also correspond to negative modes \eqref{negmodeG} of the black hole partition function. The boundary conditions are that $h_{\mu\nu}$ should be regular on the future event horizon ${\cal H}^+$ and vanishing at infinity.

The strategy for studying perturbations of the black hole will be to seek a solution of (\ref{lichnMPeq}), i.e. a negative mode of the black hole, and then vary the spin of the black hole until $k$ vanishes, i.e. the negative mode becomes a zero-mode. This strategy is motivated by the availability of numerical techniques for solving eigenvalue equations of the form (\ref{lichnMPeq}). Solutions with non-zero $k$ correspond to perturbations of black branes. Therefore our method will yield results for the Gregory-Laflamme instability of rotating black branes as well as enabling us to search for black hole instabilities.

We can Fourier analyse our perturbation in the time and $\psi$ directions, i.e. we assume that the dependence on $t$ and $\psi$ is given by
\be
\label{eqn:fourier}
 h_{\mu\nu} \propto e^{-\ii\omega t + \ii m \psi}\,,
\ee
where $m$ is an integer. As we shall explain in detail below, we can also decompose the perturbation into harmonics on $CP^N$. These can be of scalar, vector or tensor type. The tensors were considered in Ref.~\cite{Kunduri:2006qa}. We shall restrict our attention to perturbations of scalar-type, which can be expanded in terms of scalar harmonics on $CP^N$. As usual, harmonics with different eigenvalue decouple from each other. The equations satisfied by $h_{\mu\nu}$ depend only on the eigenvalue of the harmonic in question.\footnote{In the last Chapter, we referred to the different negative modes as different ``harmonics''. However, in the present case, we do have a precise harmonic structure in $CP^N$, which is why the problem has codimension 1.} Eigenvalues of the scalar Laplacian on $CP^N$ are labelled by a non-negative integer $\kappa$ (see Section~\ref{sec:decomposition}). Hence our perturbation is labelled by $(\omega,m,\kappa)$. 

Consider the (Lorentzian) negative mode equation (\ref{lichnMPeq}). As we have explained above, this arises from classical perturbations of a rotating black brane. The usual approach to this problem is to fix $(k,m,\kappa)$ and to determine $\omega$. However, our approach will be to fix $(\omega,m,\kappa)$ and determine the possible eigenvalue(s) $-k^2$. In other words, we are determining the wavenumber $k$ for which black brane perturbations with given $m$ and $\kappa$ have time-dependence associated with the given $\omega$. We shall fix the overall scale $r_M=1$ and determine the eigenvalue(s) $-k^2$ for fixed $(\omega,m,\kappa)$ as $a$ increases from $0$ to extremality. If $k$ vanishes for some value of $a$ then the associated black hole admits a zero-mode with the given values of $(\omega,m,\kappa)$ (of course it must be checked that this is not pure gauge).

In searching for an instability, we are looking for solutions of (\ref{lichnMPeq}) with ${\rm Im}(\omega)>0$. A problem with our approach is that we expect unstable modes to have complex $\omega$ in general, with the real and imaginary parts of $\omega$ related in some way. In other words, for given $m,\kappa$ and $a$, the complex quantity $\omega$ will be a function of the real quantity $k$ and hence the real and imaginary parts of $\omega$ cannot be independent. If we try to follow the above strategy for a randomly chosen complex value of $\omega$, then this will not satisfy the required relation between its real and imaginary parts, and therefore our numerical method will not output a real value of $k$. In order to locate where $k$ vanishes we would have to scan over both $a$ and, say, the real part of $\omega$. It would be difficult to do this with high accuracy.

We shall circumvent this problem by restricting attention to modes with $m=0$, i.e. modes preserving the rotational symmetry of the black hole that follows from the rigidity theorem. There are reasons to expect that unstable modes with $m=0$ will have purely imaginary $\omega$: $\omega = i \Gamma$, $\Gamma>0$, and this is confirmed by our results. Ref.~\cite{Kleihaus:2007dg} has obtained numerically non-uniform rotating black brane solutions that do indeed bifurcate from the uniform branch (based on cohomogeneity-1 MP solutions) at a point corresponding to a stationary perturbation. Hence stationary perturbations do indeed exist. We believe that the reason for this is that unstable modes will have ${\rm Re}(\omega)=0$ if they are invariant under the rotational symmetry of the black hole predicted by the theorems of Refs.~\cite{Hollands:2006rj,Moncrief:2008mr}, i.e. the symmetry generated by the Killing field $\Omega_i m_i$, where $m_i$ are the rotational Killing vector fields and $\Omega_i$ the associated angular velocities of the horizon. We do not have a proof of this, but our results, and the results of Refs.~\cite{Kleihaus:2007dg,Dias:2009iu}, indicate that it is true. In the limit $k \rightarrow k_\ast$, this gives a stationary threshold mode that preserves this symmetry. This symmetry is a necessary condition for the threshold mode to correspond to a bifurcation into a new family of non-uniform rotating black brane solutions, since presumably this new family should respect the theorems of Refs.~\cite{Hollands:2006rj,Moncrief:2008mr} (although, strictly speaking, these theorems apply only to black holes, not black branes).

In summary, we expect an instability of the black branes ($\Gamma>0$) to appear for wavenumbers $|k|<k_\ast$. If we restrict attention to modes invariant under the symmetry generated by $\Omega_i m_i$ then unstable modes will have ${\rm Re}(\omega)=0$, and the threshold unstable mode, with $k=k_\ast$, will be stationary and invariant under the same symmetry. For cohomogeneity-1 black holes, $\Omega_i m_i$ is proportional to $\partial/\partial \psi$, so modes invariant under the symmetry generated by $\Omega_i m_i$ must have $m=0$, which is why we set $m=0$ above. Notice that only $m=0$ modes are required to obey the ultraspinning conjecture in Section~\ref{sec:zeromodeultraspin}, since the conjecture only applies to perturbations which do not break the rotational symmetry generated by $\Omega_i m_i$.

In summary, we shall set $m=0$, $r_M=1$ and, for given $(\Gamma,\kappa,a)$ we shall determine the possible eigenvalues $-k^2$. Then we vary $a$ until the eigenvalue vanishes. We have then found a black hole that admits an unstable zero-mode with the given values of $\Gamma$ and $\kappa$.

\subsection{Results for $D=5$}

\begin{figure}[t]
\centerline{\includegraphics[width=.45\textwidth]{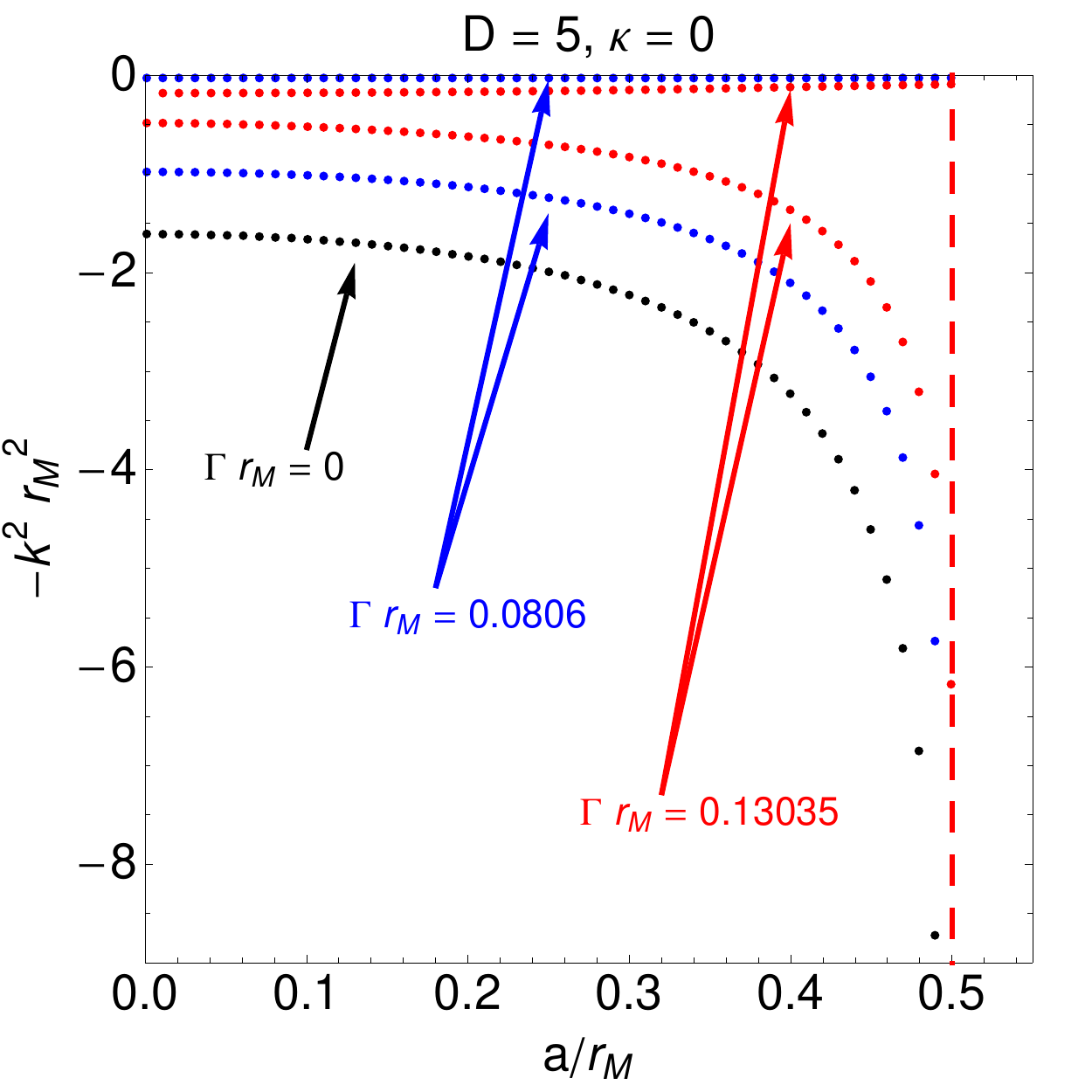}
\hspace{1cm}\includegraphics[width=.45\textwidth]{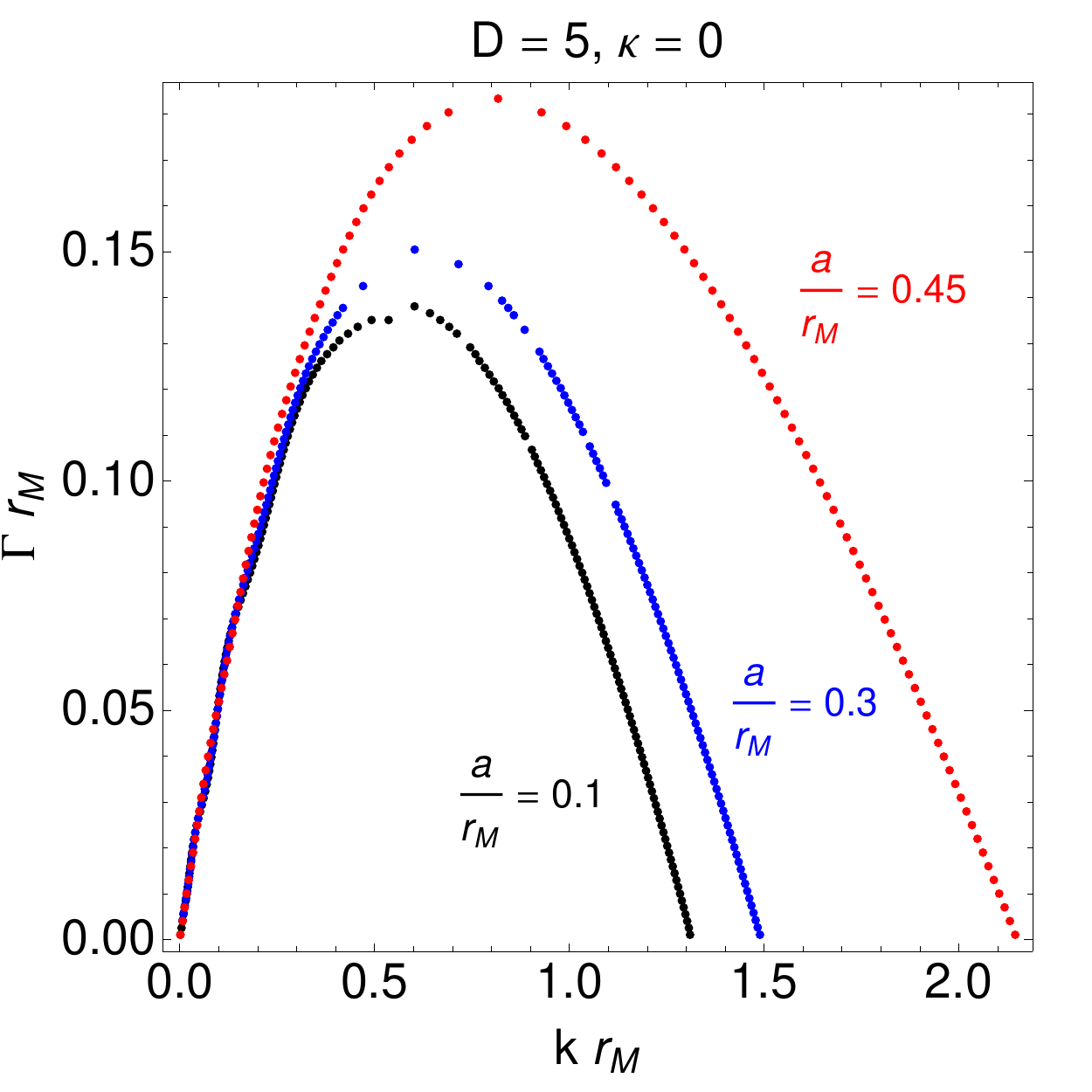}}
\caption{Results in $D=5$, $\kappa=0$. We represent this mode for fixed values of $\Gamma \,r_M$ (first graph) and $a/ r_M$ (second graph). This unstable mode of the branes corresponds to the well-known Gregory-Laflamme mode.}
\label{fig:5dkzero}
\end{figure}

Our expectation is that, for small $a$, the black hole will be classically stable but the associated black branes will suffer a Gregory-Laflamme instability~\cite{Gregory:1993vy}. Therefore, for a range of $\Gamma$, there should exist real solutions for $k$ (corresponding to unstable perturbations of the branes) but $k$ will never vanish for $\Gamma>0$, so the black hole is stable. For a static brane, the Gregory-Laflamme instability is an $s$-wave perturbation of the transverse black hole, which for us translates into a $\kappa=0$ perturbation. Hence, for a rotating brane, it is natural to expect this instability also to have $\kappa=0$.

This is indeed what we find. The left plot in Fig.~\ref{fig:5dkzero} shows our result for $-k^2$ for given $a$ and $\Gamma$, with $\kappa=0$. The plot with $\Gamma=0$ corresponds to a stationary perturbation of the black branes. This is the ``threshold unstable mode" at the critical wavelength beyond which the black branes are unstable. Note that the curves exist for all values of $a$, i.e. the Gregory-Laflamme instability is always present, it does not ``switch off" as $a$ increases. The upper curves do not extend to $k=0$ so there is no indication of any black hole instability.

In the right plot of Fig.~\ref{fig:5dkzero}, we give a more familiar plot of $\Gamma$ against $k$ for different values of $a$. For each value of $a$, we have a curve that takes the usual Gregory-Laflamme form. The maximum value of $\Gamma$ is $10-20 \%$ of $r_M$ and increases with increasing $a$. Furthermore, the range of $k$ for which there exists an instability increases, i.e. the instability persists down to shorter wavelengths as $a$ increases. Hence rotation makes the branes more unstable. As usual, $\Gamma \rightarrow 0$ as $k \rightarrow 0$ and as $k \rightarrow k_\ast>0$. The mode with $\Gamma=k=0$ has the usual interpretation of a gauge mode \cite{Gregory:1994bj}. The mode with $\Gamma=0$ and $k=k_\ast$ is the threshold unstable mode associated with the onset of instability. This marks the bifurcation of a new family of non-uniform rotating black brane solutions. These are the solutions constructed in Ref.~\cite{Kleihaus:2007dg} in the black string case. They preserve the symmetries of the cohomogeneity-1 MP black hole but break the translational symmetry along the string.

Note that the slope $\Gamma/k$ approaches a common limiting value as $k \rightarrow 0$, independently of the value of $a$. This long-wavelength limiting behaviour is captured by the blackfold approach: it follows from Ref. \cite{Emparan:2009at} that $\Gamma/k \rightarrow 1/\sqrt{D-2}$ as $k \rightarrow 0$. This is consistent with our numerical results.

We find no solution of (\ref{lichnMPeq}) with $\kappa=1$. Therefore our results are consistent with stability of these black holes, in agreement with the results of Ref.~\cite{Murata:2008yx}.

\subsection{Results for $D=7$}

\begin{figure}[t]
\centerline{\includegraphics[width=.45\textwidth]{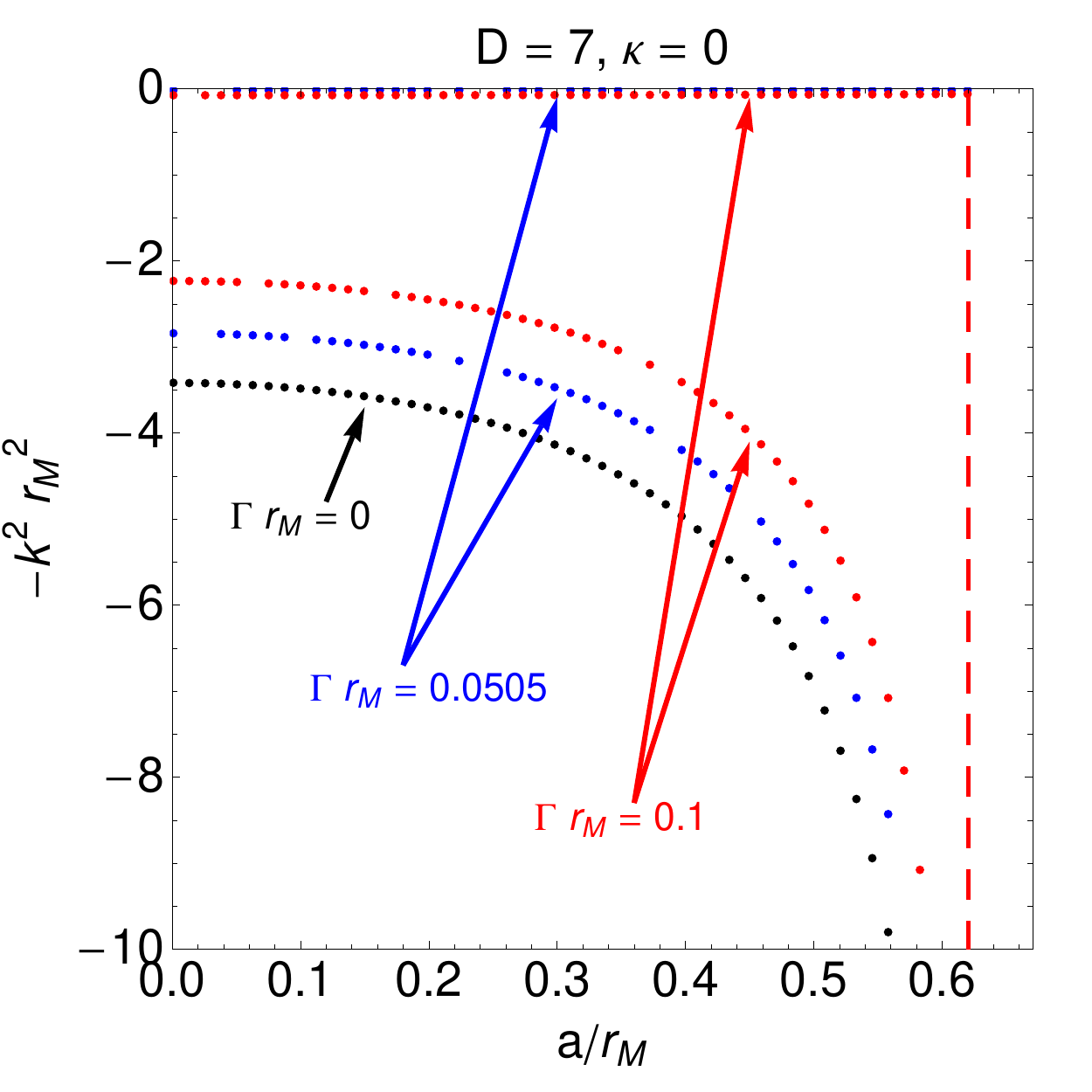}
\hspace{1cm}\includegraphics[width=.45\textwidth]{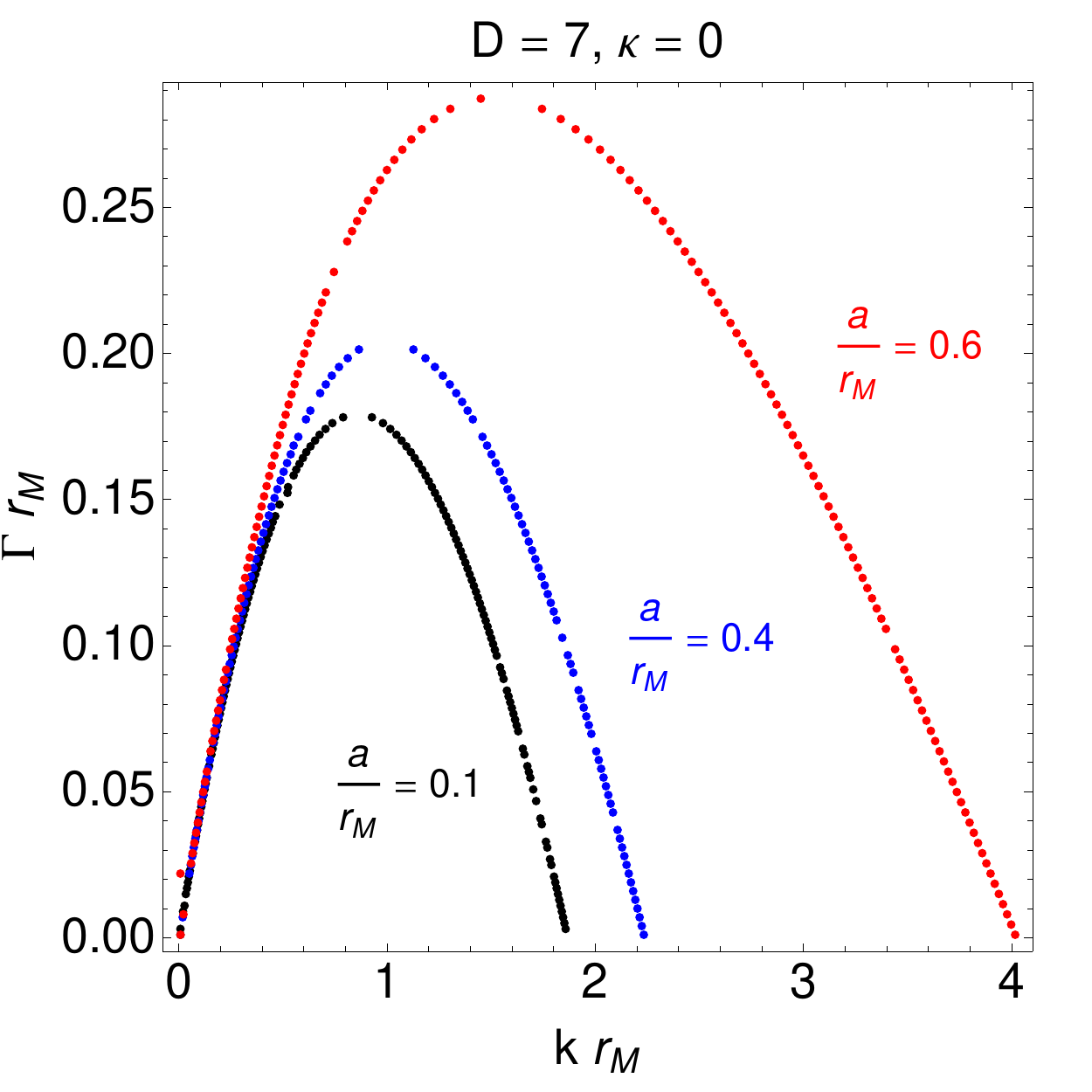}}
\caption{Results in $D=7$, $\kappa=0$. The graphs are entirely analogous to the ones in Fig.~\ref{fig:5dkzero}.}
\label{fig:7dkzero}
\end{figure}

For $\kappa=0$, we have the Gregory-Laflamme instability shown in Fig.~\ref{fig:7dkzero}. This is qualitatively the same as for $D=5$. Once again, rotation makes the branes more unstable, the behaviour as $k \rightarrow 0$ is consistent with $\Gamma/k \rightarrow 1/\sqrt{D-2}$ independently of $a$, and there is a threshold mode at a critical value of $k$ at which we expect a bifurcation of a new family of non-uniform branes analogous to the strings constructed in Ref.~\cite{Kleihaus:2007dg}.

\begin{figure}[t]
\centerline{\includegraphics[width=.45\textwidth]{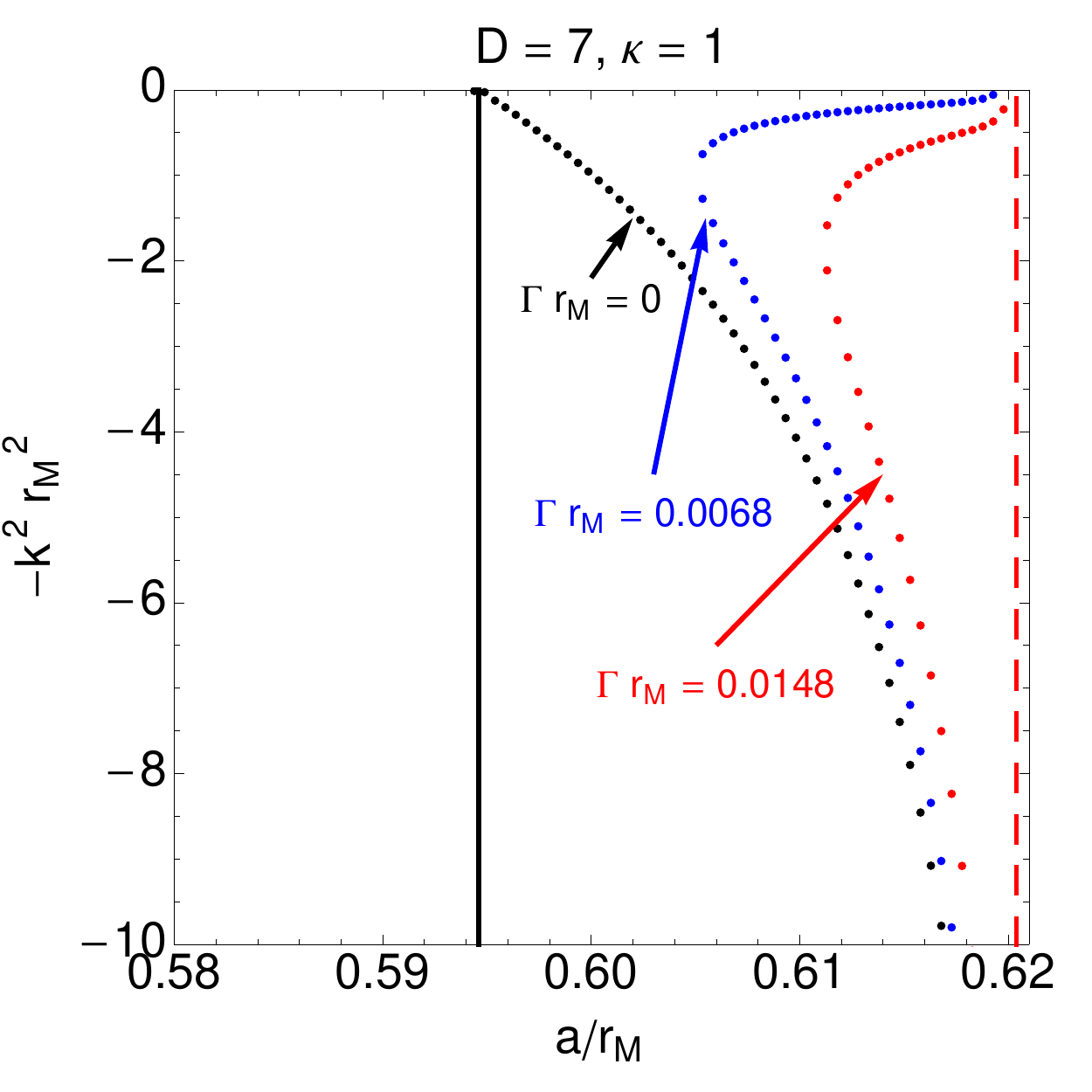}
\hspace{1cm}\includegraphics[width=.45\textwidth]{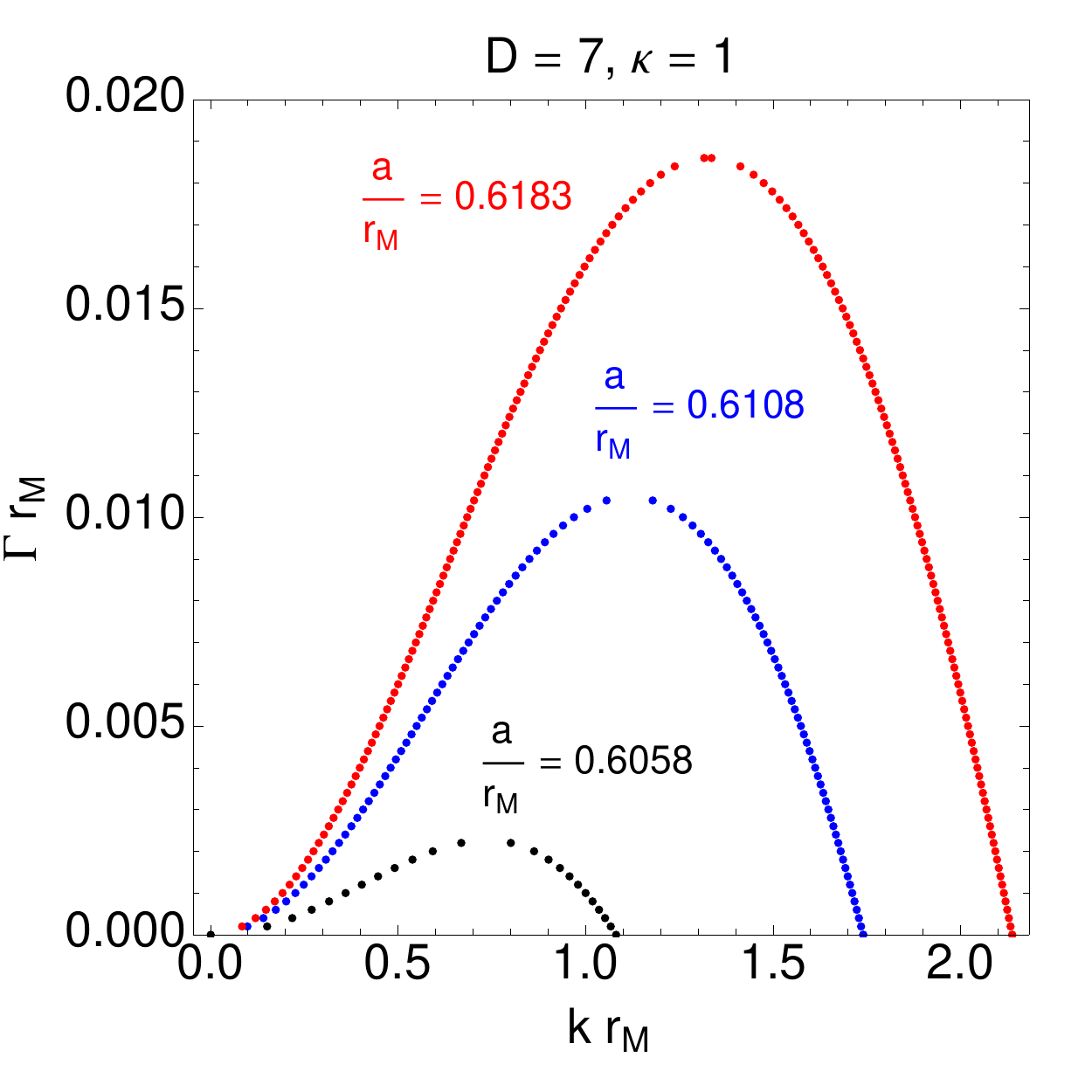}}
\caption{Results in $D=7$, $\kappa=1$. We represent this mode for fixed values of $\Gamma \,r_M$ (first graph) and $a/ r_M$ (second graph). It corresponds to a new Gregory-Laflamme instability of the rotating black branes, appearing for the numerical value $(a_1/r_M)^\mathrm{num}=0.5949$. In the first graph, the vertical line to the left corresponds to the analytical prediction of \eqref{eqn:a1}, $(a_1/r_M)=0.5946\,$, and the interrupted vertical line to the right corresponds to extremality. The second graph clearly indicates that the instability of the black branes does not extend to an instability of the black hole ($k = 0$).}
\label{fig:7d}
\end{figure}

A new feature of $D=7$ is the existence of an ultraspinning regime. This occurs for $a>a_1$, where $a_1$ was defined in equation (\ref{eqn:a1}). Just as in Ref.~\cite{Dias:2009iu} (Chapter~\ref{cha:MPsingle}), we find that a new stationary ($\Gamma=0$) negative mode of the black hole appears at this point. This negative mode has $\kappa=1$. Our numerical results are shown in  Fig.~\ref{fig:7d}. These results demonstrate that the black branes have an instability in the $\kappa=1$ sector when $a>a_1$.
This is a {\it new} Gregory-Laflamme instability of the black branes, distinct from the instability in the $\kappa=0$ sector. The plots of $\Gamma$ against $k$ have the same qualitative shape as for the $\kappa=0$ instability except that the slopes of the curves appear to vanish (for all $a$) as $k \rightarrow 0$.\footnote{It would be interesting to investigate whether this behaviour can be explained using blackfold methods.} Once again there is a threshold unstable mode at a critical value of $k$. Presumably this corresponds to a bifurcation to a new family of non-uniform black brane solutions. In addition to breaking the symmetry along the branes, this mode also breaks some of the symmetry of the black hole (typically down to that of a generic MP black hole\footnote{
This is because $\kappa=1$ harmonics are in one to one correspondence with Killing vector fields of $CP^N$: see Appendix~\ref{cpnscalarharmonicsappendix}.}) so this new family has less symmetry than the non-uniform branes associated to the threshold unstable mode with $\kappa=0$.

Note that the $\kappa=1$ black brane instability coexists with the $\kappa=0$ instability. The latter is clearly dominant since it has much larger $\Gamma$ and the instability exists for a larger range of $k$, i.e. down to shorter wavelengths.

It is important to note that there is no evidence of any instability of the black {\it hole}: none of the curves with non-zero $\Gamma$ extends to $k=0$. In the limit  $k \rightarrow 0$, solutions with non-zero $\Gamma$ approach a pure gauge mode, just as for $\kappa=0$. Additionally, the solution with $\Gamma=0$ does not approach a gauge mode as $k \rightarrow 0$. Instead, as anticipated in Chapter~\ref{cha:classthermo}, it corresponds simply to a variation of parameters within the MP family of solutions.

For $D=7$, since the black hole is ultraspinning for $a>a_1$, there is the possibility of an ultraspinning instability appearing at $a=a_2>a_1$. However, we find no solution of equation (\ref{lichnMPeq}) for $\kappa=2$ so our results are consistent with stability of $D=7$ cohomogeneity-1 MP black holes for axisymmetric perturbations ($m=0$).

\subsection{Results for $D=9$: Black hole instability}

\begin{figure}[t]
\centerline{\includegraphics[width=.45\textwidth]{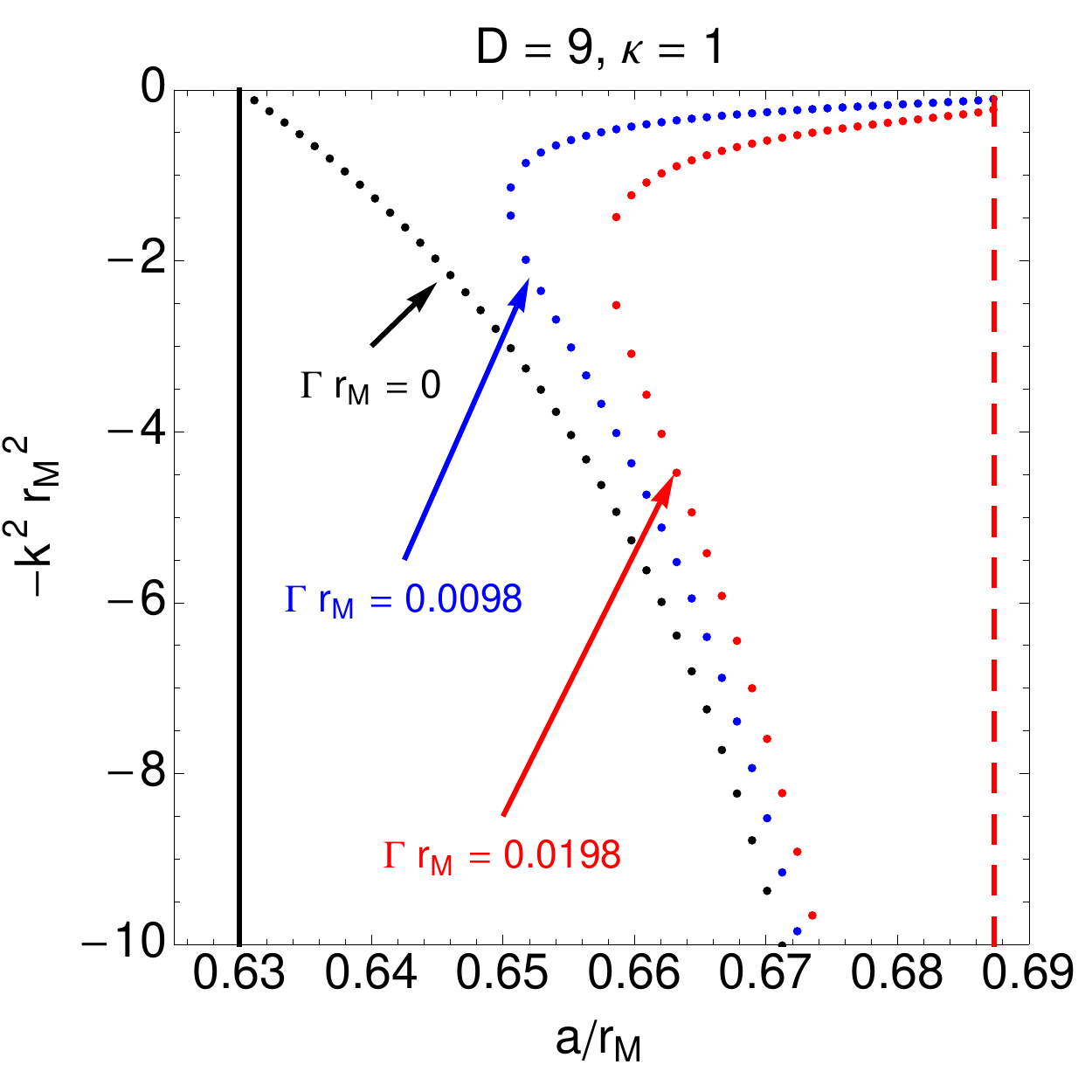}
\hspace{1cm}\includegraphics[width=.45\textwidth]{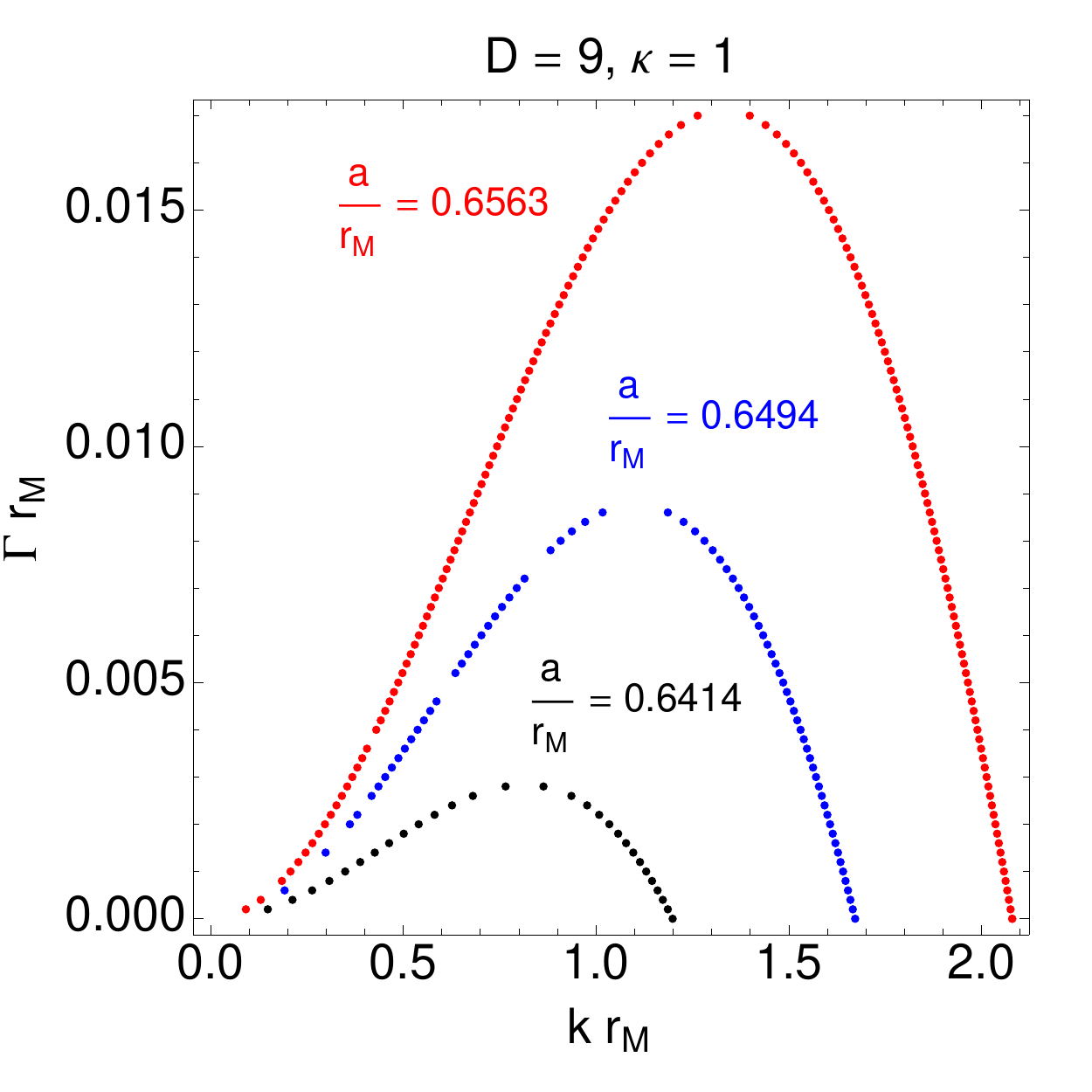}}
\caption{Results in $D=9$, $\kappa=1$. The graphs are entirely analogous to the ones in Fig.~\ref{fig:7d}.}
\label{fig:9dk1}
\end{figure}

For $\kappa=0$, we have the expected Gregory-Laflamme instability of the black branes. For $\kappa=1$, we find, as for $D=7$, a new Gregory-Laflamme instability of the black branes that appears at $a=a_1$. This is shown in figure Fig.~\ref{fig:9dk1}.

\begin{figure}[t]
\centerline{\includegraphics[width=.45\textwidth]{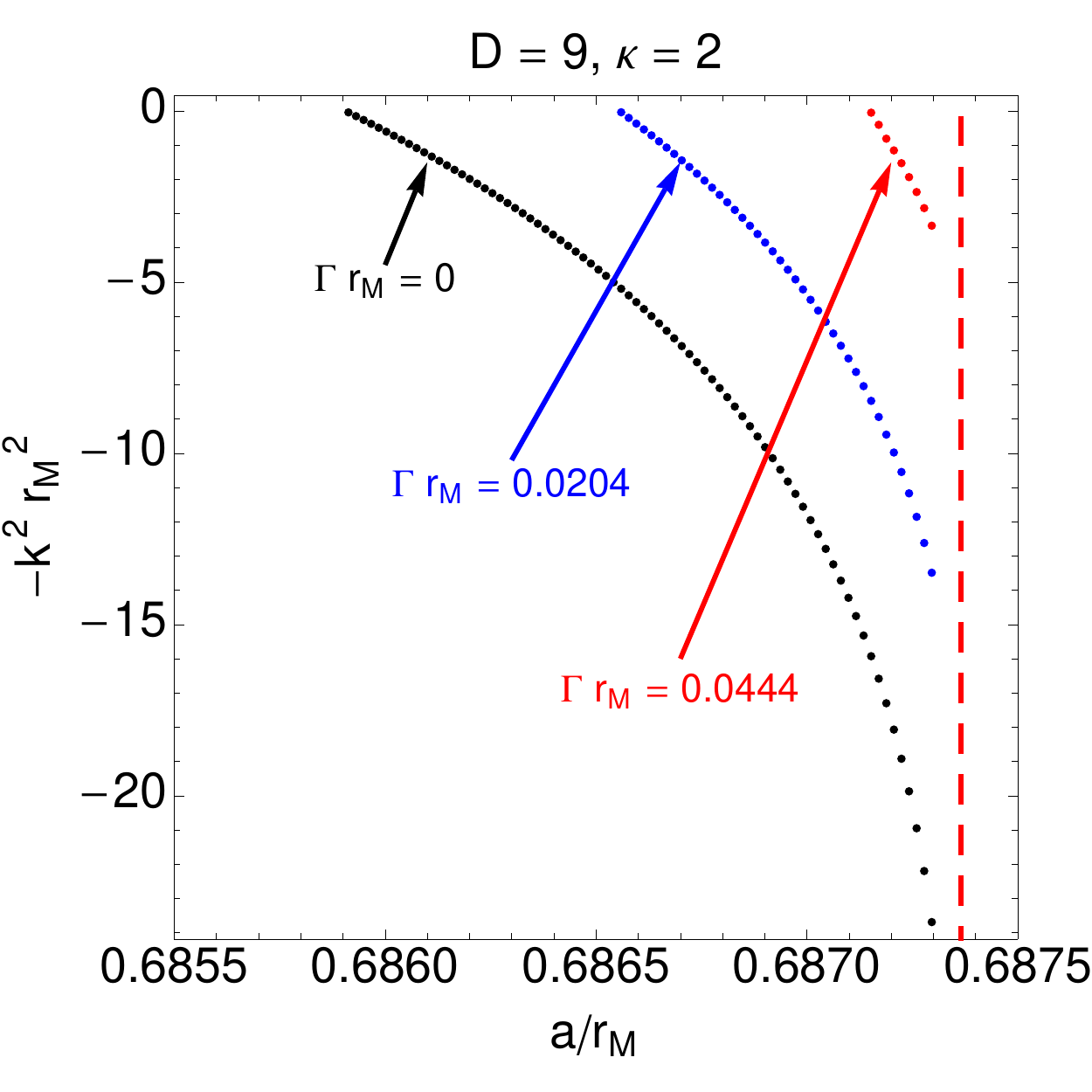}
\hspace{2cm}\includegraphics[width=.45\textwidth]{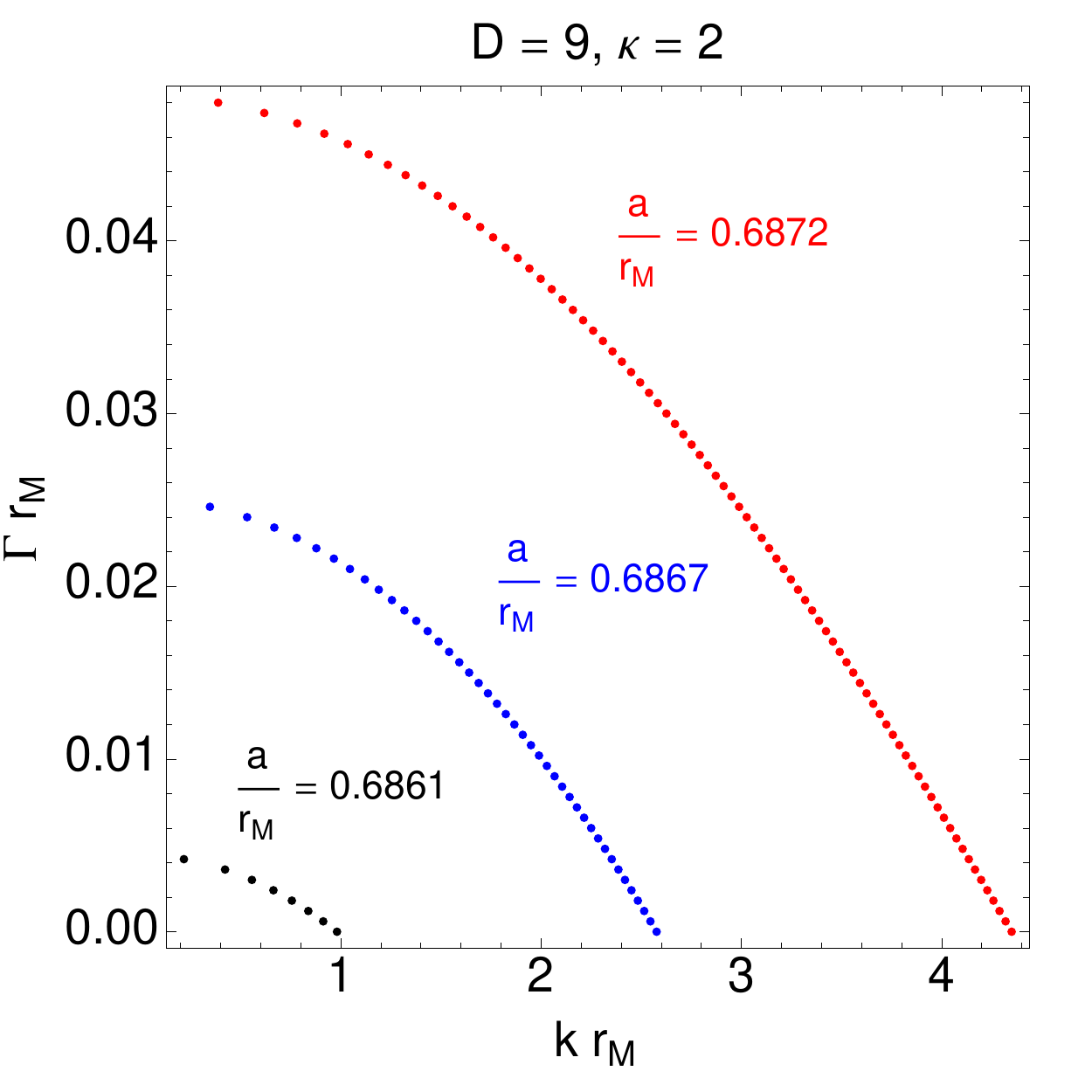}}
\caption{Results in $D=9$, $\kappa=2$. We represent this mode for fixed values of $\Gamma \,r_M$ (first graph) and $a/ r_M$ (second graph). As opposed to the $\kappa=1$ case, the time-dependent mode extends all the way to $k=0$. There is not only a new Gregory-Laflamme instability of the black branes, but also an instability of the black hole, appearing at $a=a_2$, where  $a_2/r_M=0.6858 > a_1/r_M =0.6300\,$. In the first graph, the interrupted vertical line to the right corresponds to extremality. In the second graph, the curve shrinks to the origin as $a \rightarrow a_2$.}
\label{fig:9dk2}
\end{figure}

The new feature that appears for $D=9$ is an instability in the $\kappa=2$ sector, which appears at 
$a=a_2>a_1$. This is shown in Fig.~\ref{fig:9dk2}. The left plot shows a new stationary ($\Gamma=0$) negative mode which emerges from a zero-mode at $a=a_2$. We shall prove in Appendix~\ref{sec:ChangeMJ} that this zero-mode cannot correspond to a variation of parameters within the MP family of solutions. Furthermore, in Appendix~\ref{sec:NoPureGauge}, we show that it cannot be a gauge mode.

For $a>a_2$, there is a new instability of the black branes, corresponding to the curves with $\Gamma>0$ in the plot. But there is a qualitative difference between the left plot of Fig.~\ref{fig:9dk2} and our previous plots: the curves with $\Gamma>0$ now intersect $k=0$, i.e. we have found perturbations of the black {\it hole} that grow exponentially in time, that is, a classical instability of black holes with $a>a_2$. This is our main result. 

The onset of instability is indicated by the stationary zero-mode ($\Gamma=0$, $k=0$) at $a=a_2$. This is analogous to the mode constructed for singly-spinning black holes in Ref.~\cite{Dias:2009iu} (Chapter~\ref{cha:MPsingle}). Our main achievement here is to demonstrate, for the first time, the existence of modes which grow exponentially with time when $a>a_2$.

The right plot of Fig.~\ref{fig:9dk2} shows a clear difference from our previous plots. Unlike the original Gregory-Laflamme instability, we find that $\Gamma$ is maximized at $k=0$ rather than vanishing there. Hence, for the black branes, the most unstable $\kappa=2$ modes are those with $k=0$, i.e. those corresponding to the black hole instability. For larger $k$, the black brane instability ``switches off" in the same way as the original Gregory-Laflamme instability, with a threshold mode at $k=k_\ast$ indicating a new family of non-uniform black brane solutions. 

\begin{figure}[t]
\centerline{\includegraphics[width=.5\textwidth]{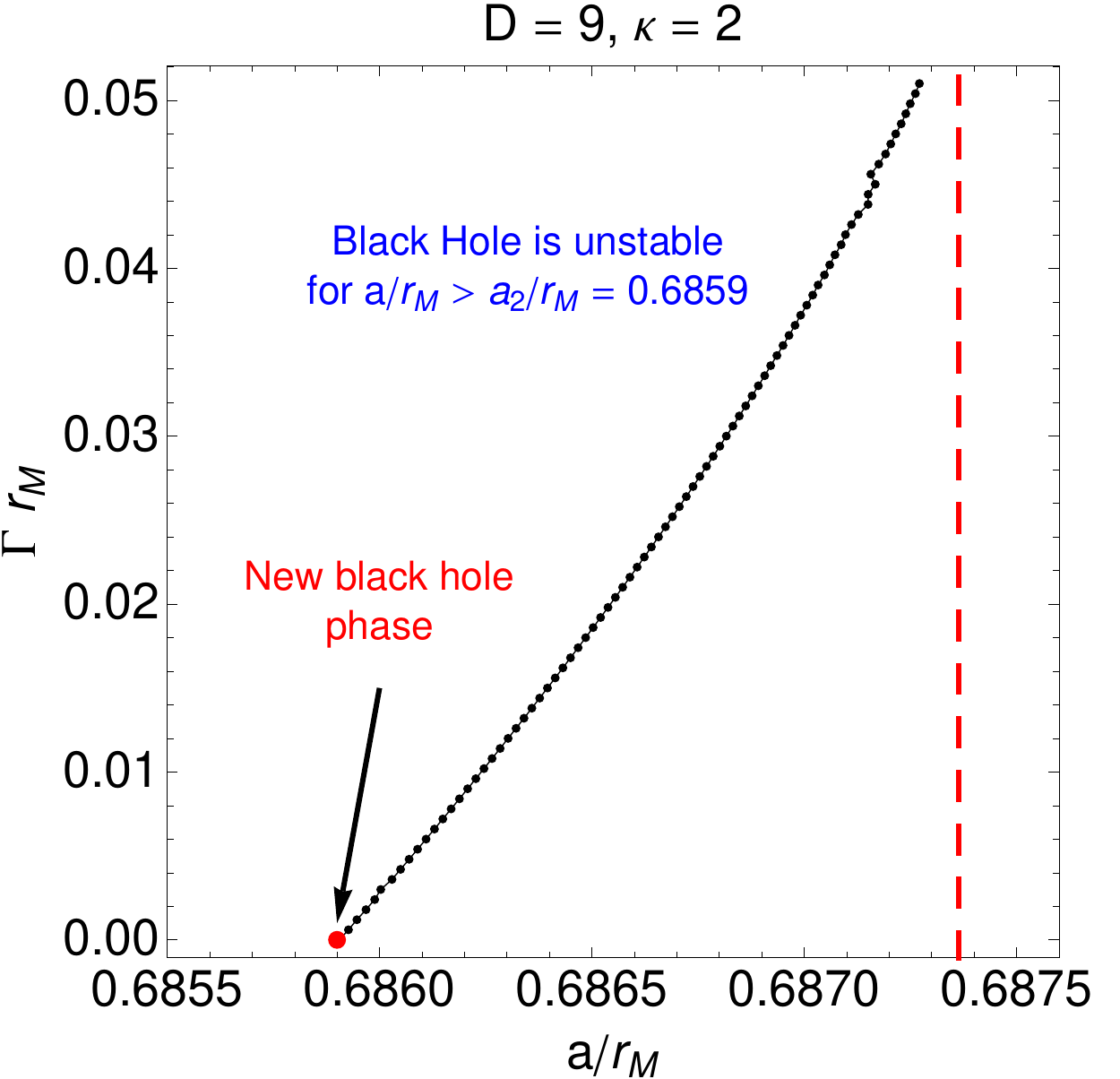}}
\caption{Results in $D=9$ for $\kappa=2$, in the limit $k=0$. We represent the black hole instability in a plot of $\Gamma \,r_M$ versus $a/ r_M$. The interrupted line corresponds to extremality. Numerical error prevents us from extending our results all the way to extremality.} 
\label{fig:9dGamma}
\end{figure}

Fig.~\ref{fig:9dGamma} presents our result for the instability time scale of the black hole as a function of its spin. For $a>a_2$, we find that $\Gamma$ increases monotonically with $a$, so extreme black holes are the most unstable.

Finally, we find no solutions of (\ref{lichnMPeq}) with $\kappa=3$.

\subsection{Rotational symmetries of higher-dimensional black holes} \label{subsec:MPeqrotsym}

As explained before, the study of perturbations of higher-dimensional black holes can be used to investigate the possible existence of new families of black hole solutions with less symmetry than the known solutions. The idea proposed in Ref.~\cite{Reall:2002bh} is to look for a stationary zero-mode of the black hole, which is interpreted as indicating the existence of a family of solutions branching off from the known solutions.

In our case, the stationary zero-modes with $\kappa=1$ are uninteresting since these correspond to variations within the MP family. However, for $D=9$, we found a stationary zero-mode with $\kappa=2$ that appears at $a=a_2$, the critical value of $a$ beyond which the black hole is unstable. Therefore, we have found evidence for a new family of black hole solutions that bifurcates from the MP family at this point.\footnote{
We expect that this new family will have unequal angular momenta in general. However, this cannot be seen from our results since $\kappa=2$ modes do not change the mass or angular momenta at the linearised level (see Appendix \ref{sec:ChangeMJ}). A second order calculation would be required to determine these changes.}

How much symmetry do these new solutions have? This can be inferred from the symmetry of the $\kappa=2$ harmonics on $CP^N$. There will, of course, be a family of degenerate scalar harmonics with $\kappa=2$. Some of these will preserve some of the symmetry of $CP^N$ whereas others break it completely (this is proved in Appendix~\ref{subsec:symmetries}).\footnote{
It is helpful to think about the case of $CP^1=S^2$, for which $\kappa=\ell$, the total angular momentum quantum number. Scalar harmonics are labelled by $\ell$ and $m$. Certain modes with $\ell=2$ (i.e. quadrupole modes) may preserve some symmetry (e.g. if $m=0$ then they are axisymmetric) but generically they break all of the continuous symmetries of $S^2$.}
For a mode of the latter type, the associated metric perturbation will break completely the $SU(N+1)$ subgroup of the $R \times U(N+1)$ isometry group of the background metric. It preserves only a $R \times U(1)$ subgroup corresponding to time-translation invariance and invariance under the rotations generated by $\partial/\partial \psi$. Hence the corresponding family of new black hole solutions will possess just a single rotational symmetry.

Very recently, Ref.~\cite{Emparan:2009vd} has constructed approximate black {\it ring} solutions with just a single rotational symmetry. Our results are the first evidence for the existence of new black hole solutions with a single rotational symmetry and horizons of spherical topology.

Note that if one used a $\kappa=2$ harmonic that does preserve some of the symmetry of $CP^N$ then presumably this would give rise to a {\it different} family of new black hole solutions, with more than one rotational symmetry. Therefore, assuming that each stationary zero-mode corresponds to a new non-linear stationary black hole solution, there must exist several new black hole solutions that bifurcate from the MP family at the same point, and these different solutions have different numbers of rotational symmetries. So how many new solutions are there?

One way of addressing this question is to determine the number of parameters in the most general $\kappa=2$ harmonic. If we take $D=9$ then $\kappa=2$ harmonics correspond to the $[2,0,2]$ representation of $SU(4)$, which is $84$-dimensional. Hence the most general $\kappa=2$ harmonic is labelled by $84$ parameters. Some such harmonics are related by acting with $SU(4)$, i.e. by rotations of the background spacetime. However, since $SU(4)$ has dimension 15, this can eliminate only 15 parameters, leaving $84-15=69$ parameters that cannot be eliminated by rotations of the background. So, up to rotations of the background, we have a family of stationary zero-modes with 69 parameters, and presumably a family of new black holes with 70 parameters, the extra parameter being the mass (or $r_M$). This is considerably more parameters than the 5 that are required to specify the $D=9$ MP solution!

\subsection{Expectations from the ultraspinning conjecture}

The predictions of the ultraspinning conjecture for MP black holes were discussed in Section~\ref{subsec:ultraspinMP}. In the equal spins case analysed here, the reduced Hessian
\be
H_{ij}\equiv -\lp{\dderf{S}{J_i}{J_j}}\rp_M
\ee
was shown to possess an eigenvalue which is always positive (with eigenvector $V_i=V\, \forall i$), and $N$ degenerate eigenvalues (with eigenvectors satisfying $\sum_i V_i=0$) which change from positive to negative as the ultraspinning surface is crossed, with no further changes of sign at larger angular momenta. Note that only the positive eigenvalue corresponds to a variation which preserves the equality of the angular momenta. We expect that new thermodynamic negative modes will emerge only at the unique value of the angular momentum corresponding to the ultraspinning surface, that there will be precisely $N$ of these, and that they will break some of the symmetries of the background. 

The ultraspinning surface corresponds to $a=a_1$ (with $r_M=1$) and we found that new stationary ($\Gamma=0$) negative modes do indeed emerge at this point in $D=7,9$. These modes correspond to $\kappa=1$ harmonics. Since these are thermodynamic negative modes, we know that the zero-mode at $a=a_1$ must be simply a variation of parameters within the MP family.

To see that there are precisely $N$ negative modes emerging at $a=a_1$, we use the fact that $\kappa=1$ harmonics are in 1-1 correspondence with Killing vector fields on $CP^N$, so there are $(N+1)^2-1$ such harmonics (see Appendix~\ref{cpnscalarharmonicsappendix}). However, some of these are related by rotations of $CP^N$, so we need to determine how many parameters can be eliminated by rotations. The counting is the same as for $SU(N+1)$ gauge theory with an adjoint Higgs field. Generically this breaks $SU(N+1)$ to $U(1)^{N}$ so we are left with $N$ parameters.\footnote{Thanks to David Tong for this argument.} Hence there are $N$ independent negative modes that emerge at $a=a_1$, in agreement with the prediction from thermodynamics.

Our numerical results confirm that the stationary negative mode that emerges at $a=a_1$ does indeed correspond to the onset of a new instability of the black branes in the $\kappa=1$ sector, in agreement with the refinement of the Gubser-Mitra conjecture discussed in Section~\ref{sec:zeromodeultraspin}.

For $D>5$, cohomogeneity-1 black holes are ultraspinning for $a>a_1$ and hence might exhibit an instability of the form anticipated in Section~\ref{sec:zeromodeultraspin}. Moreover, as explained at the end of Section~\ref{sec:background}, as we increase $D$, there is more ``space" between the ultraspinning surface and the surface of extremality for such black holes. Therefore the likelihood of an instability might be expected to increase with $D$. This is in agreement with our numerical results, which show no sign of any instability for $D=7$ but confirm that an instability is present for $D=9$. Note that this instability does indeed occur inside the ultraspinning region.

The onset of instability is associated with the appearance of a new stationary zero-mode (at $a=a_2$). This cannot correspond to a variation of parameters within the MP family (we show in Appendix~\ref{sec:ChangeMJ} that $\kappa=2$ modes cannot change the asymptotic charges). Furthermore, we prove in Appendix~\ref{sec:NoPureGauge} that it cannot be a pure gauge mode. This zero-mode is continuously connected to a stationary ($\Gamma=0$) negative mode that exists for $a>a_2$. This is an example of a non-thermodynamic negative mode, i.e. one which is not associated to a local thermodynamic instability. The same behaviour was observed in Ref.~\cite{Dias:2009iu} (Chapter~\ref{cha:MPsingle}), i.e. the onset of a classical instability of the black hole is associated with the appearance of a new stationary negative mode.

Ref.~\cite{Dias:2009iu} found that further non-thermodynamic negative modes appear as the spin of the black hole is increased still further. In the present case, extremality imposes an upper bound on the spin of the black hole and we do not find any further negative modes beyond the ones associated with the instability in the $\kappa=2$ sector. However, we believe that, for larger $D$, as well as an instability in the $\kappa=2$ sector there will be further negative modes in sectors with larger $\kappa$. These new negative modes will be associated with new instabilities of the black hole in sectors with larger $\kappa$. The stationary zero-modes marking the onset of these instabilities will provide evidence for the bifurcation of new families of black hole solutions, involving a large number of parameters, and generically with just one rotational symmetry.

\section{Scalar perturbations and $CP^N$ harmonics\label{sec:decomposition}}

\subsection{Introduction}

The rest of this Chapter is devoted to explaining the technical details of our work. We shall start by explaining the decomposition of metric perturbations into harmonics on $CP^N$.

Metric perturbations can be decomposed into scalar, vector and tensor types according to how they transform under isometries of $CP^N$. Pertubations of different type must decouple from each other. The decomposition is explained (for a different problem) in Ref.~\cite{Martin:2008pf}. Tensor perturbations are the simplest, these were discussed in Ref.~\cite{Kunduri:2006qa}. We are interested in scalar perturbations, for which the perturbation can be expanded in scalar harmonics on $CP^N$.

We shall assume that our perturbation has been Fourier decomposed as in equation (\ref{eqn:fourier}). All of our numerical results assume $m=0$. However, for the sake of completeness, we shall derive equations that are valid for non-zero $m$. In order to do this, we must address a subtlety (already encountered in Ref.~\cite{Kunduri:2006qa}), that such a perturbation couples with charge $m$ to the 1-form $A_a$ on $CP^N$ defined in Section~\ref{sec:background}. Hence we must consider {\it charged} scalar harmonics on $CP^N$. First we shall describe these harmonics and then explain how to construct gravitational perturbations from them.

\subsection{Charged scalar-derived harmonics in $CP^N$ \label{secscalars}}

We define the gauge-covariant derivative acting on a charge-$m$ tensor field on $CP^N$ as
\begin{equation}
 \CD_a=\hat{\nabla}_a-\ii\, m\,A_a\,,
\end{equation}
where $\hat \nabla$ is the metric covariant derivative on $CP^N$.

\subsubsection{Scalars}

Charged scalar fields on $CP^N$ can be expanded in terms of charged scalar harmonics defined by
\begin{equation}
 (\CD^2+\lambda)\Y=0\,.\label{eqn:chargedS}
\end{equation}
These were studied in \cite{Hoxha:2000jf} (see summary in Appendix~\ref{cpnappendix}), where it is found that
\be
\label{eqn:eigenvalues}
\lambda=\ell(\ell+2N) -m^2\,, \qquad \ell=2\kappa +|m|\,,
\ee
with $\kappa=0,1,2,\ldots$. The modulus sign guarantees that the eigenvalue is the same for positive and negative charges, the corresponding eigenfunctions being related by complex conjugation. 

Notice that the presence of the `gauge field' $A$ leads to
\begin{equation}
\label{commutator}
 [\CD_a,\,\CD_b]\Y=-\ii\, 2\,m\,J_{ab}\,\Y\,.
\end{equation}

\subsubsection{Scalar-derived 1-forms}

Given a scalar harmonic $\Y$, we can define\footnote{Note that $\lambda=0$ if, and only if, $\kappa=m=0$, in which case $\Y$ is uncharged and constant. In this case there are no scalar-derived vectors nor scalar-derived tensors.}
\begin{equation}
 \Y_a=-\frac{1}{\sqrt{\lambda}}\,\CD_a \Y\,, \label{eqn:vectorH}
\end{equation}
which transforms as a charged 1-form on $CP^N$. This can be decomposed into its $(1,0)$ and $(0,1)$ parts using the complex structure on $CP^N$. Denote these by $\Y^+_a$ and $\Y^-_a$ respectively, where 
\be
 J_a{}^b \Y^\pm_b = \mp \ii \Y^\pm_a.
\ee
We shall refer to $\Y^\pm_a$ as scalar-derived 1-form harmonics.
We find that they satisfy
\begin{equation}
\CD^2\Y_a^{\pm}=-\left[\lambda-2(N+1)\mp 4\,m\right]\Y_a^{\pm}\,\end{equation}
and
\begin{equation}
\CD^a\Y_a^\pm=
	\frac{1}{2\sqrt{\lambda}}\left(\lambda \mp 2\,m\,N\right)\,\Y\,.
\end{equation}
We shall make use of the result that Killing vectors of $CP^N$ are in one-to-one correspondence with uncharged ($m=0$) scalar harmonics with $\kappa=1$ (see e.g. \cite{Hoxha:2000jf} and our Appendix~\ref{cpnappendix}).
Given such a harmonic $\Y$, the corresponding Killing vector field is $-\ii(\Y^+_a-\Y^-_a)$.

\subsubsection{Scalar-derived tensors}

Following Ref.~\cite{Martin:2008pf}, we  decompose a symmetric tensor $\Y_{ab}$ into its Hermitian (or $(1,1)$) and anti-Hermitian components according to the eigenvalue of the map
\begin{equation}
 ({\cal J}\Y)_{ab}=J_a^{\phantom a c}J_{b}^{\phantom b d}\Y_{cd}\,.
\end{equation}
If the eigenvalue is $+1$ the corresponding eigentensor is called Hermitian, and if it is $-1$ the eigentensor is called anti-Hermitian. In the anti-Hermitian case, we can further distinguish between the $(2,0)$ and $(0,2)$ components of $\Y_{ab}\,$, which are defined by $J_a^{\phantom a c}\Y_{cb}=\mp \ii\, \Y_{ab}$ with the upper and lower signs for the $(2,0)$ and $(0,2)$ components respectively.

The following quantities form a basis for anti-hermitian scalar-derived tensors:\footnote{This follows from the scalar part of equation (39) of Ref.~\cite{Martin:2008pf}.}
\begin{equation}
 \Y_{ab}^{++}=\CD_{(a}^+\Y_{b)}^+\,,\qquad \Y_{ab}^{--}=\CD_{(a}^-\Y_{b)}^-\,.
\end{equation}
$\Y_a^{\pm}$ denotes the scalar-derived 1-form harmonics of the previous Section, and $\CD_a^{\pm}$ denotes the projection of $\CD_a$ onto its $(1,0)$ and $(0,1)$ components. Notice that the correspondence between $m=0$, $\kappa=1$ scalar harmonics and Killing vector fields on $CP^N$ implies that $\Y^{\pm\pm}_{ab}$ vanish for such harmonics.

Hermitian scalar-derived tensors can be written in terms of a trace and a traceless part, for which the following quantities give a basis:\footnote{
In Ref.~\cite{Martin:2008pf}, hermitian tensors were converted into $(1,1)$-forms by contracting with $J^a{}_b$. The two quantities written here correspond to terms of the form $J \Y$ and (the primitive part of) $dd^c \Y$ in equation (47) of Ref.~\cite{Martin:2008pf}.}
\begin{equation}
\hat{g}_{ab} \Y, \qquad \Y_{ab}^{+-}=\CD_{(a}^+\Y_{b)}^-+\CD_{(a}^-\Y_{b)}^+-\frac{1}{2\,N}\,\hat{g}_{ab}\,(\CD\cdot \Y)\,.
\end{equation}

These tensor harmonics satisfy 
\begin{equation}
\begin{aligned}
\CD^2\Y_{ab}^{\pm\pm}&=-\left[\lambda-4(N+3) \mp 8\,m\right]\Y_{ab}^{\pm\pm}\,,\\
\CD^2\Y_{ab}^{+-}&=-\left(\lambda-4\,N\right)\Y_{ab}^{+-}\,,
\end{aligned}
\end{equation}
with
\begin{equation}
\begin{aligned}
\CD^c\Y_{ca}^{\pm\pm}&=-\frac{\lambda-4(N+1) \mp 2\,m(N+2)}{2}\,\Y_a^\pm \,,\\
\CD^c\Y_{ca}^{+-}&=-\frac{N-1}{2\,N}\left[(\lambda+2\,m\,N)\Y_a^+
	+(\lambda-2\,m\,N)\Y_a^-\right]\,.
\end{aligned}
\end{equation}
 

\subsection{Decomposition of perturbations in scalar-derived harmonics}

Let us now consider the perturbations of the full spacetime metric. We introduce the orthonormal basis
\begin{equation}
\label{vielbein}
 e^{(0)}=f\,dt\,,\quad e^{(1)}=g\,dr\,,\quad e^{(2)}=h\,\left(d\psi+A-\Omega\,dt\right)\,,\quad
e^{(i)}=r\,\hat e^{(i)}\,,
\end{equation}
where $\hat e^{(i)}$ is the tetrad of the $CP^N$ manifold. The dual basis is then
\begin{equation}
 e_{(0)}=\frac{1}{f}\left(\partial_t+\Omega\,\partial_\psi\right)\,,\quad 
e_{(1)}=\frac{1}{g}\,\partial_r\,,\quad e_{(2)}=\frac{1}{h}\,\partial_\psi\,,\quad
e_{(i)}=\frac{1}{r}\left[\hat e_{(i)}-\langle A,\hat e_{(i)}\rangle\,\partial_\psi\right]\,.
\end{equation}
Take $e^{(A)}=\{e^{(0)},e^{(1)},e^{(2)}\}$ and a coordinate basis $dx^a$ on $CP^N$. The components $h_{AB}$ of the metric perturbation transform as scalars under isometries of $CP^N$ and can therefore be decomposed using scalar harmonics on $CP^N$. Similarly, since we are restricting attention to scalar-type perturbations, components of the form $h_{Aa}$ and $h_{ab}$ can be decomposed using scalar-derived 1-forms and scalar-derived tensors on $CP^N$:
\begin{equation}
\begin{aligned}
h_{AB}&=f_{AB}\,\Y\,,\\
h_{Aa}&=r\left(f^+_A\,\Y^+_a+f^-_A\,\Y^-_a\right)\,,\\
h_{ab}&=-\frac{r^2}{\sqrt\lambda}\left(
H^{++}\,\Y^{++}_{ab}+H^{--}\, \Y^{--}_{ab}+H^{+-}\,\Y^{+-}_{ab}\right)
+r^2\,H_L\,\hat{g}_{ab}\Y\,,
\end{aligned}
\label{eqn:hmetric}
\end{equation}
where $f^\pm_A=\{W^\pm,X^\pm,Z^\pm\}$, and the functions multiplying the harmonics depend only on ($t,r,\psi$) and not on the coordinates of $CP^N$. The real spacetime metric perturbation is given by $\textrm{Re}\left(h_{\mu\nu}\right)$. Since $\partial_t$ and $\partial_\psi$ are Killing vectors of the background solution, we will Fourier expand all of these functions in $t$ and $\psi$, i.e. we assume a dependence $e^{-\ii\omega t+\ii m\psi}$. It remains to determine the dependence of these functions on $r$. The stability problem will thus be reduced to a system of linear ordinary differential equations.

\subsection{Boundary conditions \label{secBC}}

The metric perturbations must be regular on the future event horizon ${\cal H}^+$. This boundary condition can be imposed by considering a basis which is regular on ${\cal H}^+$, since the components of the perturbation in that basis must be regular. Let us change to the ingoing Eddington-Finkelstein coordinates that are regular at ${\cal H}^+$:
\begin{equation}
 dt\to dv-\frac{g}{f}\,dr\,,\qquad d\psi\to d\varphi -\frac{\Omega\,g}{f}\,dr\,,
\end{equation}
and consider the basis $\{ dv, dr, d\varphi +A -\Omega dv, dx^a \}$. Denote the components of the metric perturbation with respect to this new basis with a bar (e.g. $f_{\bar{0}\bar{0}}$, $\bar{W}^+$). Our boundary condition is that these components should be smooth functions of $(v,r,\phi,x^a)$ at the horizon.\footnote{In other words, we demand that the tensor field $h_{\mu\nu}$ should be regular at ${\cal H}^+$. This is stronger than the statement that the metric perturbation should be regular at the horizon, e.g. it excludes the possibility that $h_{\mu\nu}$ is singular at ${\cal H}^+$ in a certain gauge but can be made regular by a gauge transformation. For example, we show in Appendix \ref{sec:PreserveTAngVel} that, in the traceless-transverse gauge, a perturbation with $\omega=m=0$ that satisfies our boundary condition cannot change the temperature or angular velocity of the black hole. It follows that a variation in the parameters of the MP solution that {\it does} change $T_H$ or $\Omega_H$ will not give a perturbation $h_{\mu\nu}$ that is regular at the horizon in this gauge.}

For this class of black holes, the horizon is located at the largest real root $r=r_+$ of $\Delta=g(r)^{-2}$. For a non-extreme black hole, near the horizon,
$\Delta(r)=\Delta'(r_+)(r-r_+)+O[(r-r_+)^2]$, with $\, \Delta'(r_+)>0$.
Using the relation $f(r)=r/(g(r)\,h(r))$, we find that, near the horizon, the metric components in the original basis are related to the components in the new basis by
\begin{equation}
\label{bchor}
\begin{aligned}
& f_{00}\approx \frac{h(r_+)^2}{r_+^2\Delta'(r_+)} \,\frac{f_{\bar{0}\bar{0}}}{r-r_+}\,,\qquad f_{01}-f_{00} = \frac{h(r_+)}{r_+} \, f_{\bar{0}\bar{1}}\,, \\ 
&  f_{00}-2 f_{01}+ f_{11}\approx \Delta'(r_+) \, f_{\bar{1}\bar{1}}\; (r-r_+)\,, \\
& f_{02} \approx \frac{1}{r_+\sqrt{\Delta'(r_+)}} \, \frac{f_{\bar{0}\bar{2}}}{\sqrt{r-r_+}}\,,\qquad 
 f_{12}-f_{02}\approx \frac{\sqrt{\Delta'(r_+)}}{h(r_+)}  \, f_{\bar{1}\bar{2}} \sqrt{r-r_+}\,, \\
& f_{22}= \frac{1}{h(r_+)^2} \, f_{\bar{2}\bar{2}}\,, \qquad Z^{\pm}= \frac{1}{r_+ h(r_+)} \, \bar{Z}^{\pm} \,,  \\
&W^\pm \approx \frac{h(r_+)}{r_+\sqrt{\Delta'(r_+)}} \,\frac{\bar{W}^{\pm}}{\sqrt{r-r_+}}\,,\qquad
X^\pm - W^\pm \approx \frac{\sqrt{\Delta'(r_+)}}{r_+} \, \bar{X}^{\pm} \sqrt{r-r_+}\,.
\end{aligned}
\end{equation}
The functions $H^{++}$, $H^{--}$, $H^{+-}$, $H_L$ associated with the components of the metric perturbation on $CP^N$ are the same in the two bases.

Since the components in the new basis should be regular at the horizon, the above expressions give us boundary conditions on the behaviour of the components in the old basis. In imposing these boundary conditions, it is important to remember that, near the horizon,
\be
\label{expohor}
e^{-\ii \omega v + \ii m \varphi} \approx e^{-\ii \omega t + \ii m \psi} \left( \frac{r-r_+}{r_+} \right)^{-\ii \alpha (\omega-m\Omega_H)},
\ee
where $\alpha \equiv h(r_+)/(r_+ \Delta'(r_+))$ is positive (for non-extreme black holes). Hence, for example, the radial dependence of $f_{00}$ near the horizon must be
\be
 f_{00} \propto  \left( \frac{r-r_+}{r_+} \right)^{-1-\ii \alpha (\omega-m\Omega_H)}F(r),
\ee
where $F(r)$ is smooth at $r=r_+$.

When numerically solving the stability equations, it will be necessary to work with the combinations that maximize the information on the boundary conditions. For instance, one should work with $f_{02}$ and $f_{12}-f_{02}$, instead of considering only the leading behaviour of $f_{02}$ and $f_{12}$, otherwise the information that $f_{12}(r_+)-f_{02}(r_+)=0$ is lost.

As for the behaviour of the perturbations at spatial infinity $r \to \infty$, we are interested in boundary conditions that preserve the asymptotic flatness of the spacetime. For perturbations of the black branes, the equations of motion \eqref{lichnMPeq} then imply that all the functions vanish exponentially for large $r$.

\section{The eigenvalue problem \label{sec:eigenvalue}}

The ansatz for the metric perturbation $h_{\mu\nu}$ is given by Eq.~\eqref{eqn:hmetric}. Ref.~\cite{Dias:2010eu} lists in a long Appendix the components of the Lichnerowicz eigenvalue equation \eqref{lichnMPeq} in the tetrad basis \eqref{vielbein}. These consist of sixteen coupled second order ordinary differential equations, each one being second order only in one of the perturbation functions. However, six of these functions can be solved for in terms of the ten remaining functions and their first derivatives when we impose the traceless-transverse (TT) gauge conditions, listed in another Appendix of \cite{Dias:2010eu}. The procedure is analogous to the one applied in Chapters~\ref{cha:KerrAdS} and \ref{cha:MPsingle}.

Notice that the TT conditions completely fix the gauge in Eq.~(\ref{lichnMPeq}) when $k>0$, since the action of the Lichnerowicz operator on a gauge mode is trivial, $\Delta_L \nabla_{(\mu} \xi_{\nu)}=0$. As for $k=0$, there are two distinct cases. The first is the limit $k \to 0$ for which $\Gamma \to 0$, as can be seen on the right plots of Figs.~\ref{fig:5dkzero}, \ref{fig:7dkzero}, \ref{fig:7d} and \ref{fig:9dk1}. The limiting perturbation $k=0$ is an unphysical pure gauge mode, as happens in the original Gregory-Laflamme case \cite{Gregory:1994bj}. The second is the much more interesting stationary perturbation $k_\ast=0$ marking the onset of a new Gregory-Laflamme instability when the rotation increases. In the left plots of Figs.~\ref{fig:7d}, \ref{fig:9dk1} and \ref{fig:9dk2}, this is the threshold mode of the curve $\Gamma=0$. In the right plots of the same Figures, it would correspond to the graph squeezing into the origin for a critical value of the rotation. These stationary perturbations are physical since there is no gauge ambiguity. The TT conditions require that any gauge vector $\xi_\mu$ is a harmonic 1-form, satisfying $\nabla^\mu \xi_\mu =0$ and $\nabla^\rho \nabla_\rho \xi_{\mu} =0$. In Appendix~\ref{sec:NoPureGauge}, we show that no regular harmonic 1-forms exist in this case: they lead to pure gauge metric perturbations that diverge either at the boundary $r=r_+$ or at infinity $r \to \infty$. A proof along the same lines, but much more cumbersome, can be given for the modes which represent the actual instability of the black holes, i.e. the exponential growth of the perturbations with time ($k=0$ and $\Gamma >0$).

The tetrad basis \eqref{vielbein} is very convenient in the explicit derivation of the TT gauge conditions and the Lichnerowicz equations. However, for the actual implementation of the numerical problem, it is convenient to choose perturbation functions that make the final equations more amenable to numerics, e.g. it is useful to avoid using expressions involving square roots. It is also helpful to define combinations of the original perturbation functions which can be solved for algebraically through the TT gauge conditions. Both features are respected if we consider the perturbations in a related basis such that:
\be
\label{eq:newfunc}
\begin{array}{l}
\mathfrak{f}_{00} = f_{00}\,f^2-2 f_{02}\,f\, h\, \Omega +f_{22} \,h^2\, \Omega^2\,, \qquad  \qquad
\mathfrak{f}_{01} = f_{01}\, f \,g-f_{12}\, g\, h\, \Omega\,, \\ \\
\mathfrak{f}_{02} = f_{02}\, f\, h-f_{22}\, f^2\, \Omega \,, \qquad
\mathfrak{f}_{11} = f_{11}\, g^2 \,, \qquad
\mathfrak{f}_{12} = f_{12}\, g\,h\,, \qquad
\mathfrak{f}_{22} = f_{22}\, h^2\,, \\ \\
\mathfrak{f}_{0} = -\frac{1}{2} \,r\, \big( (W^+ +W^-)\,f -(Z^+ +Z^-)\,h\, \Omega \big) \,, \\ \\
\tilde{\mathfrak{f}}_{0} =- \ii \, \frac{1}{2} \,r\, \big( (W^+ -W^-)\,f -(Z^+ -Z^-)\,h\, \Omega \big) \,, \\ \\
\mathfrak{f}_{1} = -\frac{1}{2} \,r \, g\, (X^+ +X^-)\,, \qquad
\tilde{\mathfrak{f}}_{1} =- \ii \, \frac{1}{2} \,r \, g\, (X^+ -X^-)\,, \\ \\
\mathfrak{f}_{2} = -\frac{1}{2} \,r \, h \,(Z^+ +Z^-)\,, \qquad
\tilde{\mathfrak{f}}_{2} = -\ii \, \frac{1}{2} \,r \, h\, (Z^+ -Z^-)\,, \\ \\
P = \frac{1}{4}\, (2 H^{+-}-H^{++}-H^{--})\,, \qquad
Q = \ii\,\frac{1}{2}\, (H^{++}-H^{--})\,,
\end{array}
\ee
\bea
\begin{array}{l}
U = \frac{1}{4}\, (2 H^{+-}+H^{++}+H^{--})\,, \qquad
V = H_L + \frac{1}{4 N}\, \left(H^{+-}-\frac{1}{2}\, (H^{++}+H^{--}) \right)\,.
\end{array} \nonumber
\eea
Now, we solve for six of these functions ($\mathfrak{f}_{00}$, $\mathfrak{f}_{0}$, $\mathfrak{f}_{2}$, $Q$, $U$, $V$) in terms of the ten remaining functions and their first derivatives by imposing the TT gauge conditions. Upon this substitution, the corresponding second order Lichnerowicz equations will become third order. The ten equations which are second order in $\mathfrak{f}_{01}$, $\mathfrak{f}_{02}$, $\mathfrak{f}_{11}$, $\mathfrak{f}_{12}$, $\mathfrak{f}_{22}$, $\tilde{\mathfrak{f}}_{0}$, $\mathfrak{f}_{1}$, $\tilde{\mathfrak{f}}_{1}$, $\tilde{\mathfrak{f}}_{2}$, $P$, will remain second order. They constitute the system of equations to be solved numerically. A non-trivial consistency check on the gauge choice procedure is that the ten final second order equations must solve the six third order equations (e.g. a third order equation is a derivative of a second order one). We verified explicitly that this is the case.

The final system will be solved using a spectral numerical method, briefly described in the Appendix at the end of this thesis. The application of the method is simpler for Dirichlet boundary conditions. We then consider the following perturbation functions:
\be
\label{eq:qs}
\begin{array}{l}
\displaystyle{q_1 =  \left( 1-\frac{r_+}{r} \right)^{\ii \alpha(\omega-m\Omega_H)+3}}\, \mathfrak{f}_{11} \,, \qquad \qquad
\displaystyle{q_2 =  \left( 1-\frac{r_+}{r} \right)^{\ii \alpha(\omega-m\Omega_H)+1}}\, \mathfrak{f}_{22} \,, \\
\displaystyle{q_3 =  \left( 1-\frac{r_+}{r} \right)^{\ii \alpha(\omega-m\Omega_H)+2}}\, \mathfrak{f}_{01} \,, \qquad \qquad
\displaystyle{q_4 =  \left( 1-\frac{r_+}{r} \right)^{\ii \alpha(\omega-m\Omega_H)+2}}\, \mathfrak{f}_{1} \,, \\
\displaystyle{q_5 =  \left( 1-\frac{r_+}{r} \right)^{\ii \alpha(\omega-m\Omega_H)+2}}\, \tilde{\mathfrak{f}}_{1} \,, \qquad \qquad \;\,
\displaystyle{q_6 =  \left( 1-\frac{r_+}{r} \right)^{\ii \alpha(\omega-m\Omega_H)+1}}\, \tilde{\mathfrak{f}}_{2} \,, \\ 
\displaystyle{q_7 =  \left( 1-\frac{r_+}{r} \right)^{\ii \alpha(\omega-m\Omega_H)+1}}\, P \,, \\
\displaystyle{q_8 =  \left( 1-\frac{r_+}{r} \right)^{\ii \alpha(\omega-m\Omega_H)}}\, \left\{\mathfrak{f}_{02}+\Omega_H\,\mathfrak{f}_{22} +
\frac{r_+}{\alpha\,\Omega_H} \left( 1-\frac{r_+}{r} \right) \left[ \mathfrak{f}_{01}-\frac{r_+}{\alpha} \left( 1-\frac{r_+}{r} \right)\mathfrak{f}_{11} \right]\right\} \,, \\
\displaystyle{q_9 =  \left( 1-\frac{r_+}{r} \right)^{\ii \alpha(\omega-m\Omega_H)+1}}\, \left\{\mathfrak{f}_{12}+ \frac{1}{\Omega_H}\,\left[ \mathfrak{f}_{01} - \frac{r_+}{\alpha} \left( 1-\frac{r_+}{r} \right) \mathfrak{f}_{11} \right]\right\} \,, \\
\displaystyle{q_{10} =  \left( 1-\frac{r_+}{r} \right)^{\ii \alpha(\omega-m\Omega_H)}}\, \left\{ \tilde{\mathfrak{f}}_{0} + \Omega_H\, \tilde{\mathfrak{f}}_{2} -\frac{r_+}{\alpha} \left( 1-\frac{r_+}{r} \right) \tilde{\mathfrak{f}}_{1} \right\}\,, \\
\end{array}
\ee
which vanish linearly at the horizon location $r=r_+$. This behaviour can be verified in the expressions \eqref{bchor} and \eqref{expohor}. The particular combinations chosen for $q_8$, $q_9$ and $q_{10}$ encode the total information about the boundary conditions imposed by regularity, as argued in the end of Section~\ref{secBC}. For the numerical implementation, it is convenient to use the variable
\be
y=1-\frac{r_+}{r}
\ee
instead of the radial coordinate $r$, since $y$ is dimensionless and bounded, $0\leq y \leq 1\,$. The functions represented above vanish at infinity $r=\infty$ ($y=1$) since the large $r$ behaviour of the solutions of \eqref{lichnMPeq} is exponential, $e^{\pm k r}$, and regularity at infinity imposes the minus sign. The $k=0$ case will be obtained as the limit $k \to 0$.

The system of ten second order ordinary differential equations is ready to be solved numerically. The results were presented in Section~\ref{sec:strategyresults}.


\begin{subappendices}

\section{Appendix: The geometry of $CP^N$ \label{cpnappendix}}

\subsection{The Fubini-Study construction}

We review here the Fubini-Study construction of the Einstein-K\"ahler metric and K\"ahler potential on $CP^N$ \cite{Hoxha:2000jf}.\footnote{We use the coordinates $\{R_N,\Psi_N\}$ that are related to the coordinates $\{\xi,\widetilde{\tau} \}$ of \cite{Hoxha:2000jf} through the coordinate transformation $\sin^2\xi=R_N^2/(1+R_N^2)$ and $\widetilde{\tau}=\Psi_N/2$.} This construction allows us to iteratively generate the $CP^N$ metric and potential from the knowledge of the metric and potential of $CP^{N-1}$.

Take the $\mathbb{C}^{N+1}$ manifold with complex coordinates $Z^A$ and flat metric
\be
ds_{2N+2}^2 = dZ^A\, d\overline{Z}_A\,,\label{FlatMetric}
\ee
where the index $A$ runs as $A=(0,\alpha)$, with $1\leq\alpha\leq N$. Introduce $N$ inhomogeneous coordinates $\zeta^\alpha = Z^\alpha/Z^0$ in the patch where $Z^0\neq 0\,$, such that
\begin{align}
& Z^0=e^{\ii \tau}\, |Z^0|\,,\quad Z^\alpha=Z^0 \,\zeta^\alpha =R_N\, u^\alpha\,, \nonumber \\
& Z^A\, \overline{Z}_A=r^2\,,\quad
f=1+\zeta^\alpha\, \overline{\zeta}^{\bar\alpha}=1+R_N^2
\,.\label{Def:zeta}
\end{align}
Furthermore, introduce a new set of $(N-1)$ inhomogeneous coordinates $v^i$ ($0\leq i\leq N-1$) such that
\be
u^N=e^{\ii \Psi_N/2}\, |u^N|\,,\qquad  u^i=u^N v^i \qquad \hbox{with}\qquad u^\alpha \, \bar u^{\bar\alpha}= 1 \,.\label{Def:v}
\ee
The flat metric on $\mathbb{C}^{N+1}$ can then be written as
\bea
ds_{2N+2}^2 = dr^2 + r^2\, d\Omega_{2N+1}^2\,, \qquad
\hbox{where}\qquad d\Omega_{2N+1}^2 = (d\tau + A_{(N)})^2 +
d\Sigma_N^2\, \label{Hopf}
\eea
is the metric on the unit sphere $S^{2N+1}$, and $d\Sigma_N^2$ is the unit $CP^N$ metric. Written in this way we see that $S^{2N+1}$ is a Hopf fibration of $S^1$ over $CP^N$. In \eqref{Hopf}, $A_{(N)}$ is the $CP^N$ K\"ahler potential. Explicitly, the  $CP^N$  metric and K\"ahler potential are given by ($R_N\geq 0$ and $0\leq \Psi_N \leq  4\pi$)
\begin{eqnarray}
&& d\Sigma_{N}^2 = \hat{g}_{ab} dx^a dx^b = \frac{dR_N^2}{\lp 1+R_N^2\rp^2}
  +\frac{1}{4}\frac{R_N^2}{\lp 1+R_N^2\rp^2}\,\lp d\Psi_N+2A_{(N-1)} \rp^2 +
\frac{R_N^2}{ 1+R_N^2}\, d\Sigma_{N-1}^2\,, \nonumber \\
&&
 A_{(N)}=\frac{1}{2}\frac{R_N^2}{1+R_N^2}\lp d\Psi_N + 2
 A_{(N-1)} \rp\,,
\label{iterativeCPn}
\end{eqnarray}
in terms of the Fubini-Study metric, $d\Sigma_{N-1}^2$, and K\"ahler potential, $A_{(N-1)}$, on the unit $CP^{N-1}$,
\begin{eqnarray}
&& d\Sigma_{N-1}^2 = f_{N-1}^{-1}\, dv^i\, d\bar v^{\bar\imath}\, -
f_{N-1}^{-2}\, |\bar v^{\bar \imath}\, dv^i|^2\,, \nonumber \\
&& A_{(N-1)}= \frac{1}{2} \, \ii\, f_{N-1}^{-1}\, \lp v^i\, d\bar
v^{\bar \imath} - \bar v^{\bar \imath}\, dv^i \rp\,, \qquad f_{N-1} = 1 +
v^i\, \bar v^{\bar \imath} \,. \label{iterativeCPn:2}
\end{eqnarray}

By definition, the K\"ahler form on $CP^{N}$, $J_{N}=\frac{1}{2}dA_{N}\,$, is covariantly conserved, $\hat{\nabla}_a J_{(N)}^{bc}=0\,$, and satisfies $J_a^{\:\:b}J_{bc}=-\hat{g}_{ac}\,$.

The lesson from this analysis is that starting from the $CP^1$ fields we can iteratively construct the $CP^N$ geometry as well as the complex coordinates $Z^A$ that define the embedding of $CP^N$ in $\mathbb{C}^{N+1}$. $CP^1$ is isomorphic to the 2-sphere $S^2$, its metric and K\"ahler potential being given by
\be
d\Sigma_{1}^2=\frac{1}{4} \lp d\theta^2 +\sin^2\theta d\phi^2\rp \qquad \mathrm{and} \qquad
A_{(1)}=\frac{1}{2} \cos\theta \,d\phi\,.
\ee
Examples of the embedding in $\mathbb{C}^{N+1}$ may be elucidative. For $CP^2$, parameterized by $(\theta,\phi,R_2,\Psi_2)$, the map is given by
\be
\lp Z^0,Z^1,Z^2\rp= \frac{r  e^{i\tau}}{\sqrt{1+R_2^2}}\lp 1,R_2 \cos\frac{\theta}{2} \,e^{i\,\frac{1}{2}\,(\Psi_2+\phi)},
R_2 \sin\frac{\theta}{2} \,e^{i\,\frac{1}{2}\,(\Psi_2-\phi)} \rp,
\ee
while for $CP^3$, parameterized by $(\theta,\phi,R_2,\Psi_2,R_3,\Psi_3)$, the map is
\begin{align}
&\lp Z^0,Z^1,Z^2,Z^3\rp= \nonumber \\
&~~~~\frac{r  e^{i\tau}}{\sqrt{1+R_3^2}} \lp 1,
\frac{R_3 e^{i\,\frac{1}{2}\,\Psi_3}}{\sqrt{1+R_2^2}} R_2\,\cos\frac{\theta}{2} \,e^{i\,\frac{1}{2}\,(\Psi_2+\phi)},
\frac{R_3 e^{i\,\frac{1}{2}\,\Psi_3}}{\sqrt{1+R_2^2}} R_2 \sin\frac{\theta}{2} \,e^{i\,\frac{1}{2}\,(\Psi_2-\phi)} \rp.
\end{align}

In order to reproduce the results in Section~\ref{secscalars}, it is useful to recall that, for $CP^N$,
\be
\hat{R}_{abcd} = \hat{g}_{ac} \hat{g}_{bd} - \hat{g}_{ad} \hat{g}_{bc} + J_{ac} J_{bd} - J_{ad} J_{bc} + 2 J_{ab} J_{cd}\,.
\ee

\subsection{Scalar harmonics and Killing vectors \label{cpnscalarharmonicsappendix}}

We review here the systematic way to construct all scalar harmonics and Killing vector fields on $CP^N$ \cite{Hoxha:2000jf}. The isometry group of $CP^N$ is $SU(N+1)$. Let $T_{A_1\cdots A_p}{}^{B_1\cdots B_q}$ be a constant Hermitian $SU(N+1)$ tensor, which is symmetric in the index set $\{A_1,\ldots,A_p\}$ and the index set $\{B_1,\ldots,B_q\}$, and traceless in any contraction between an $A_i$ and a $B_i$ index. This defines the $(p,q)$ representation of $SU(N+1)$. The charged scalar harmonics are then given by
\be
\mathbb{Y}=T_{A_1\cdots A_p}{}^{B_1\cdots B_q}\, Z^{A_1}\cdots Z^{A_p}\, \overline{Z}_{B_1}\cdots \overline{Z}_{B_q}\,,
\label{Harmonics}
\ee
and satisfy the Laplacian \eqref{eqn:chargedS} for $\lambda=2[2pq+N(p+q)]$. We have $\kappa=\mathrm{max}\{p,q\}$ and $m=p-q$. Uncharged scalar harmonics have $\kappa=p=q$ and $\lambda=4\kappa(\kappa+N)$.

The Killing vectors on an Einstein-K\"ahler space can be constructed from the uncharged scalar harmonics with $\kappa=1\,$, which have eigenvalue $\lambda=4(1+N)$. Indeed all the Killing vectors $\xi_{(i)}$ of $CP^N$ are generated by the relation
\be 
\xi^a_{(i)} = J_{(N)}^{ab}\, \partial_b \mathbb{Y}_{\kappa=1,(i)}^{m=0}\,\,,\label{Killing}
\ee
where $i=1,\ldots,N(N+2)\,$. That is, setting to zero all but one of the constant components of the arbitrary Hermitian traceless tensor $T_A^{\:\:B}$ we get a $\kappa=1$, $m=0$ scalar harmonic on $CP^N$ through \eqref{Harmonics}. Repeating the exercise for all possible combinations, we generate the $N(N+2)$ uncharged scalar
harmonics $\mathbb{Y}_{\kappa=1,(i)}^{m=0}\,$, and the associated $(N+1)^2-1=N(N+2)$ Killing vectors through \eqref{Killing}. There are $N$ linearly independent Killing vectors which commute with all the others, thus generating the Cartan subgroup $U(1)^N$ of $SU(N+1)$.

\subsection{Symmetries of $\kappa=2$ harmonics} \label{subsec:symmetries}

The symmetry group $SU(N+1)$ is broken, at least partially, by any linear perturbation $h_{ab}$ satisfying $\mathcal{L}_\xi h_{ab}\neq 0\,$, where $\xi$ is one of the $N(N+2)$ Killing vectors of $CP^N$. For scalar type perturbations, if $\mathcal{L}_\xi \Y \neq 0$ for some $\xi$ then the symmetry associated with $\xi$ is broken by the perturbation.

We explained above how to construct the Killing vectors from the uncharged $\kappa=1$ scalar harmonics. Consider now the most general linear combination of Killing vectors $K=\sum_i c_i\, \xi_{(i)}$, with $i=1,\ldots,N(N+2)\,$. The entire symmetry group $SU(N+1)$ is broken by $h_{ab}$ if the only solution to $\mathcal{L}_K \mathbb{Y} =0$ is  $c_i=0$ for all $i$, i.e. $K=0$. 
There are (uncharged) $\kappa=2$ harmonics on $CP^3$ ($D=9$) for which this is true. For reference, we present here a particular example in the coordinate system used above: a family of $m=0$, $\kappa=2$ harmonics with three non-zero continuous parameters, $\beta_1$, $\beta_2$ and $\beta_3$,
\bea
& \displaystyle{ \Y_{\kappa=2}^{m=0} \;=\; \beta_1 \; \frac{R_3 R_2}{\left(1+R_3^2\right)^2} \sqrt{\frac{1+\cos\theta}{1+R_2^2}} \left[1-\frac{R_3^2 R_2^2 (1+\cos\theta)}{2 \left(1+R_2^2\right)}\right] e^{\frac{1}{2} \ii (\Psi_3 +\Psi_2+\phi)} } \qquad \qquad \nonumber \\
& \displaystyle{ + \beta_2 \; \frac{R_3^2 R_2 \sqrt{1-\cos\theta}}{\left(1+R_3^2\right)^2 \left(1+R_2^2\right)} \left[1-\frac{R_3^2}{2 \left(1+R_2^2\right)}\right] 
e^{\frac{1}{2} \ii (\phi-\Psi_2)} + \beta_3 \; \frac{ R_3^2  \;e^{i \Psi_3 }}{\left(1+R_3^2\right)^2 \left(1+R_2^2\right)} }\,. \nonumber \\
\eea
If one of $\beta_1,\beta_2,\beta_3$ vanishes then this is still a $\kappa=2$ harmonic but it preserves some symmetry.

\section{Appendix: Properties of stationary axisymmetric modes   \label{sec:stationaryModes}}

In the main body of the Chapter, we have presented our numerical
results for general axisymmetric  time dependent scalar
perturbations. Our numerical code has a continuous limit as
$\omega=\ii\, \Gamma \rightarrow 0$. Therefore this particular case
was already included in our discussion. However, in this appendix we
want to have a closer look at stationary axisymmetric modes.
The reasons are: (i) we developed an independent code
for this particular case which confirms the results from
our general code with time dependence; (ii) we proved that the
stationary zero-modes (with $\omega=0$ and $k=k_\ast=0$) are not pure gauge
modes; (iii) we confirmed that our stationary perturbations
preserve the angular velocity and temperature of the background
geometry; (iv) we determined which stationary zero-mode perturbations
can change the mass and angular momenta of the background solution;
finally (v) we want to give special attention to the stationary modes,
and not just to the time dependent instability, since these may
indicate bifurcation points to new branches of black hole solutions.

\subsection{Stationary perturbations sub-sector \label{sec:statModes2}}

As described in Section~\ref{sec:eigenvalue}, the stability problem
of axisymmetric  perturbations with time dependence consists of a
system of 16 Lichnerowicz eigenvalue equations for 16 unknown
functions. Choosing the TT gauge reduces the problem to a system of
6 TT gauge conditions and 10 Lichnerowicz equations. The procedure
is consistent because, as explained in that Section, the latter 10
equations automatically imply, through the 6 gauge conditions, that
the other 6 Lichnerowicz eigenvalue equations are satisfied.

When we consider the axisymmetric stationary sub-sector of the perturbations, i.e. $m=0$, $\omega=\ii\, \Gamma=0$ we find
that the initial system of  16 Lichnerowicz equations decouples into
a subsystem of 10 equations for 10 functions and another subsystem
with 6 equations involving only the remaining 6 perturbation
functions. Moreover, we can use the stationarity condition, $\omega=\ii\, \Gamma=0$, to further simplify our system of equations. To see how this is accomplished, let us introduce
the harmonics associated with the time and azimuthal Killing
directions as $\mathbb{S}=e^{-i\omega t}e^{im \psi}$. We can then
decompose the perturbations according to how they transform under the $\{t,\psi\}$ Killing
isometries. For example, the scalar-derived vector perturbations satisfy
$h_{A\bar{b}}\sim f_A \,\partial_b\mathbb{S}$ for $b=t,\psi$ and the index $A$
running over the radial and $CP^N$ coordinates (the bar in $\bar{b}$
denotes that the 1-form basis is really $\{dt,d\psi+A-\Omega dt\}$).
These perturbations, $h_{At}$ and $h_{A \bar{\psi}}$, must
then vanish when we set $\omega=0$ and $m=0$. This amounts to requiring that $\mathfrak{f}_{01}=
\mathfrak{f}_{12}=\mathfrak{f}_0=\mathfrak{f}_2=0$. The original 10
time dependent Lichnerowicz equations then imply that
$\tilde{\mathfrak{f}}_1=0$ when we set $\omega=0$. Similarly, the
original 6 TT gauge conditions imply that $Q=0$. We are then led to the following conditions
\be
\mathfrak{f}_{01}=\mathfrak{f}_{12}=\mathfrak{f}_0=\mathfrak{f}_2=\tilde{\mathfrak{f}}_1=Q=0\,,
\label{killed}
\ee
or, using the map \eqref{eq:newfunc}: $f_{01}=f_{12}=0$, $W^-=-W^+$,
$X^-=X^+$, $Z^-=-Z^+$,  and $H^{--}=H^{++}$.

With (\ref{killed}) the original system reduces to a closed system
of 3 TT gauge conditions and 7 Lichnerowicz equations. In this
axisymmetric stationary case, the boundary conditions at the horizon
(\ref{bchor}) reduce to
\begin{equation}
\label{BCsStationary}
\begin{aligned}
& f_{00}= -\frac{h(r_+)}{r_+} \, f_{\bar{0}\bar{1}}\,,
\qquad  
f_{00}+f_{11}\approx 
\Delta'(r_+) \, f_{\bar{1}\bar{1}}\; (r-r_+)\,, \\
& f_{02} \approx -\frac{\sqrt{\Delta'(r_+)}}{h(r_+)}  \,
f_{\bar{1}\bar{2}} \sqrt{r-r_+}\,, \qquad 
f_{22}= \frac{1}{h(r_+)^2} \, f_{\bar{2}\bar{2}}\,, \qquad Z^{\pm}= \frac{1}{r_+ h(r_+)} \, \bar{Z}^{\pm} \,,  \\
&W^\pm \approx \frac{h(r_+)}{r_+\sqrt{\Delta'(r_+)}}
\,\bar{W}'^{\pm}\sqrt{r-r_+}\,,\qquad X^\pm - W^\pm \approx
\frac{\sqrt{\Delta'(r_+)}}{r_+} \, \bar{X}^{\pm} \sqrt{r-r_+}\,,
\end{aligned}
\end{equation}
where we used $\bar{W}^{\pm}=\bar{W'}^{\pm}(r-r_+)$ and one further
has $\bar{W'}^{-}=-\bar{W'}^{+}$, $\bar{X}^{-}=\bar{X}^{+}$ and
$\bar{Z}^{-}=-\bar{Z}^{+}$. It is important to emphasize that
(\ref{killed}) already encode the information that the perturbations
are in the TT gauge, and that the boundary conditions
(\ref{BCsStationary}) are compatible with the TT gauge.

We have done an explicit search of the stationary modes using only
the subsystem of 3 TT gauge conditions and 7 Lichnerowicz equations
described above subject to (\ref{BCsStationary}). We recover
independently the same results that we obtain when we set $\omega=0$ in
our time dependent code.

\subsection{Stationary zero-modes are not pure gauge  \label{sec:NoPureGauge}}

As discussed in Section~\ref{sec:eigenvalue}, perturbations with
$k>0$ in the TT gauge all have the gauge freedom fixed. However, TT
perturbations with $k=0 \neq k_\ast$ and $\Gamma=0$ are pure gauge modes \cite{Gregory:1994bj}.
In this subsection, we want to confirm that our stationary
axisymmetric zero-modes with $k=k_\ast=0$ cannot be pure gauge modes. Given that for any residual gauge
freedom the gauge parameter would be constrained to be a harmonic 1-form, we will
prove that there is no regular harmonic 1-form that could generate our perturbations.

Consider the effect of a scalar gauge transformation on the metric perturbations. The most general scalar-type gauge parameter can be decomposed as
\begin{equation}
 \xi=e^{-\ii\omega t+\ii m\psi}\big[\xi_0(r)\Y\,e^{(0)}+\xi_{1}(r)\Y\,e^{(1)}+\xi_{2}(r)\Y\,e^{(2)}+ r \big(\xi^+(r)
 \Y^+_a +\xi^-(r)\Y^-_a \big) dx^a\big]\,.
\end{equation}
Under a gauge transformation,
\begin{equation}
 h_{\mu\nu}\to h_{\mu\nu}+2\,\nabla_{(\mu}\xi_{\nu)}\,,
\end{equation}
the tetrad components of the metric perturbations transform as
\begin{equation}
\begin{aligned}
&f_{00}\to
f_{00}-2\bigg[\frac{\ii(\omega-m\,\Omega)}{f}\,\xi_0+\frac{f'}{f\,g}\,\xi_1\bigg]\,,
\quad
f_{01} \to f_{01}+\frac{1}{g}\bigg(\dr-\frac{f'}{f}\bigg)\xi_0-\frac{\ii(\omega-m\,\Omega)}{f}\,\xi_1\,,\\
&f_{02}\to
f_{02}+\frac{\ii\,m}{h}\,\xi_0-\frac{h\,\Omega'}{f\,g}\,\xi_1-\frac{\ii(\omega-m\,\Omega)}{f}\,\xi_2\,,
\qquad
f_{11}\to f_{11}+\frac{2}{g}\,\dr\,\xi_1\,,\\
&f_{12}\to
f_{12}-\frac{h\,\Omega'}{f\,g}\,\xi_0+\frac{\ii\,m}{h}\,\xi_1+\frac{1}{g}\bigg(\dr-\frac{h'}{h}\bigg)\xi_2\,,\qquad
f_{22}\to f_{22}+2\bigg[\frac{h'}{h\,g}\,\xi_1+\frac{\ii\,m}{h}\,\xi_2\bigg]\,,
\end{aligned} \nonumber
\end{equation}
\begin{equation}
\begin{aligned}
&W^+\to
W^+-\bigg[\frac{\sqrt\lambda}{r}\,\xi_0+\frac{\ii(\omega-m\,\Omega)}{f}\,\xi^+\bigg]\,,\quad
W^-\to
W^--\bigg[\frac{\sqrt\lambda}{r}\,\xi_0+\frac{\ii(\omega-m\,\Omega)}{f}\,\xi^-\bigg]\,,\\
&X^+\to X^+-\bigg[\frac{\sqrt\lambda}{r}\,\xi_1-\frac{1}{g}\bigg(\dr-\frac{1}{r}\bigg)\xi^+\bigg]\,,\quad X^-\to X^--\bigg[\frac{\sqrt\lambda}{r}\,\xi_1-\frac{1}{g}\bigg(\dr-\frac{1}{r}\bigg)\xi^-\bigg]\,,\\
&Z^+\to
Z^+-\bigg[\frac{\sqrt\lambda}{r}\,\xi_2-\ii\bigg(\frac{m}{h}+\frac{2\,h}{r^2}\bigg)\xi^+\bigg]\,,\quad
Z^-\to Z^--\bigg[\frac{\sqrt\lambda}{r}\,\xi_2-\ii\bigg(\frac{m}{h}-\frac{2\,h}{r^2}\bigg)\xi^-\bigg]\,,\\
&H_L\to H_L+\frac{2}{r}\bigg[\frac{1}{g}\,\xi_1+\frac{1}{4\,N\,\sqrt\lambda}\big(\xi^+(\lambda-2\,m\,N)+\xi^-(\lambda+2\,m\,N)\big)\bigg]\,,\\
&H^{+-}\to H^{+-}-\frac{\sqrt\lambda}{r}(\xi^++\xi^-)\,,\quad
H^{++}\to H^{++}-\frac{2\,\sqrt\lambda}{r}\,\xi^+\,,\quad H^{--}\to
H^{--}-\frac{2\,\sqrt\lambda}{r}\,\xi^-\,.
\end{aligned}
\end{equation}
Our stationary axisymmetric perturbations must satisfy (\ref{killed}) which requires that a potentially dangerous gauge parameter $\xi$ must satisfy
\begin{equation}
 \xi_0(r)=\xi_{2}(r)=0\,, \quad\hbox{and} \qquad
 \xi^-(r)=\xi^+(r)\,. \label{KillComp}
\end{equation}
We now prove that a parameter $\xi$ obeying these conditions cannot generate a pure gauge metric perturbation that is regular. By regularity we mean that the gauge transformation cannot diverge at the horizon $r=r_+$ nor at the asymptotic boundary $r \to \infty$.

A TT gauge perturbation generated by $\xi$ must satisfy the conditions $\nabla_\mu\xi^\mu=0$ and $\Box\xi_\nu=0$. If we introduce the antisymmetric tensor $F_{\mu\nu}=\nabla_{[\mu}\xi_{\nu]}$, for a Ricci flat background, these conditions reduce to $\nabla_\mu\xi^\mu=0$ and $\nabla_\mu F^{\mu\nu}=0$ which read simply
  \be
 \partial_\mu\lp \sqrt{-g}\xi^\mu \rp=0\,,\qquad \partial_\mu\lp \sqrt{-g}F^{\mu\nu}
 \rp=0\,. \label{NoPureGaugeEqs}
 \ee
Using $\sqrt{-g}=r^{2N+1}\sqrt{\hat{g}}$ and Eq.~(\ref{eqn:chargedS}), the first of the equations above requires that
 \be
 \xi^+(r)=\frac{1}{\lambda r^{2N-1}}\,\partial_r\left[ r^{2N+1} g(r)^{-2}\xi_1(r)
 \right]\,,
 \ee
where the background function $g(r)$ is defined in (\ref{background}) and $\lambda$ is the $CP^N$ eigenvalue (\ref{eqn:eigenvalues}). The second family of equations in (\ref{NoPureGaugeEqs}) further demands that $\left(\xi^+\right)^{\prime}(r)=\xi_1(r)$. Introducing the new
variable
 \be
 \xi_1(r)= r^{-(N+3/2)g^2(r)}\,\chi(r)\,,
 \ee
the solution of (\ref{NoPureGaugeEqs}) must then solve
 \be
 \chi^{\prime\prime}(r)=V(r)\chi(r)\,,  \qquad \hbox{with}\quad V(r)=\frac{1}{r^2}\left[\lp N^2-\frac{1}{4}\rp+\lambda
 g^2(r)\right]>0\,. \label{Qeq}
 \ee

To study the regularity of the associated gauge transformation, we need the asymptotic behavior of $\chi(r)$ at the horizon and at infinity. The solution to Eq.~\eqref{Qeq} can be obtained in these regions by considering the dominant contributions of $V(r)$ or by doing a Frobenius analysis. Using (\ref{eqn:eigenvalues}), we find that
\begin{eqnarray}
&&\chi(r){\bigl|}_h\sim a_0(r-r_+)\,\quad \mathrm{or} \quad \chi(r){\bigl|}_h\sim a_0 \,, \label{BC:Q1} \\
&& \chi(r){\bigl|}_{\infty}\sim b_0 r^{\frac{1}{2}\pm
(2\kappa+N)}\,.
 \label{BC:Q2}
\end{eqnarray}
Recall that $\delta h_{\mu\nu}=-\mathcal{L}_\xi g_{\mu\nu}$. We have
to  discard the second possibility in \eqref{BC:Q1} because it would
generate a dependence $f_{11}{\big|}_h \sim (r-r_+)^{-2}$, not
compatible with the TT boundary conditions \eqref{BCsStationary}. On
the other hand, we have to discard the solution with the positive
sign in \eqref{BC:Q2} because it generates a perturbation that grows
faster than the unperturbed background metric as $r\to 0$. The
appropriate boundary conditions for \eqref{Qeq} are thus
 \be
 \chi(r){\bigl|}_h\sim a_0(r-r_+)\,,\qquad \chi(r){\bigl|}_{\infty}\sim b_0 r^{\frac{1}{2}- (2\kappa+N)}
 \,. \label{BC:Qfinal}
 \ee

We can now complete our proof. Notice that
\begin{eqnarray}
 0\leq
\int_{r_+}^{\infty}\chi^{\prime}(r)^2&=&\chi(r)\chi^{\prime}(r){\biggl|}_{r_+}^{\infty}-\int_{r_+}^{\infty}\chi(r)\chi^{\prime\prime}(r)
\nonumber\\
 &=& -\int_{r_+}^{\infty}V(r)\chi(r)^2 \leq 0 \,, \label{nullQ}
\end{eqnarray}
where we used \eqref{BC:Q2} and \eqref{Qeq}. But these relations can be satisfied only for \be
 \chi(r)=0 \qquad \Rightarrow \qquad \xi_1(r)=0\,, \quad \hbox{and}
 \quad \xi^+(r)=0 \,, \label{BC:Qend}
 \ee
which in addition to (\ref{KillComp}) proves that our regular zero-mode perturbations in the TT gauge cannot be pure gauge modes.

\subsection{Temperature and angular velocity preserved  \label{sec:PreserveTAngVel}}

In Section~\ref{sec:zeromodeultraspin}, we discussed the connection between the
classical instability of the black hole and its thermodynamics. This
connection was built on the claim that the stationary and axisymmetric
modes that we study preserve the temperature and the
angular velocities of the background solution. Here we will prove
that this is indeed the case.

We will first compute the angular velocity and the temperature of
the unperturbed background solution using standard Euclidean
methods. Our strategy is then to check that our TT metric
perturbation $h_{\mu\nu} dx^\mu dx^\nu$ is a regular symmetric 2-tensor, when expressed
in coordinates where the background metric is regular, which confirms
that they preserve the angular velocity and temperature.

We start with the computation of the background angular velocity and
temperature. This is done performing the standard three steps in the
background solution: i) a coordinate transformation to coordinates
$(t,\tilde{\psi})$ that corotate with the black horizon, ii) a Wick rotation
of the time coordinate so that we work with the Euclidean solution,  and iii) a change to a new radial
coordinate that zooms the geometry in the near-horizon region. That
is, we perform the coordinate transformations
 \be
 \tilde{\psi}=\psi-\Omega_H t\,, \qquad t=-i\tau \,, \qquad
 r=r_+ +\frac{\Delta'(r_+)}{4}\,\rho^2\,,
  \ee
with $\Omega_H$ being the angular velocity of the background
solution \eqref{angvel}. A final coordinate
transformation,
 \be
 \tilde{\tau}=2\pi T_H \tau\,, \qquad \hbox{with} \qquad T_H= \frac{r_+ \Delta'(r_+)}{4\pi
 h(r_+)}\,,
  \ee
sets the period of $\tau$ to be the inverse of the horizon
temperature and avoids a conical singularity at the horizon. The
Euclidean sector of the near horizon region of the background
solution (\ref{background}) then reads
 \be
\label{NHbackground} ds^2_E \simeq \rho^2 \,d\tilde{\tau}^2 +
d\rho^2 + h(r_+)^2[d\tilde{\psi} +A_a dx^a]^2 + r_+^2 \hat{g}_{ab}
dx^a dx^b\,,
 \ee
which is a manifestly regular geometry. Indeed the polar coordinate
singularity can be removed by a coordinate transformation into
cartesian coordinates, $\tilde{\tau}=\hbox{Arctan}(y/x)$ and
$\rho=\sqrt{x^2+y^2}$. This concludes our computation of the angular
velocity and temperature of the background black hole.

To study the regularity of the perturbations, start by introducing the
manifestly regular 1-forms,
 \be \label{NHperturbations}
E^{\tilde{\tau}}=\rho^2 d\tilde{\tau}=x\,dy-y\,dx\,,\qquad E^{\rho}=\rho\,
d\rho=x\,dx+y\,dy\,.
 \ee
Consider now our TT metric perturbation $h_{\mu\nu} dx^\mu dx^\nu$.
After using the boundary conditions (\ref{BCsStationary}), which
satisfy the TT gauge conditions, we get
\begin{equation}
\begin{aligned}
h_{\mu\nu}\,dx^\mu\,dx^\nu&\simeq
 \frac{h(r_+)}{r_+} f_{\bar{0}\bar{1}}\Y \lp \rho^2 \,d\tilde{\tau}^2
 + d\rho^2\rp +f_{\bar{2}\bar{2}}\Y  \lp d\tilde{\psi} +A_a dx^a \rp^2 \\
&~~~+ {\rm i}\frac{\Delta'(r_+)}{h(r_+)} f_{\bar{1}\bar{2}}\Y E^{\tilde{\tau}} \lp d\tilde{\psi} +A_a dx^a \rp
+ \frac{\Delta'(r_+)^2}{4}  f_{\bar{1}\bar{1}}\Y (E^{\rho})^2 \\
&~~~-4 {\rm i}\frac{r_+ h(r_+)}{\Delta'(r_+)}E^{\tilde{\tau}}\lp \bar{W}'^{+}_a + \bar{W}'^{-}_a \rp dx^a + 2 r_+E^{\rho} \lp \bar{X}_a^+ + \bar{X}_a^-\rp dx^a\\
&~~~+ 2 \lp d\tilde{\psi} +A_a dx^a \rp \lp \bar{Z}_a^+ +
\bar{Z}_a^-\rp dx^a\\
&~~~+r_+^2\left[-\frac{1}{\sqrt{\lambda}}\left(H_{ab}^{++}+H_{ab}^{--}+H_{ab}^{+-}\right)+
\tilde H_L\,\hat g_{ab}\right]dx^adx^b\,.
\end{aligned}
\label{eqn:Regularhmetric}
\end{equation}
Notice that the first term on the right-hand side ensures that there is no conical singularity if $\tilde{\tau}$ has the same periodicity as the background metric. Furthermore, the remaining dependence on $\tilde{\tau}$ and $\rho$ is given by the manifestly regular 1-forms $E^{\tilde{\tau}}$ and $E^{\rho}$. Hence, $h_{\mu\nu} dx^\mu dx^\nu$ is a regular 2-tensor in the
cartesian coordinates $(x,y,\tilde{\psi},x^a)$ where the background
metric is regular, which confirms that the perturbations indeed preserve the
angular velocity and temperature.

\subsection{Perturbations of the asymptotic charges  \label{sec:ChangeMJ}}

The stationary zero-modes can potentially change the mass and angular momenta of the geometry. In this subsection, we determine which of our perturbations with $\omega=0$, $m=0$ and $k=0$ can change these conserved charges.

The change on the conserved charges associated to a Killing generator $\xi$ introduced by a perturbation $h_{\mu\nu}$ can be defined via a surface boundary integral as (for TT perturbations) \cite{Abbott:1981ff,Barnich:2001jy}
\begin{eqnarray}
   Q_\xi[h,g]=-\frac{1}{32 \pi G} \int_{\partial \Sigma} \epsilon_{\alpha\beta\mu\nu} \big[
  \xi_\sigma \nabla^\nu h^{\mu\sigma}
 - h^{\nu\sigma} \nabla_\sigma \xi^\mu  
  + \frac{1}{2} h^{\sigma\nu}(\nabla^\mu\xi_\sigma
+ \nabla_\sigma\xi^\mu)\big] dx^\alpha\wedge dx^\beta \,. \nonumber \\
\label{chargeXi}
\end{eqnarray}
The conserved charges of interest are the energy, for $\xi=-\partial/\partial t$, and the angular momenta associated with the $\lfloor (D-1)/2 \rfloor$ $U(1)$ Killing vectors $\xi=\partial/\partial \Psi_i$. The corresponding changes are denoted by $\mathcal{E}$ and $\mathcal{J}_i$, respectively. The surface integral is over a constant time hypersurface at asymptotic infinity, $\partial \Sigma$.

To compute the charges of our perturbations, we need the asymptotic behaviour of our solutions. This can be obtained from a Frobenius analysis of the Lichnerowicz equations at $r \to \infty$. We are interested in boundary conditions that preserve the asymptotic flatness of the spacetime, i.e. that decay (strictly) faster than the background geometry. We find that the behaviour of the regular perturbations is such that they decay at infinity according to
\begin{eqnarray}
 &&  \mathfrak{f}_{11}\approx \mathcal{O}\lp r^{-4-2\kappa}\rp\,, \quad \mathfrak{f}_{12}\approx\mathcal{O}\lp r^{-2\kappa}\rp\,, \quad
\mathfrak{f}_{22}\approx\mathcal{O}\lp r^{-2\kappa}\rp\,, \quad
\mathfrak{f}_1\approx\mathcal{O}\lp r^{-3-2\kappa}\rp\,, \nonumber\\
&& \tilde{\mathfrak{f}}_0\approx\mathcal{O}\lp r^{-2\kappa}\rp\,,
\quad \tilde{\mathfrak{f}}_2\approx\mathcal{O}\lp r^{-2\kappa}\rp\,,
\quad P\approx\mathcal{O}\lp r^{-2-2\kappa}\rp\,.
\label{FroebInfinity}
\end{eqnarray}

We then find the generic behaviour of the changes in the conserved charges,
\begin{eqnarray}
 && \mathcal{E}= E_0 \, r^{-2\kappa}\lp 1+ \mathcal{O}\lp r^{-1}\rp \rp\,,\nonumber\\ \label{ChargesPerturb}
 && \mathcal{J}_{\Psi_i}= A_i\, r^{2-2\kappa} + B_i\, r^{-2\kappa}
+ \mathcal{O}\lp r^{-1-2\kappa}\rp \,,
\end{eqnarray}
where  $E_0,A_i,B_i$ are functions of $\kappa$ and $r_+$ and, in particular, $A_i=B_i=0$ for $\kappa=0$.

We thus conclude that the $\kappa=0$ modes are the only ones that decay sufficiently slowly to change the mass of the geometry. Moreover these modes cannot change the angular momenta. The only modes that change the angular momenta are those with $\kappa=1$. No modes with $\kappa\geq 2$ can change the mass or the angular momenta. We have done these computations explicitly for the $D=7$ case.

\end{subappendices}

%% file: Conclusion.tex
\chapter{Conclusion and outlook} \label{cha:conclusion}

The original work described in this thesis can be divided into two parts. In the first, we studied the black hole partition function of Euclidean quantum gravity. We extended the study of negative modes, which represent pathologies in the one-loop quantum corrections, to black holes which are charged \cite{Monteiro:2008wr} (Chapter~\ref{cha:RN}) or rotating \cite{Monteiro:2009tc} (Chapter~\ref{cha:KerrAdS}, and also \cite{Dias:2009iu,Dias:2010eu}, Chapters~\ref{cha:MPsingle}--\ref{cha:MPequal}). We found that local thermodynamic instabilities are always signalled (individually) by the existence of a negative mode. The results strengthen the claim that the gravitational partition function indeed describes semiclassical quantum gravity at low energies, even beyond the leading order instanton approximation. In the charged case, a trick based on a Kaluza-Klein reduction allowed for the decoupling of the unphysical divergent sector, the analogue of the conformal sector for pure-gravity. However, it would be convenient to have a more general procedure to deal with gravity-matter instantons, applicable to the charged rotating case.

In the second part (Refs.~\cite{Dias:2009iu,Dias:2010eu}, Chapters~\ref{cha:classthermo}--\ref{cha:MPequal}), we explored the connection between classical stability and local thermodynamic stability of black holes, numerically analysing perturbations of Myers-Perry solutions with a single spin and with equal spins (cohomogeneity-1 in odd $D$). The connection is that negative modes can have implications for the classical stability problem. We started by refining the Gubser-Mitra conjecture, showing that one Gregory-Laflamme-type instability of uniform black branes appears for each individual local thermodynamic instability, and in fact for each negative mode whether it is related to a thermodynamic instability or not. The onsets of these instabilities should be associated with bifurcations to new non-uniform black brane families. Moreover, we showed that \emph{all} asymptotically vacuum black holes possess a local thermodynamic instability, which implies that the associated uniform black branes are classically unstable \cite{Dias:2010eu}.

The main achievement in this thesis was to show that rapidly-rotating Myers-Perry black holes can be classically unstable, as first conjectured for the $D\geq 6$ singly-spinning sector in Ref.~\cite{Emparan:2003sy}. We presented in Ref.~\cite{Dias:2009iu} (Chapter~\ref{cha:MPsingle}) the first evidence for this instability by showing that, as high rotations are considered, additional negative modes of the partition function arise which are not associated with the standard local thermodynamic instabilities. The zero-modes marking the appearance of each new negative mode are instead the thresholds of classical instabilities of the black hole, and not just of the black branes. In Ref.~\cite{Dias:2010eu} (Chapter~\ref{cha:MPequal}), we went further and, analysing cohomogeneity-1 MP black holes, verified explicitly the existence of such an instability in $D=9$, by determining its timescale at the linear level.

The non-thermodynamic zero-modes also mark the bifurcation of new families of stationary black holes from the MP family. In the singly-spinning MP case \cite{Dias:2009iu}, there should be infinite new families of this type. The first may interpolate between a MP black hole and a black ring through a horizon topology transition, and the second may interpolate between a MP black hole and a black Saturn, etc., as conjectured in Refs.~\cite{Emparan:2003sy,Emparan:2007wm}. In the equal spins case \cite{Dias:2010eu}, these zero-modes indicate the existence of higher-dimensional black holes in $D=9$ (and, we believe, higher odd $D$) with a single rotational symmetry. This is the first known example with spherical horizon topology (see Ref.~\cite{Emparan:2009vd} for the very first example, the helical black rings). Furthermore, we argue that this family of black holes which generically has a single rotational symmetry is determined by 70 parameters, while the MP solution has only 5 parameters. These results are a stark measure of the challenge of classifying higher-dimensional black holes.

The negative modes found obey a certain harmonic structure, with the thermodynamic negative modes corresponding to the lowest harmonics (s-wave and p-wave).\footnote{Notice that, in general, there may exist different harmonic structures corresponding to different perturbation subsectors, e.g. there may exist unrelated p-waves.} Therefore, we conjectured that classical instabilities for perturbations preserving the rotational symmetry, corresponding to higher harmonics (d-wave, etc.), can only be excited for rotations higher than the first thermodynamic zero-mode (p-wave). This is the ultraspinning conjecture of Ref.~\cite{Dias:2009iu}, consistent with the posterior results of Ref.~\cite{Dias:2010eu}. The preservation of the rotational symmetry is required in our argument in order to make the connection with regular Euclideanised negative modes. It was recently found in Refs.~\cite{Shibata:2009ad,Shibata:2010wz} that MP black holes can also be unstable after a critical value of the rotation for perturbations breaking this symmetry. This occurs even in $D=5$, where there is no instability in the sector that we consider in this thesis.

It would be important to focus also on different higher-dimensional solutions, such as black rings. What is perhaps the most important question of higher-dimensional black hole physics, along with the closely related classification problem, remains open: is there uniqueness of stable solutions?

A major part of the work reported in this thesis, from Refs.~\cite{Monteiro:2009tc,Dias:2009iu,Dias:2010eu}, is based on a numerical analysis of coupled linear second order differential equations, ODEs or PDEs. The spectral method used to solve these equations is briefly described in the Appendix A. To the best of our knowledge, this powerful method was used in the context of general relativity for the first time in Ref.~\cite{Monteiro:2009tc}. It would be interesting to also apply spectral methods to non-linear black hole problems, such as the construction of the new solutions whose existence we conjectured here, or of solutions relevant for applications of the holographic correspondence. Closely connected is the recent progress -- likely to make an impact in the future -- in extending to high energy physics problems the well-developed numerical techniques available for general relativity in four dimensions, e.g. \cite{Zilhao:2010sr,Witek:2010qc}.

Let us conclude on a different note. The plethora of higher-dimensional black holes and its seemingly impossible classification somewhat remind us of the situation in nuclear/particle physics before the quark model was proposed. The challenge presented by higher dimensions contrasts with the simplicity of the four-dimensional case, where the Kerr black hole is the unique asymptotically flat vacuum solution. In the same way, the myriad of particles detected in collision experiments contrasted with the previous situation where only a handful of more familiar particles were known. The quark model put some order in that particle zoo. Our hope is that there is also an underlying structure connecting the several phases of black holes. This thesis is a small step towards unveiling that structure. Or one may ask the same question for the black ring that Rabi asked for the muon: ``Who ordered that?''

%% file: Spectralmethod.tex
\appendix

\chapter{Spectral numerical method} \label{app:spectral}

In this Appendix, we briefly describe the spectral numerical method that we employed to solve the linear ordinary differential equations (ODEs) and the coupled partial differential equations (PDEs) in this thesis, namely in Chapters~\ref{cha:KerrAdS}, \ref{cha:MPsingle} and \ref{cha:MPequal}. See Ref.~\cite{Trefethen} for more details.

Spectral methods can solve a system of coupled ODEs or PDEs to high accuracy on a finite domain, as long as the system allows for analytic solutions. To our knowledge, the first application of these methods to general relativity was given in \cite{Monteiro:2009ke}, followed by \cite{Dias:2009iu,Dias:2010eu}. We will start by considering the case of ODEs, and in the end we will discuss how to generalise the procedure for PDEs.

The goal is to approximate a given function, defined on a finite domain, as a finite sum of algebraic polynomials $p(y) = \sum_{i=0}^{\mathcal N} a_i y_i$. Performing the polynomial interpolation in an equidistant grid with ${\mathcal N}+1$ points turns out to be catastrophic in many cases due to the oscillation of high-degree polynomials, the so-called Runge phenomenon. In general, this approximation does not converge as ${\mathcal N}\to\infty$, and may get worse at a rate as large as $2^{\mathcal N}$. The correct approach is to perform the interpolation in a non-uniform grid, distributed more densely near the edges of the interpolation interval. We use the Chebyshev grid, whose points are the extrema of Chebyshev polynomials,
\begin{equation}
y_j = \frac{a+b}{2}+\frac{a-b}{2}\cos \left(\frac{j \pi}{{\mathcal N}}\right),\qquad
j\in\{0,1,\ldots,{\mathcal N}\}\,,
\label{eq:ap1}
\end{equation}
for $y_j\in[a,b]$. Not only does this grid avoid the Runge phenomenon, by clustering points near the boundary, but it has another well-known advantage over uniform grids: it typically leads to an exponential accuracy of the approximantion as ${\mathcal N}\to\infty$. However, because we are approximating a function as a sum of polynomials, we must restrict to analytic functions. The reason for the exponential accuracy is that such functions have rapidly decaying Fourier transforms. Since spectral methods, as the name indicates, act in Fourier space in a certain sense, the common difficulties of having singular points in the equations can be avoided, as long as analytic solutions exist.

The procedure to solve differential equations is in the same spirit as standard quantum mechanics. Consider an eigenvalue system of $n$ coupled linear ODEs with variable coefficients,
\begin{equation}
\sum_{\beta=1}^n H_{\alpha\beta}\; q_\beta^{(\lambda)} = \lambda \sum_{\beta=1}^n
T_{\alpha\beta}\;q_\beta^{(\lambda)},\qquad \alpha\in\{0,1,\ldots,n\}\,,
\label{eq:ap2}
\end{equation}
where each $H_{\alpha\beta}$ is a second order operator in $y$, each $T_{\alpha\beta}$ is a scalar function and $\{\lambda,q_\beta^{(\lambda)}\}$ are the eigenvalues and eigenfunctions that we want to determine. We now perform our approximation and discretise the $[a,b]$ interval according to the grid (\ref{eq:ap1}). Each $q_\beta^{(\lambda)}$ is then approximated by a vector, $\vec{q}_\beta^{(\lambda)}$, whose entries are the values at $y_j$ of the eigenfunctions we want to determine. Following this procedure, one represents derivatives with respect to $y$ by matrices, $D_{\mathcal N}$, that act on the vectors $\vec{q}_\beta^{(\lambda)}$, mixing adjacent points (see p.53 of \cite{Trefethen} for an explicit construction of such matrices). After this approach is complete, each $H_{\alpha\beta}$ and $T_{\alpha\beta}$ are transformed into square matrices $\vec{\vec{H}}_{\alpha\beta}$  and $\vec{\vec{T}}_{\alpha\beta}$, respectively, of dimension $({\mathcal N}+1)\times ({\mathcal N}+1)$.

We conveniently chose to work with Dirichlet boundary conditions by considering the rescaled functions \eqref{eq:KAdSbc}, \eqref{MPs:newfunc} and (\ref{eq:qs}). The ellipticity of the perturbation operator is consistent with a boundary value problem. Dirichlet boundary conditions are imposed by setting the first and last elements of each $\vec{q}_\beta^{(\lambda)}$ to zero, and by eliminating the first and last columns and rows of each matrix $\vec{\vec{H}}_{\alpha\beta}$ and $\vec{\vec{T}}_{\alpha\beta}$ \cite{Trefethen}. We are then left with the following system of linear algebraic equations
\begin{equation}
\sum_{\beta=1}^n\hat{\vec{\vec{H}}}_{\alpha\beta}\; \hat{\vec{q}}_\beta^{(\lambda)} =
\lambda
\sum_{\beta=1}^n\hat{\vec{\vec{T}}}_{\alpha\beta}\;\hat{\vec{q}}_\beta^{(\lambda)}\,,
\label{eq:ap3}
\end{equation}
where $\hat{\vec{\vec{H}}}_{\alpha\beta}$ and $\hat{\vec{\vec{T}}}_{\alpha\beta}$ are obtained from $\vec{\vec{H}}_{\alpha\beta}$  and $\vec{\vec{T}}_{\alpha\beta}$ by deleting their first and last columns and rows, and thus are square $({\mathcal N}-1)\times ({\mathcal N}-1)$ matrices. The system of equations (\ref{eq:ap3}) can be written as
\begin{equation}
\left[
\begin{array}{ccc}
\hat{\vec{\vec{H}}}_{11} & \ldots & \hat{\vec{\vec{H}}}_{1n}
\\
\vdots & \ddots & \vdots
\\
\hat{\vec{\vec{H}}}_{n1} & \ldots & \hat{\vec{\vec{H}}}_{nn}
\end{array}\right]
\left[\begin{array}{c}
\hat{\vec{q}}_1^{(\lambda)}
\\
\vdots
\\
\hat{\vec{q}}_{n}^{(\lambda)}
\end{array}\right]=\lambda\left[
\begin{array}{ccc}
\hat{\vec{\vec{T}}}_{11} & \ldots & \hat{\vec{\vec{T}}}_{1n}
\\
\vdots & \ddots & \vdots
\\
\hat{\vec{\vec{T}}}_{n1} & \ldots & \hat{\vec{\vec{T}}}_{nn}
\end{array}\right]
\left[\begin{array}{c}
\hat{\vec{q}}_1^{(\lambda)}
\\
\vdots
\\
\hat{\vec{q}}_{n}^{(\lambda)}
\end{array}\right]\,,
\end{equation}
which is just a standard generalised eigenvalue problem of dimension $n ({\mathcal N}-1)$.

The generalisation to the cases where we have a system of PDEs is straightforward. Say we have two coordinates $y$ and $x$. Then we can work with a two-dimensional grid where each coordinate is discretised according to (\ref{eq:ap1}). In our works, we have used the same number of points $\mathcal N$ for both $y$ and $x$, although this is not a requirement. The vectors $\vec{q}_\beta^{(\lambda)}$ which approximate the eigenfunctions $q_\beta^{(\lambda)}$ will have $({\mathcal N}+1)^2$ components instead of ${\mathcal N}+1$, one component for each grid point, and the matrices $\vec{\vec{H}}_{\alpha\beta}$  and $\vec{\vec{T}}_{\alpha\beta}$ will accordingly become $({\mathcal N}+1)^2\times ({\mathcal N}+1)^2$ matrices, increasing the computational demands.